\RequirePackage{ifpdf}
\documentclass[paper]{JHEP3}
\pdfoutput=1 
\usepackage{epsfig,subfigure}
\usepackage{morefloats}
\usepackage{graphicx}
\usepackage{amssymb}
\usepackage{amsmath}
\usepackage{booktabs,multirow,tabularx}
\usepackage{slashed}
\usepackage{cite}
\usepackage{caption}
\usepackage{float}
\usepackage{placeins}
\usepackage{rotating}
\usepackage{lscape}
\usepackage[Q=yes,pverb-linebreak=no]{examplep}

\DeclareGraphicsExtensions{.eps}
\graphicspath{{./}}

\newcommand\sss{\scriptscriptstyle}

\def\beq{\begin{equation}}
\def\beqn{\begin{eqnarray}}
\def\eeq{\end{equation}}
\def\eeqn{\end{eqnarray}}
\def\beal{\begin{align}}
\def\endal{\end{align}}
\def\abs#1{\left|#1\right|}
\def\remove#1#2{#1\hspace{-#2truecm}\backslash}

\newcommand\as{\alpha_{\sss S}}
\newcommand\aem{\alpha_{\sss EM}}

\newcommand\epem{e^+e^-}
\newcommand\mpmm{\mu^+\mu^-}
\newcommand\muF{\mu_{\sss F}}
\newcommand\muR{\mu_{\sss R}}

\newcommand\bt{\bar{t}}

\newcommand\aNLO{{\sc\small MadGraph5\_aMC@NLO}}
\newcommand\MadGraphf{{\sc\small MadGraph5}}

\newcommand\HWpp{{\sc\small Herwig++}}
\newcommand\PYe{{\sc\small Pythia8}}
\newcommand{\pt}{p_{\sss T}}
\newcommand{\kt}{k_{\sss T}}
\newcommand{\mt}{m_{\sss T}}
\newcommand{\Ht}{H_{\sss T}}
\newcommand\ident{{\cal I}}
\newcommand\Zjs{Z\!+\!{\rm jets}}
\newcommand\Wjs{W\!\!+\!{\rm jets}}
\newcommand\Vjs{V\!\!+\!{\rm jets}}
\newcommand\prompt{{\tt MG5\_aMC>}}
\newcommand\Etmiss{\remove{E}{0.18}_{\sss T}}
\newcommand\Njet{N_{jet}}

\author{R.~Frederix$^a$, S.~Frixione$^b$, A.~Papaefstathiou$^c$, 
S.~Prestel$^d$, P.~Torrielli$^e$\\
$^a$ Physik Department T31, Technische Universit\"at M\"unchen, 
James-Franck-Str.~1,\\
$\phantom{^a}$ D-85748 Garching, Germany\\
$^b$ INFN, Sezione di Genova, Via Dodecaneso 33, I-16146, Genoa, Italy\\
$^c$ PH Department, TH Unit, CERN, CH-1211 Geneva 23, Switzerland\\
$^d$ SLAC, National Accelerator Laboratory\\
$\phantom{^d}$ 2575 Sand Hill Road, Menlo Park, CA 94025-7090, USA\\
$^e$ Dipartimento di Fisica, Universit\`a di Torino and INFN,
 Sezione di Torino,\\
$\phantom{^e}$ Via P.~Giuria~1, I-10125, Turin, Italy\\
}
\title{A study of multi-jet production in association with
an electroweak vector boson}
\abstract{We consider the production of a single $Z$ or $W$ boson
in association with jets at the LHC. We compute the corresponding
cross sections by matching NLO QCD predictions with the {\sc\small Herwig++} 
and {\sc\small Pythia8} parton showers, and by merging all of the underlying 
matrix elements with up to two light partons at the Born level. We compare 
our results with several 7-TeV measurements by the ATLAS and 
CMS collaborations, and overall we find a good agreement between 
theory and data.}
\preprint{
 CERN-PH-TH-2015-255\\
 SLAC-PUB-16421\\
 TUM-HEP-1025/15\\
 MCNET-15-22\\
 }
\begin{document}

\section{Introduction and motivation\label{sec:intro}}
The production of jets in association with a $Z$ or a $W$ boson
plays a role of primary importance at high-energy hadron colliders.
On the one hand, when regarded as backgrounds either or both of these 
processes are relevant to the determination of the properties of the Standard 
Model (SM) Higgs, of the top quark, and of the gauge sector of
the SM, as well as playing a major role in many searches for new
physics beyond the SM. On the other hand, they are also 
interesting in their own right, because they constitute an
excellent testing ground for theoretical predictions. This stems 
from a variety of reasons: the high statistical accuracy of the 
measurements performed and foreseen by the LHC collaborations;
the fact that these processes, especially if leptonic channels 
are considered, are simpler from a theoretical viewpoint than 
other SM processes with a similar number of final-state particles
(e.g.~$t\bt\!+\!{\rm jets}$); and because their complicated kinematics 
allow one to probe phase-space regions sensibly described either
by Monte Carlo event generators (MCs henceforth), or by fixed-order
calculations.

The features just mentioned are what render $\Zjs$ and $\Wjs$ 
production the primary processes for the validation of multi-jet
merging approaches. As a technical aside, we remind the reader
that a matching formalism aims at including, in a consistent manner
and in particular without double counting, the results of matrix 
elements computations into MCs, at a given order in perturbation theory 
and at a given number of final-state particles at the Born level. 
Conversely, a merging formalism combines several matched samples, which are
identified by different parton multiplicities at the Born level. This 
implies that, by merging, one extends the scope of matched results towards
kinematic regions defined by many, well-separated jets. These
are indeed well studied by LHC collaborations: even in Run I,
jet multiplicities exceeding five have been measured, with transverse 
momenta reaching the TeV range; in Run II, these impressive achievements 
will be easily surpassed. A good understanding of these characteristics, 
through the capability of merged predictions to reproduce data, is vital 
for many new-physics searches, which are very often heavily reliant on
theoretical predictions, and categorised in terms of jet multiplicities.
Theoretically, a significant amount of information is involved in
merged results, obviously including the merging formalisms themselves;
detailed comparisons to data will help tell the various proposals 
apart, and ultimately suggest improvements. One also expects 
to become sensitive to large-logarithm effects of electroweak 
origin~\cite{Ciafaloni:1998xg,Ciafaloni:2000df,Denner:2000jv,Denner:2001gw},
whence the necessity of going beyond the present results, which are
based on QCD computations.

Until recently, merging formalisms have relied on underlying computations
of LO (i.e.~tree-level only) accuracy. Techniques such as CKKW, CKKW-L, MLM,
and their variants~\cite{Catani:2001cc,Lonnblad:2001iq,Krauss:2002up,
Mrenna:2003if,Lavesson:2005xu,Alwall:2007fs,Hoeche:2009rj,Hamilton:2009ne,
Lonnblad:2011xx,Lonnblad:2012ng} have been systematically compared
to each other (see e.g.~ref.~\cite{Alwall:2007fs} for a study focusing
on $\Wjs$ production), and to data. The extension of merging to NLO 
accuracy is a challenging theoretical problem~\cite{Lavesson:2008ah,
Hamilton:2010wh,Hoche:2010kg,Giele:2011cb,Alioli:2011nr,Hoeche:2012yf,
Frederix:2012ps,Platzer:2012bs,Alioli:2012fc,Lonnblad:2012ix,Hamilton:2012rf,
Alioli:2013hqa}. Comprehensive comparisons among different approaches
are so far lacking, and it is only lately that experimental collaborations
have started to employ these results, with the idea of using them to
replace gradually the LO-accurate ones. Of course, this assumes that
mergings at the NLO will improve the description of the data given
by their LO counterparts.

The aim of this paper is to use $\Zjs$ and $\Wjs$ LHC data
for a phenomenology validation of the FxFx NLO-merging
formalism~\cite{Frederix:2012ps}. In order to rely on well-understood
measurements, and to avoid making mistakes in the involved definitions
of cuts and observables as defined by the experiments, we have limited
ourselves to considering the 7-TeV analyses by ATLAS and CMS based
on the full dataset (stemming from an integrated luminosity of 
about 5~fb$^{-1}$), and associated with officially-supported
Rivet~\cite{Buckley:2010ar} routines. Specifically, the focus of this 
paper will be the $\Zjs$ results of ATLAS~\cite{Aad:2013ysa}
and CMS~\cite{Chatrchyan:2013oda,Chatrchyan:2013tna}, and the $\Wjs$ 
results of ATLAS~\cite{Aad:2014qxa} and CMS~\cite{Khachatryan:2014uva}.
We shall also briefly deal with the inclusive and the underlying-event
analyses of refs.~\cite{Aad:2012wfa,Chatrchyan:2012tb}, in order
to raise some points of potential future interest that concern
the interplay of tunable parameters in MCs with
(NLO) matching and merging. In all cases,
our predictions are obtained by means of the fully automated 
\aNLO\ framework~\cite{Alwall:2014hca}; parton showers are 
simulated with \HWpp~\cite{Bahr:2008pv,Gieseke:2011na,Arnold:2012fq,
Bellm:2013lba} and \PYe~\cite{Sjostrand:2007gs,Sjostrand:2014zea}.

We conclude this section by giving the briefest possible introduction
to the FxFx approach; the interested reader can find fuller details
in the original paper~\cite{Frederix:2012ps} as well as in
ref.~\cite{Alwall:2014hca}. FxFx is based on the MC@NLO matching
procedure~\cite{Frixione:2002ik}. MC@NLO samples are constructed
for the processes whose Born-level contributions are:
\beq
\ident_1+\ident_2\;\longrightarrow\;S+i~{\rm partons}\,,
\label{partproc}
\eeq
with $i\ge 0$, and 
where $S$ is a set of $p$ particles which does not contain any QCD 
massless partons. In the present case, $S=\ell^+\ell^-$ or $S=\ell\nu_\ell$
(with $\ell=e$ or $\mu$)
for $\Zjs$ and $\Wjs$ production respectively; by considering lepton 
pairs already at the matrix-element level, one takes into account 
exactly spin-correlation and off-shell effects. The samples corresponding
to eq.~(\ref{partproc}) are defined in such a way that hard emissions
are suppressed by means of a function that can be parametrised in
terms of a hard mass scale (called the (FxFx) merging scale), 
except in the case of the largest multiplicity (i.e.~the
largest $i$) considered. In other words, hard partons are mostly seen as 
entering a Born-level contribution (say, at $i=i_0$), rather than the 
real-emission correction to the previous multiplicity ($i=i_0-1$). Such a 
suppression, enforced at the level of matrix elements, must be accompanied 
by suitable choices of shower scales. On top of this, matrix elements
are also multiplied by appropriate Sudakov factors. Showered
events are subject to an MLM-type rejection, quite analogous to the
original one~\cite{Alwall:2007fs}, except for the fact that a $\kt$ 
jet-finding algorithm~\cite{Catani:1993hr} is adopted, as is already the 
case in the LO \MadGraphf\ implementation~\cite{Alwall:2008qv}. The only
difference w.r.t.~an LO treatment is the fact that, at the NLO, the
rejection procedure must be based on a jet-jet matching, rather 
than on a parton-jet one, in order to preserve IR safety.
Finally, we point out that FxFx is non-unitary.
In other words, the total rate resulting from an FxFx-merged sample
is not necessarily equal to the total rate of the inclusive sample
of lowest multiplicity, even before any final-state cuts are applied.
While the rate thus obtained is, like any other observable, merging-scale
dependent, it constitutes a genuine prediction of the method, in that
it incorporates some of the contributions to the inclusive $K$ factor
at orders higher than NLO.

This paper is organised as follows: technical details are reported
in sect.~\ref{sec:tech}; we present our predictions and the comparisons
to data in sect.~\ref{sec:res} (sect.~\ref{sec:resZ} for $\Zjs$,
sect.~\ref{sec:resW} for $\Wjs$, and sect.~\ref{sec:incl} for
inclusive and small-$\pt$ observables); we conclude in sect.~\ref{sec:conc}.

\section{Technicalities and settings\label{sec:tech}}
The hard-event samples have been obtained with
\aNLO v2.2.1~\cite{Alwall:2014hca}. The whole procedure is fully
automated: for either $\Zjs$ or $\Wjs$ production, a single file of 
{\em unweighted} hard events is generated, and is subsequently showered 
by an MC. No post-processing (e.g.~rescaling and combining samples relevant
to different parton multiplicities) is necessary. The MLM rejection
is performed on the fly, on a event-by-event basis, by the MCs.
Event files are created with the following commands
(here given for $\Zjs$ production in order to be definite):

\noindent
{\tt ~./bin/mg5\_aMC}

\noindent
~\prompt\ {\tt ~import model loop\_sm-no\_b\_mass}

\noindent
~\prompt\ {\tt ~define p = p b b\~{}; define j = p}

\noindent
~\prompt\ {\tt ~define l+ = e+ mu+; define l- = e- mu-}

\noindent
~\prompt\ {\tt ~generate  p p > l+ l- [QCD] @ 0}

\noindent
~\prompt\ {\tt ~add~process  p p > l+ l- j [QCD] @ 1}

\noindent
~\prompt\ {\tt ~add~process  p p > l+ l- j j [QCD] @ 2}

\noindent
~\prompt\ {\tt ~output; launch} 

\noindent
with {\tt ickkw}$=3$ in {\tt run\_card.dat}\footnote{This
setting tells the code that FxFx merging is to be used. For
UNLOPS~\cite{Lonnblad:2012ix} one sets {\tt ickkw}$=4$.}.
The commands in the first two lines instruct \aNLO\ to
perform a five-flavour computation, where the $b$ quark
is treated as massless and may appear in the initial states
of partonic subprocesses. In the case of $W^+ \!+ W^-$ production
(i.e.~the two $W$'s are generated simultaneously), one simply 
replaces the {\tt l+ l-} pair above with {\tt l vl}, with
{\tt l  = e+ mu+ e- mu-} and {\tt vl = ve vm ve\~{} vm\~{}}.
As these command lines imply, all of our simulations are
based on underlying matrix elements computed at the NLO accuracy,
with up to two extra partons at the Born level (i.e.~the highest
final-state multiplicity considered at tree-level is equal
to two leptons and three QCD partons): we do not consider any 
LO-accurate matrix elements for multiplicities larger than those that 
enter our NLO computations (i.e.~four or higher). In the case of 
$\Zjs$ production, both $Z$ and $\gamma^\star$ contributions, 
as well as their interference, are included. For showering,
\PYe.210~\cite{Sjostrand:2014zea} and \HWpp 2.7.1~\cite{Bellm:2013lba}
have been used. As far as \PYe\ is concerned, v8.210 embeds FxFx-related
modules\footnote{FxFx merging in \PYe\ takes advantage of the available 
{\tt UserHooks} facilities, meaning that the merging is (almost) completely 
decoupled from the event generator; note that all MLM-inspired merging 
methods are handled in such a way.}, and includes all of the fixes to 
previous un-validated versions of those. In particular, the
prescription on how to match the pseudo-jets of the short-distance cross
section to jets after parton showering has now been aligned to that
of \HWpp. Furthermore, the numerical stability of the 
beam-remnant handling was improved to cope with some very small 
Bjorken-$x$ values that are induced by NNPDF-derived tunes. 
In \HWpp, all FxFx modules are treated as an external 
plugin\footnote{These modules will be available in the forthcoming
release of \HWpp. In the meanwhile, they can be obtained upon
request to the authors of this paper.}, fully analogous to a generic 
user-defined analysis; thus, such modules must be compiled and loaded 
through the standard \HWpp\ input file. 

The hard event files, as customary in \aNLO, contain additional weights 
that allow one to compute the hard-scale and PDF uncertainties according 
to the reweighting procedure introduced in ref.~\cite{Frederix:2011ss};
such weights are stored as dictated by the LHA v3.0
format~\cite{Butterworth:2014efa}. The FxFx-specific
modules of both \PYe\ and \HWpp\ are fully compatible with such a
structure, and thus can handle all of the additional weights at
the same time as the main event weight. Unfortunately, due to
limitations in the present Rivet release, that package can only
deal with one weight at a time; hence, at the level of analysis we
have been forced to launch a separate Rivet run for each of the 
additional weights (therefore repeating exactly the same operations
multiple times, since the kinematics of the event does not change).
Although these operations are highly parallelisable, they could
be avoided simply by giving Rivet the possibility of filling
multiple histograms with the same kinematics and different weights.

In the computations {\em of the matrix elements} we have set the
most relevant parameters as follows: 
\begin{itemize}
\item $m_Z = 91.188$~GeV, $m_W = 80.419$~GeV.
\item $\Gamma_Z = 2.441$~GeV, $\Gamma_W = 2.0476$~GeV.
\item Lepton masses ($e^\pm$ and $\mu^\pm$) set equal to 
zero\footnote{The MCs will then put the charged leptons
on their physical mass shells, but genuine lepton-mass effects are
not included. We expect them to be utterly negligible for the
observables we have considered.}.
\item NNPDF2.3~\cite{Ball:2012cx} NLO PDFs ({\tt LHAGLUE} number 230000), 
which also set the value of $\as(m_Z)=0.119$. We work in a 
five-flavour scheme.
\item Central hard-scale choice: $\mu_0=\Ht/2$, with $\Ht$ the scalar sum
of the transverse masses $\sqrt{\pt^2+m^2}$ of all final state particles.
In merged samples, this is computed by using the $2\to V+1$ process 
reconstructed by clustering back the given kinematic 
configuration\footnote{The scale $\mu_0$ may not vanish, owing to 
the hard cuts imposed on the final-state leptons.}; 
on top of that, an $\as$ reweighting is performed -- see 
ref.~\cite{Frederix:2012ps} for more details.
\item Hard-scale variations: independent, $1/2 \mu_0 < \muR, \muF < 2\mu_0$.
\item FxFx merging scales\footnote{We note that the merging scale need not 
 be smaller than the minimal jet transverse momentum imposed in the analysis, 
 and it is actually desirable that its range includes the latter. More details 
 can be found at pages 56-57 of ref.~\cite{Alwall:2014hca}, as well as at 
 pages 18-19 of ref.~\cite{Frederix:2012ps}.}: $\mu_Q=15$, $25$, and $45$~GeV.
\item $1/\aem=132.507$.
\item Diagonal CKM matrix.
\end{itemize}
Most of the above settings may be checked by inspecting the headers
of the hard event files, where they are included through copies
of the relevant input cards. The PDFs have been obtained by means 
of the {\tt LHAPDF6}~\cite{Buckley:2014ana} package.
Our simulations are based on 15M unweighted events for the sum of the two 
leptonic channels in $\Zjs$ and in $\Wjs$ production. Given that lepton-mass 
effects are ignored here, the electron and muon channels basically account for 
half of those events each. In order to have a benchmark independent of merging,
a further 5M events for the sum of the two decay channels have been 
generated for the inclusive MC@NLO sample of lowest multiplicity 
($i=0$ in eq.~(\ref{partproc})). 
We shall refer to these events as the ``(fully) inclusive'' samples.
We finally recall that FxFx makes use of a $\kt$-clustering 
algorithm~\cite{Catani:1993hr}, with $R=1$, both at the short-distance level 
and after parton showering. A jet at short distance is said to match a jet 
after parton showering if the $\kt$ jet algorithm with minimal separation 
$\mu_Q$ would combine two such objects into a single pseudo-jet.

In the MCs, the parameters have been set to their defaults\footnote{FxFx 
dictates shower-scale choices, as explained in ref.~\cite{Frederix:2012ps}. 
As far as fully-inclusive samples are concerned, the shower scales are 
assigned according to the \aNLO\ defaults (see pages 48-49 of 
ref.~\cite{Alwall:2014hca}); we point out that the latter are partly under 
the user's control through the input parameter {\tt shower\_scale\_factor}.}
as defined by the \aNLO\ interface (see {\tt amcatnlo.cern.ch},
``Special settings for the parton shower'' under the ``Help and FAQ''
item), or by the native defaults. We point out that this implies,
in particular, that the values of $\as(M_Z)$ adopted by \PYe\
(equal to 0.1365 for initial- and final-state radiation, and to 0.130 
for multiparton interactions) and by \HWpp\ (equal to 0.118,
and internally evolved at two loops with a native scheme), as
dictated by the respective tunes chosen here (see below),
have not been modified. On the other hand, in the showers
the same PDFs as in the hard-matrix element computations have
been used, in order to guarantee that the formal NLO accuracy
of the underlying MC@NLO simulation be respected. The underlying
event tunes are Monash~2013~\cite{Skands:2014pea} in \PYe,
and UE-EE-3-CTEQ6L1~\cite{Gieseke:2011na} in \HWpp.
Since neither of them has been obtained with NNPDF2.3 NLO PDFs,
the quality of the comparisons to low-$\pt$ data might be degraded
w.r.t.~one performed by using the MCs standalone in conjuction with the same 
PDFs as in the tunes. While this is a common issue with NLO+PS simulations,
and one that will be addressed possibly only in the context of a genuine
NLO tune, the impact on $\Zjs$ and $\Wjs $ observables is expected,
and will be shown, to be rather minimal. We shall discuss this point 
in more details in sect.~\ref{sec:incl}.

\section{Comparison to data\label{sec:res}}
This section contains the main results of this paper. We present
$\Zjs$ observables in sect.~\ref{sec:resZ}, $\Wjs$ observables in 
sect.~\ref{sec:resW}; inclusive and underlying-event results are reported
in sect.~\ref{sec:incl}. In the remainder of this preamble we give
some general information, and discuss features common to all analyses.

\vskip 0.4truecm
\noindent
{\bf $\blacklozenge$ Meaning of leptons}

\noindent
In several of the Rivet routines we have used, results are obtained
for both the bare and the dressed leptons. In the following, we
shall restrict ourselves to considering only the latter, in view
of their being more inclusive in QED radiation, and thus less
sensitive to the modeling of QED showers in MCs; in general, however,
the differences between bare and dressed predictions are fairly
small. Consistently with such a choice, in our simulations both \HWpp\ 
and \PYe\ do feature QED showers; the input parameters that control 
them have been left equal to their default values as given by the MC authors.

\vskip 0.4truecm
\noindent
{\bf $\blacklozenge$ Event generation and efficiencies}

\noindent
\begin{table}
\begin{center}
\begin{tabular}{c|ccc}
\toprule
 &  {$\mu_Q=15$~GeV} & {$\mu_Q=25$~GeV} & 
    {$\mu_Q=45$~GeV} \\\hline
\HWpp\ 
    & 44\%(2.7)
    & 38\%(3.2)
    & 35\%(3.5)
    \\
\PYe\ 
    & 45\%(4)
    & 37\%(4)
    & 32\%(4)
    \\
\bottomrule
\end{tabular}
\end{center}
\caption{\label{tab:rej}
Efficiencies of the MLM-type rejection in FxFx merging, rounded
to the percent; in brackets, we report the corresponding oversampling 
factors (see the text for details).
}
\end{table}
The hard jet cuts and widely different kinematic configurations
relevant to $\Zjs$ and $\Wjs$ production imply that it is generally
time-consuming to accumulate sufficient statistics in MC simulations.
We point out that, in the context of merging approaches that feature
an MLM-type rejection procedure, the efficiency for finding a jet-jet
match and hence not to reject an event is larger the larger are the
generation cuts (with the obvious and usual condition that such cuts
do not bias the physical results). When one makes several choices
for the merging scale, there is therefore a trade-off between
generating as many event samples as merging-scale choices with
maximally-efficient generation cuts, and generating a single event
sample, where the generation cuts are so that they do not bias any
of the physics results obtained with the different merging scales.
In order to reduce the hard-event generation time, and the number of
associated files, we have adopted the latter option. However, we point
out that the former option is a perfectly reasonable one too, which may 
actually be more convenient with large-scale computer clusters
(where multiple hard-event generations can be launched in parallel).
In the generation of our hard events, we have imposed $\pt(j)\ge 8$~GeV,
without any restrictions on jet rapidities\footnote{Jets are obtained 
with an $R=1$ $\kt$-clustering algorithm. We remark that in principle 
minimum-$\pt$ cuts are not mandatory in the Sudakov-reweighted version of 
the FxFx merging~\cite{Frederix:2012ps} (which is the default, used here and
recommended for phenomenology applications), although they help increase 
significantly the generation efficiency.}; the invariant mass of
opposite-charge lepton pairs has been constrained to be larger than 40~GeV.
The efficiencies (i.e.~the probability that events not be rejected in the 
MLM procedure) resulting from these cuts that we have measured in our 
simulations are reported in table~\ref{tab:rej}. On top of those, one 
needs to take into account the fact that a further loss of statistics is
entailed by the presence of negatively-weighted events. The fraction
of these is equal to about 25\% in the FxFx samples for both $\Zjs$
and $\Wjs$ production (for comparison, it is less than 10\% in the
fully-inclusive samples). In order to improve the statistics of
the final physics plots, we have {\em oversampled} our hard events.
In other words, the same events have been showered more
than once, with care being taken to change the seeds that govern
parton showers -- in this way, a given hard event MLM-rejected 
with a certain shower configuration might be accepted by generating
a different shower configuration. Of course, in order not to bias
the final results the oversampling must not be too large; in this
paper, we have used what is reported in round brackets in
table~\ref{tab:rej} (a non-integer value implies that only a
fraction of the hard-event file is considered; note that such a file
is randomised at generation time).

\vskip 0.4truecm
\noindent
{\bf $\blacklozenge$ Choice of parameters in MC simulations}

\noindent
Before turning to presenting our predictions, we briefly return 
to the fact that, as discussed in sect.~\ref{sec:tech}, there 
are certain small inconsistencies in the parameter choices made at the
level of hard matrix elements and in the parton showers; this is common
in the context of NLO+PS simulations. In particular, we are concerned 
here with $\as$ and the PDFs. For parton showering, we have set the
$\as(m_Z)$ input value equal to that relevant to the MC tunes, rather
than as prescribed by the PDFs (note that either choice guarantees
the NLO accuracy of the results). In the case of \HWpp, the difference
is negligible (0.118 vs 0.119), while it is larger\footnote{However, 
note that \HWpp\ uses the Monte Carlo scheme~\cite{Catani:1990rr}
that, through a rescaling of $\Lambda_{\sss QCD}$ as obtained from
the input $\as(m_Z)$, effectively gives larger values of $\as(Q)$,
which are thereby closer to those of \PYe.} in the case of
\PYe\ (0.1365 vs 0.119). We have verified that, by setting 
$\as(m_Z)=0.130$ in the \HWpp\ showers, the effects on some
observables are sizable, and the agreement with data worsens.
A similar degradation of the theory-data agreement is seen by choosing 
$\as(m_Z)=0.119$ in the \PYe\ showers, with lower $\as$ values 
leading to lower jet rates and softer $\pt$ spectra than in data. 
Thus, the conclusion of this heuristic study is that it appears to be 
best to set $\as(m_Z)$ in the MCs equal to the value(s) that result(s)
from tuning. As far as the PDFs are concerned, we have also showered events
by choosing the LO version of the NNPDF2.3 sets (which formally spoils
the NLO accuracy of the computations); some differences w.r.t.~our
standard simulations do appear also in this case, restricted to 
small-$\pt$ regions, but are generally much smaller than in the case
of $\as$ variations (with the exception of the underlying event
analysis of ref.~\cite{Chatrchyan:2012tb}). While this is reassuring, we 
point out that, given the correlations between $\as$, the PDFs, and the 
low-energy parameters that are set when tuning an MC, a more consistent
theoretical treatment might necessitate beyond-LO tunes;
we shall further comment on this point in sect.~\ref{sec:incl}.

\vskip 0.4truecm
\noindent
{\bf $\blacklozenge$ Total inclusive rates}

\noindent
\begin{table}
\begin{center}
\begin{tabular}{c|ccc|c|l}
\toprule
 &  {$\mu_Q=15$~GeV} & {$\mu_Q=25$~GeV} & 
    {$\mu_Q=45$~GeV}   & inclusive \\\hline\hline
\multirow{2}{*}{$\Zjs$}
    & 2.055($-$0.9\%)
    & 2.074
    & 2.085($+$0.5\%)
    & 2.012($-$3.0\%)
    & {\sc\small HW++}
    \\
    & 2.168($+$0.8\%)
    & 2.150
    & 2.117($-$1.5\%)
    & 2.011($-$6.5\%)
    & {\sc\small PY8}
    \\\hline
\multirow{2}{*}{$\Wjs$}
    & 20.60($-$0.9\%)
    & 20.78
    & 20.87($+$0.4\%)
    & 19.96($-$3.9\%)
    & {\sc\small HW++}
    \\
    & 21.71($+$1.0\%)
    & 21.50
    & 21.18($-$1.5\%)
    & 19.97($-$7.1\%)
    & {\sc\small PY8}
    \\
\bottomrule
\end{tabular}
\end{center}
\caption{\label{tab:totFxFx}
Total rates (in nb) for the three different choices of the FxFx merging
scale, as well as those for the inclusive (i.e.~non-merged) samples,
obtained with \HWpp\ (upper rows) and \PYe\ (lower rows).
Relative differences w.r.t.~the FxFx results obtained with the central
merging scale are also reported in brackets.
}
\end{table}
In table~\ref{tab:totFxFx} we report the predictions for the total
rates (hence, independent of final-state cuts and jet definitions)
that result from the FxFx-merged and the fully-inclusive samples; the
latter are, by construction in the MC@NLO formalism, equal to those
one obtains with fixed-order computations -- indeed, the \PYe\ and
\HWpp\ results in the last column agree to a 0.05\% level, which is the 
statistical inaccuracy one expects from a 5M-event sample. There are two 
features in table~\ref{tab:totFxFx} which are particularly
worth remarking. Firstly, the merged 
results obtained with different merging scales are very close to each other.
This gives one confidence on the fact that merging-scale systematics
is under control, in spite of the large range chosen for 
$\mu_Q$ variations. Secondly, the merged rates are a few percent
larger than the fully-inclusive one, with the exact amount depending
on the MC adopted for showering. This is a manifestation of the non-unitary 
behaviour of FxFx, and the MC-dependent amount of ``unitarity violation''
should be seen as an actual prediction associated with the given MC.
On the other hand, the differences w.r.t.~the fully-inclusive cross sections
are not large, which is perfectly compatible with the moderate NNLO $K$ 
factors for $Z$ and $W$ hadroproduction.
We shall see that the small increase of the merged cross sections
w.r.t.~the inclusive ones is beneficial in terms of the comparisons to data.

\vskip 0.4truecm
\noindent
{\bf $\blacklozenge$ Normalisation of results}

\noindent
The features just mentioned, and the predictivity they underpin, help us
stress the following point. All of our predictions are reported
with their native normalisation: in other words, {\em no rescaling}
has been performed. While an overall re-normalisation
by a constant (e.g.~the NNLO/NLO fully-inclusive $K$ factor) common 
to all observables is acceptable, we believe that the practice of rescaling 
theoretical results by factors that depend on the jet multiplicity
leads to confusion, and especially when such a multiplicity 
is understood in the inclusive sense.
Although by so doing one generally makes theory-data comparisons
look better, one also tends to neglect the fact that merged 
results, especially at the NLO, are supposed to be predictive
for both shapes and rates. At the very least, a rescaling dependent 
on the jet multiplicity renders it more difficult to understand
the strengths and weaknesses of a given merging approach, and to 
assess the overall predictivity of different merging techniques.
The latter problem is clearly more acute in the case where the rescaling
factors exhibit a non-negligible dependence on the jet multiplicity,
and/or on the particular MC considered. As an example of
both of these issues, we refer the reader to
table~7, appendix~A, of ref.~\cite{Aad:2014qxa}, where the results
of several state-of-the-art simulations are reported. From a purely
theoretical viewpoint, the effects of a multiplicity-dependent
rescaling have been considered e.g.~in ref.~\cite{Lonnblad:2012ix};
fig.~6 there, in particular, shows that such a rescaling may induce
a large increase of the theoretical systematics, for cross sections
characterised by large $K$ factors.

\vskip 0.4truecm
\noindent
{\bf $\blacklozenge$ Generalities concerning differential observables}

\noindent
All of the figures we shall present below have the same layout.
In the main frame, the data are displayed as full black circles,
with bars representing the associated errors as quoted by the experiments.
The FxFx results are shown as green bands (labelled by ``Var''
in the plots), which give the full span of the theoretical uncertainties 
considered here: these are due to the variations of the hard scales, 
of the PDFs, and of the merging scale. Such bands have been obtained
as follows. For a given merging scale, the envelope is constructed 
for the hard-scale dependence bin-by-bin, by taking the maximum and 
the minimum in each bin among the predictions associated with all 
possible hard-scale choices. Similarly, one constructs the PDF-dependence
envelope according to the prescription of the PDF authors~\cite{Ball:2012cx}.
The two envelopes thus obtained are combined in quadrature, and 
the result is therefore the theoretical uncertainty band associated
with the given merging scale. Finally, the envelope of these theoretical
uncertainties relevant to the three merging scales is constructed.
In the main frame we also report the central MC@NLO result for the 
fully-inclusive samples (i.e.~non-merged) as a solid red histogram 
(labelled by ``inc''). Below the main frame
there are two insets. The upper one displays the fractional experimental
errors as a yellow band. The ratios of the theoretical predictions over
data are given as green bands (for the FxFx results) and as a solid
red histogram (for the fully-inclusive MC@NLO results). The central
FxFx predictions (i.e.~obtained with reference hard scales and PDFs,
and $\mu_Q=25$~GeV), divided by data,
are displayed as full green triangles. Finally, dark-green
error bars represent the fractional statistical uncertainties,
in the case of $\mu_Q=25$~GeV. In the lower inset we show the ratios
of the upper and lower end of the hard-scale-plus-PDF envelopes,
over the central result; these are
therefore theoretical fractional uncertainties, displayed as
solid blue, dashed magenta, and dot-dashed light blue histograms
for $\mu_Q=15$, $25$, and $45$~GeV, respectively.

In order to facilitate the comparisons with published results,
the observables in the plots are strictly labelled as in the original
papers (i.e.~the labels are inherited from the relevant Rivet 
routines). We warn the reader that this might imply some
inconsistencies in the notations of different analyses.

Before turning to commenting the comparisons to $\Vjs$ data, we stress the
following point: the observables we are interested in are characterised
by a number of jets ($\Njet$) that accompany the vector boson equal to or 
larger than one. For these, the MC@NLO fully-inclusive predictions are 
at most of LO accuracy, when $\Njet=1$; for $\Njet>1$, they are MC-induced,
hence of LL accuracy. This implies that for the present observables the
comparison between data and fully-inclusive predictions is not particularly
meaningful: hence, from a phenomenological viewpoint we believe it should
not be considered. Note that when $\Njet=1$ the fully-inclusive MC@NLO results
may actually be farther from data than those obtained from a $V\!+\!1$-parton 
LO sample, owing to the different choices of shower scales in the two 
simulations. Therefore, the {\em sole} reason for including the
fully-inclusive MC@NLO predictions in the plots below is that of 
giving a benchmark against which the changes due to the inclusion
of higher-multiplicity matrix elements in the FxFx procedure can be
gauged in a purely theoretical way.

\subsection{$\Zjs$\label{sec:resZ}}
This section is devoted to the comparison between data and 
theoretical predictions relevant to $\Zjs$ production.
We shall make use of the ATLAS measurements of ref.~\cite{Aad:2013ysa},
and of the CMS measurements of refs.~\cite{Chatrchyan:2013oda,
Chatrchyan:2013tna}. For each of these, a (non-exhaustive) summary 
of the characteristics of the analysis is given; however, we encourage 
the reader to check the original experimental papers for more details.
In both the experimental results and in our simulations (see the
beginning of sect.~\ref{sec:tech}), a ``$Z$'' is a shorthand
for a lepton pair.

\vskip 0.4truecm
\noindent
$\bullet$ ATLAS~\cite{Aad:2013ysa}
({\tt arXiv:1304.7098}, Rivet analysis {\tt ATLAS\_2013\_I1230812}).

\noindent
Study of jet, $Z$, inclusive properties (the latter two defined 
by requiring the presence of at least one jet in the final state), 
and jet-jet correlations. Based on an integrated luminosity
of 4.6 fb$^{-1}$, using both $\epem$ and $\mpmm$ pairs, with $R=0.4$ 
anti-$\kt$ jets~\cite{Cacciari:2008gp} within $\pt(j)>30$ GeV
and $\abs{y(j)}<4.4$. Further cuts ($\ell$ denotes either
an electron or a muon): $\pt(\ell)\ge 20$~GeV,
$66\le M(\ell\ell)\le 116$~GeV, $\Delta R(j\ell)\ge 0.5$,
$\Delta R(\ell\ell)\ge 0.2$, $\abs{\eta(\mu)}\le 2.4$,
$\abs{\eta(e)}\le 1.37$ and $1.52\le\abs{\eta(e)}\le 2.47$.

\begin{figure}[!ht]
  \includegraphics[width=0.499\linewidth]{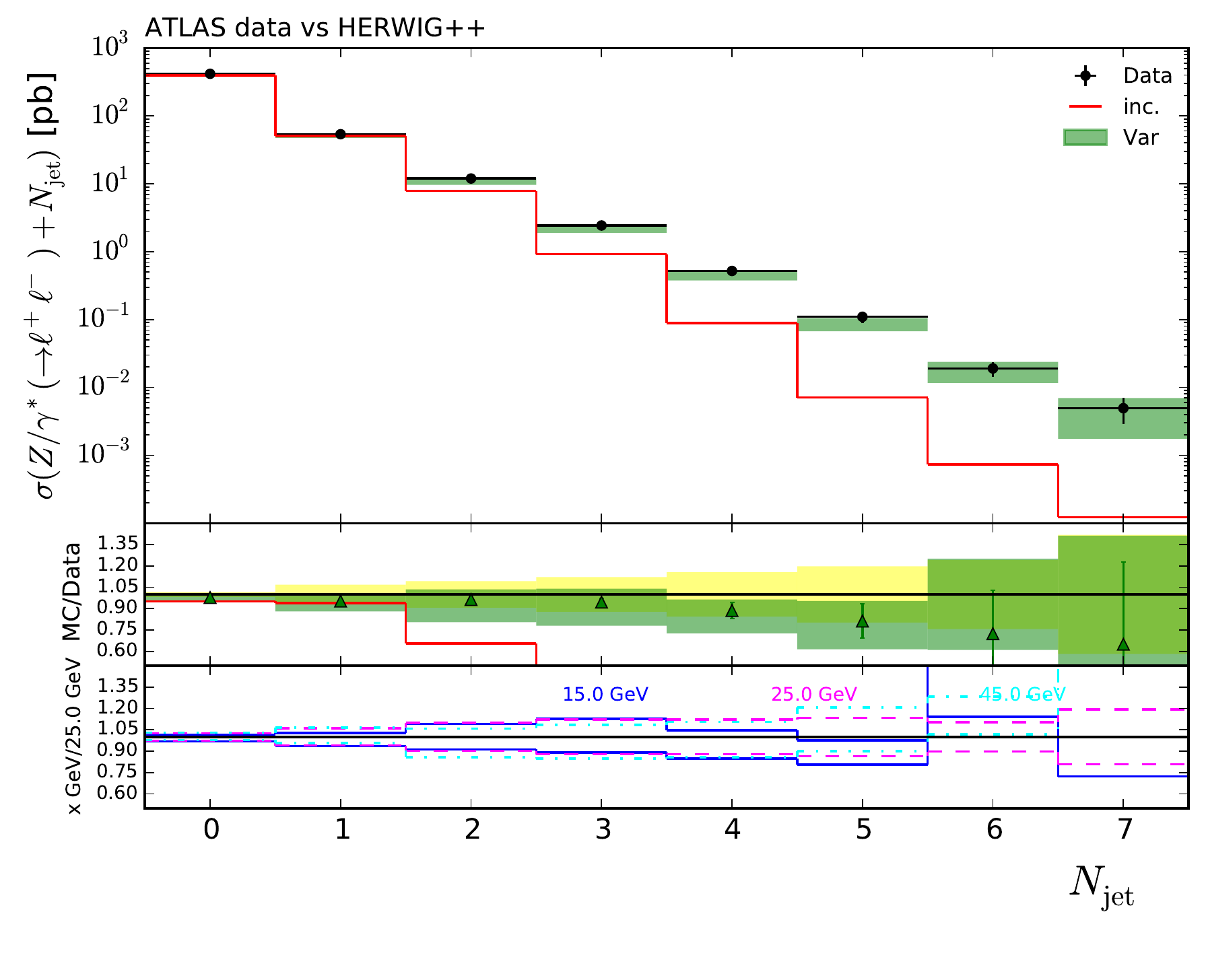}
  \includegraphics[width=0.499\linewidth]{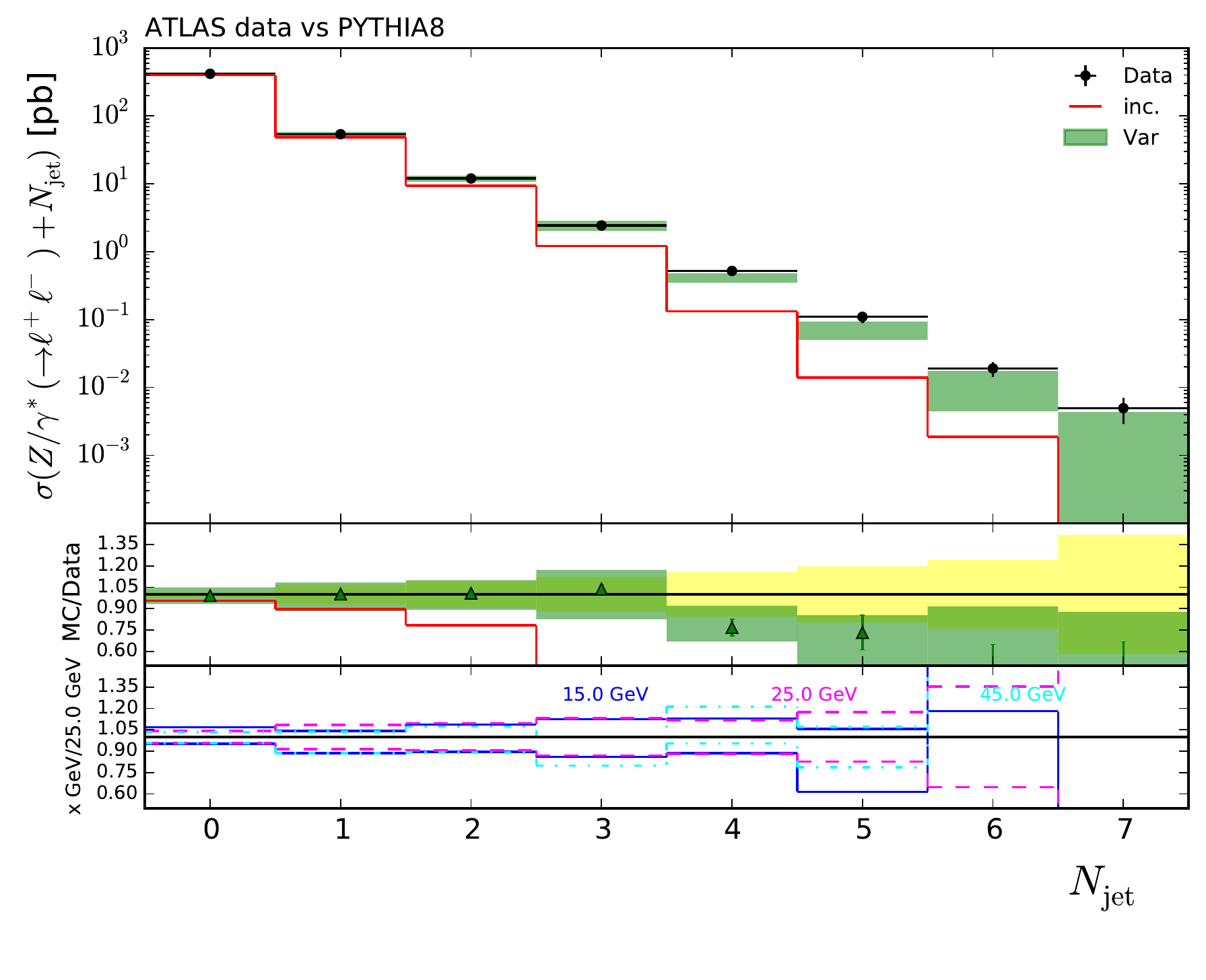}
  \caption{Exclusive jet multiplicity. 
 Data from ref.~\cite{Aad:2013ysa}, compared to \HWpp\ (left panel) 
 and \PYe\ (right panel) predictions. The FxFx uncertainty envelope 
 (``Var'') and the fully-inclusive central result (``inc'') are shown
 as green bands and red histograms respectively. See the end of
 sect.~\ref{sec:tech} for more details on the layout of the plots.
}
  \label{fig:Z.1304.7098:03}
\end{figure}

\begin{figure}[!ht]
  \includegraphics[width=0.499\linewidth]{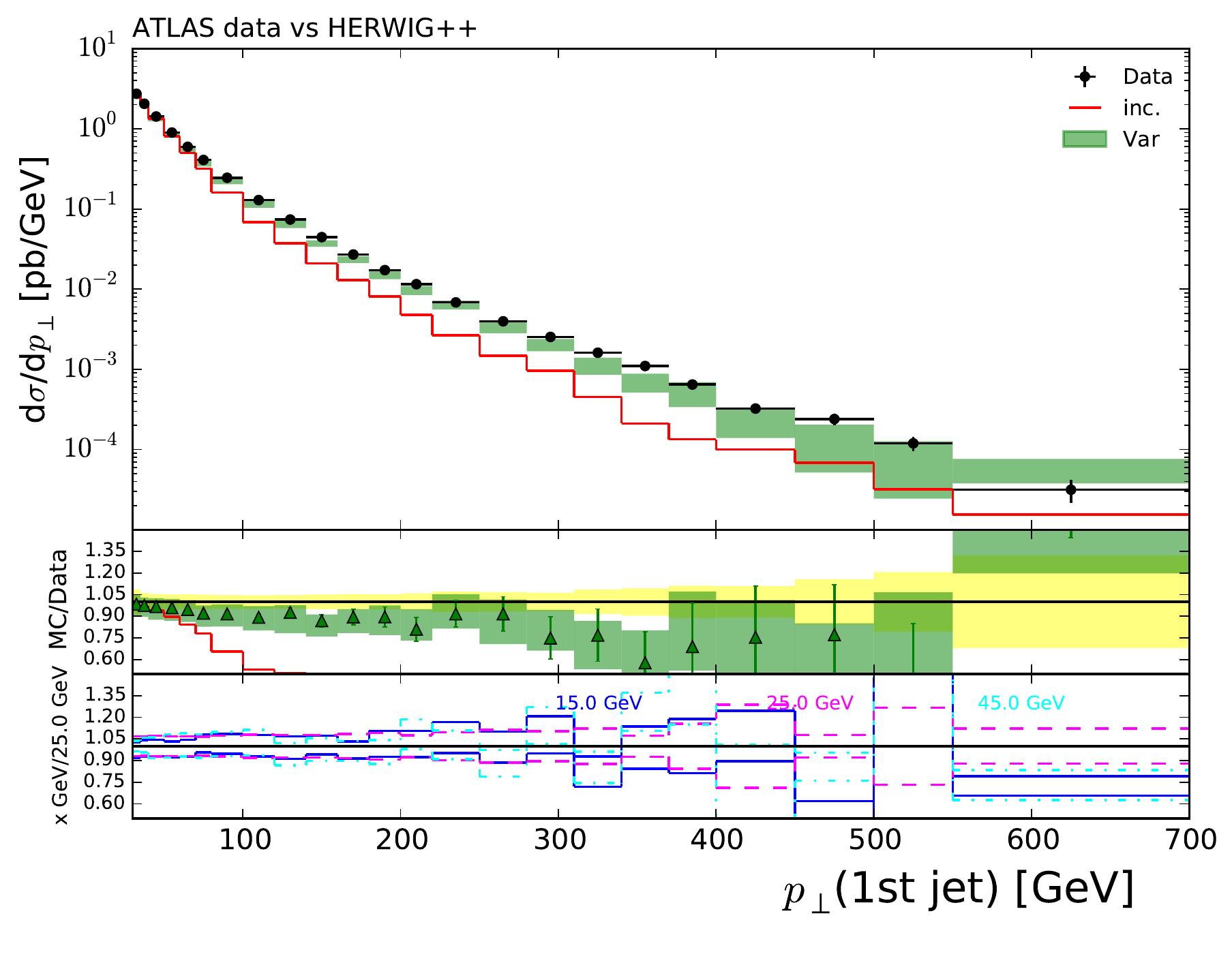}
  \includegraphics[width=0.499\linewidth]{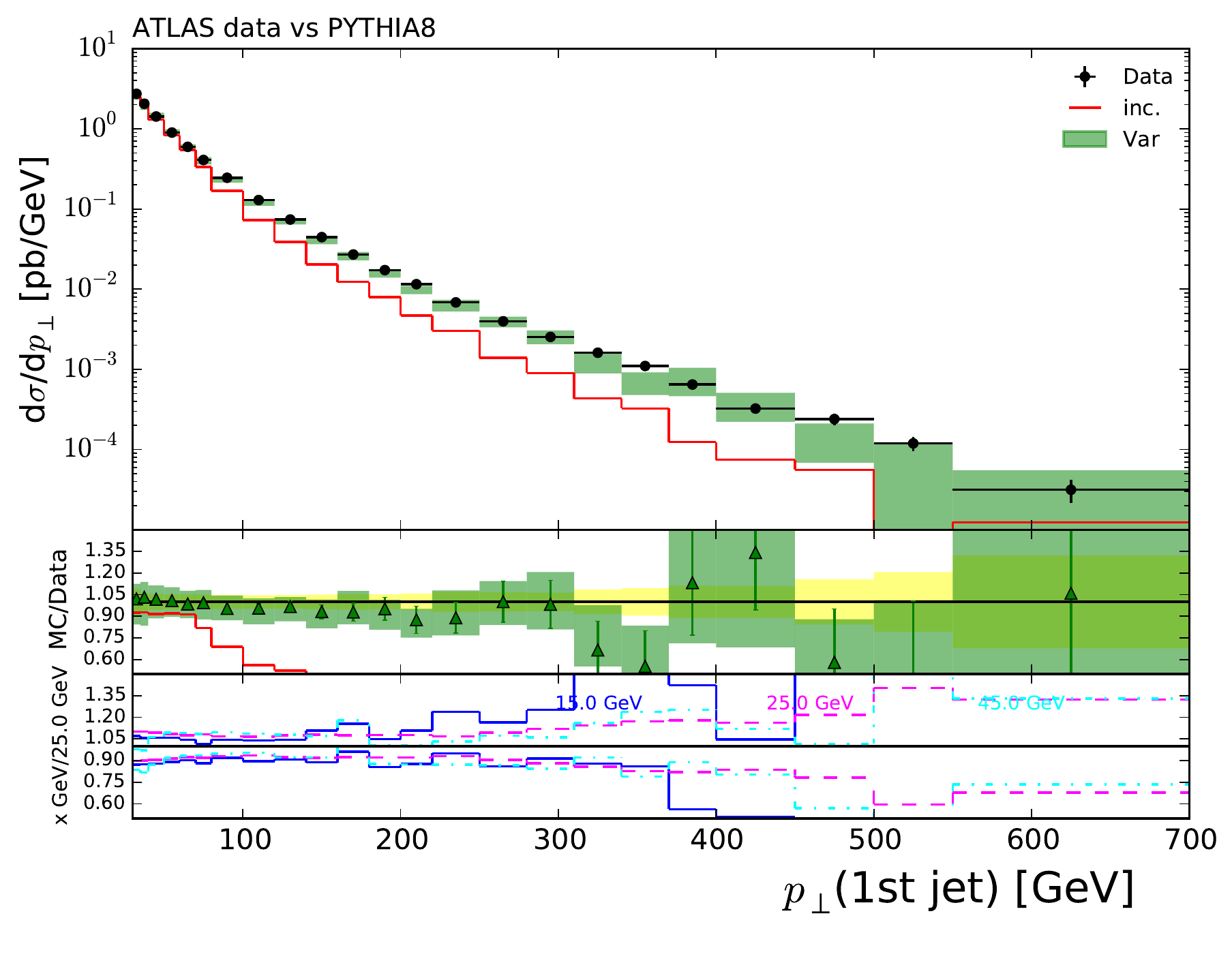}
  \caption{As in fig.~\ref{fig:Z.1304.7098:03},
 for the transverse momentum of the 1$^{st}$ jet.
}
  \label{fig:Z.1304.7098:09}
\end{figure}

\begin{figure}[!ht]
  \includegraphics[width=0.499\linewidth]{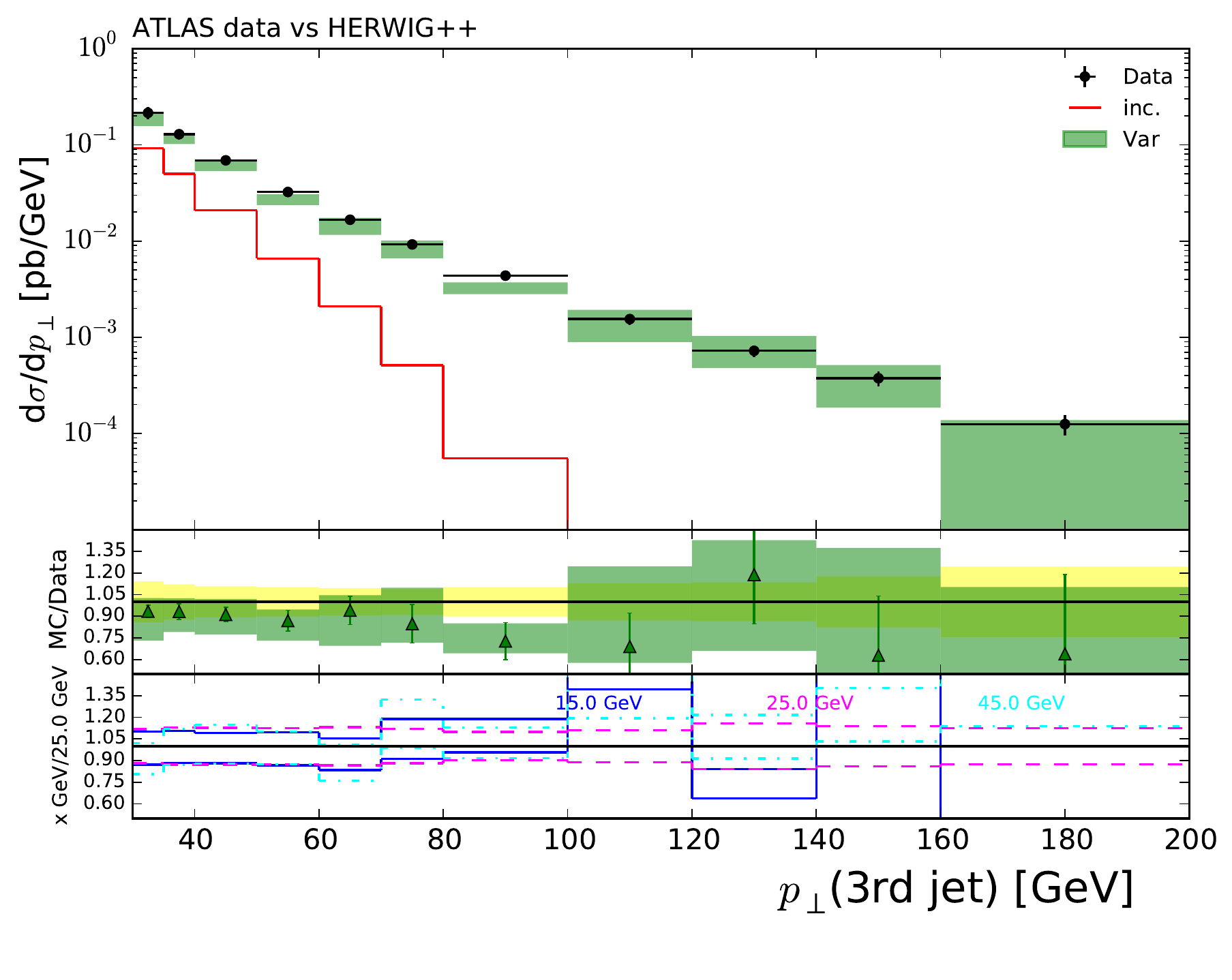}
  \includegraphics[width=0.499\linewidth]{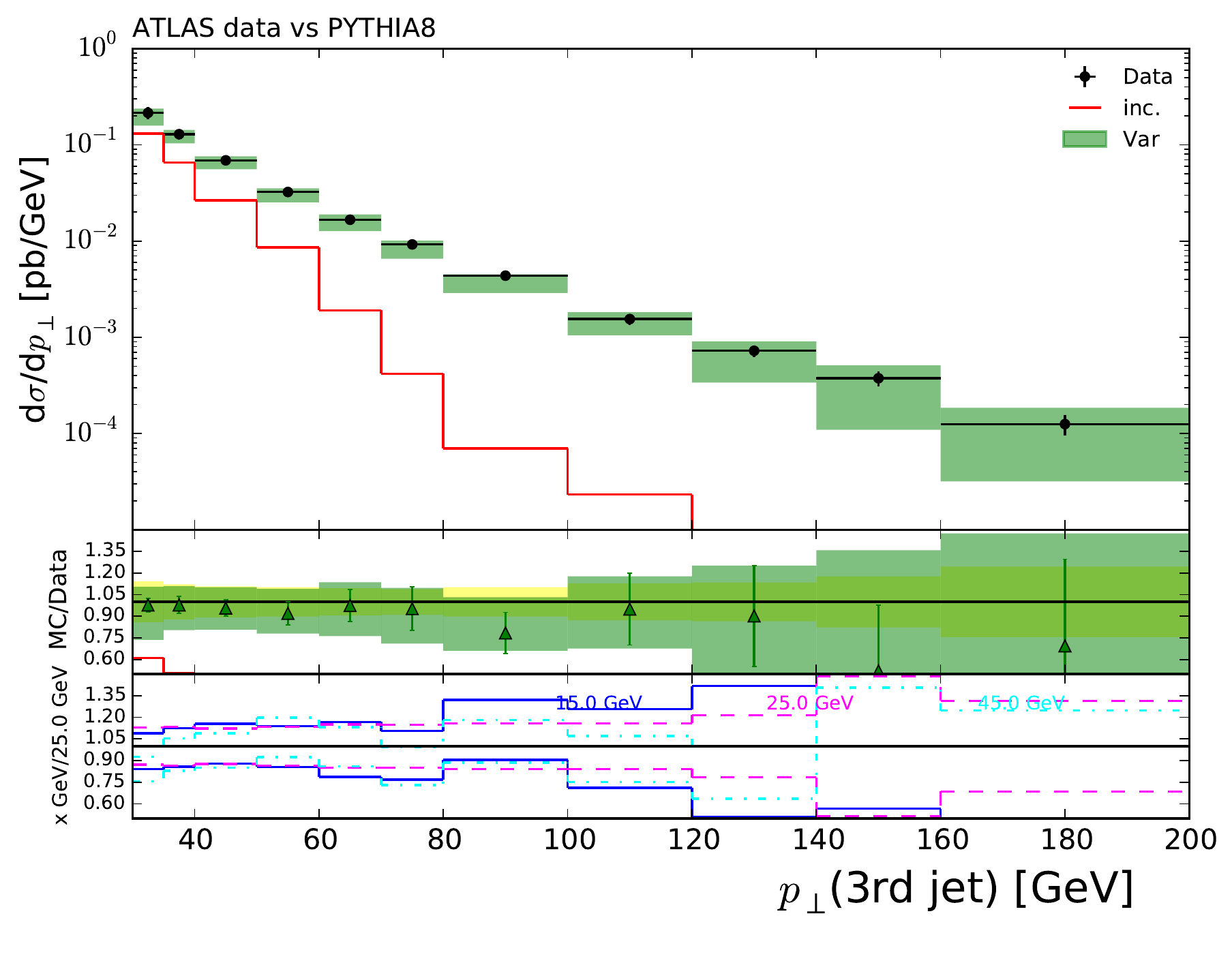}
  \caption{As in fig.~\ref{fig:Z.1304.7098:03},
 for the transverse momentum of the 3$^{rd}$ jet.
}
  \label{fig:Z.1304.7098:11}
\end{figure}

\begin{figure}[!ht]
  \includegraphics[width=0.499\linewidth]{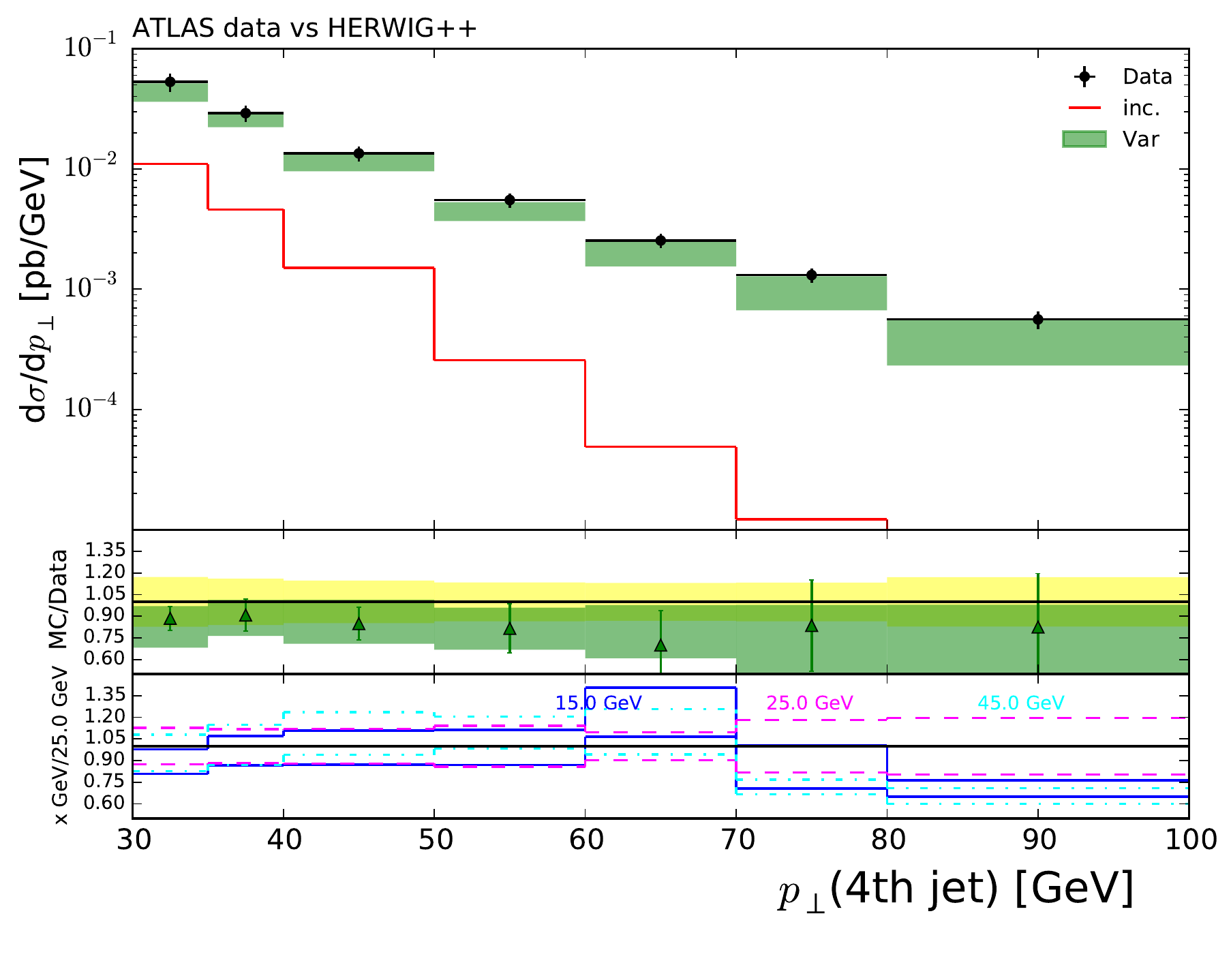}
  \includegraphics[width=0.499\linewidth]{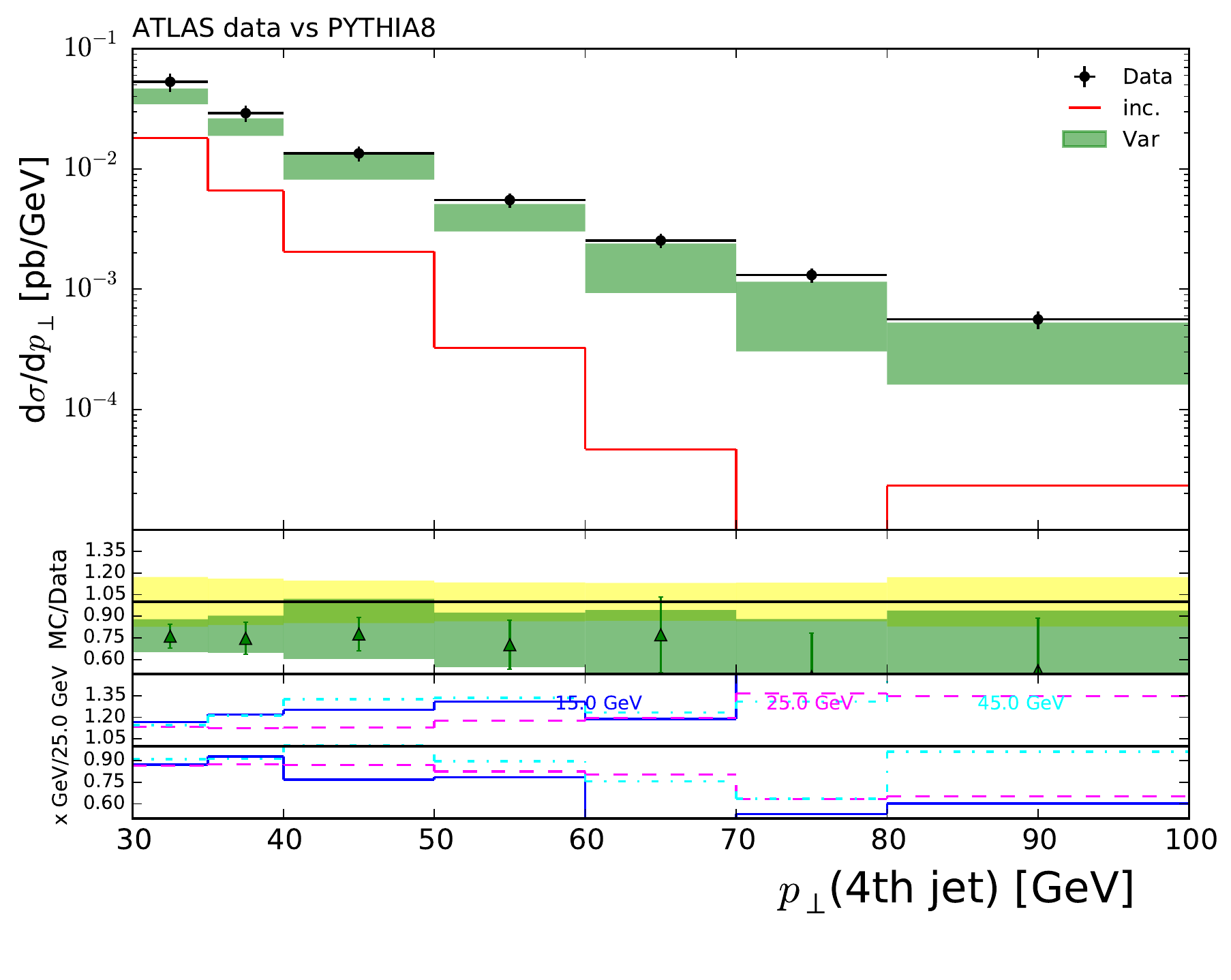}
  \caption{As in fig.~\ref{fig:Z.1304.7098:03},
 for the transverse momentum of the 4$^{th}$ jet.
}
  \label{fig:Z.1304.7098:12}
\end{figure}

\begin{figure}[!ht]
  \includegraphics[width=0.499\linewidth]{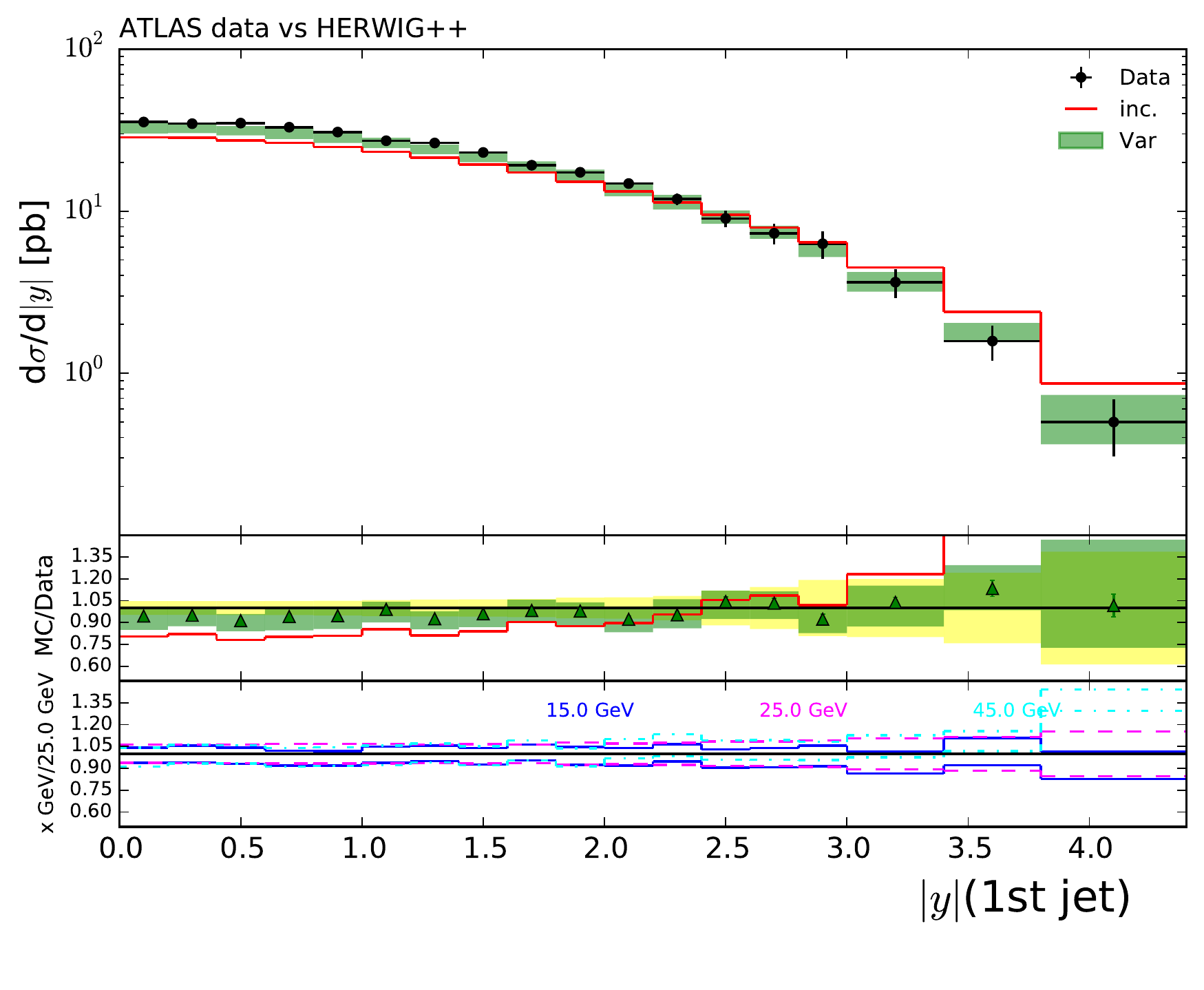}
  \includegraphics[width=0.499\linewidth]{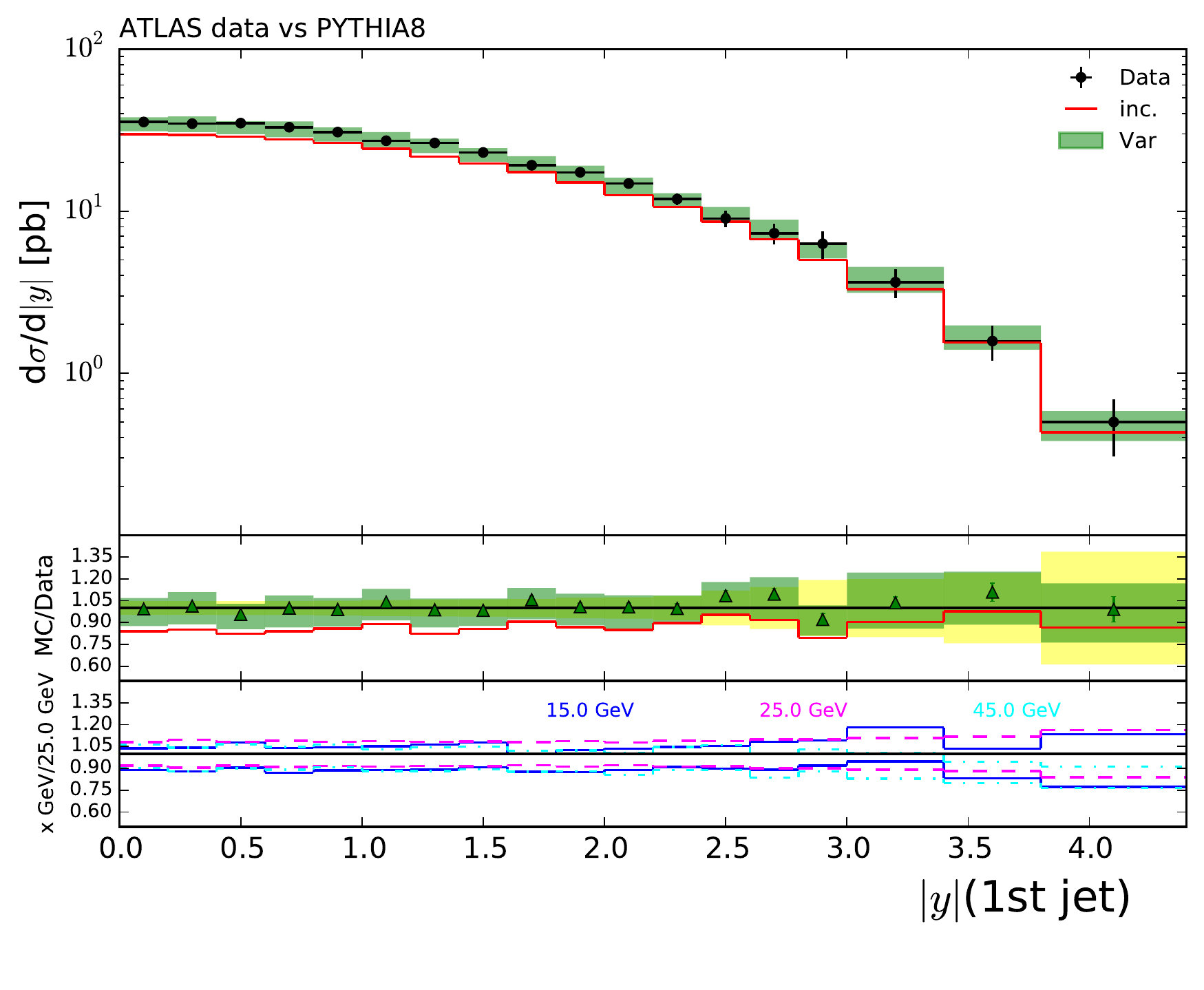}
  \caption{As in fig.~\ref{fig:Z.1304.7098:03},
 for the rapidity of the 1$^{st}$ jet.
}
  \label{fig:Z.1304.7098:17}
\end{figure}

\begin{figure}[!ht]
  \includegraphics[width=0.499\linewidth]{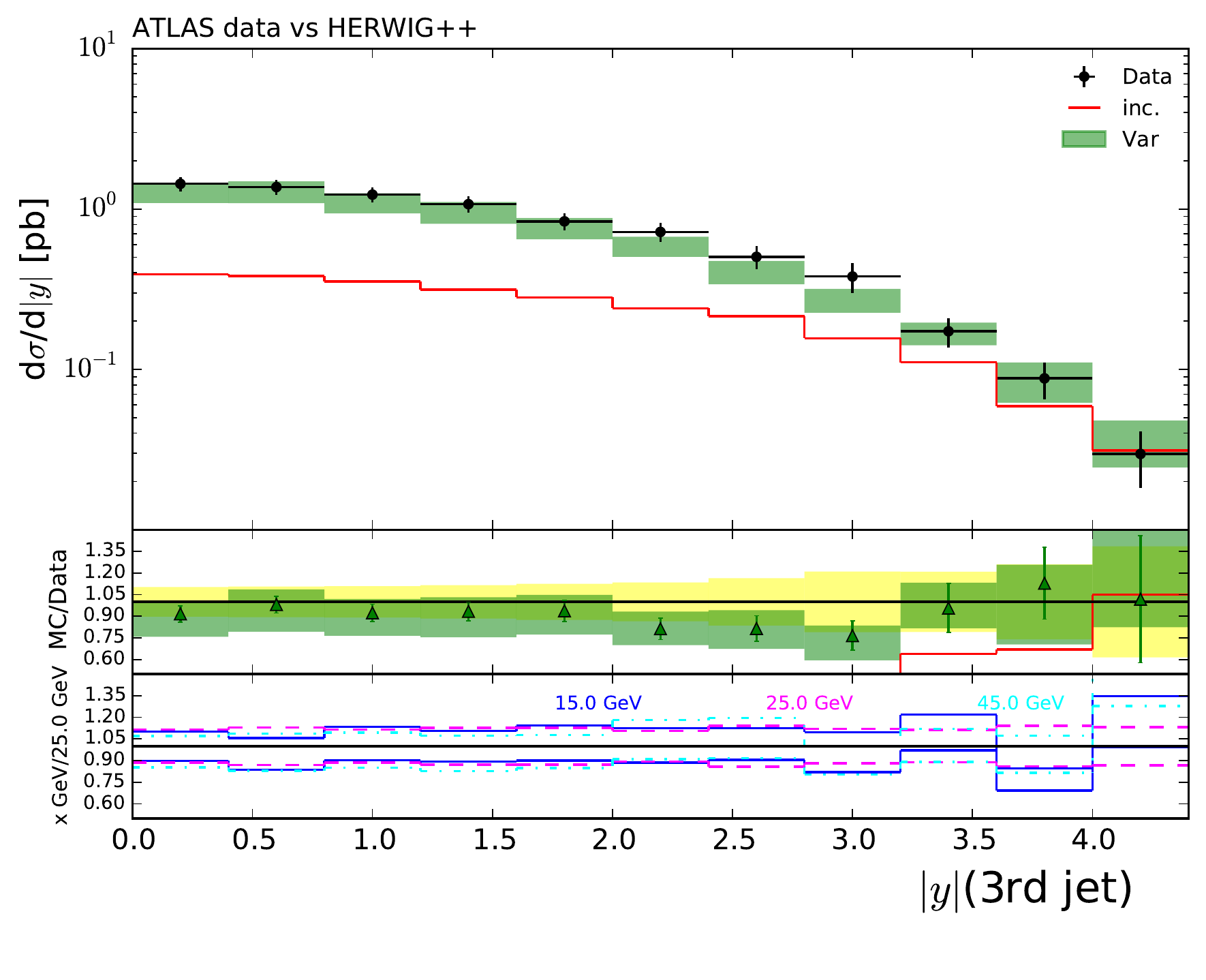}
  \includegraphics[width=0.499\linewidth]{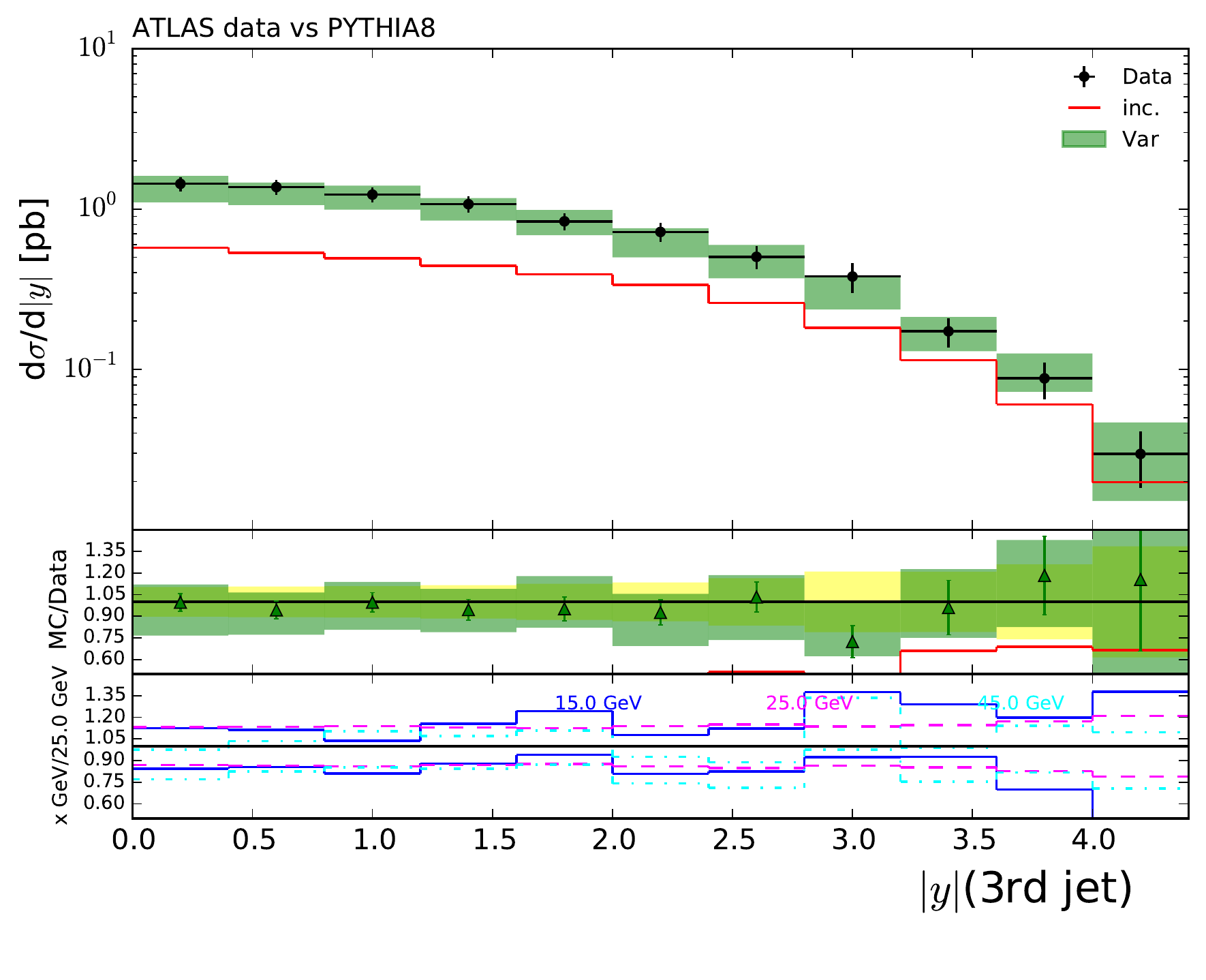}
  \caption{As in fig.~\ref{fig:Z.1304.7098:03},
 for the rapidity of the 3$^{rd}$ jet.
}
  \label{fig:Z.1304.7098:19}
\end{figure}

\begin{figure}[!ht]
  \includegraphics[width=0.499\linewidth]{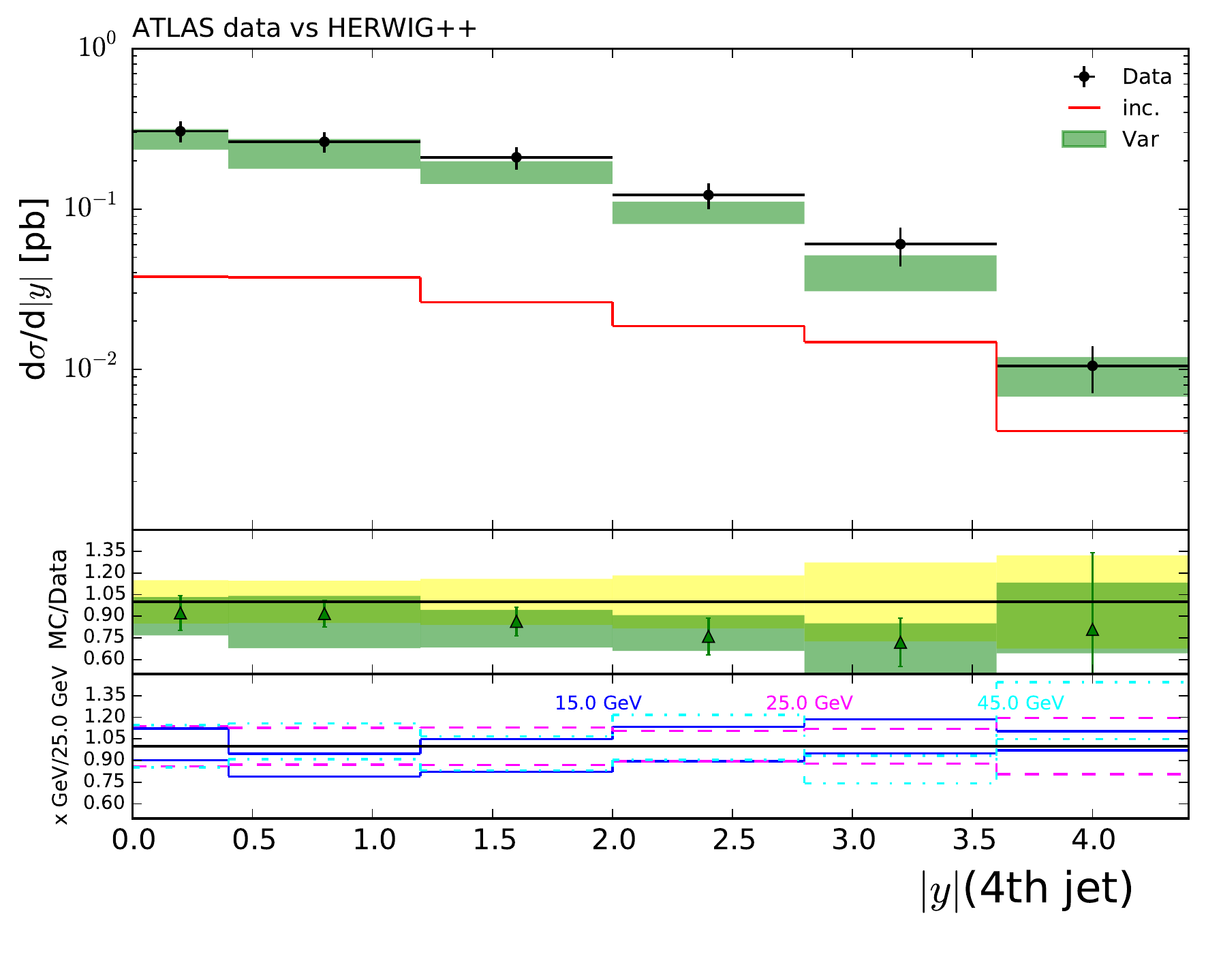}
  \includegraphics[width=0.499\linewidth]{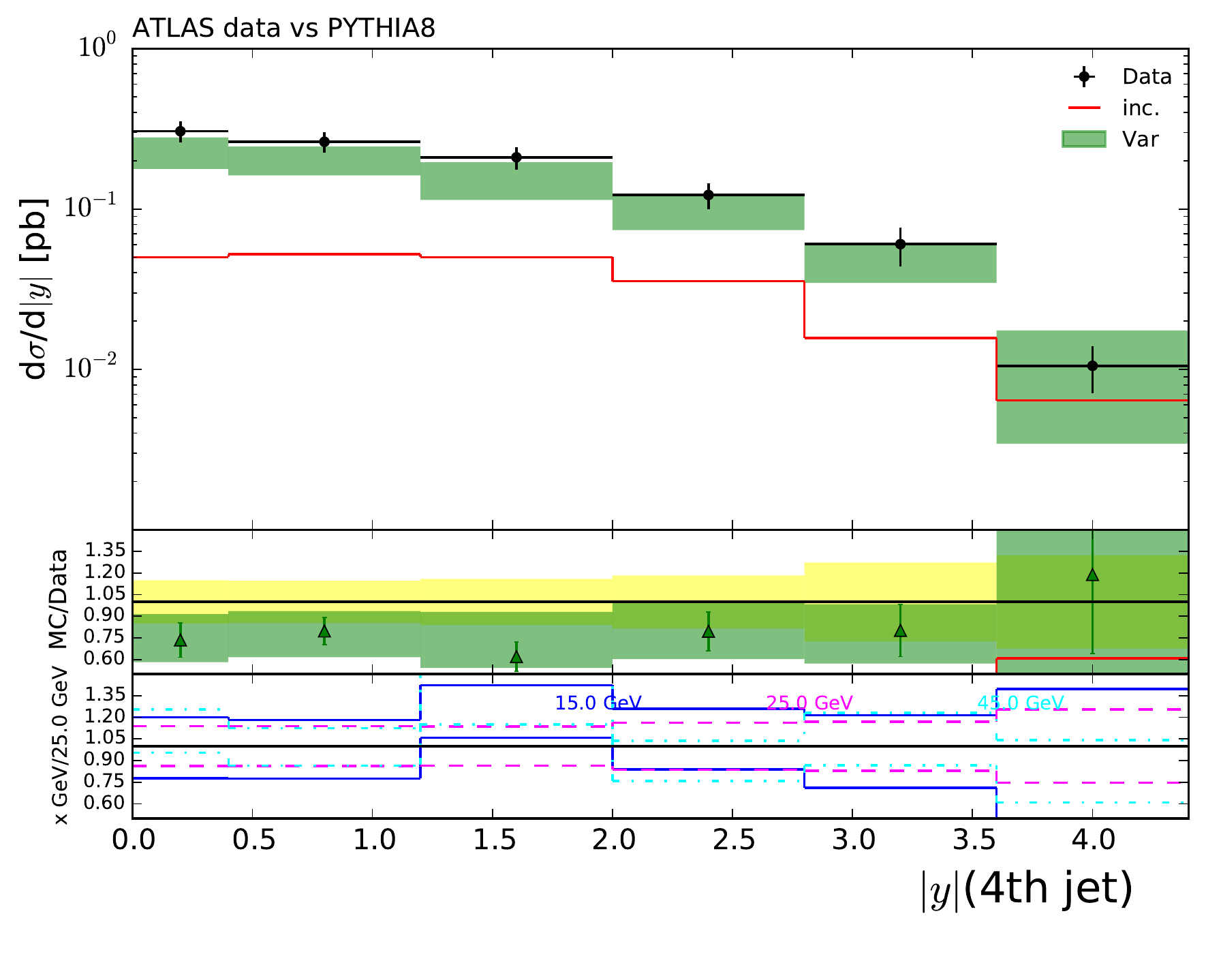}
  \caption{As in fig.~\ref{fig:Z.1304.7098:03},
 for the rapidity of the 4$^{th}$ jet.
}
  \label{fig:Z.1304.7098:20}
\end{figure}

\begin{figure}[!ht]
  \includegraphics[width=0.499\linewidth]{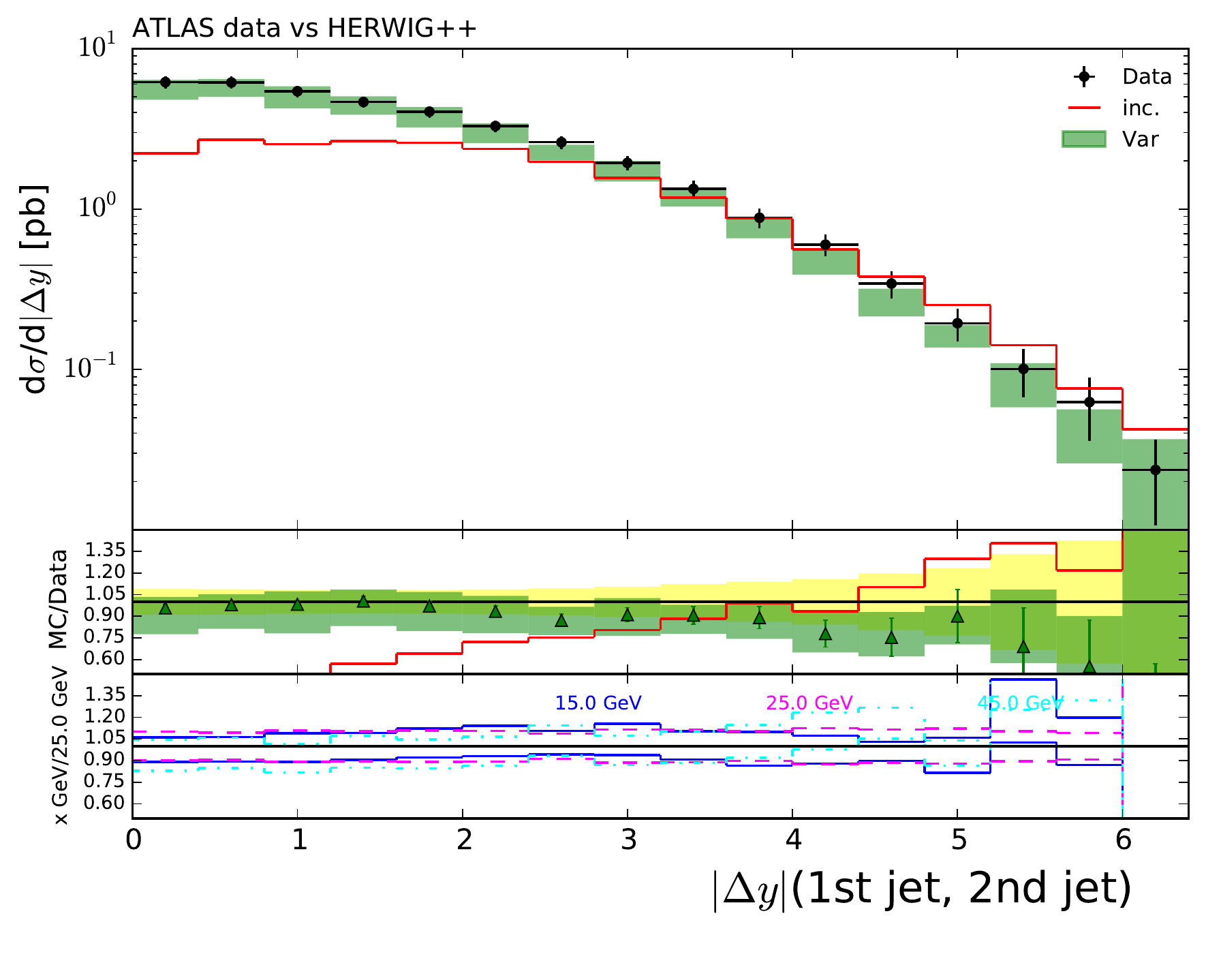}
  \includegraphics[width=0.499\linewidth]{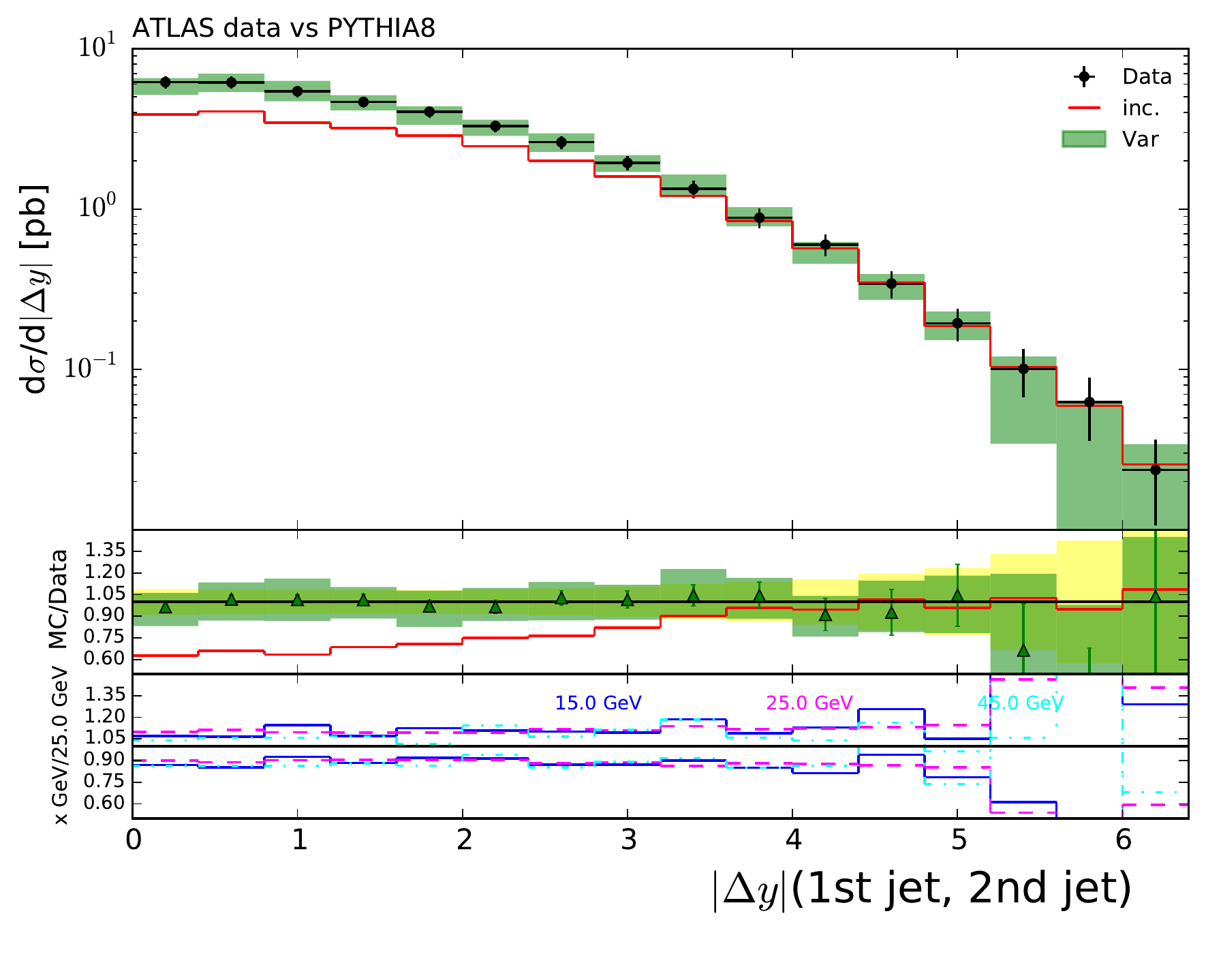}
  \caption{As in fig.~\ref{fig:Z.1304.7098:03},
 for the rapidity distance between the two hardest jets.
}
  \label{fig:Z.1304.7098:21}
\end{figure}

\begin{figure}[!ht]
  \includegraphics[width=0.499\linewidth]{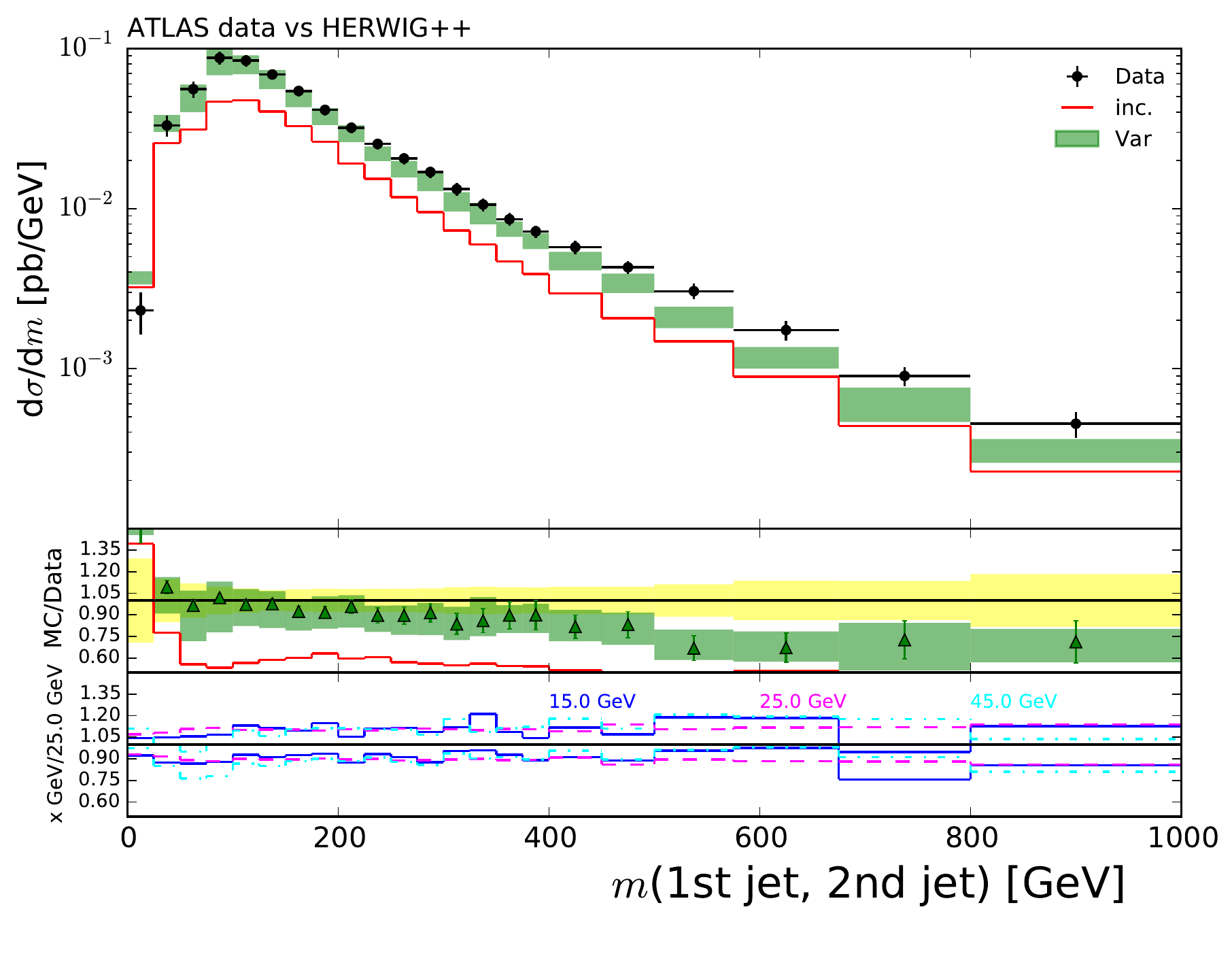}
  \includegraphics[width=0.499\linewidth]{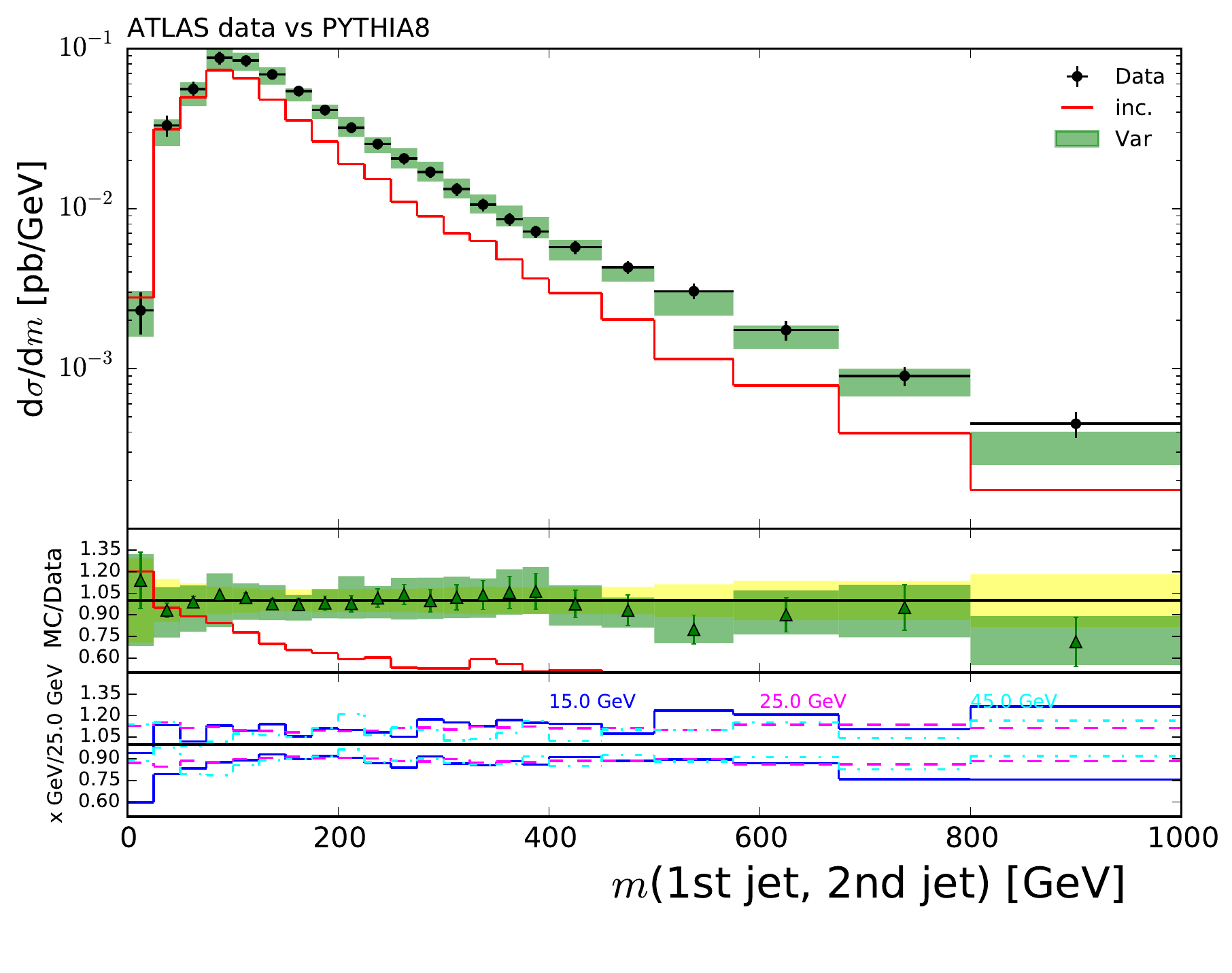}
  \caption{As in fig.~\ref{fig:Z.1304.7098:03},
 for the invariant mass of the two hardest jets.
}
  \label{fig:Z.1304.7098:22}
\end{figure}

\begin{figure}[!ht]
  \includegraphics[width=0.499\linewidth]{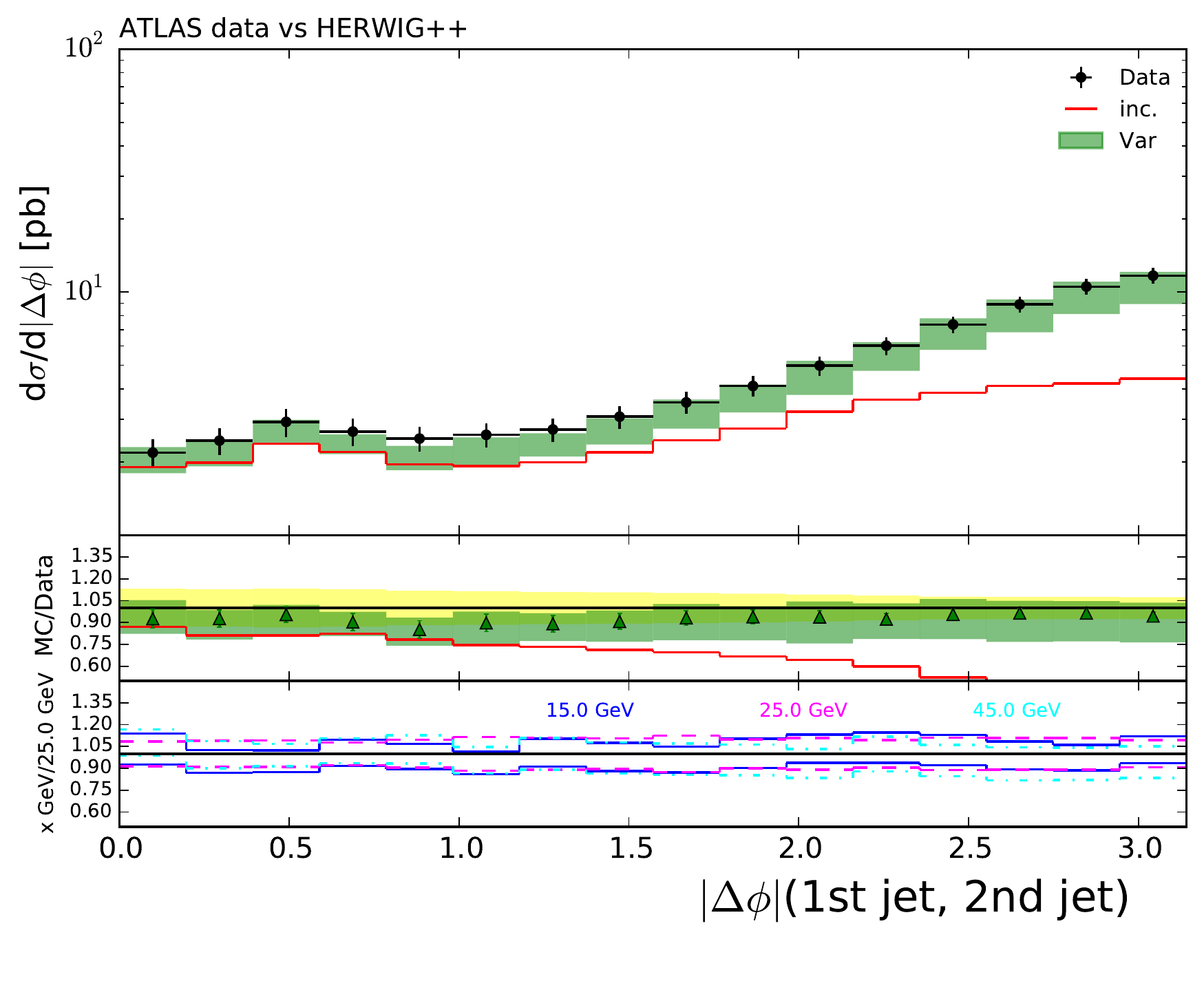}
  \includegraphics[width=0.499\linewidth]{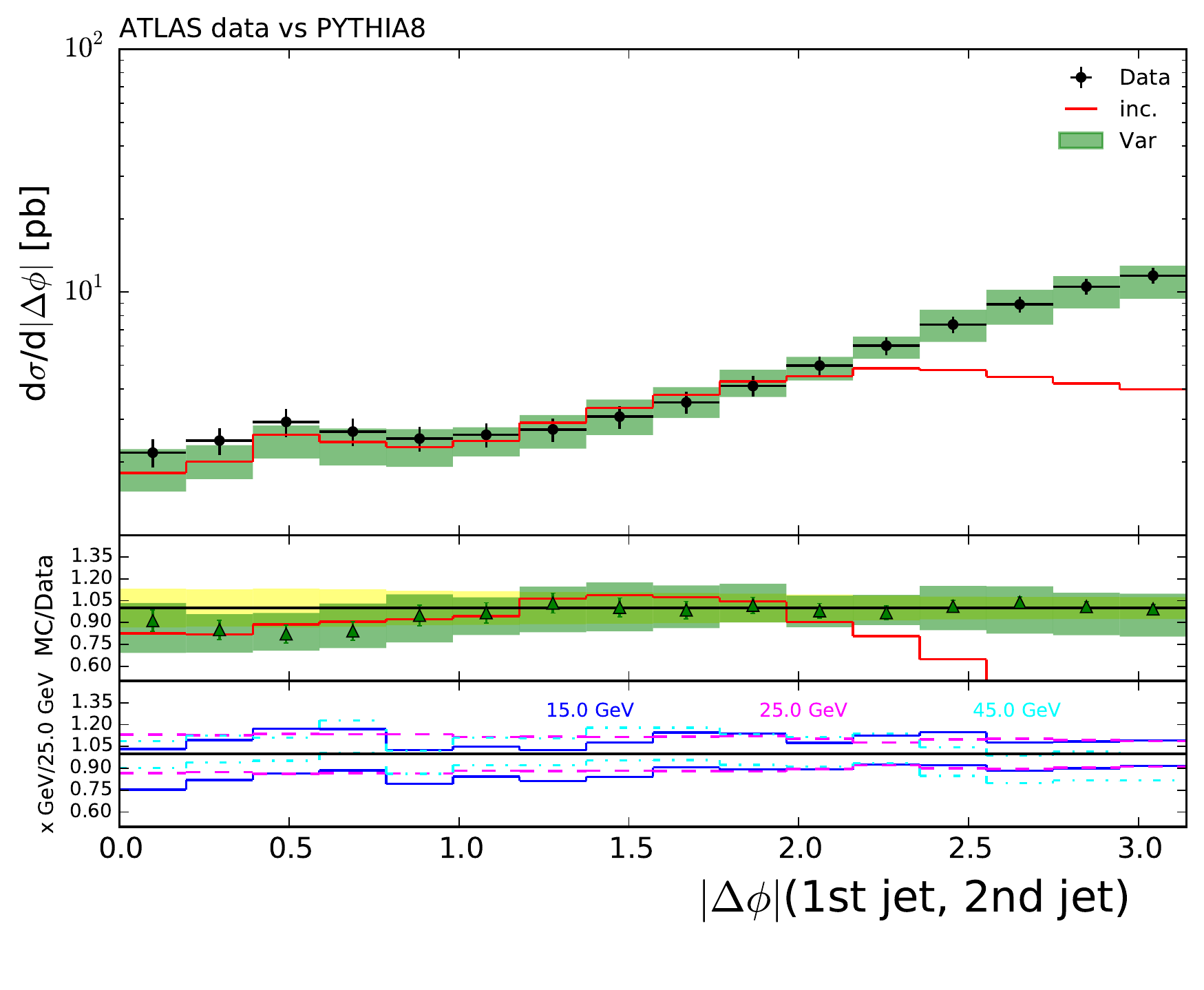}
  \caption{As in fig.~\ref{fig:Z.1304.7098:03},
 for the azimuthal distance between the two hardest jets.
}
  \label{fig:Z.1304.7098:23}
\end{figure}

\begin{figure}[!ht]
  \includegraphics[width=0.499\linewidth]{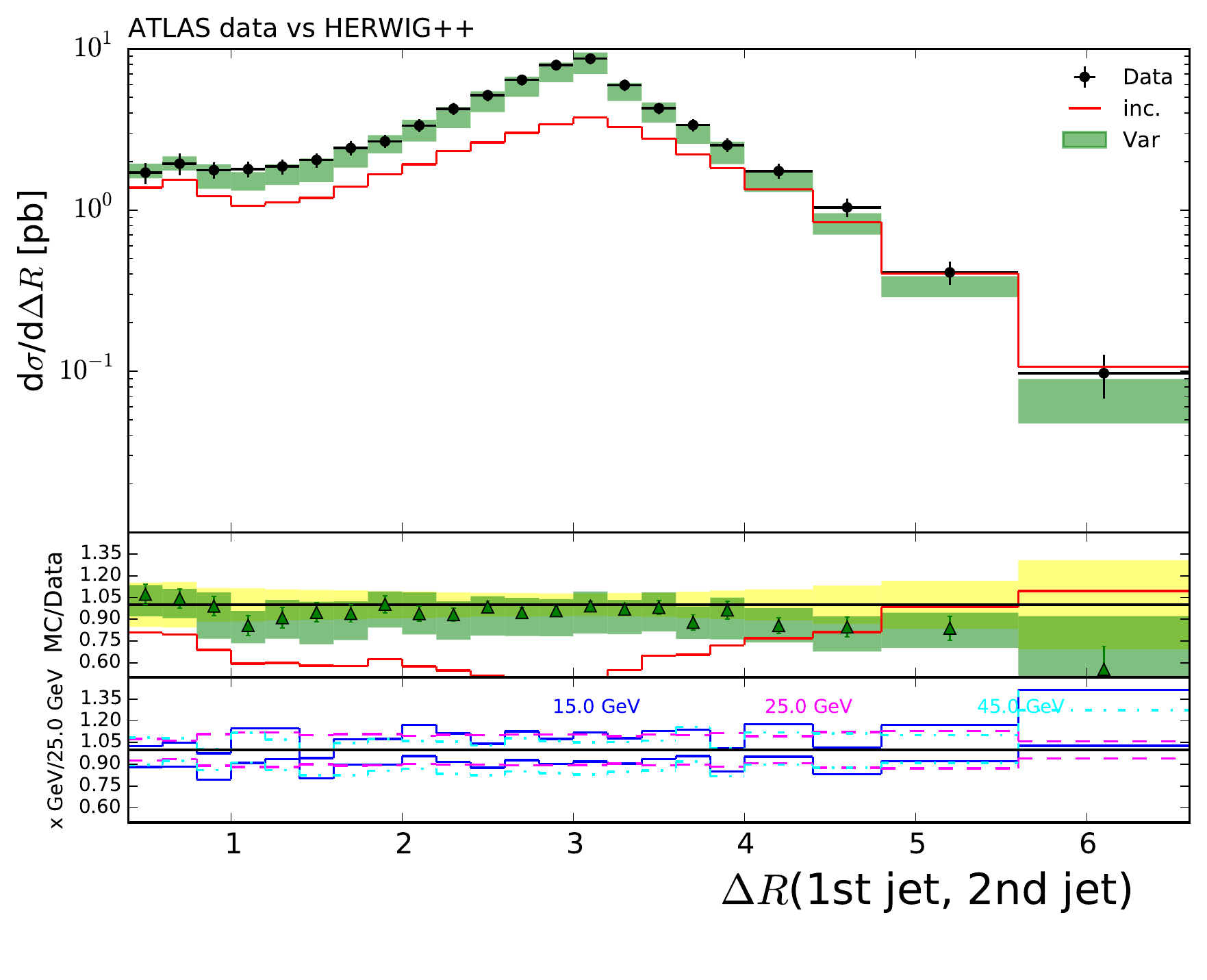}
  \includegraphics[width=0.499\linewidth]{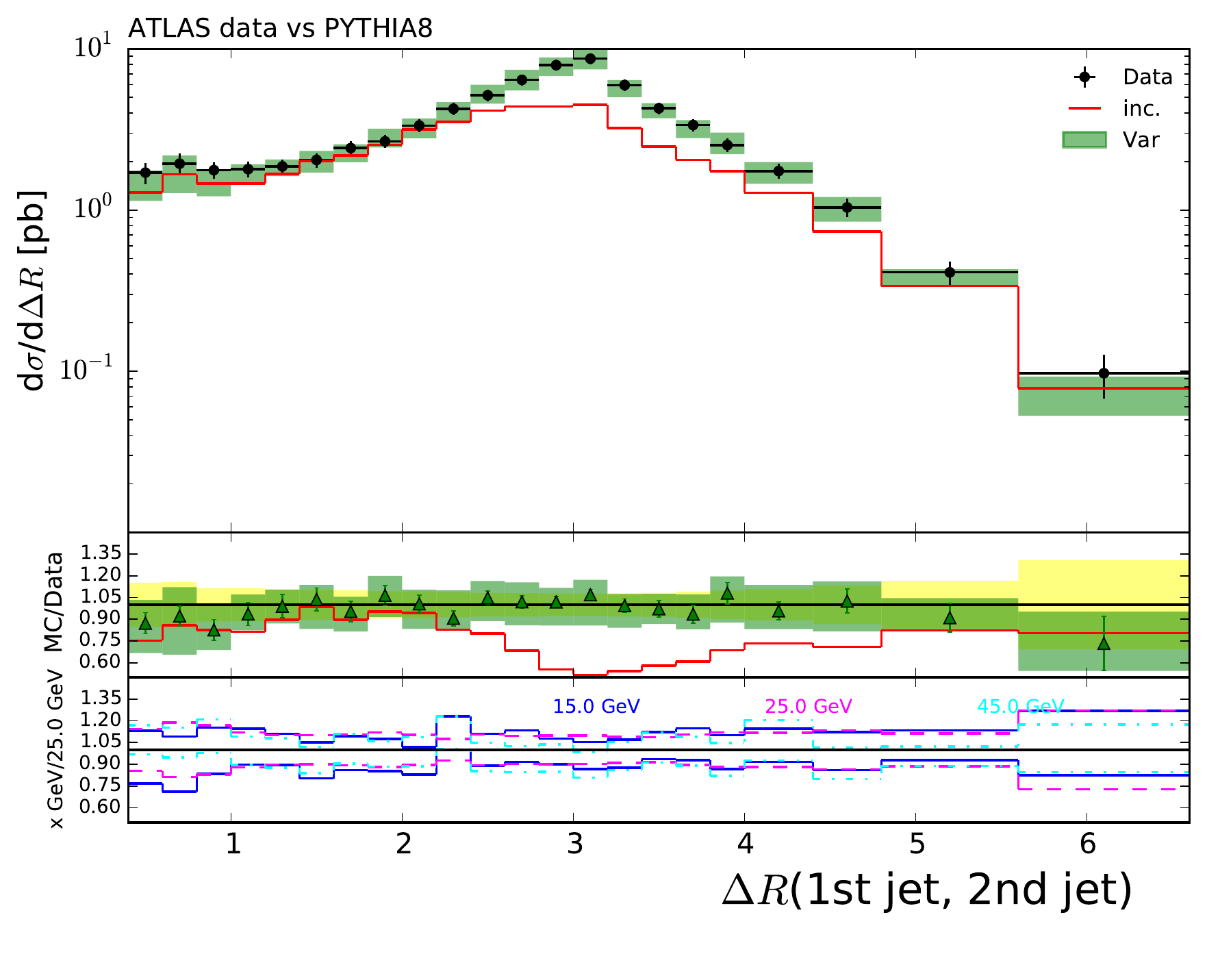}
  \caption{As in fig.~\ref{fig:Z.1304.7098:03},
 for the $\Delta R$ between the two hardest jets.
}
  \label{fig:Z.1304.7098:24}
\end{figure}

\begin{figure}[!ht]
  \includegraphics[width=0.499\linewidth]{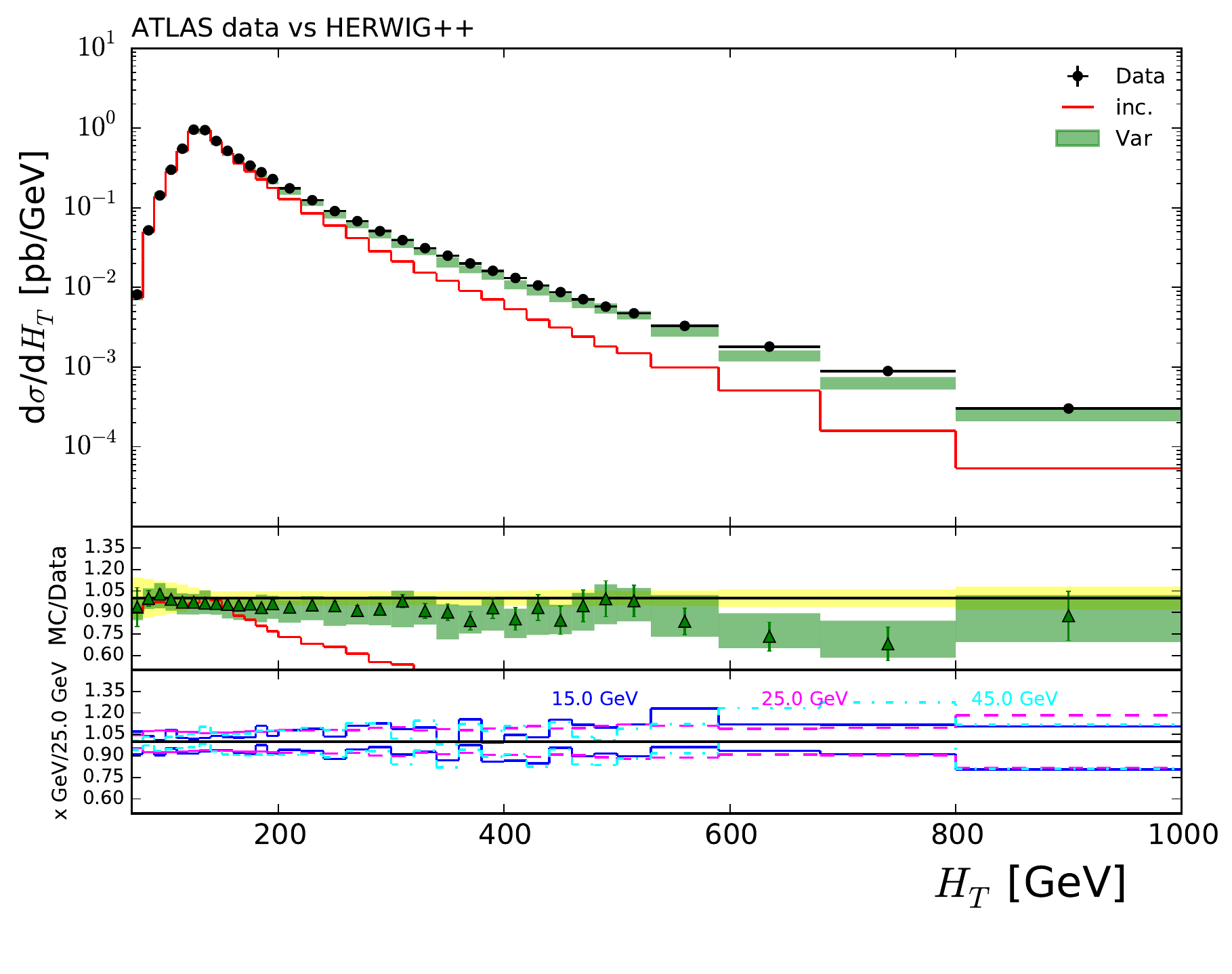}
  \includegraphics[width=0.499\linewidth]{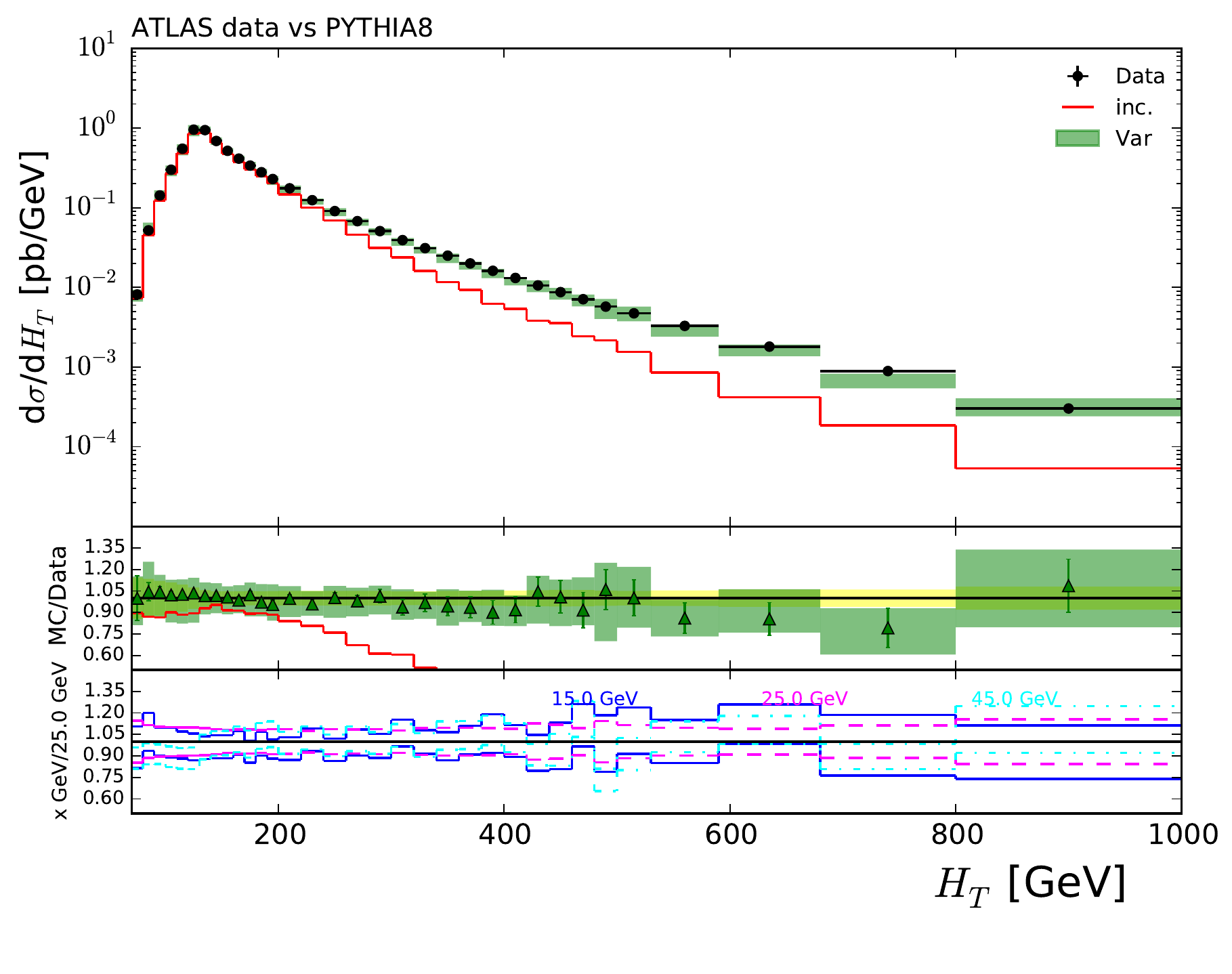}
  \caption{As in fig.~\ref{fig:Z.1304.7098:03},
 for the $H_{\sss T}$ of leptons and jets.
}
  \label{fig:Z.1304.7098:27}
\end{figure}

Each of the figures that follow is relevant to one observable
and includes two panels, with identical data but different
theoretical predictions (obtained by showering with \HWpp\ 
and with \PYe). We present, in particular:
fig.~\ref{fig:Z.1304.7098:03}: exclusive jet multiplicity;
fig.~\ref{fig:Z.1304.7098:09}: transverse momentum of the 1$^{st}$ jet;
fig.~\ref{fig:Z.1304.7098:11}: transverse momentum of the 3$^{rd}$ jet;
fig.~\ref{fig:Z.1304.7098:12}: transverse momentum of the 4$^{th}$ jet;
fig.~\ref{fig:Z.1304.7098:17}: rapidity of the 1$^{st}$ jet;
fig.~\ref{fig:Z.1304.7098:19}: rapidity of the 3$^{rd}$ jet;
fig.~\ref{fig:Z.1304.7098:20}: rapidity of the 4$^{th}$ jet;
fig.~\ref{fig:Z.1304.7098:21}: rapidity distance between the two hardest jets;
fig.~\ref{fig:Z.1304.7098:22}: invariant mass of the two hardest jets;
fig.~\ref{fig:Z.1304.7098:23}: azimuthal distance between the two hardest jets;
fig.~\ref{fig:Z.1304.7098:24}: $\Delta R$ between the two hardest jets;
fig.~\ref{fig:Z.1304.7098:27}: $H_{\sss T}$ of leptons and jets.
Among the observables measured in ref.~\cite{Aad:2013ysa}, those
chosen here constitute a subset which is sufficiently representative,
both in terms of kinematic characteristics and for the comparison
with the two different MCs used in the present paper. The hardest,
third-hardest, and fourth-hardest jets have been selected since
formally they are predicted, given the matrix elements we have
employed, at the NLO, LO, and LL accuracy respectively, and thus
they cover all possible situations as far as the nominal predictivity
of the simulations is concerned. The second-hardest jet, whose 
single-inclusive observables are not shown here,
is expected to have a similar behaviour as the leading one, which is what 
we have indeed explicitly verified.


The exclusive jet multiplicity (fig.~\ref{fig:Z.1304.7098:03})
is very well predicted by both MCs, up to $\Njet=3$. Although in a
statistically non-significant way, the central \HWpp\ prediction
slightly undershoots the data, at variance with the \PYe\ one;
this very minor difference between the two MCs is basically an 
overall effect, and can be accounted for by the total-rate results
of table~\ref{tab:totFxFx}. The lack of high-multiplicity matrix
elements starts to be visible for $\Njet\ge 4$, with \PYe\
dropping faster than \HWpp\ (whose central prediction is at the
border of the data error band up to $\Njet=7$); it must be
kept in mind that this multiplicity region is entirely dominated 
by MC effects, and formally of LL accuracy. The impact of multi-parton
matrix elements, measured by the distance between the FxFx and the
inclusive predictions, is dramatic.

The predictions for the single-jet transverse momenta of 
figs.~\ref{fig:Z.1304.7098:09}--\ref{fig:Z.1304.7098:12} tend
to be marginally softer than data, although this trend is hardly
statistically significant, except perhaps for the leading-jet
distribution in the case of \HWpp. It is worth remarking that,
shape-wise, the agreement between theory and data is rather
good even for the $4^{th}$ jet, in spite of this being beyond
matrix-element accuracy; for such a jet, the only difference
between predictions and measurements is one of rates, whose values
can be read off the $\Njet=4$ bin of fig.~\ref{fig:Z.1304.7098:03}.
As expected, the differences between merged and inclusive predictions
increase, in shape and rate, with the jet multiplicity.
The situation for the single-jet rapidities
(figs.~\ref{fig:Z.1304.7098:17}--\ref{fig:Z.1304.7098:20})
is quite analogous to that of the $\pt$'s: they are described
very well by both \HWpp\ and \PYe\ (bar the rate effect on the
$4^{th}$ jet). The almost identical FxFx predictions of the two MCs
have no analogue at the inclusive level, where \HWpp\ and \PYe\
behave differently (with only the latter in reasonable agreement
with data, and only for the leading jet); this is another clear
indication of the benefits of the merging procedure. We finally
point out that the theory-data comparison for the $2^{nd}$ jet
largely follows the same pattern as for the hardest jet,
with reduced differences in hardness for the $\pt$ distribution
w.r.t.~the latter case.

Figures~\ref{fig:Z.1304.7098:21}--\ref{fig:Z.1304.7098:24}
display four correlations, constructed with the two leading jets.
Overall, the agreement between theory and data is good,
and especially so in the case of \PYe; the only exception is
the invariant mass of the pair as predicted by \HWpp, which is
clearly softer than what is measured. Smaller discrepancies
(i.e.~within uncertainties) can be seen in the tail of the
$\Delta y$ distribution (slightly larger in the case of \HWpp), 
at small $\Delta\phi$ for \PYe, and at large $\Delta R$; note
that the statistical errors start to be non-negligible in the
tails of the $\Delta y$ and $\Delta R$ distributions.
For all of these correlations, the differences between the
merged and inclusive predictions are extremely significant.

Finally, the scalar transverse momentum sum of jets and
leptons is shown in fig.~\ref{fig:Z.1304.7098:27}. Both
MCs do well, with only a marginal tendency to be softer
than data, more marked in the case of \HWpp.

A general conclusion that can be drawn from the present
theory-data comparison is that the benefits of merging extend
beyond what one might naively expect on the basis of the
multiplicities employed at the matrix element level.
Even for observables that are entirely MC dominated,
the presence of a few hard partons in the final state allows
the MCs to stay within their natural range of validity, 
so that their underpinning approximations are not overstretched.
While some of the $\pt$-related theoretical predictions are
slightly softer than data, the trend is generally not statistically
significant; thus, there is no evidence for the necessity
of including $Z\!+\!3j$ matrix elements in the case of the observables 
studied here (bar the $4^{th}$-jet rate), since small 
differences could also be induced by e.g.~different choices of 
input parameters in the matrix elements and/or the MCs.


\vskip 2.0truecm
\noindent
$\bullet$ CMS~\cite{Chatrchyan:2013oda}
({\tt arXiv:1310.3082}, Rivet analysis {\tt CMS\_2013\_I1258128}).

\noindent
Study of rapidity distributions in $Z\!+\!1$~jet events (i.e.~exactly one jet). 
Based on an integrated luminosity of 5 fb$^{-1}$, using both $\epem$ and 
$\mpmm$ pairs, with $R=0.5$ anti-$\kt$ jets,
within $\pt(j)>30$ GeV and $\abs{\eta(j)}<2.4$. Further cuts:
$\pt(\ell)\ge 20$~GeV, $76\le M(\ell\ell)\le 106$~GeV,
$\abs{\eta(\ell)}\le 2.1$, $\pt(\ell\ell)\ge 40$~GeV,
$\Delta R(j\ell)\ge 0.5$.

\begin{figure}[!ht]
  \includegraphics[width=0.499\linewidth]{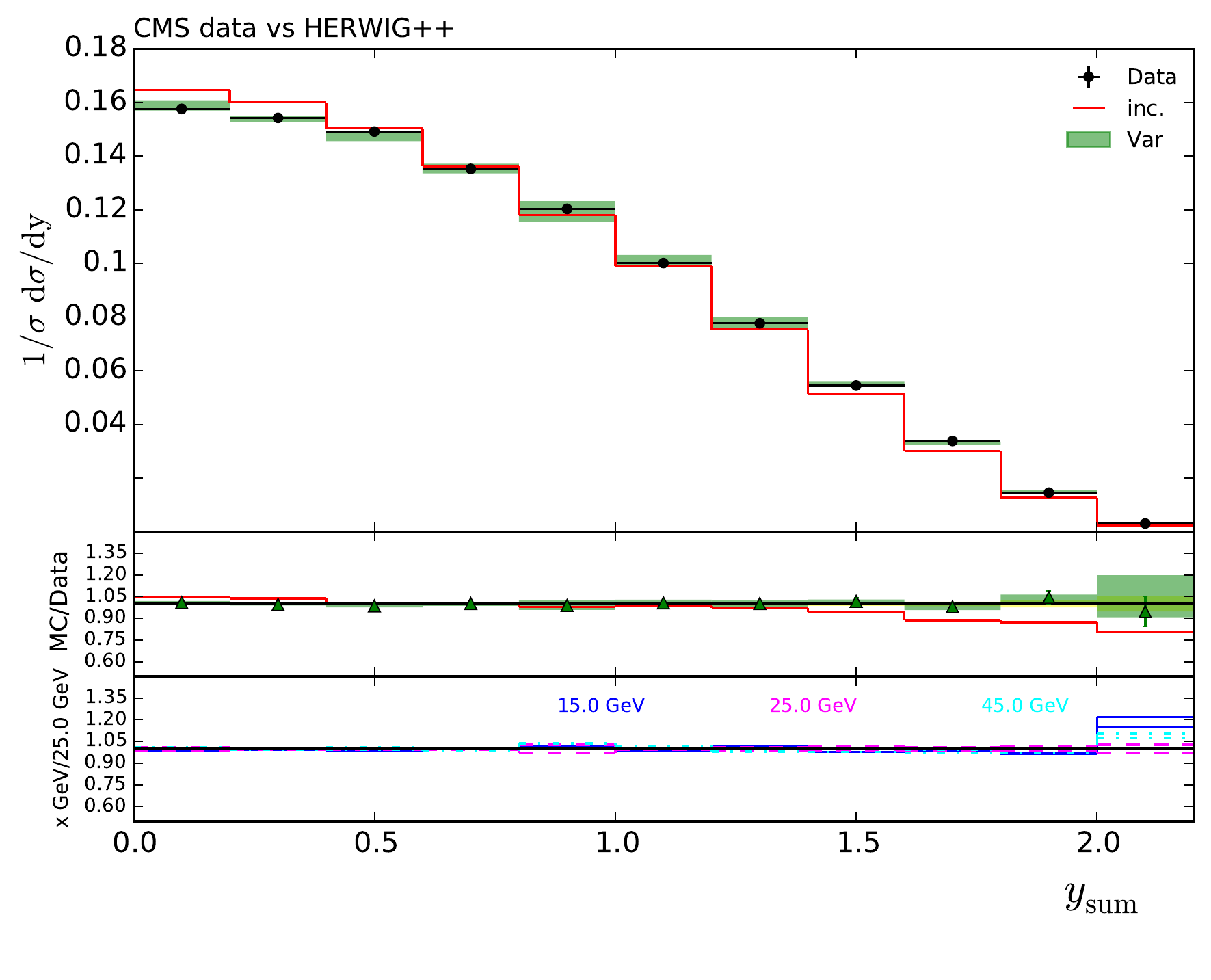}
  \includegraphics[width=0.499\linewidth]{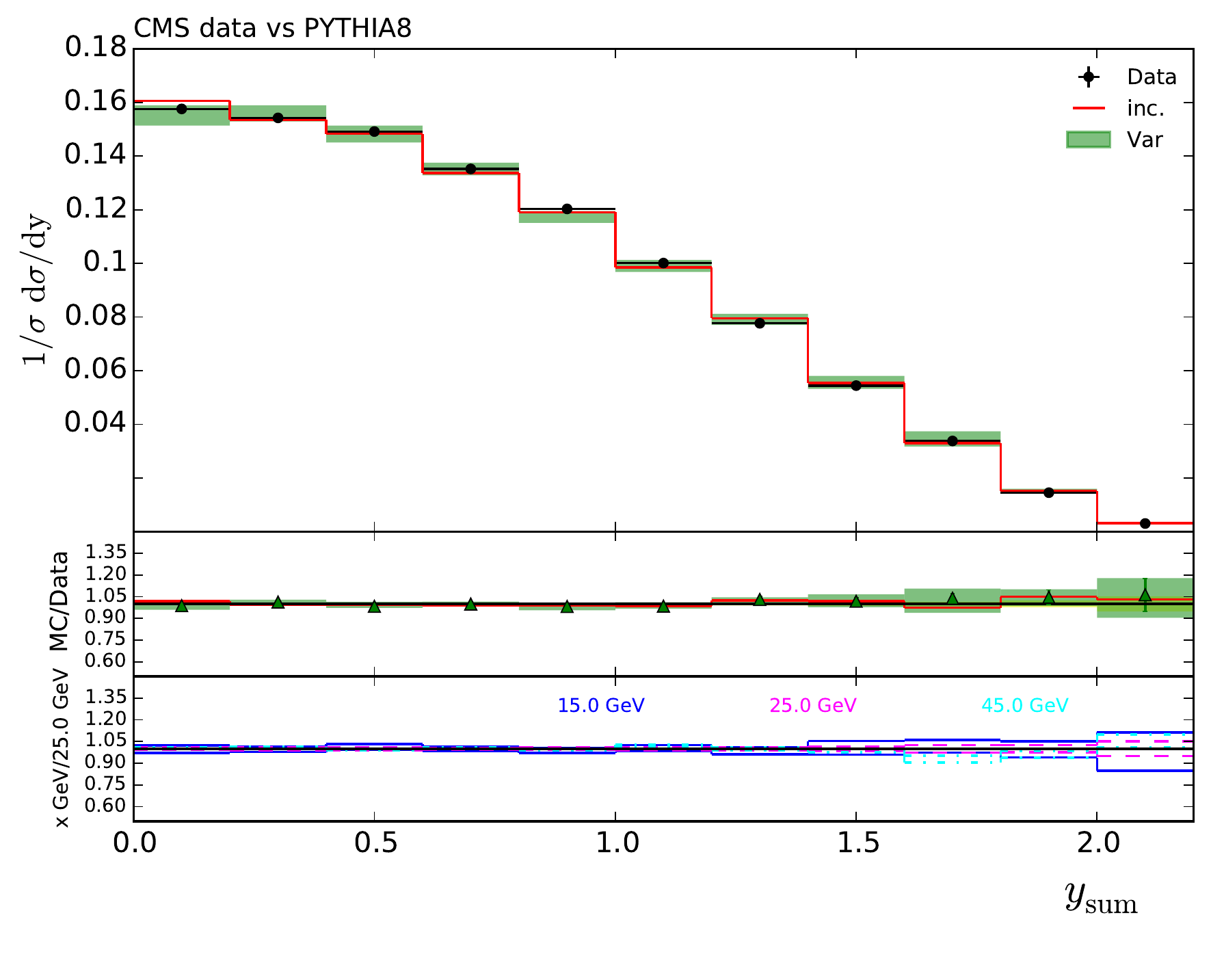}
  \caption{Sum of the rapidities of the $Z$ and the 1$^{st}$ jet.
 Data from ref.~\cite{Chatrchyan:2013oda}, compared to \HWpp\ (left panel) 
 and \PYe\ (right panel) predictions.
 The FxFx uncertainty envelope 
 (``Var'') and the fully-inclusive central result (``inc'') are shown
 as green bands and red histograms respectively. See the end of
 sect.~\ref{sec:tech} for more details on the layout of the plots.
}
  \label{fig:Z.1310.3082:03}
\end{figure}

\begin{figure}[!ht]
  \includegraphics[width=0.499\linewidth]{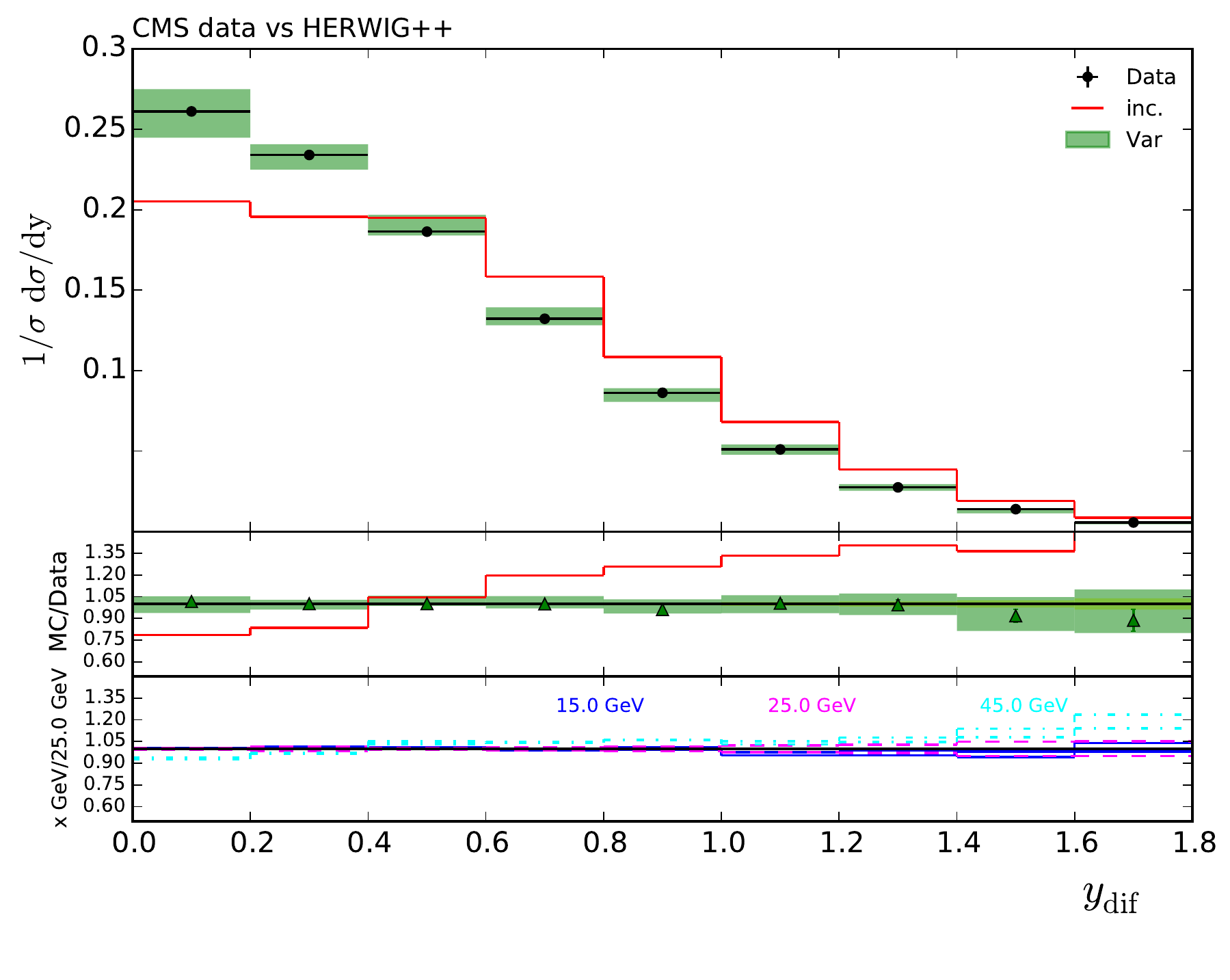}
  \includegraphics[width=0.499\linewidth]{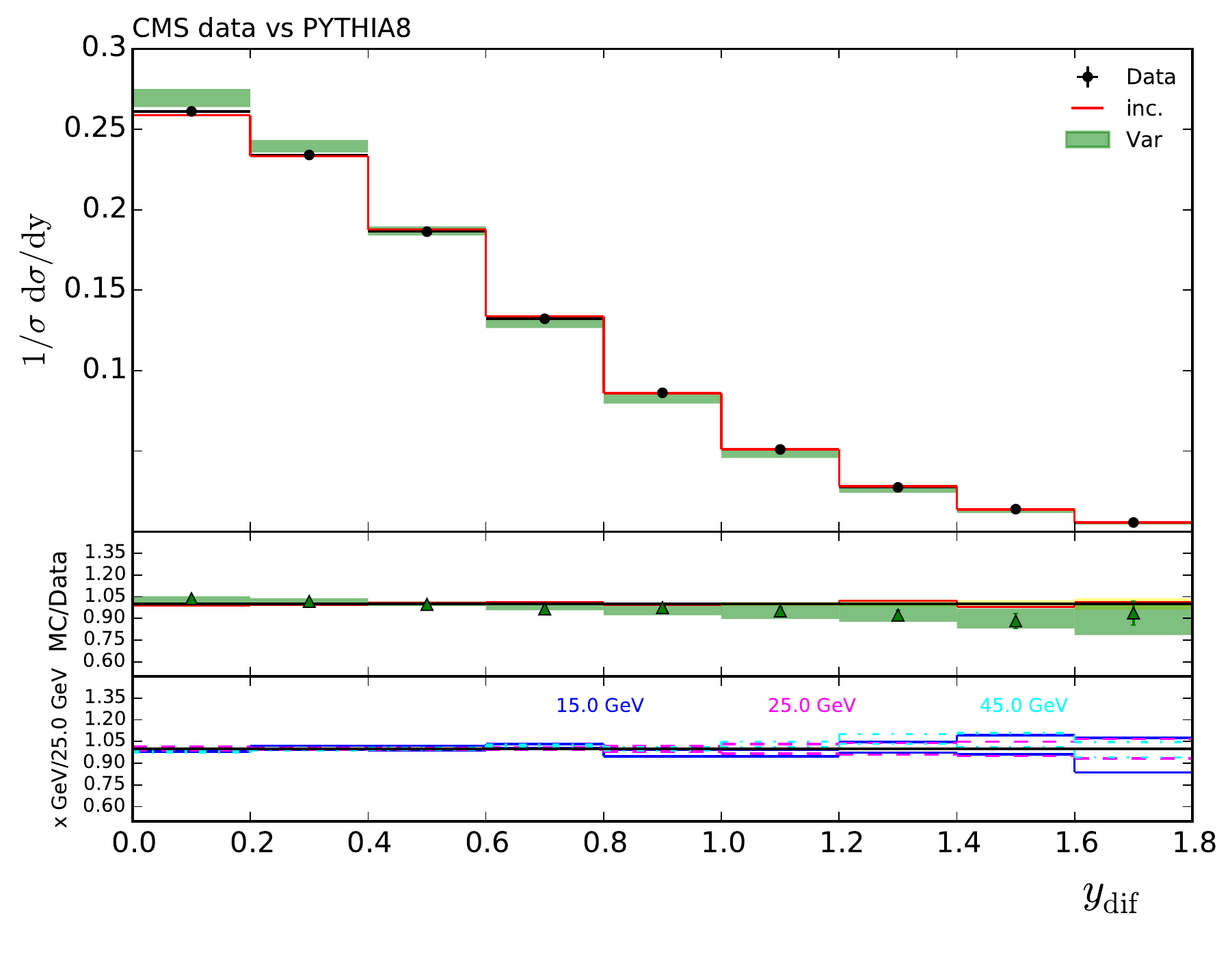}
  \caption{As in fig.~\ref{fig:Z.1310.3082:03}, for the 
 difference of the rapidities of the $Z$ and the 1$^{st}$ jet.
}
  \label{fig:Z.1310.3082:04}
\end{figure}

We present here two observables: in fig.~\ref{fig:Z.1310.3082:03} 
and fig.~\ref{fig:Z.1310.3082:04} the sum and the difference, 
respectively, of the rapidities of the $Z$ and of the jet; these 
we have chosen for being the most involved cases among the measurements
in ref.~\cite{Chatrchyan:2013oda}, and because their comparison
with the theoretical LO+PS predictions considered by CMS 
was not entirely satisfactory.

As one can see from the figures, the agreement between merged predictions
and data is excellent for both MCs. This result appears to be strongly
driven by matrix-element effects, given the very significant differences
between the \HWpp\ and \PYe\ predictions which result from the inclusive
samples. This is especially true in the case of the rapidity difference,
which in inclusive simulations is known to be affected by large MC
systematics -- for a detailed discussion on this point, see
refs.~\cite{Torrielli:2010aw,Frederix:2012ps}.
We point out that we have found a level of agreement identical 
to that of figs.~\ref{fig:Z.1310.3082:03} and~\ref{fig:Z.1310.3082:04}
also in the case of the single-inclusive rapidities (of the $Z$ and
the jet) measured in ref.~\cite{Chatrchyan:2013oda}.
\enlargethispage*{50pt}


\vskip 0.4truecm
\noindent
$\bullet$ CMS~\cite{Chatrchyan:2013tna}
({\tt arXiv:1301.1646}, Rivet analysis {\tt CMS\_2013\_I1209721}).

\noindent
Study of event shapes and azimuthal correlations.
Based on an integrated luminosity of 5 fb$^{-1}$,
using both $\epem$ and $\mpmm$ pairs, with $R=0.5$ 
anti-$\kt$ jets within $\pt(j)>50$ GeV
and $\abs{\eta(j)}<2.5$. Further cuts:
$\pt(\ell)\ge 20$~GeV, $71\le M(\ell\ell)\le 111$~GeV,
$\abs{\eta(\ell)}\le 2.4$, $\Delta R(j\ell)\ge 0.4$.
We point out that, owing to the hardness of the jets defined
by CMS in this analysis (which is significantly larger than
that relevant to refs.~\cite{Aad:2013ysa,Chatrchyan:2013oda}),
and given the generation cuts we have chosen (see the beginning
of sect.~\ref{sec:res}), our simulations have statistical accuracies
relatively lower than those accumulated for the other measurements
considered in this paper. For this reason, for certain observables
we have limited ourselves to presenting only \HWpp\ results, after 
having checked the statistical compatibility of these with their \PYe\ 
counterparts, which have slightly larger fluctuations.

\begin{figure}[!ht]
  \includegraphics[width=0.499\linewidth]{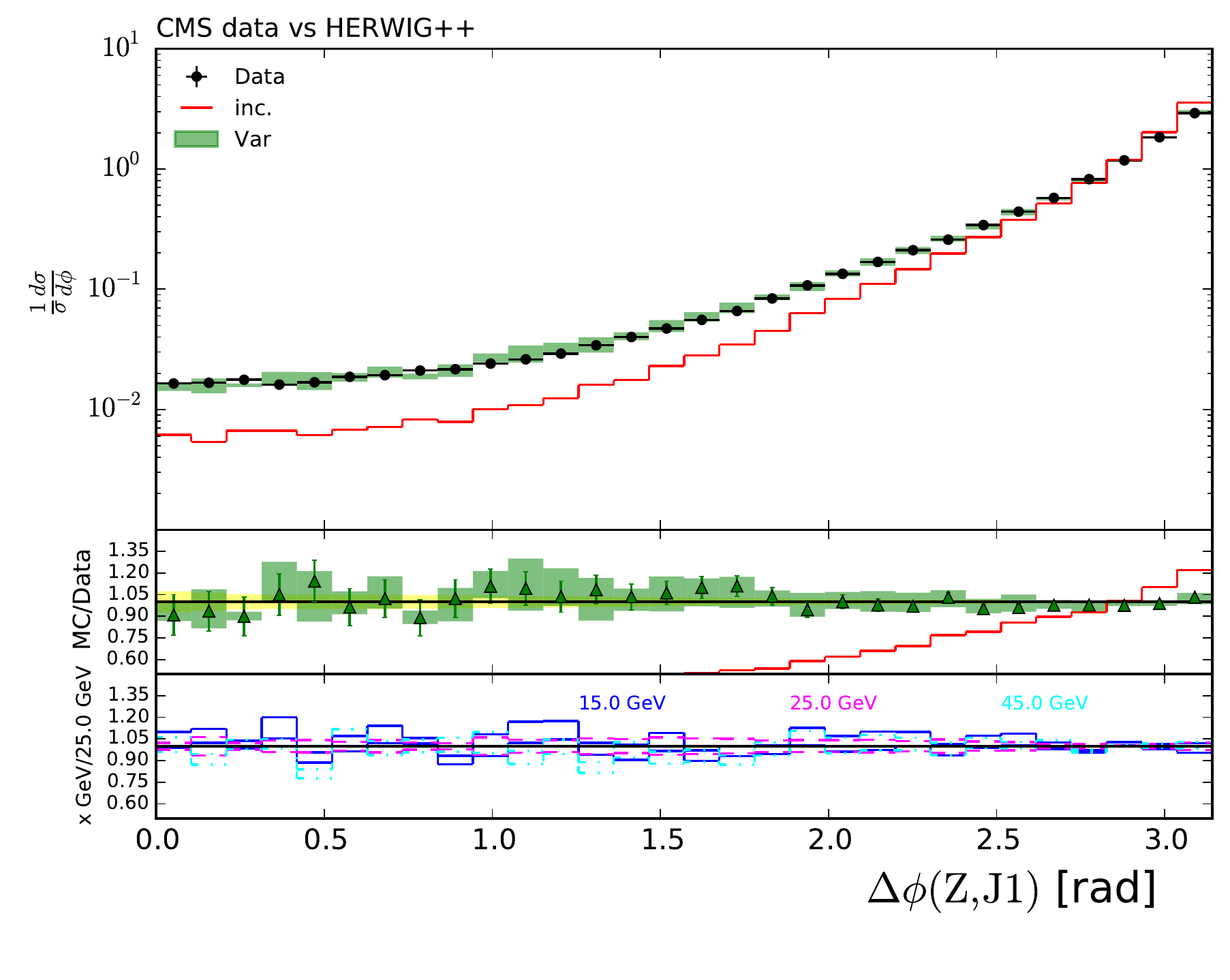}
  \includegraphics[width=0.499\linewidth]{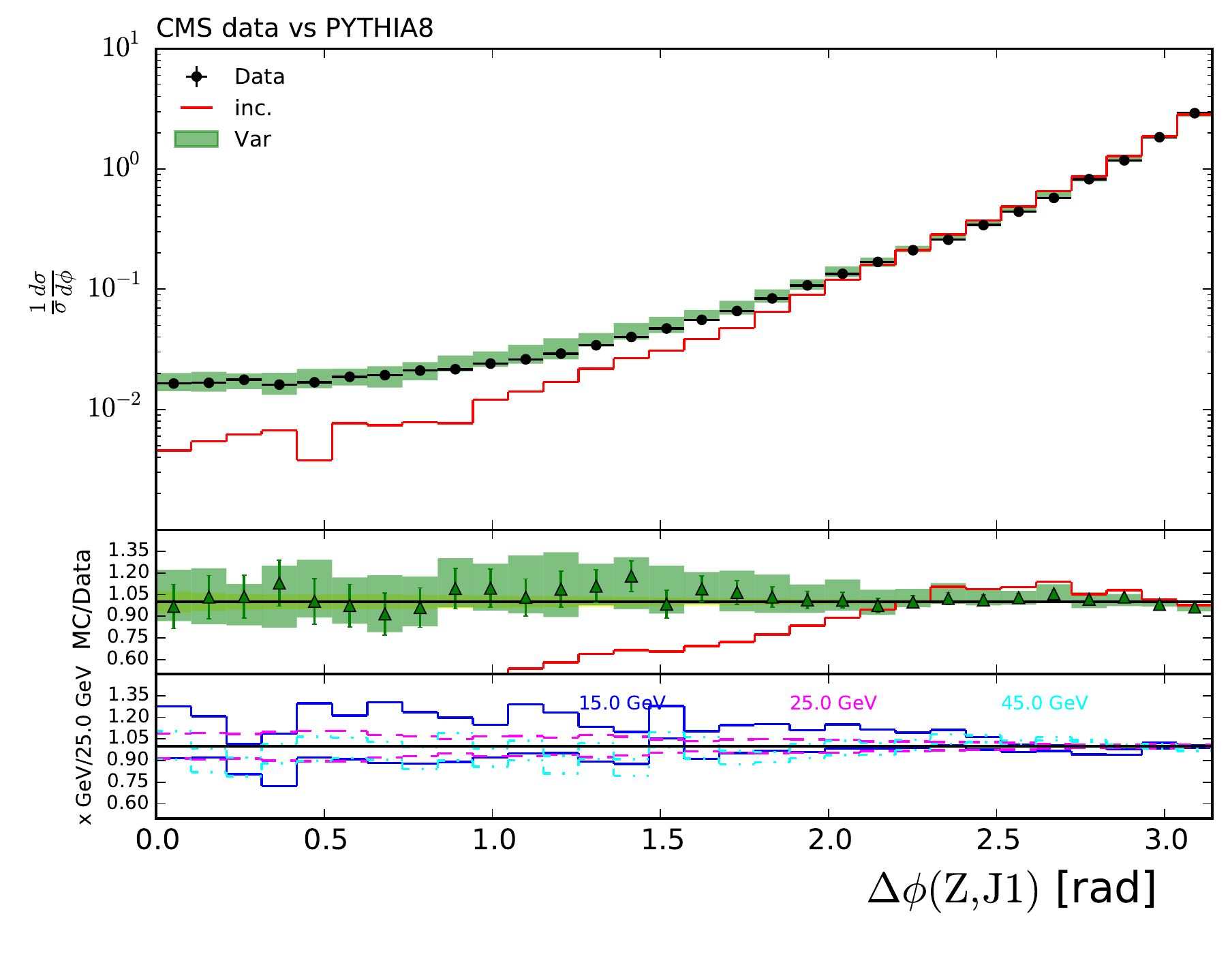}
  \caption{Azimuthal distance between the $Z$ and the 1$^{st}$ jet.
 Data from ref.~\cite{Chatrchyan:2013tna}, compared to \HWpp\ (left panel) 
 and \PYe\ (right panel) predictions.
 The FxFx uncertainty envelope 
 (``Var'') and the fully-inclusive central result (``inc'') are shown
 as green bands and red histograms respectively. See the end of
 sect.~\ref{sec:tech} for more details on the layout of the plots.
}
  \label{fig:Z.1301.1646:01}
\end{figure}

\begin{figure}[!ht]
  \includegraphics[width=0.499\linewidth]{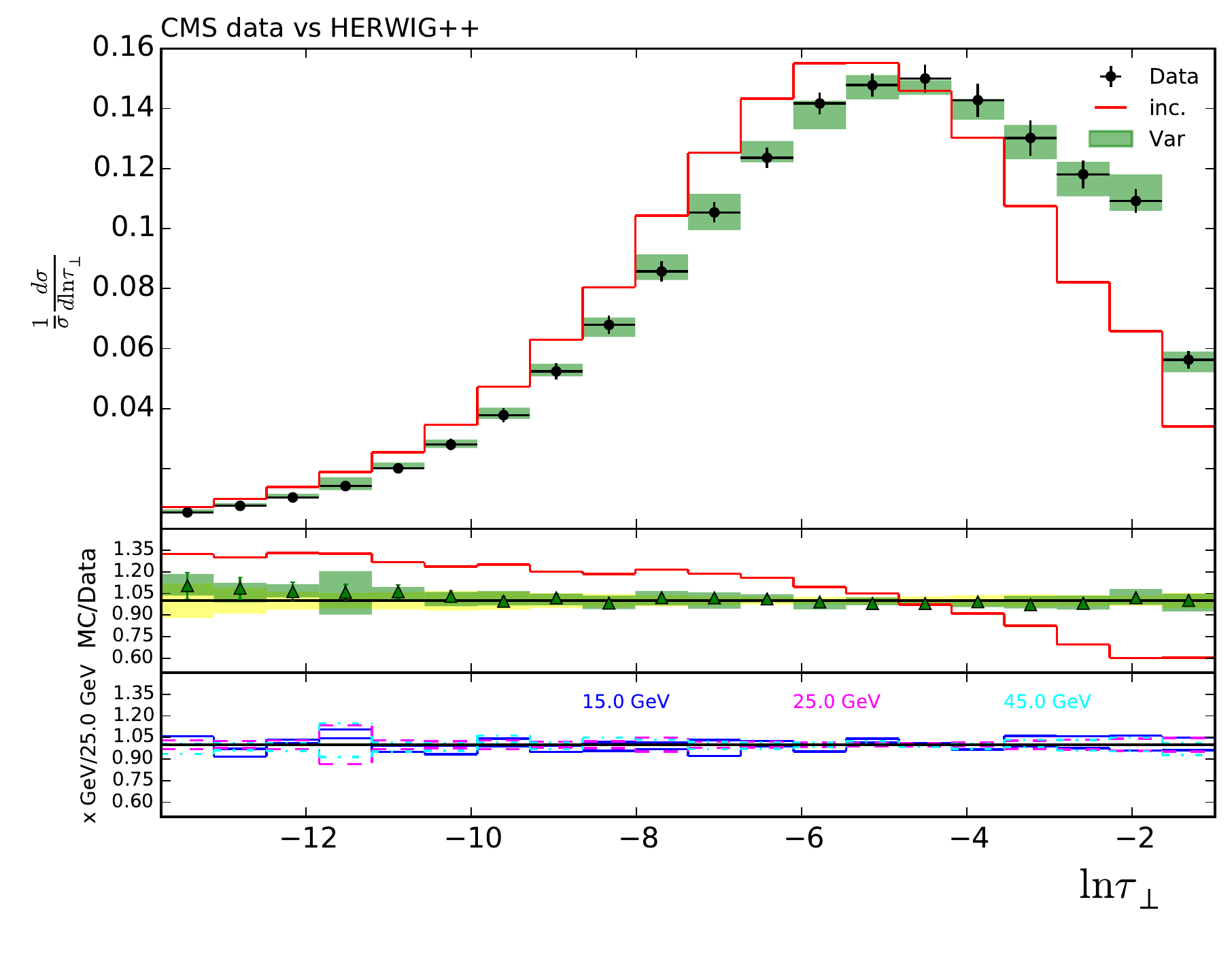}
  \includegraphics[width=0.499\linewidth]{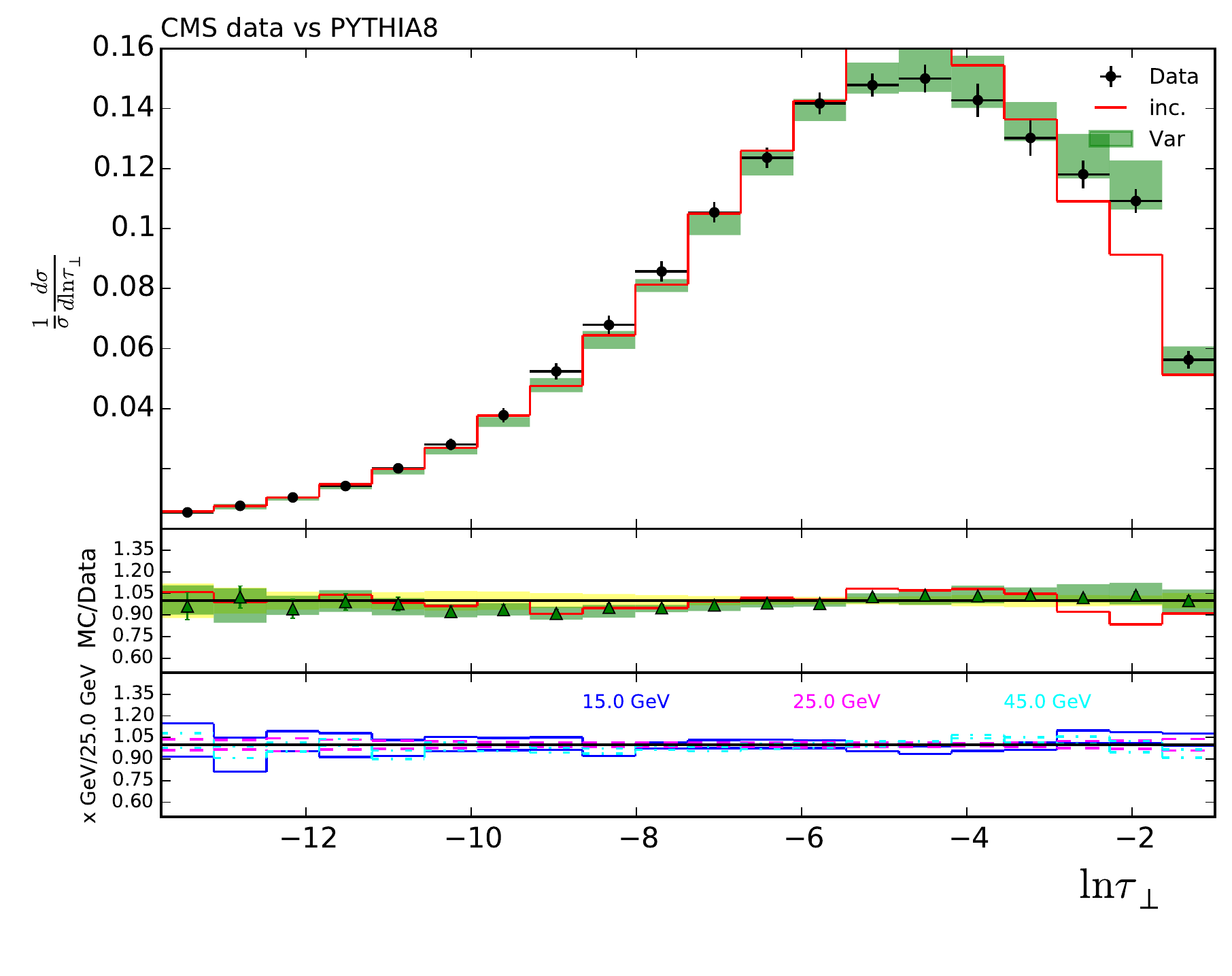}
  \caption{As in fig.~\ref{fig:Z.1301.1646:01}, for the 
 transverse thrust.
}
  \label{fig:Z.1301.1646:09}
\end{figure}

\begin{figure}[!ht]
  \includegraphics[width=0.499\linewidth]{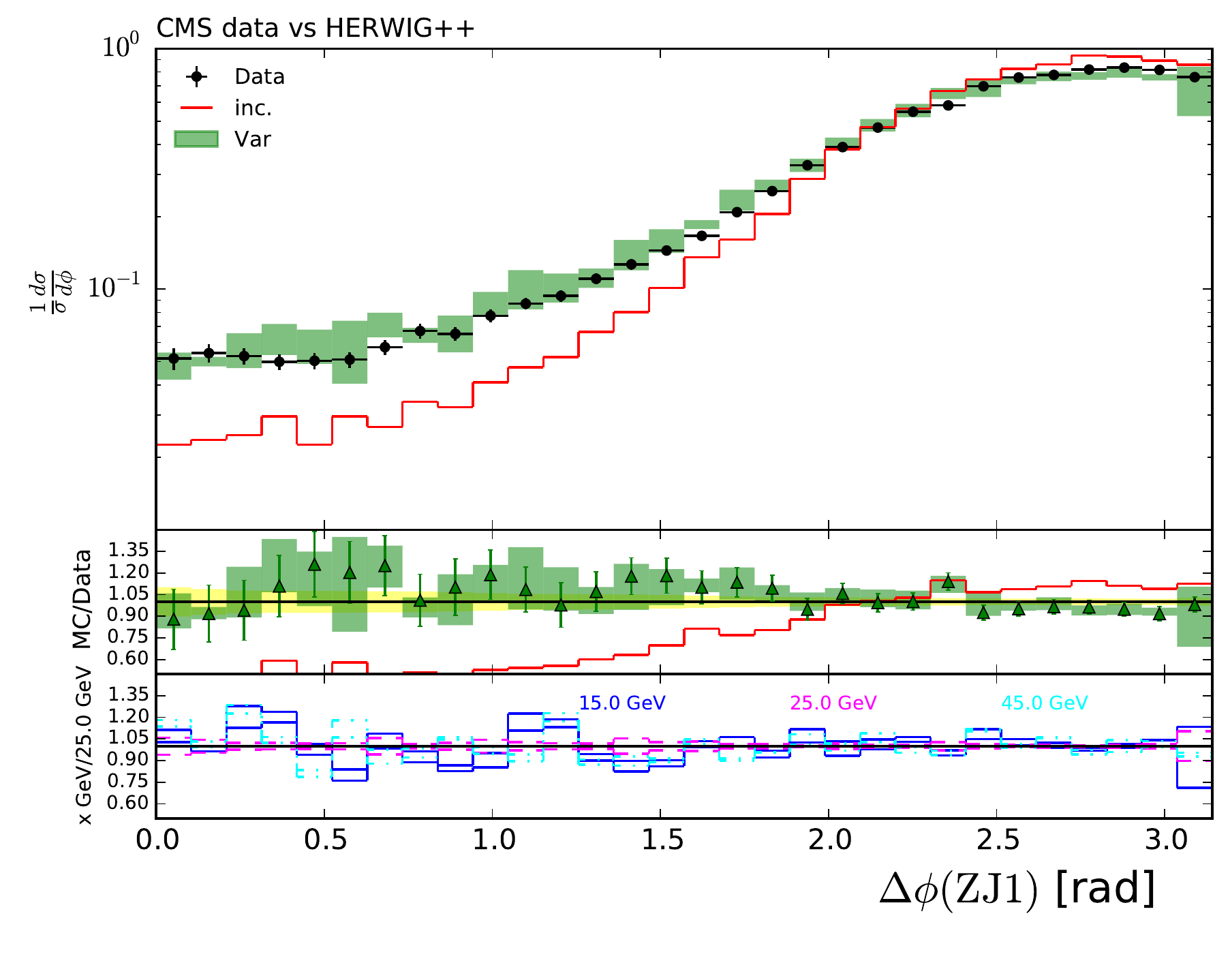}
  \includegraphics[width=0.499\linewidth]{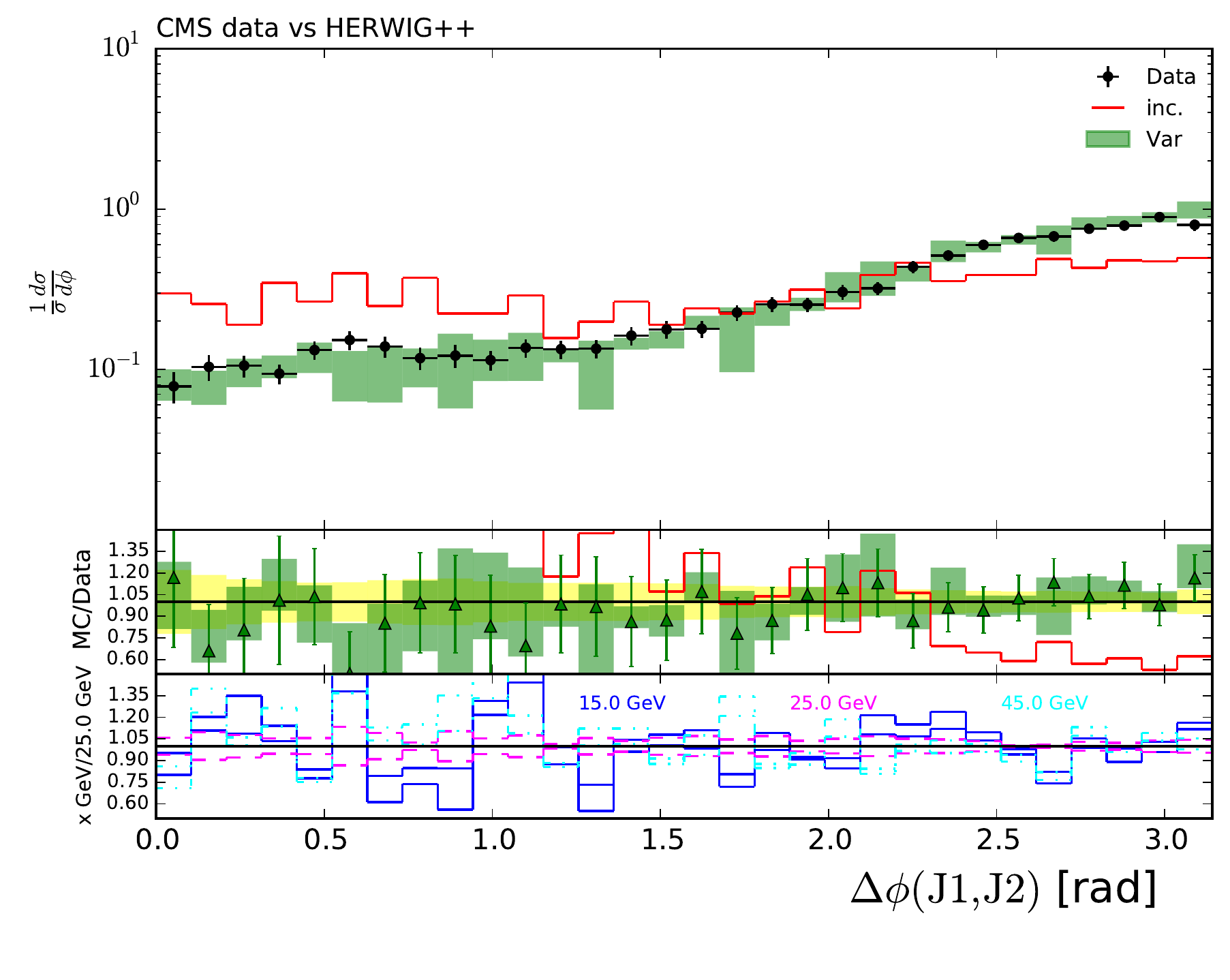}
  \caption{As in fig.~\ref{fig:Z.1301.1646:01}, for the 
 azimuthal distance between the $Z$ and the 1$^{st}$ jet,
 in events with at least two jets (left panel), and the
 azimuthal distance between the 1$^{st}$ and the 2$^{nd}$ jet,
 in events with at least three jets (right panel), both
 compared to \HWpp\ predictions.
}
  \label{fig:Z.1301.1646:02+06}
\end{figure}

\begin{figure}[!ht]
  \includegraphics[width=0.499\linewidth]{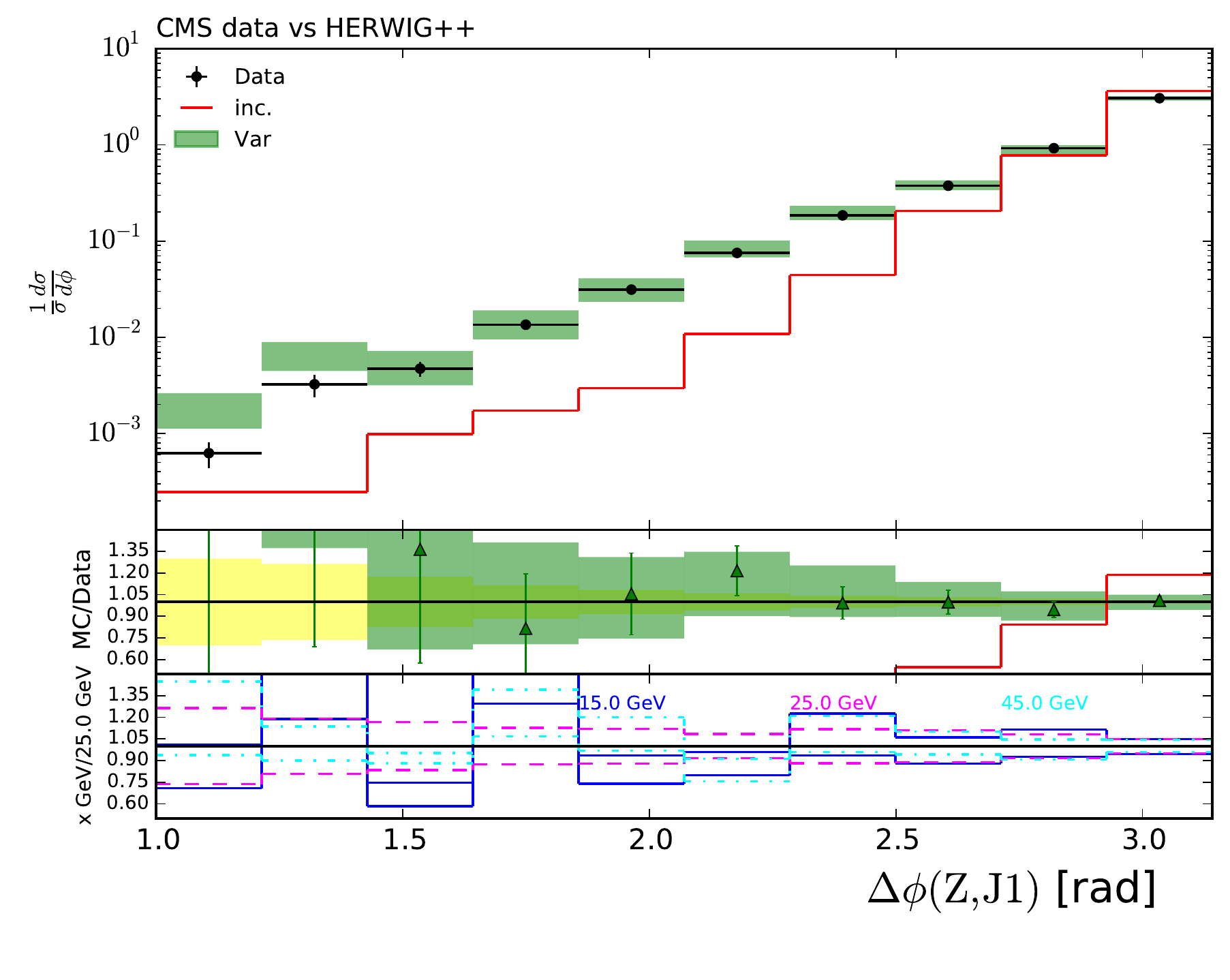}
  \includegraphics[width=0.499\linewidth]{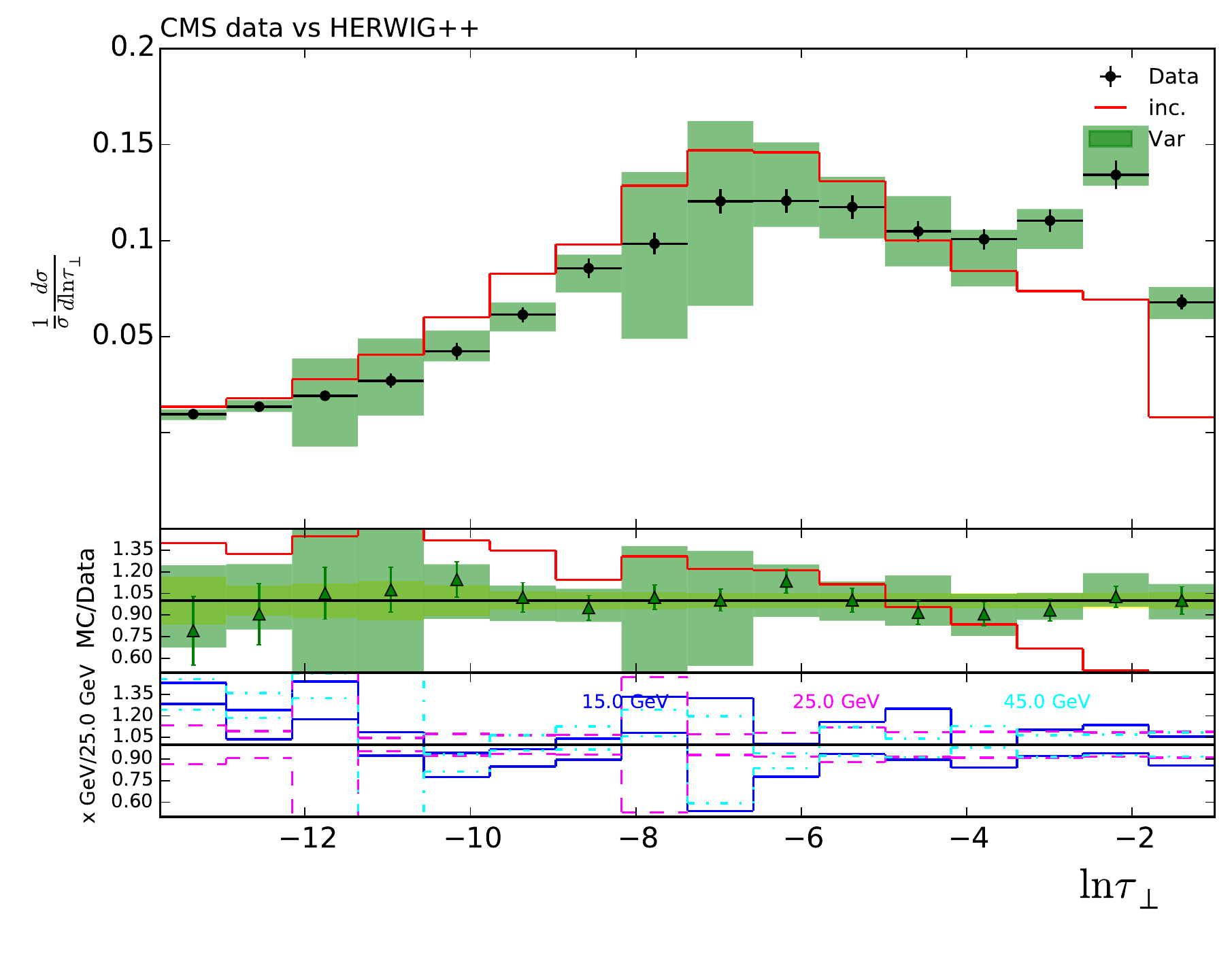}
  \caption{As in fig.~\ref{fig:Z.1301.1646:02+06}, for the 
 azimuthal distance between the $Z$ and the 1$^{st}$ jet
 (left panel), and the transverse thrust (right panel), both 
 in events with $\pt(Z)\ge 150$~GeV. 
}
  \label{fig:Z.1301.1646:10+18}
\end{figure}

We have selected the following observables.
Figure~\ref{fig:Z.1301.1646:01}: azimuthal distance between the $Z$ 
and the 1$^{st}$ jet;
fig.~\ref{fig:Z.1301.1646:09}: transverse thrust;
fig.~\ref{fig:Z.1301.1646:02+06}: azimuthal distance between the $Z$ 
and the 1$^{st}$ jet, in events with at least two jets, and azimuthal 
distance between the 1$^{st}$ and the 2$^{nd}$ jet, in events with 
at least three jets;
fig.~\ref{fig:Z.1301.1646:10+18}: azimuthal distance between the $Z$ 
and the 1$^{st}$ jet, and transverse thrust, both in events with 
$\pt(Z)\ge 150$~GeV.


Two $\Njet\ge 1$ observables are presented in fig.~\ref{fig:Z.1301.1646:01}
(distance between the $Z$ and the leading jet in the azimuthal angle)
and in fig.~\ref{fig:Z.1301.1646:09} (transverse thrust), and compared
to both \HWpp\ and \PYe. When considered in their full ranges, these 
observables are very sensitive to the interplay between matrix elements, 
parton showers, and possibly soft physics modelling\footnote{Multi-parton 
scatterings are conjectured to have a non-negligible impact on observables 
such as the transverse thrust~\cite{Gaunt:2014ska}.}, and thus ultimately 
to the merging procedure. When increasing either the minimal jet multiplicity, 
or the $Z$ transverse momentum, final-state objects are in fact less 
correlated (see ref.~\cite{Chatrchyan:2013tna}). The agreement between 
the measurements and the FxFx theoretical predictions is excellent for 
both MCs, indicating that the cross-talk between merging and soft-physics 
modelling does not deteriorate the overall quality of 
event generation. On average, the merging systematics is larger
with \PYe\ than with \HWpp. Inspection of the lower insets shows that this
is basically due to the behaviour of the $\mu_Q=15$~GeV \PYe\ sample, which 
has the largest fluctuations; however, from the upper insets one sees that 
the central predictions are quite consistent between the two MCs, and 
almost on top of the data in the whole ranges.

Two azimuthal distances ($\Delta\phi(Z,j_1)$ with $\Njet\ge 2$, and
$\Delta\phi(j_1,j_2)$ with $\Njet\ge 3$) are shown in
fig.~\ref{fig:Z.1301.1646:02+06}, together with the corresponding
\HWpp\ results. As expected, these are flatter than those of
fig.~\ref{fig:Z.1301.1646:01}, i.e.~more de-correlated. However,
a certain amount of matrix-element-induced effect is still
present, as one can infer by comparing the FxFx with the inclusive results, 
with the latter predicting less events in the $\Delta\phi(Z,j_1)\to 0$
tail and a flatter $\Delta\phi(j_1,j_2)$ distribution. The agreement
of the theoretical predictions with data is again quite good, albeit 
in presence of larger systematics.

Finally, the same observables as in figs.~\ref{fig:Z.1301.1646:01}
and~\ref{fig:Z.1301.1646:09}, but subject to a $\pt(Z)\ge 150$~GeV cut, 
are compared to \HWpp\ predictions in fig.~\ref{fig:Z.1301.1646:10+18}.
The only discrepancy with data is seen in the two leftmost bins of
the azimuthal correlation, which are however affected by extremely
large statistical errors (note, furthermore, that the bin with the 
smallest $\Delta\phi(Z,j_1)$ value is not shown in fig.~4 of 
ref.~\cite{Chatrchyan:2013tna}). Observe that, in the case of the transverse
thrust, the central FxFx result (green triangles) is in good agreement 
with data, and affected by a relatively small scale-and-PDF uncertainty
(magenta dashed band); the large overall systematics is again dominated
by the behaviour of the $\mu_Q=15$~GeV sample. We point out that
in the context of the present analysis, which is characterised
by very hard jets, such a choice for the merging scale is extreme,
and might in fact be safely neglected.

As was the case for the two $\Zjs$ analyses discussed previously,
the differences between the FxFx and the fully-inclusive results
are always very significant. The latter also exhibit a strong
MC dependence, which is essentially absent in the former.

\clearpage

\subsection{$\Wjs$\label{sec:resW}}
This section is devoted to the comparison between data and 
theoretical predictions relevant to $\Wjs$ production.
We shall make use of the ATLAS measurements of ref.~\cite{Aad:2014qxa},
and of the CMS measurements of ref.~\cite{Khachatryan:2014uva}. 
For each of these, a (non-exhaustive) summary of the characteristics 
of the analysis is given; however, we encourage the reader to check 
the original experimental papers for more details.

\vskip 0.4truecm
\noindent
$\bullet$ ATLAS~\cite{Aad:2014qxa}
({\tt arXiv:1409.8639}, Rivet analysis {\tt ATLAS\_2014\_I1319490}).

\noindent
Study of jet and inclusive properties (the latter defined 
by requiring the presence of at least one jet in the final state), 
and of jet-jet correlations. Based on an integrated luminosity
of 4.6 fb$^{-1}$, using both the electron and muon channels,
with $R=0.4$ anti-$\kt$ jets within $\pt(j)>30$ GeV and $\abs{y(j)}<4.4$. 
Further cuts: $\pt(\ell)\ge 25$~GeV, $\abs{\eta(\mu)}\le 2.4$,
$\abs{\eta(e)}\le 1.37$ and $1.52\le\abs{\eta(e)}\le 2.47$,
$\Etmiss>25$~GeV, $\mt(\ell\nu)>40$~GeV (see ref.~\cite{Aad:2014qxa} for 
the definition of the missing energy and the neutrino transverse momentum).
Events are rejected if a second electron or muon that passes
the cuts is present. Note that ATLAS treat $W\to\tau\nu$ events as 
background, and consistently with that we have not generated this
channel.

\begin{figure}[!ht]
  \includegraphics[width=0.499\linewidth]{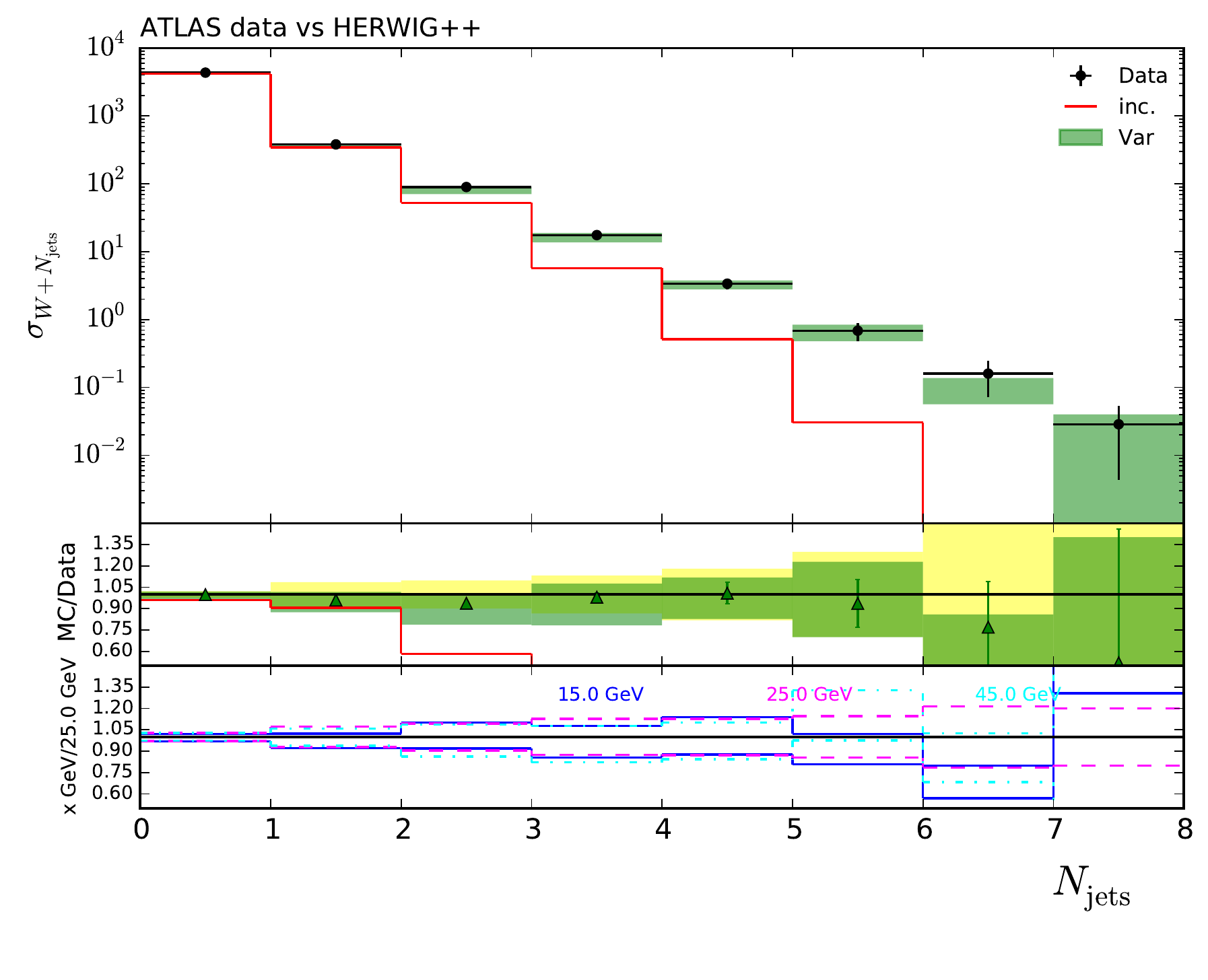}
  \includegraphics[width=0.499\linewidth]{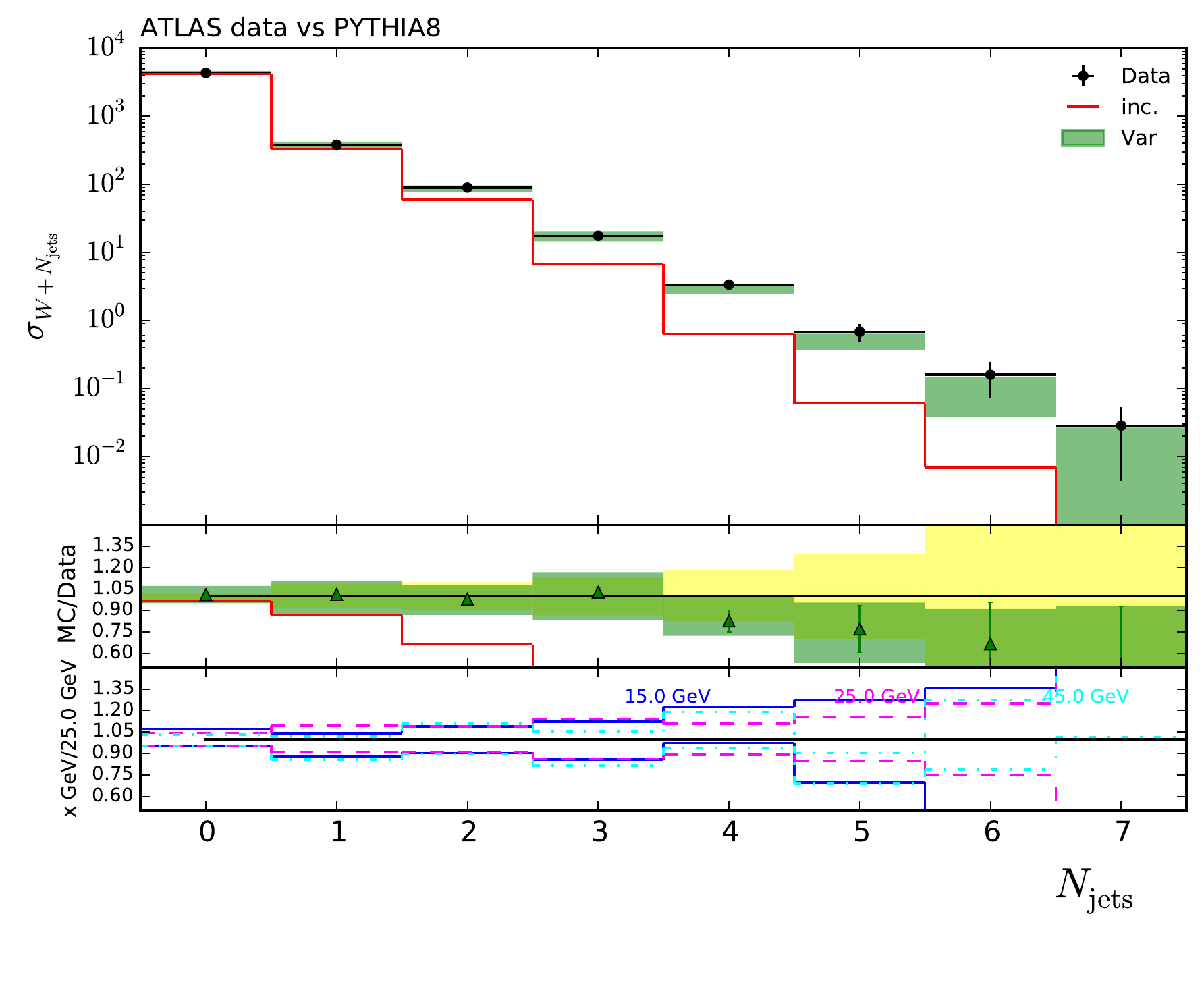}
  \caption{Exclusive jet multiplicity. 
 Data from ref.~\cite{Aad:2014qxa}, compared to \HWpp\ (left panel) 
 and \PYe\ (right panel) predictions.
 The FxFx uncertainty envelope 
 (``Var'') and the fully-inclusive central result (``inc'') are shown
 as green bands and red histograms respectively. See the end of
 sect.~\ref{sec:tech} for more details on the layout of the plots.
}
  \label{fig:W.1409.8639:04}
\end{figure}

\begin{figure}[!ht]
  \includegraphics[width=0.499\linewidth]{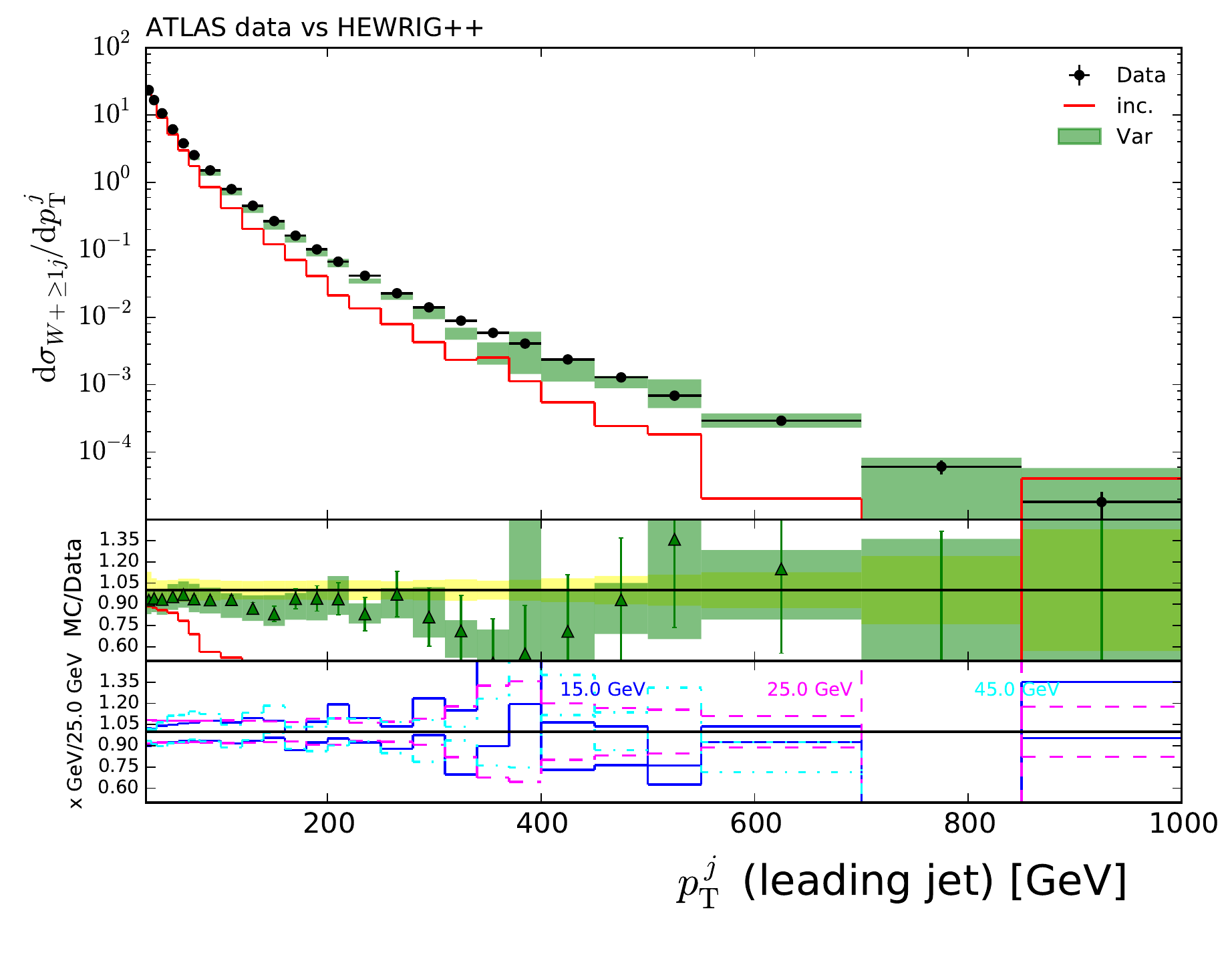}
  \includegraphics[width=0.499\linewidth]{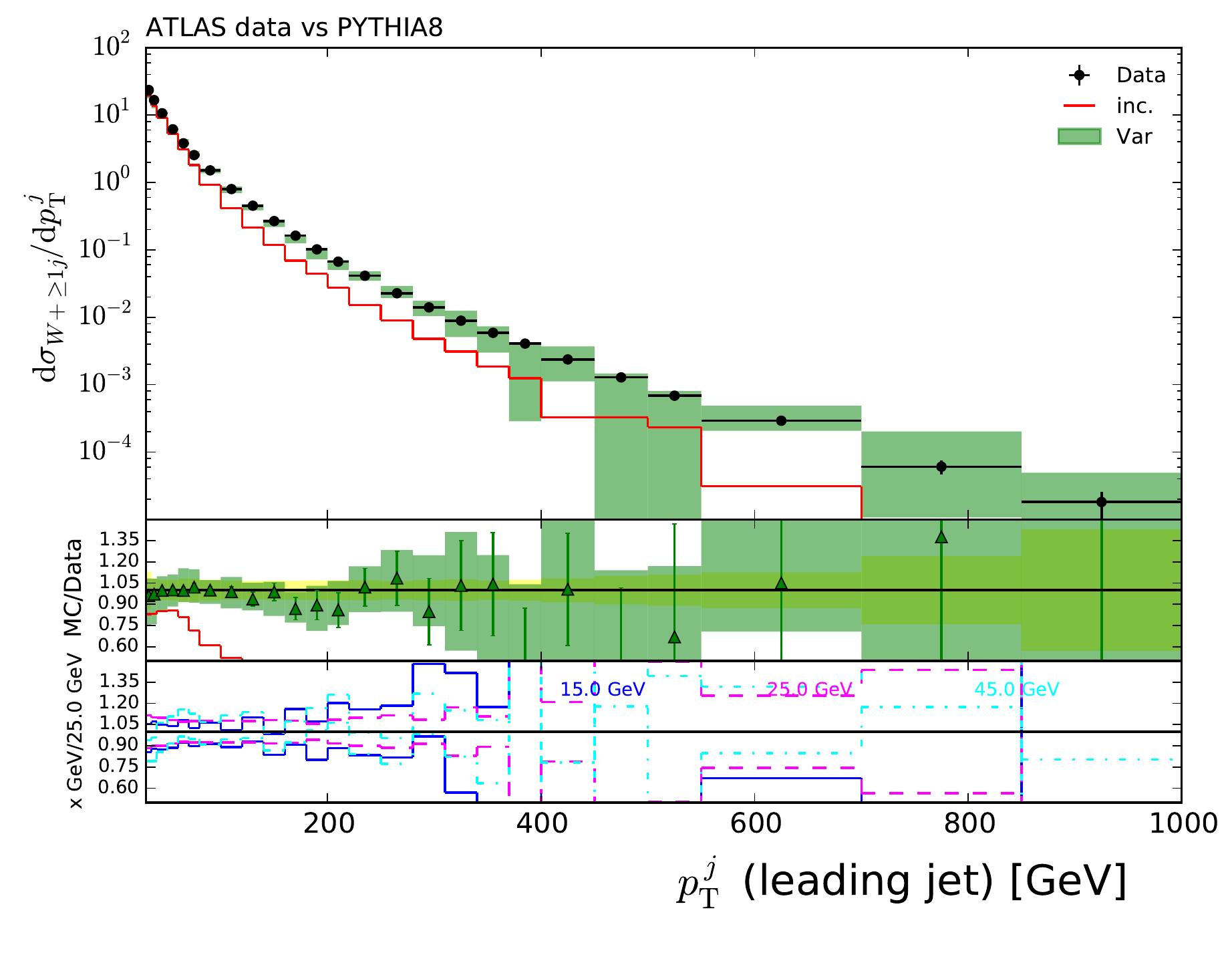}
  \caption{As in fig.~\ref{fig:W.1409.8639:04}, for the 
 transverse momentum of the 1$^{st}$ jet.
}
  \label{fig:W.1409.8639:05}
\end{figure}

\begin{figure}[!ht]
  \includegraphics[width=0.499\linewidth]{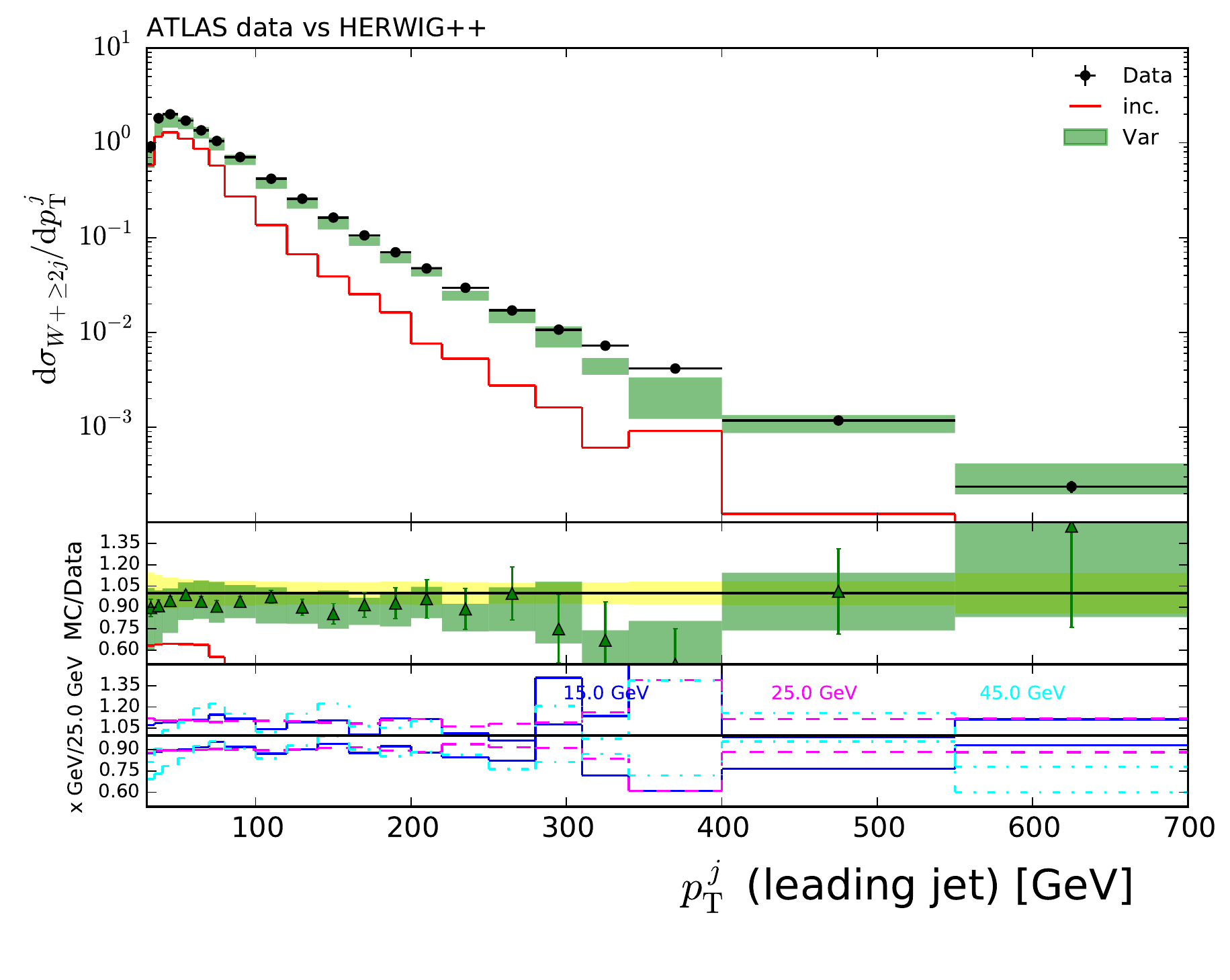}
  \includegraphics[width=0.499\linewidth]{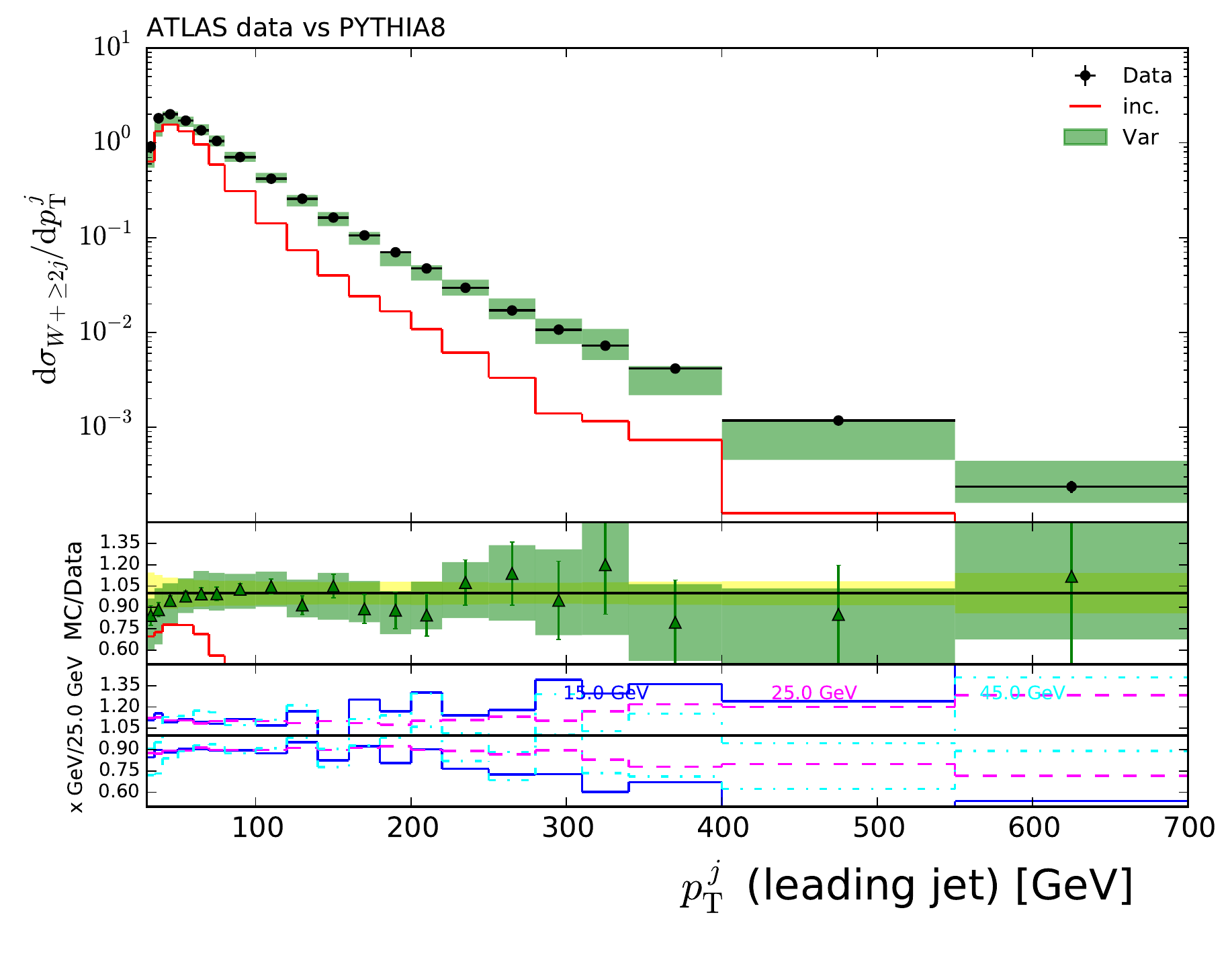}
  \caption{As in fig.~\ref{fig:W.1409.8639:04}, for the 
 transverse momentum of the 1$^{st}$ jet, in events
 with at least two jets.
}
  \label{fig:W.1409.8639:07}
\end{figure}

\begin{figure}[!ht]
  \includegraphics[width=0.499\linewidth]{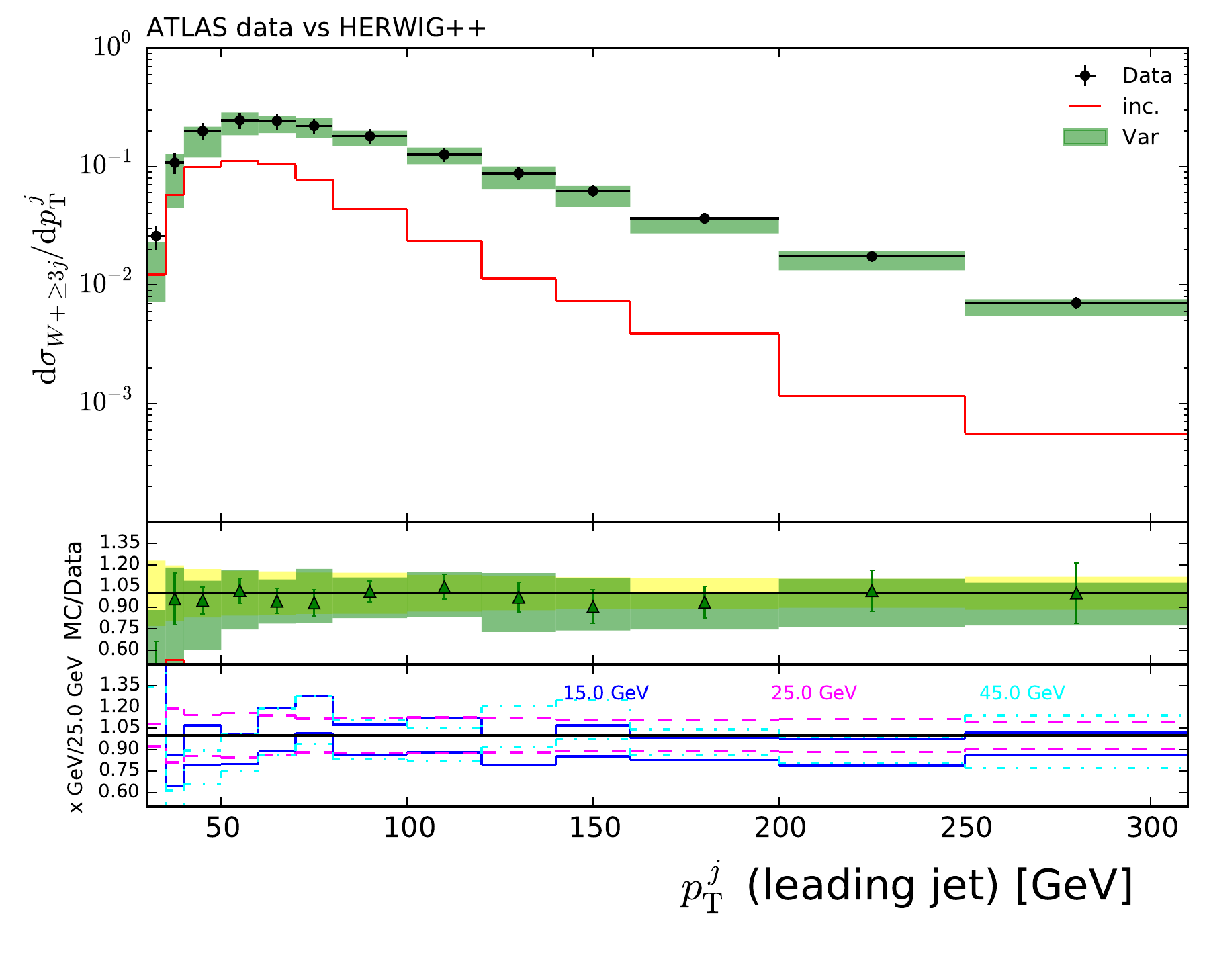}
  \includegraphics[width=0.499\linewidth]{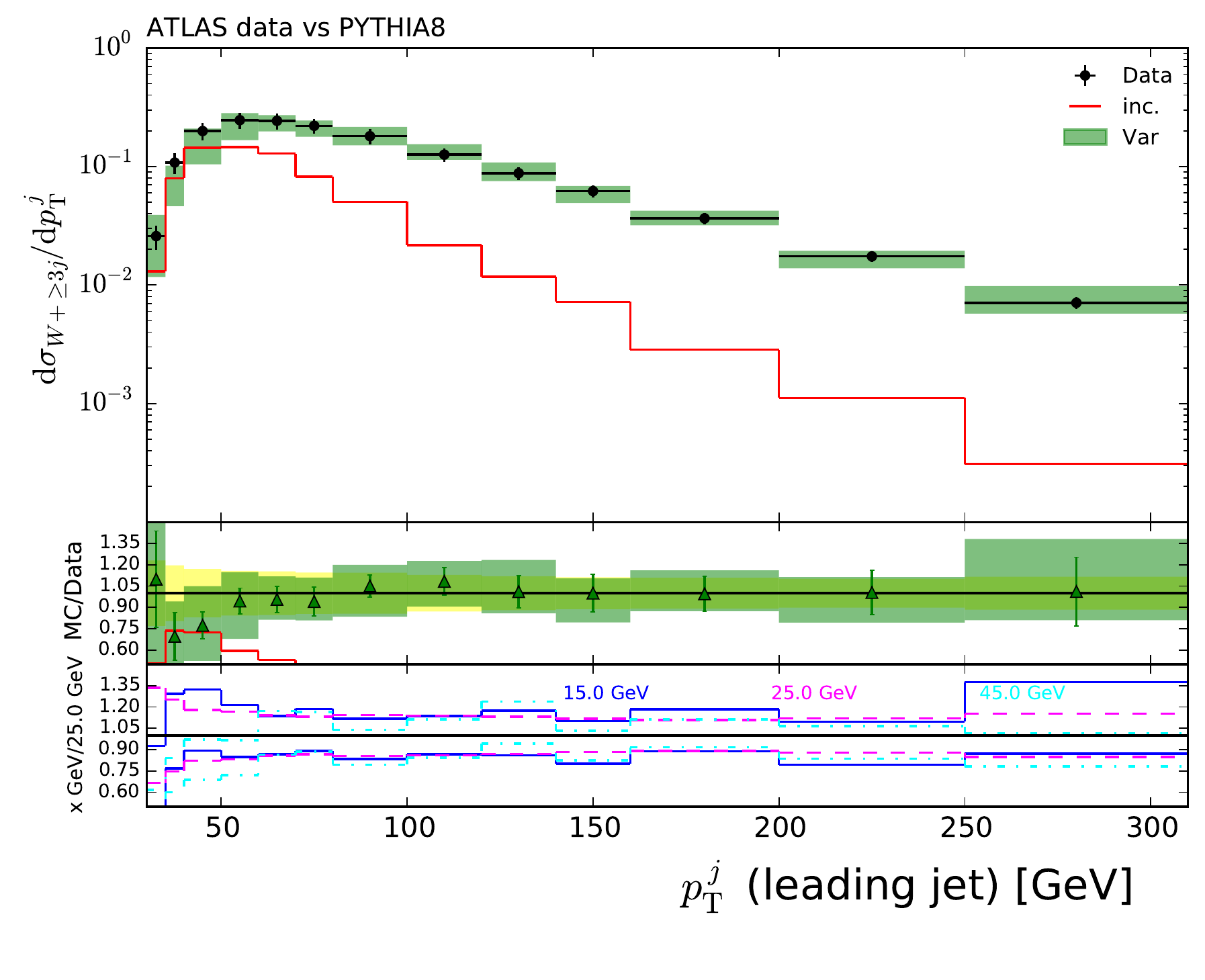}
  \caption{As in fig.~\ref{fig:W.1409.8639:04}, for the
 transverse momentum of the 1$^{st}$ jet, in events
 with at least three jets.
}
  \label{fig:W.1409.8639:08}
\end{figure}

\begin{figure}[!ht]
  \includegraphics[width=0.499\linewidth]{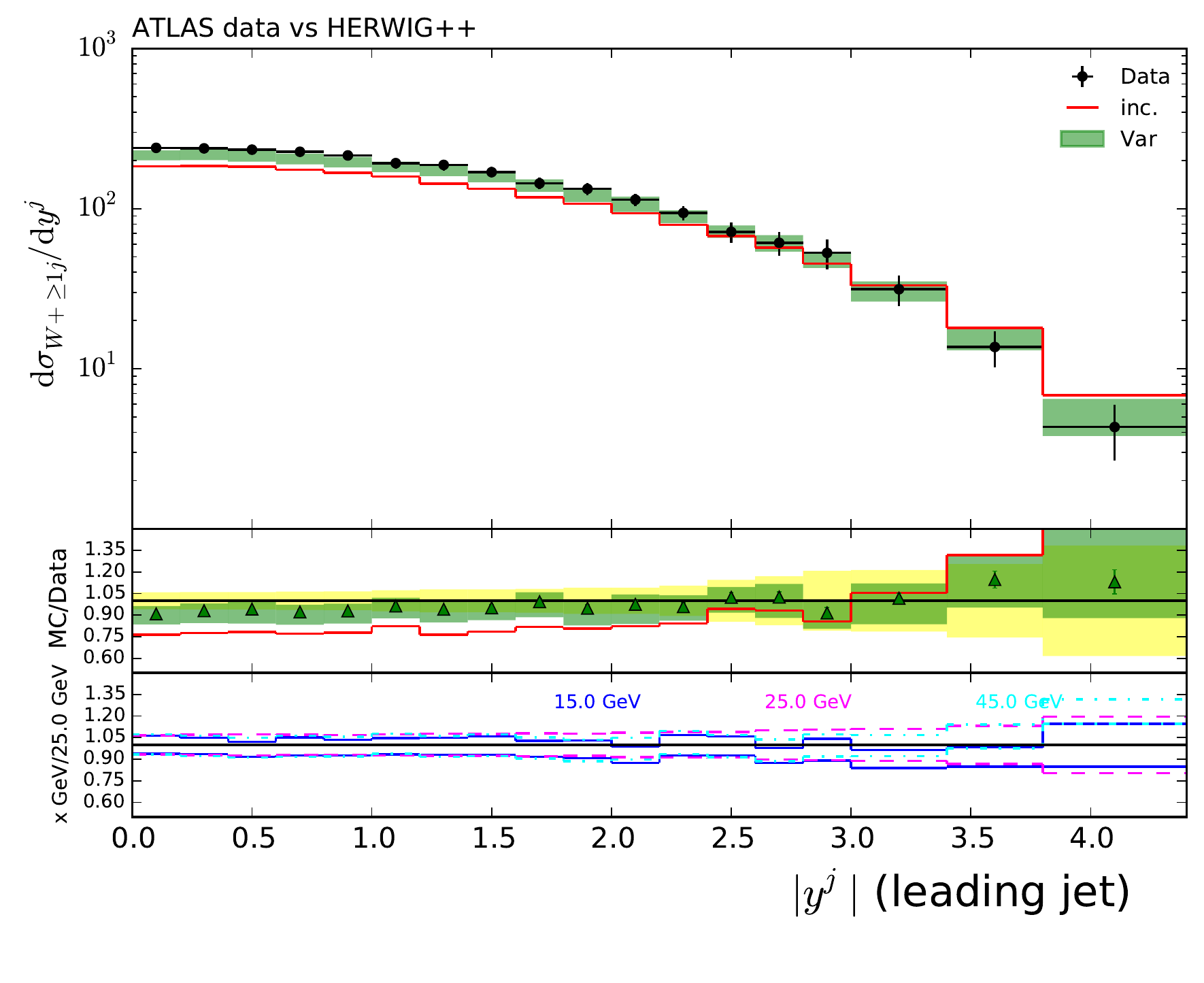}
  \includegraphics[width=0.499\linewidth]{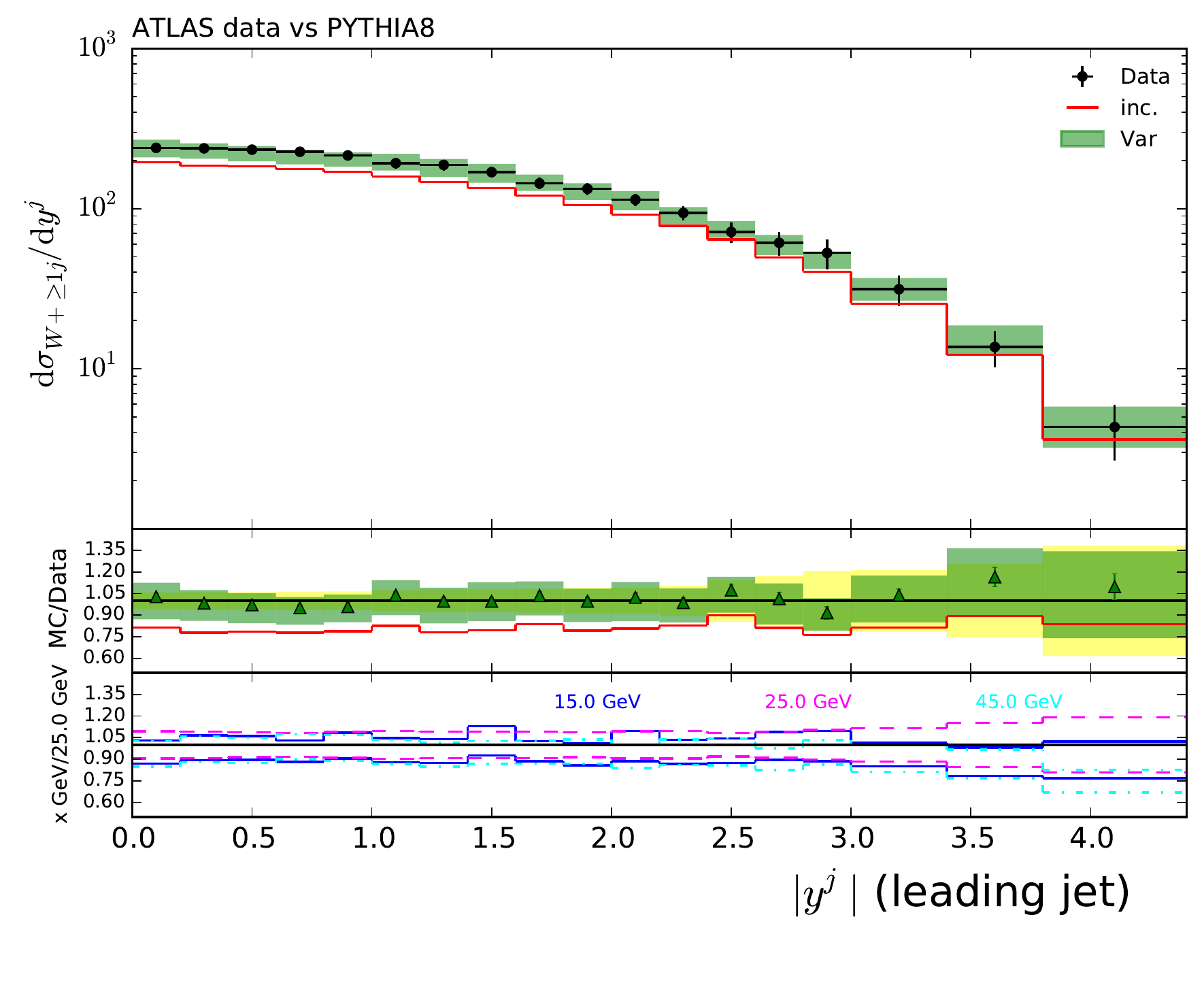}
  \caption{As in fig.~\ref{fig:W.1409.8639:04}, for the 
 rapidity of the 1$^{st}$ jet.
}
  \label{fig:W.1409.8639:13}
\end{figure}

\begin{figure}[!ht]
  \includegraphics[width=0.499\linewidth]{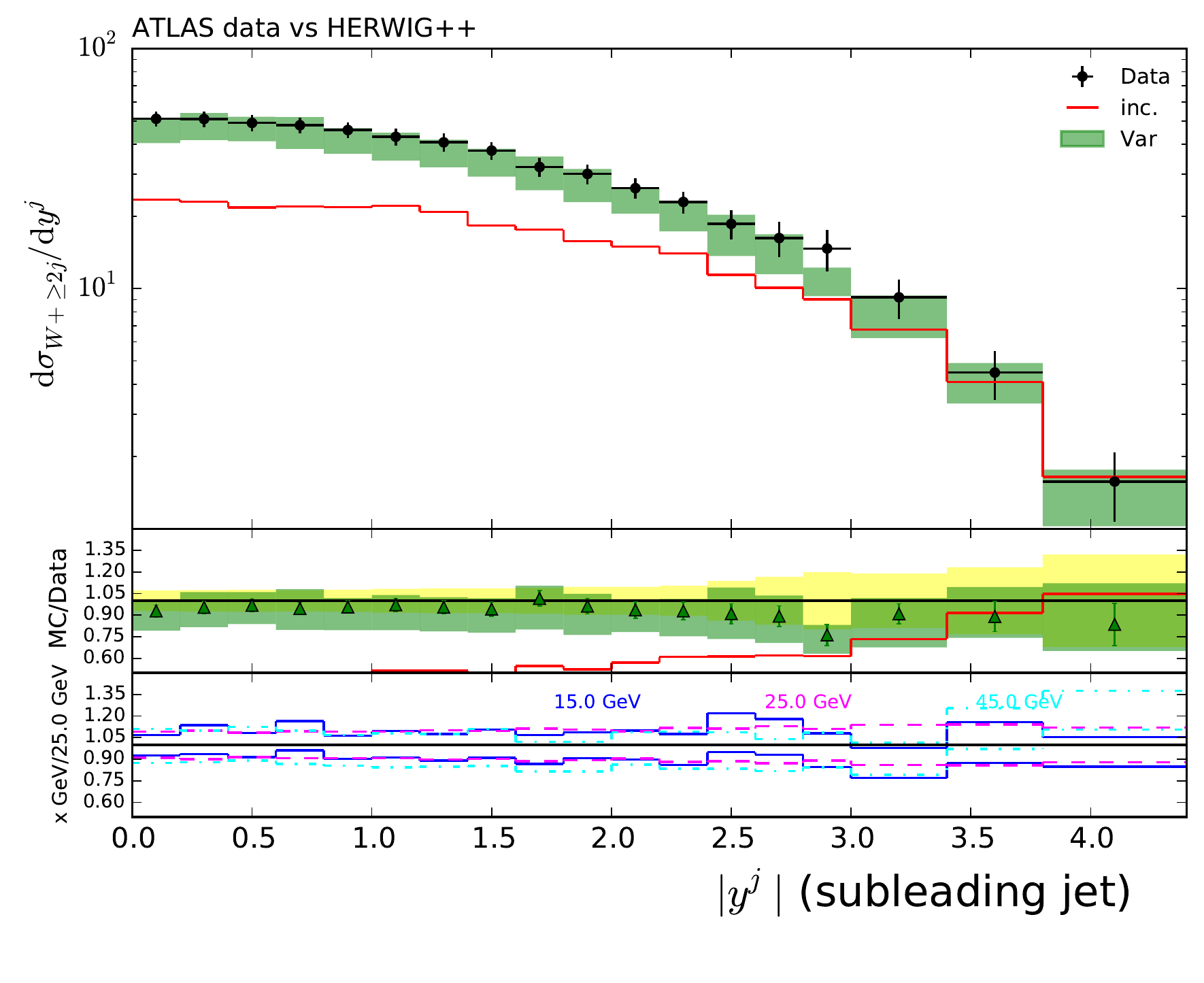}
  \includegraphics[width=0.499\linewidth]{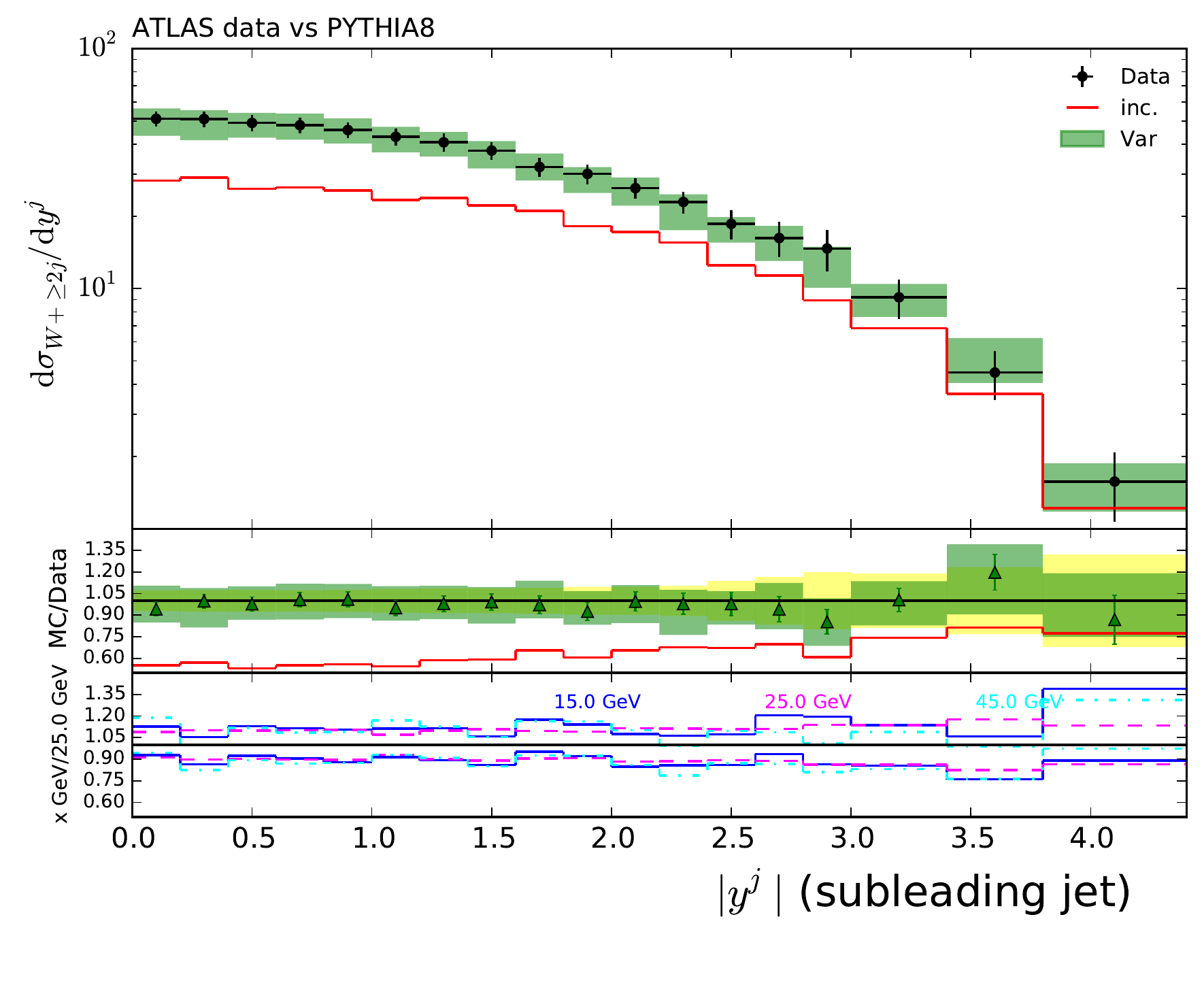}
  \caption{As in fig.~\ref{fig:W.1409.8639:04}, for the 
 rapidity of the 2$^{nd}$ jet.
}
  \label{fig:W.1409.8639:14}
\end{figure}

\begin{figure}[!ht]
  \includegraphics[width=0.499\linewidth]{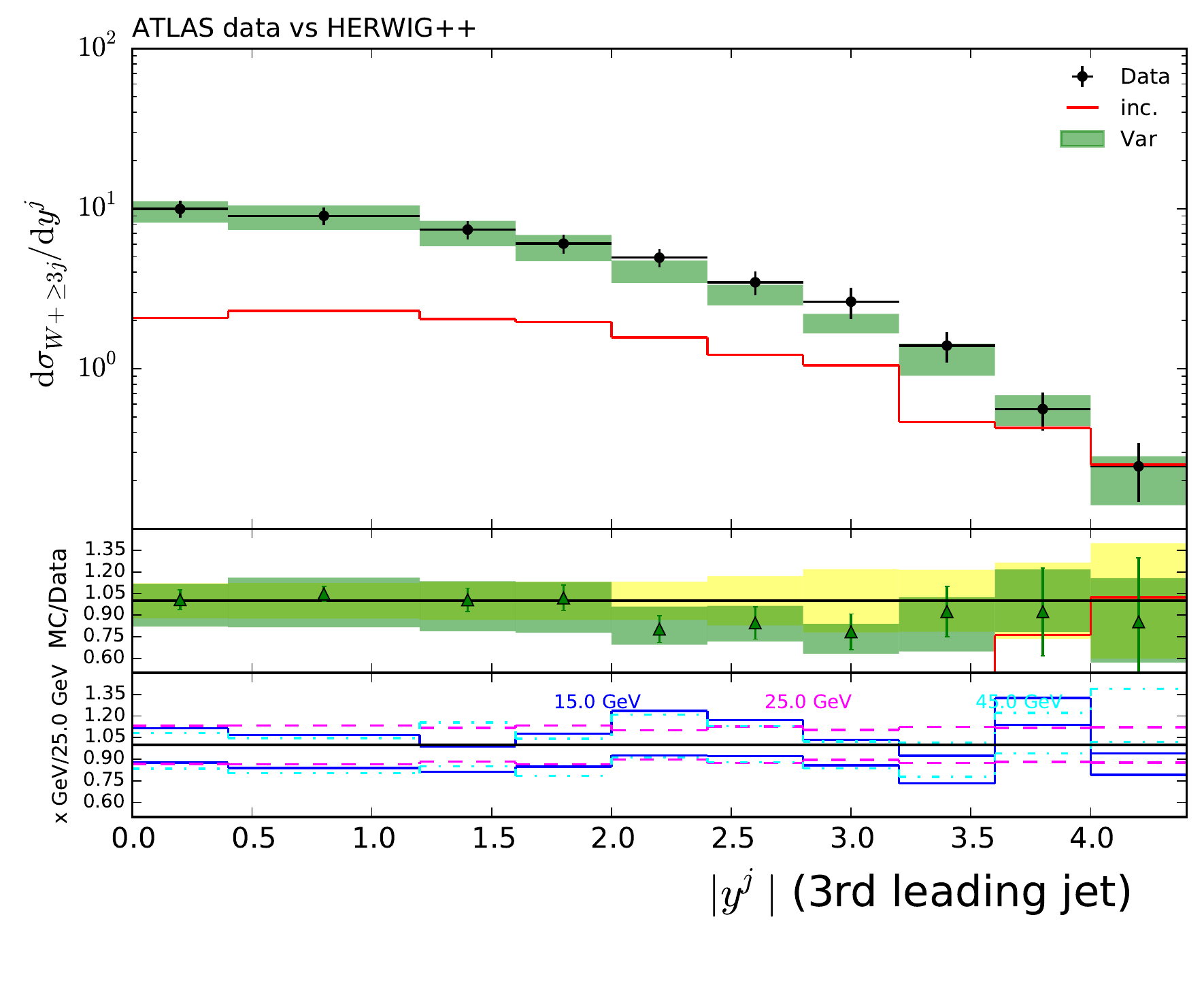}
  \includegraphics[width=0.499\linewidth]{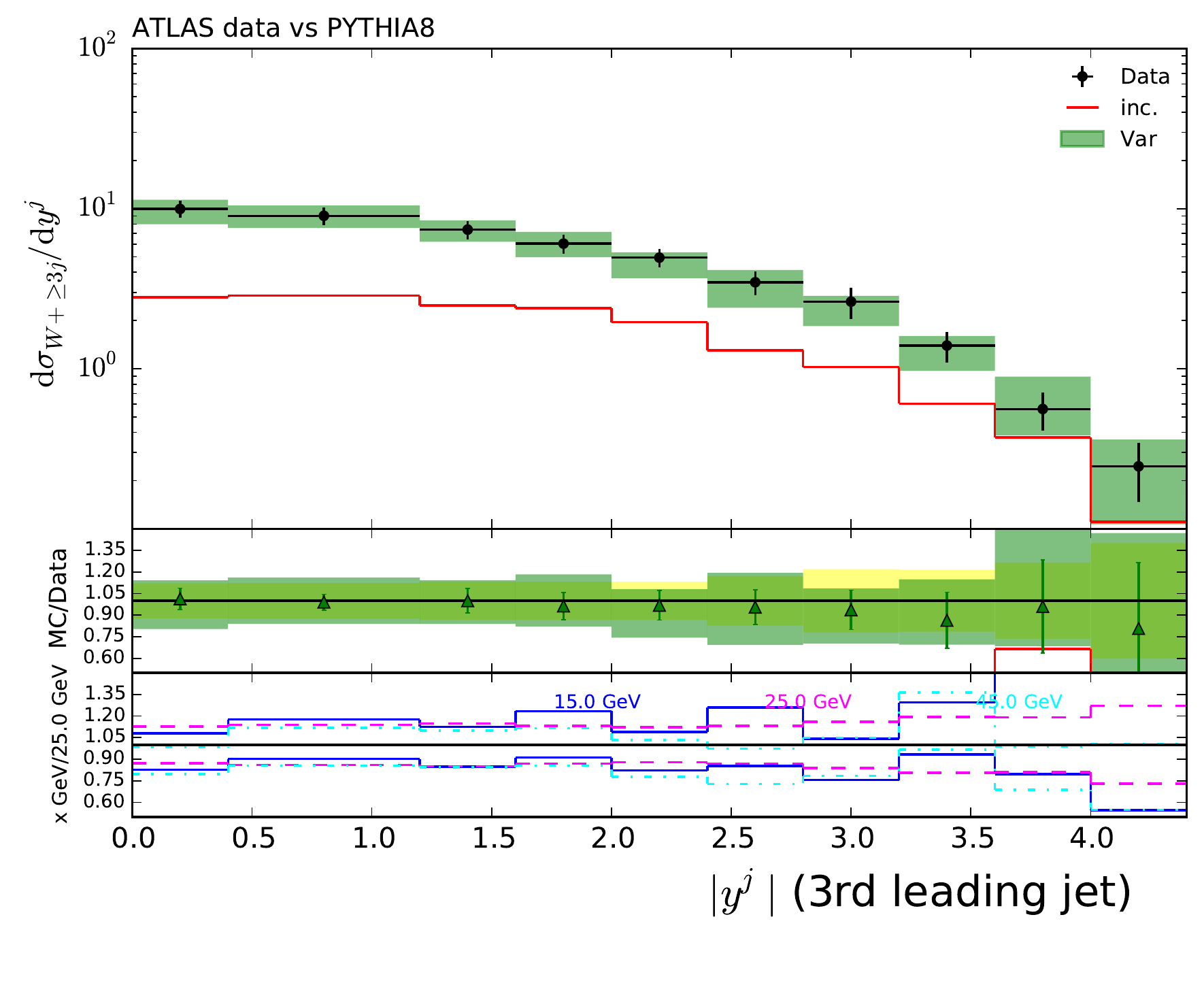}
  \caption{As in fig.~\ref{fig:W.1409.8639:04}, for the 
 rapidity of the 3$^{rd}$ jet.
}
  \label{fig:W.1409.8639:27}
\end{figure}

\begin{figure}[!ht]
  \includegraphics[width=0.499\linewidth]{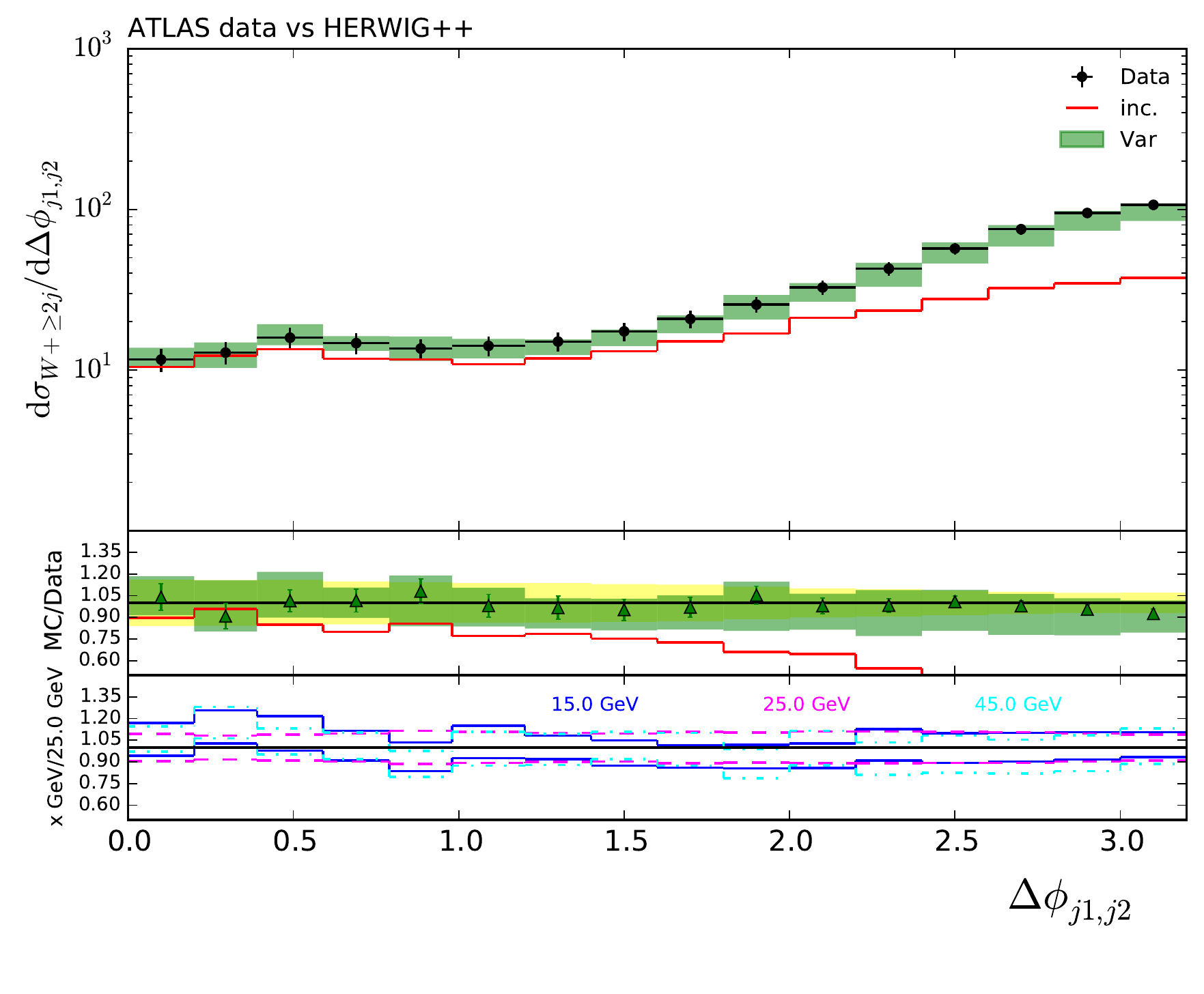}
  \includegraphics[width=0.499\linewidth]{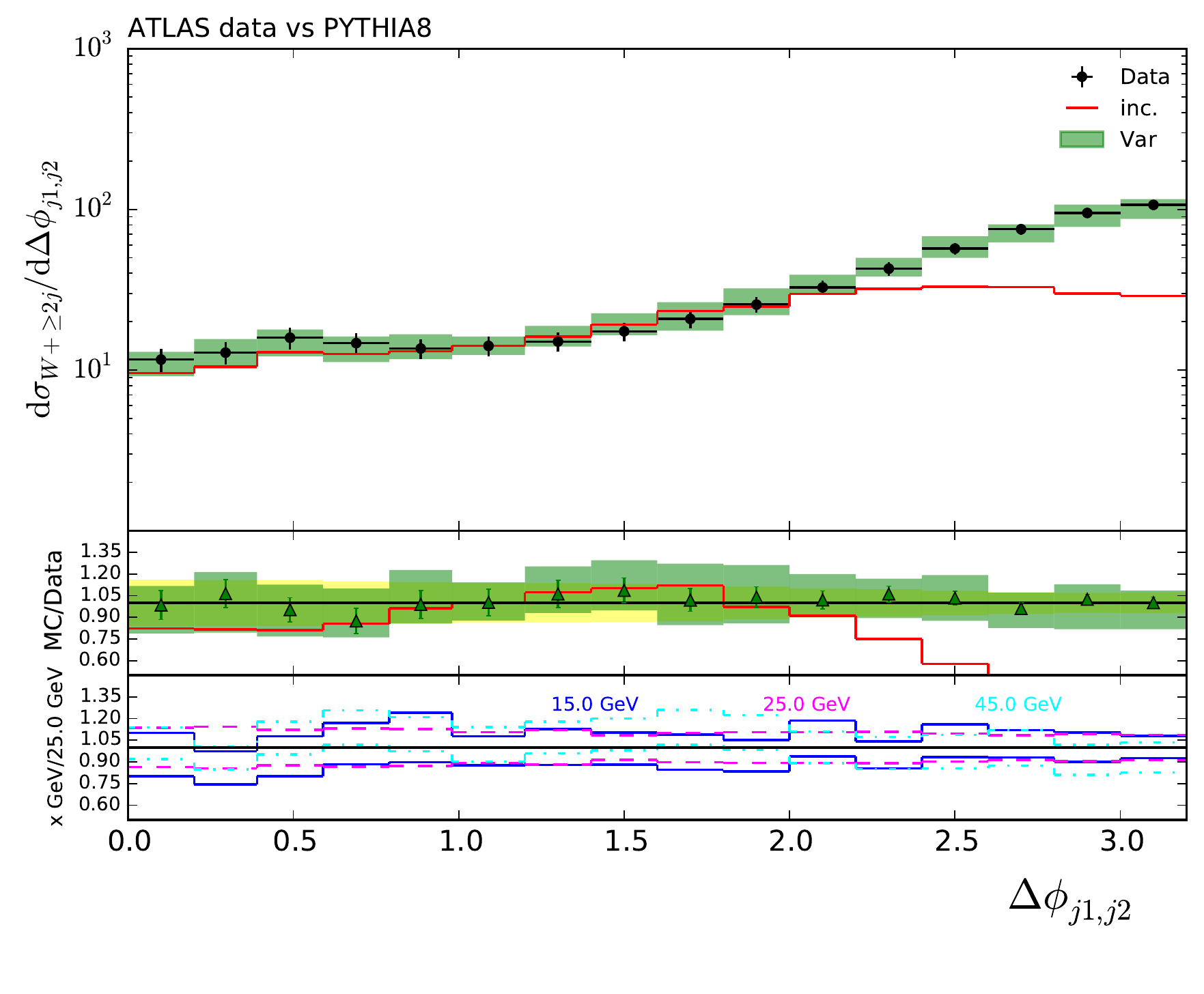}
  \caption{As in fig.~\ref{fig:W.1409.8639:04}, for the 
 azimuthal distance between the two hardest jets.
}
  \label{fig:W.1409.8639:23}
\end{figure}

\begin{figure}[!ht]
  \includegraphics[width=0.499\linewidth]{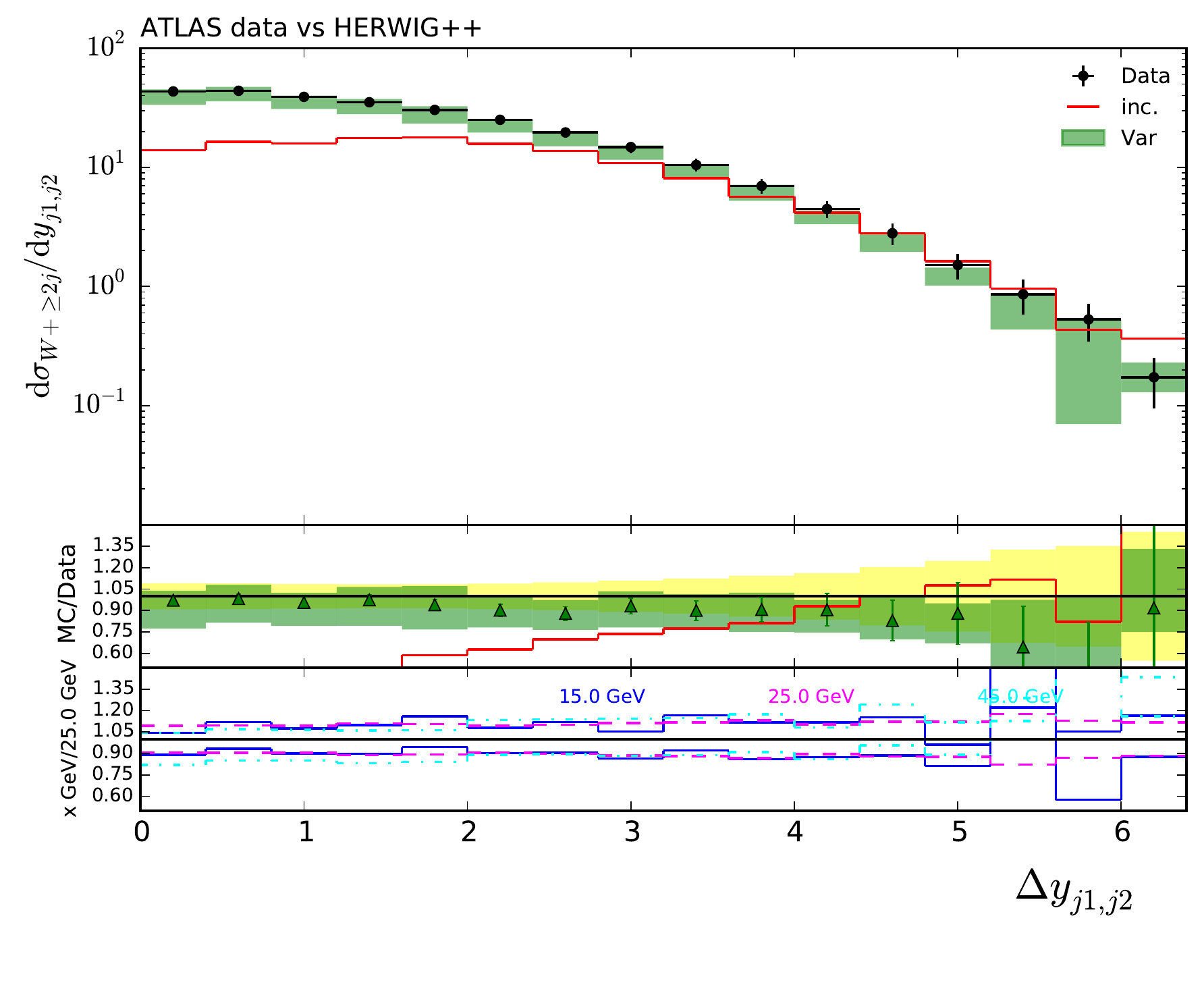}
  \includegraphics[width=0.499\linewidth]{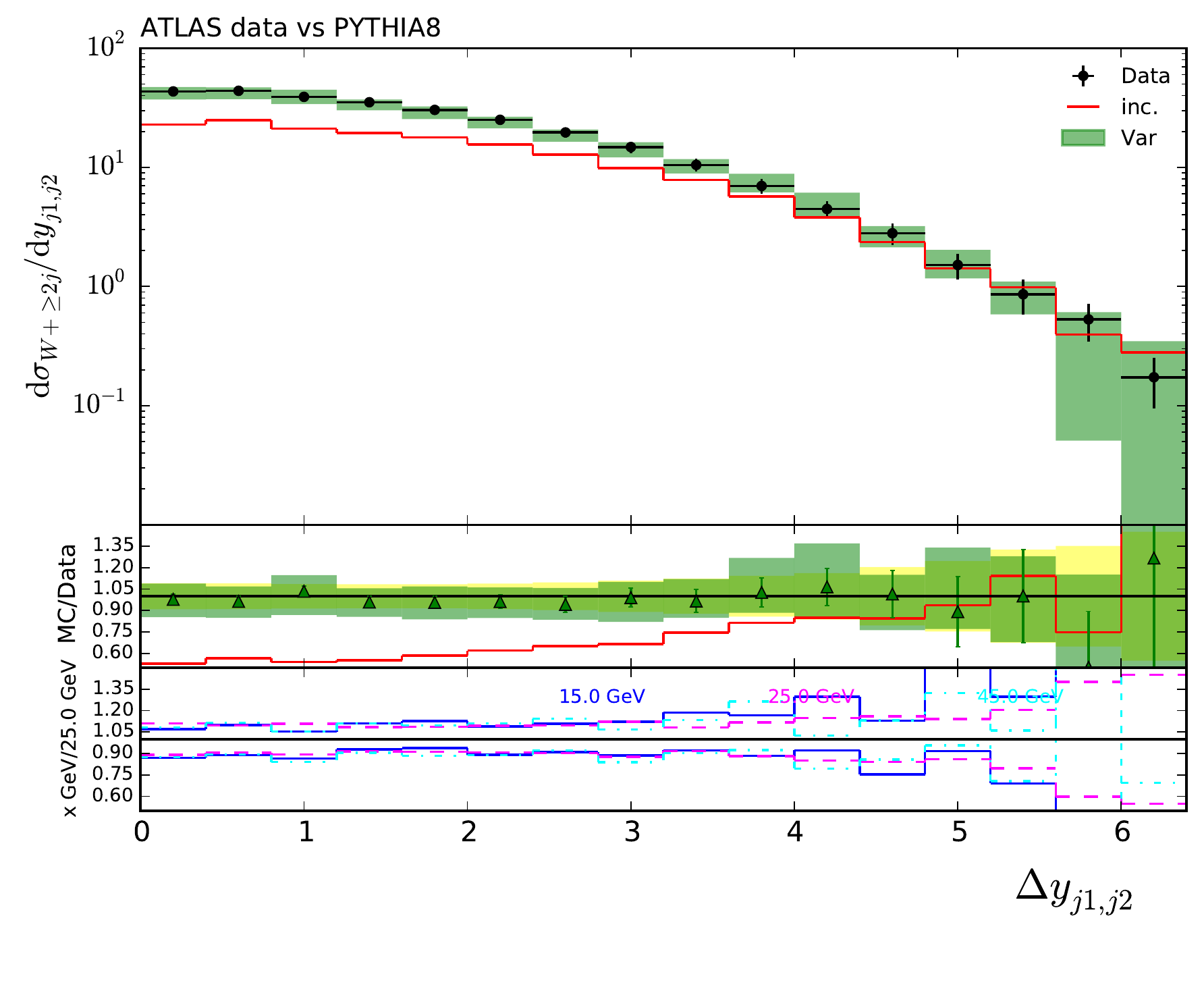}
  \caption{As in fig.~\ref{fig:W.1409.8639:04}, for the 
 rapidity distance between the two hardest jets.
}
  \label{fig:W.1409.8639:24}
\end{figure}

\begin{figure}[!ht]
  \includegraphics[width=0.499\linewidth]{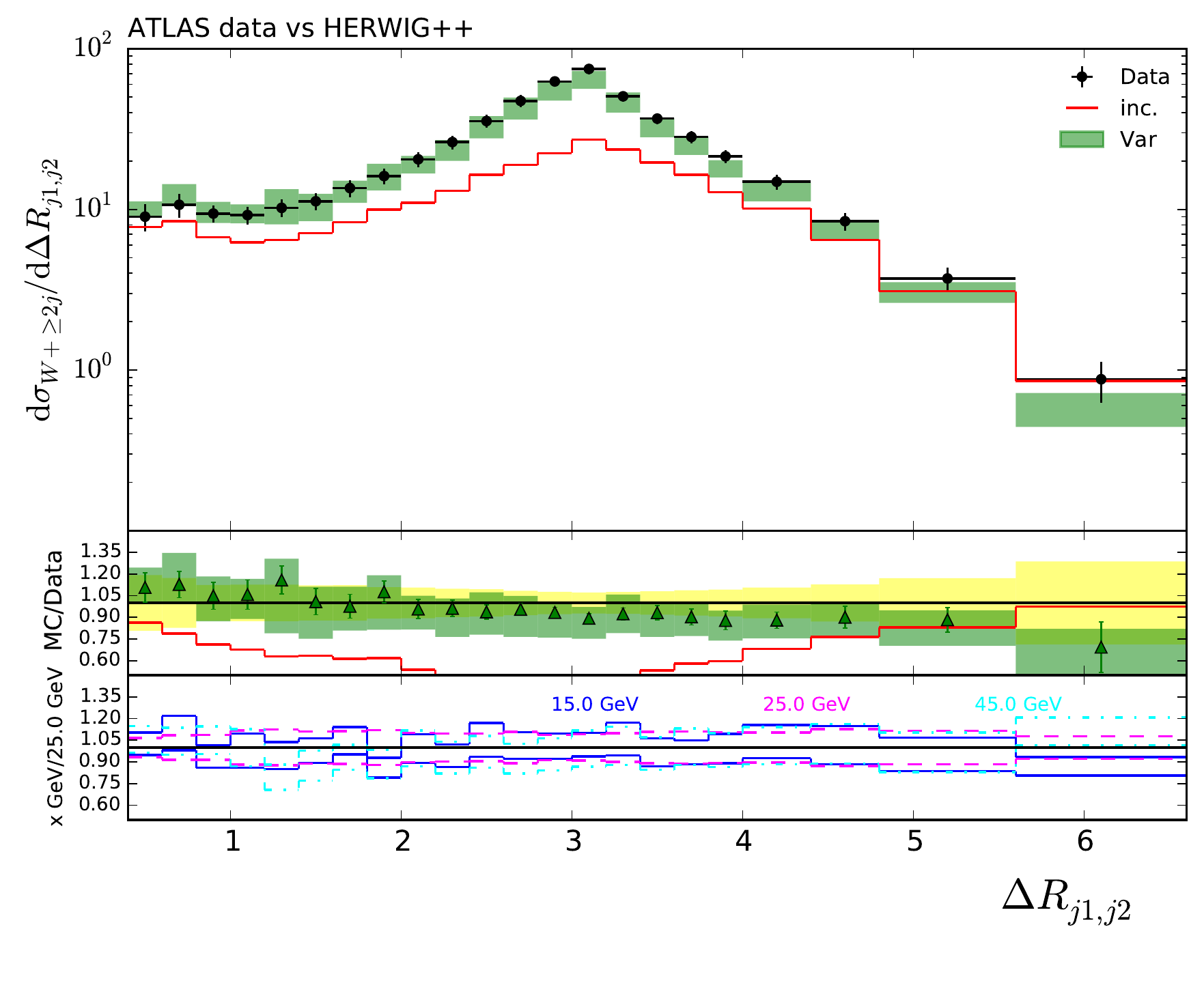}
  \includegraphics[width=0.499\linewidth]{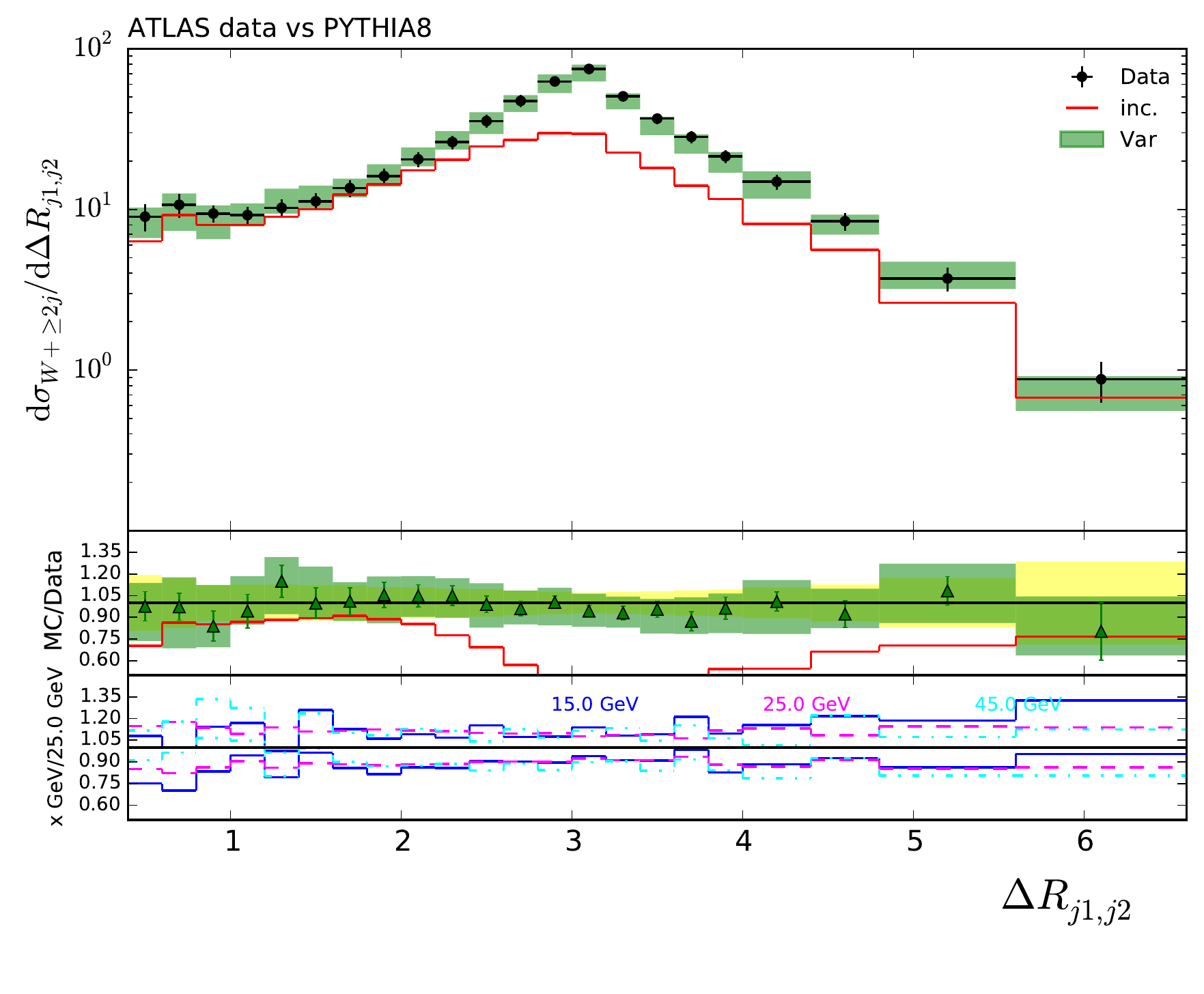}
  \caption{As in fig.~\ref{fig:W.1409.8639:04}, for the 
 $\Delta R$ between the two hardest jets.
}
  \label{fig:W.1409.8639:25}
\end{figure}

\begin{figure}[!ht]
  \includegraphics[width=0.499\linewidth]{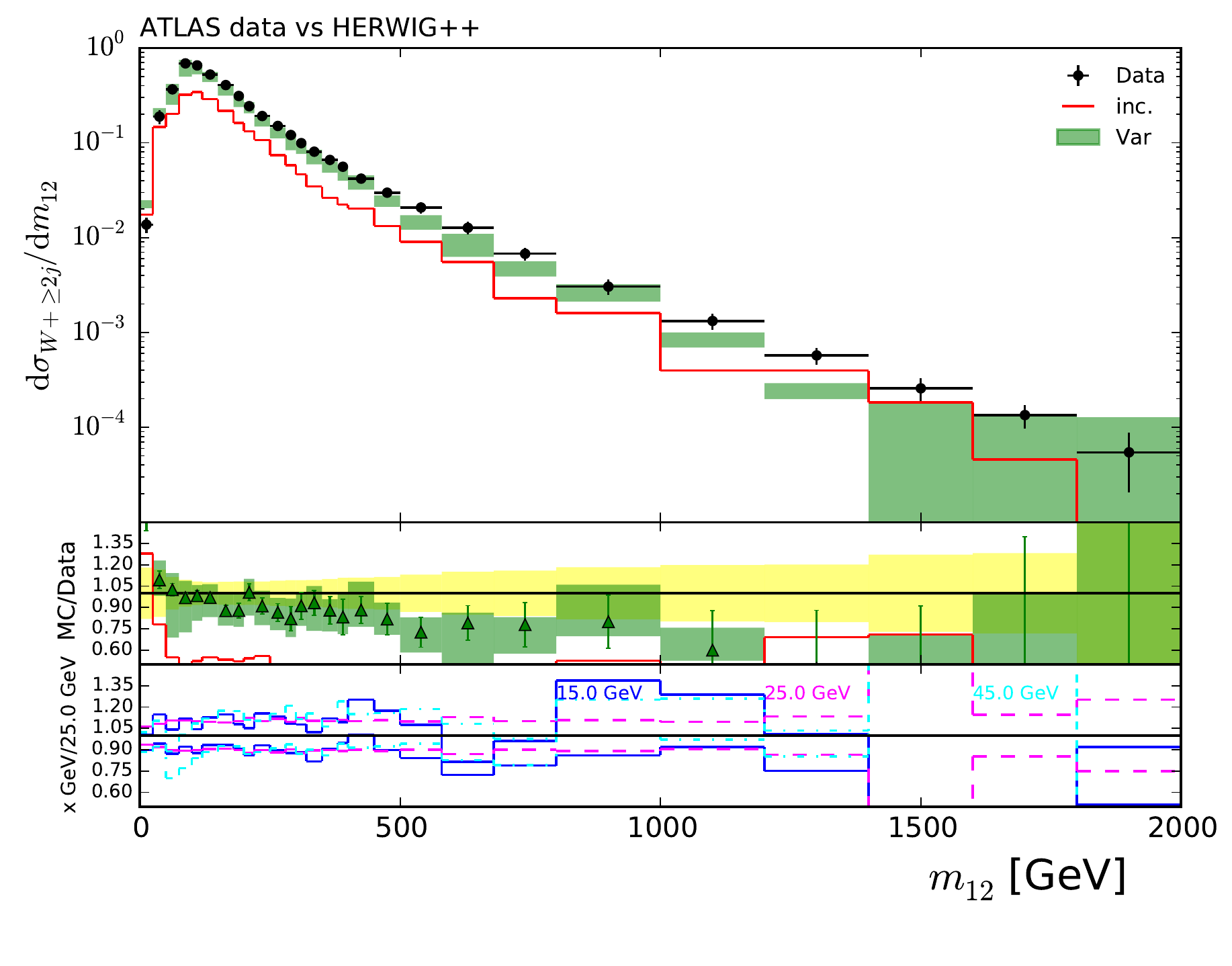}
  \includegraphics[width=0.499\linewidth]{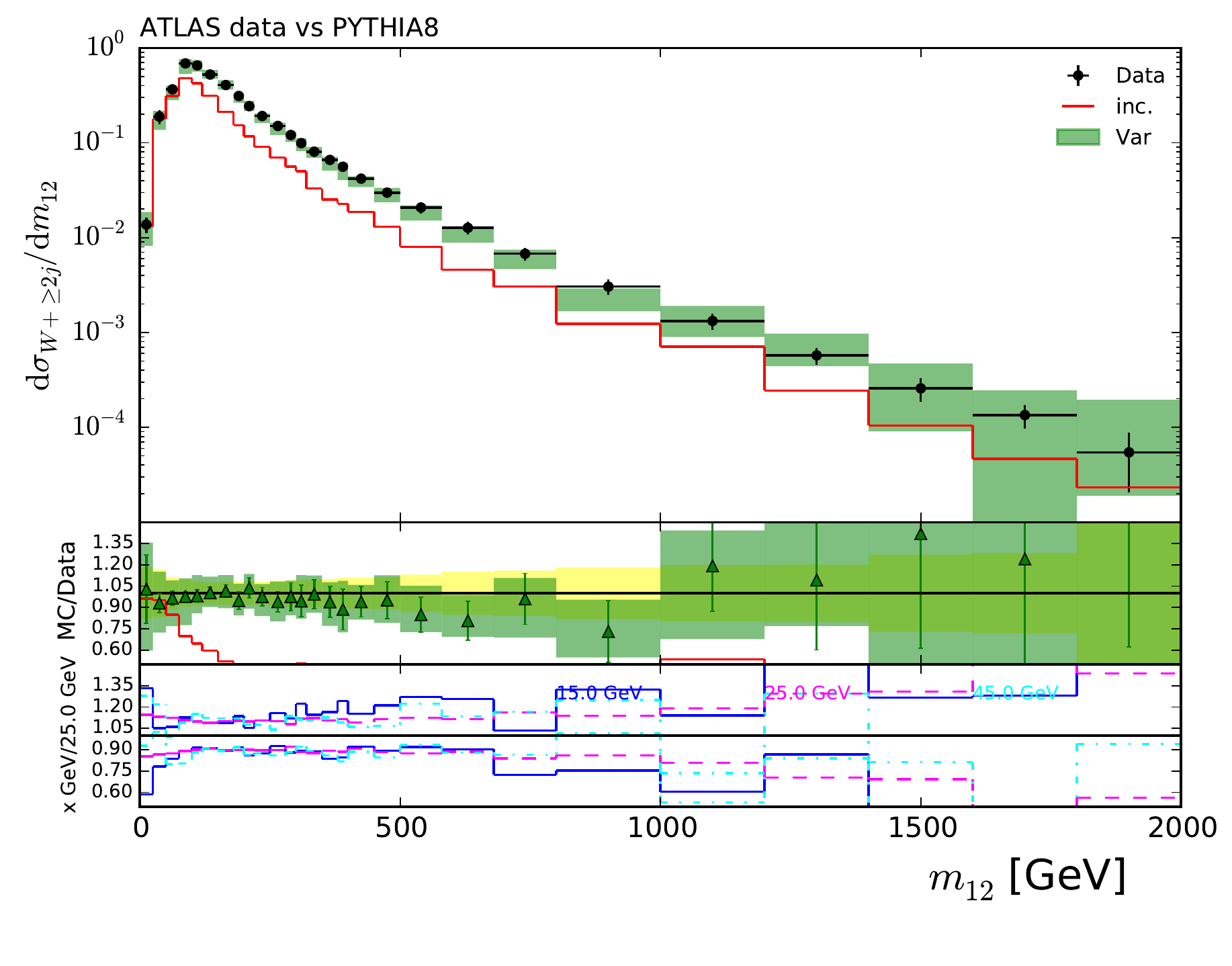}
  \caption{As in fig.~\ref{fig:W.1409.8639:04}, for the 
 invariant mass of the two hardest jets.
}
  \label{fig:W.1409.8639:26}
\end{figure}

\begin{figure}[!ht]
  \includegraphics[width=0.499\linewidth]{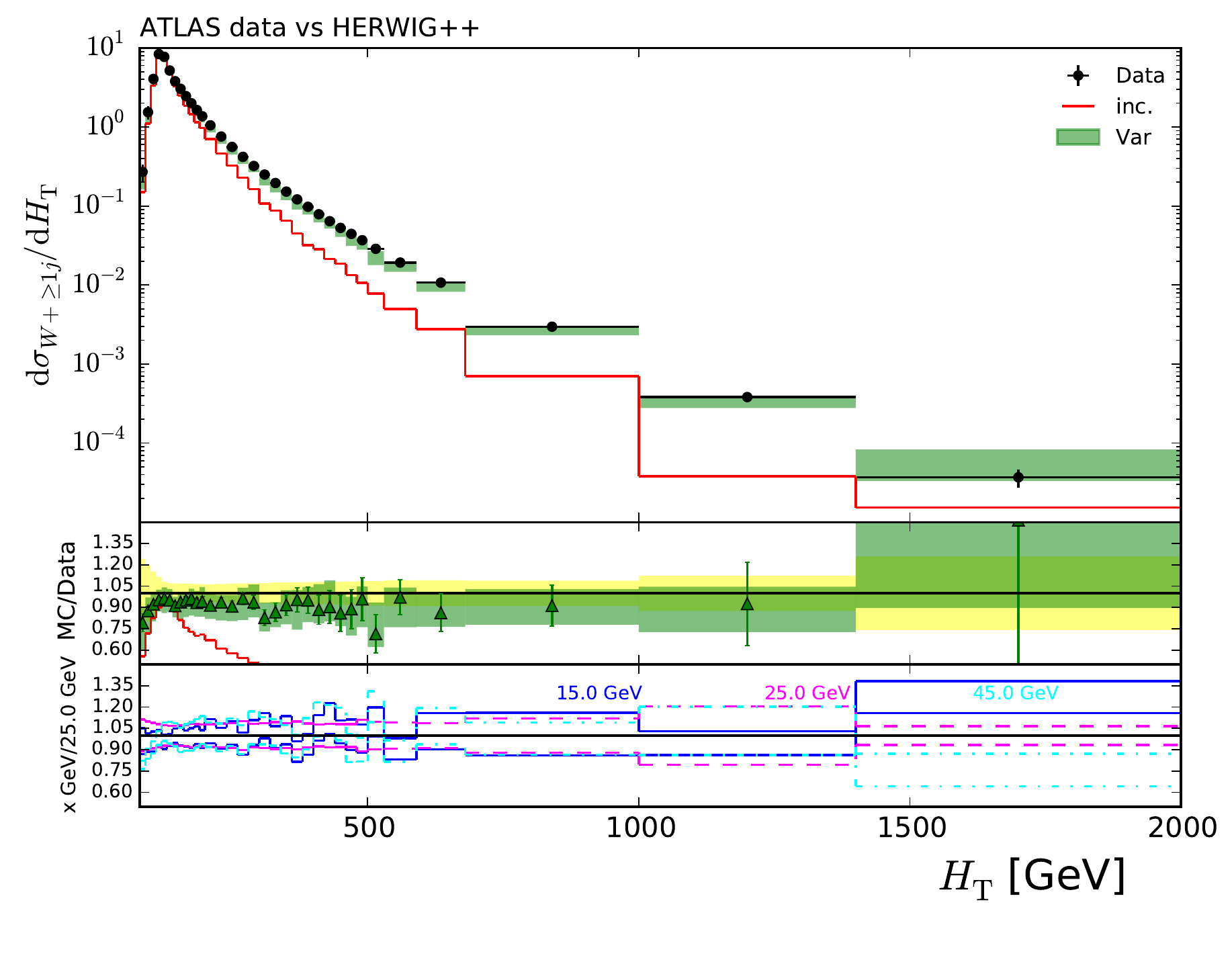}
  \includegraphics[width=0.499\linewidth]{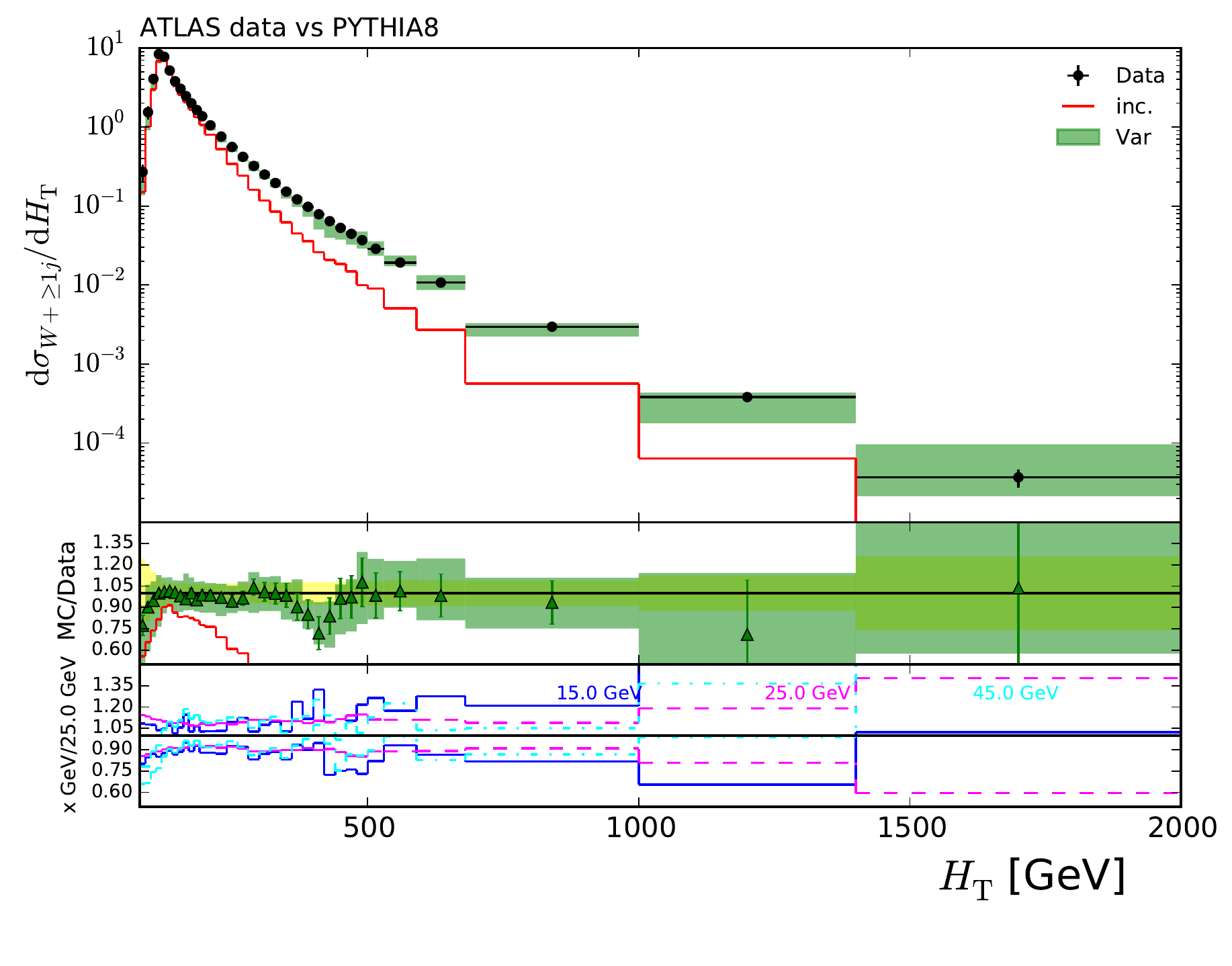}
  \caption{As in fig.~\ref{fig:W.1409.8639:04}, for $H_{\sss T}$.
}
  \label{fig:W.1409.8639:15}
\end{figure}

\begin{figure}[!ht]
  \includegraphics[width=0.499\linewidth]{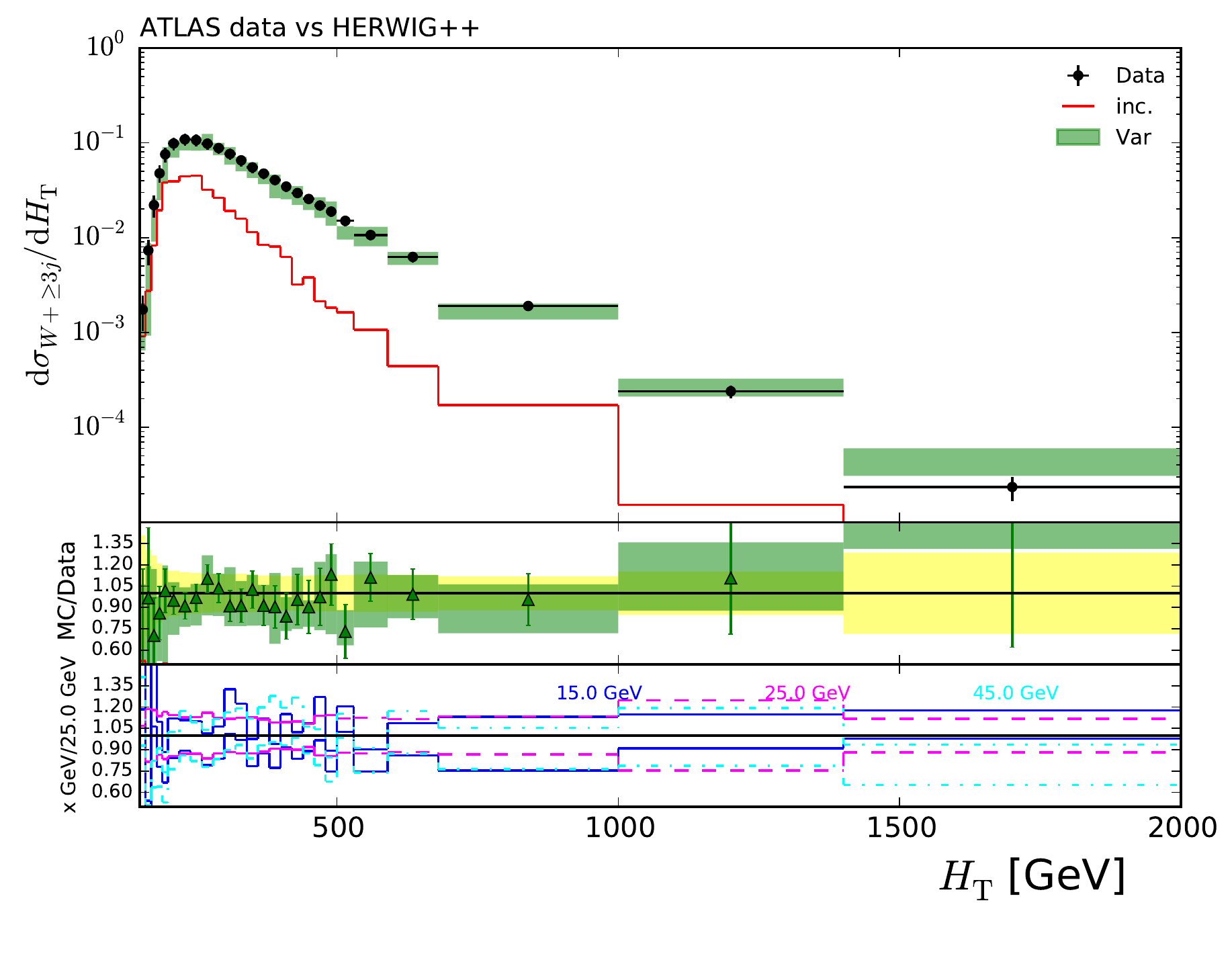}
  \includegraphics[width=0.499\linewidth]{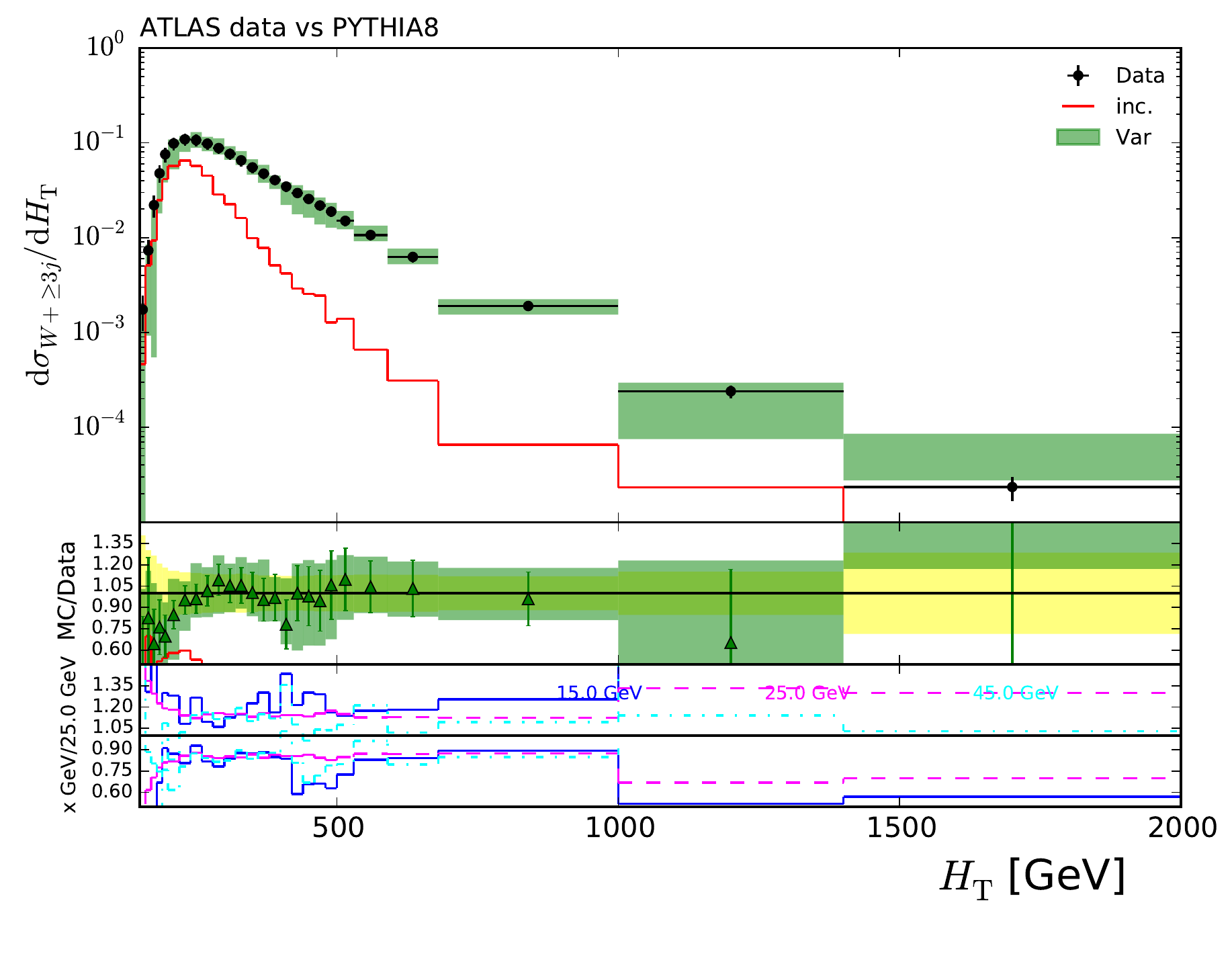}
  \caption{As in fig.~\ref{fig:W.1409.8639:04}, for $H_{\sss T}$ 
 in events with at least three jets.
}
  \label{fig:W.1409.8639:19}
\end{figure}

The observables that we have selected in this analysis are displayed
in the following plots.
Figure~\ref{fig:W.1409.8639:04}: exclusive jet multiplicity;
fig.~\ref{fig:W.1409.8639:05}: transverse momentum of the 1$^{st}$ jet;
fig.~\ref{fig:W.1409.8639:07}: transverse momentum of the 1$^{st}$ jet, 
in events with at least two jets;
fig.~\ref{fig:W.1409.8639:08}: transverse momentum of the 1$^{st}$ jet, 
in events with at least three jets;
fig.~\ref{fig:W.1409.8639:13}: rapidity of the 1$^{st}$ jet;
fig.~\ref{fig:W.1409.8639:14}: rapidity of the 2$^{nd}$ jet;
fig.~\ref{fig:W.1409.8639:27}: rapidity of the 3$^{rd}$ jet;
fig.~\ref{fig:W.1409.8639:23}: azimuthal distance between the two hardest jets;
fig.~\ref{fig:W.1409.8639:24}: rapidity distance between the two hardest jets;
fig.~\ref{fig:W.1409.8639:25}: $\Delta R$ between the two hardest jets;
fig.~\ref{fig:W.1409.8639:26}: invariant mass of the two hardest jets;
fig.~\ref{fig:W.1409.8639:15}: $H_{\sss T}$;
fig.~\ref{fig:W.1409.8639:19}: $H_{\sss T}$ in events with at least three jets.


The comparison between theory and data for the exclusive jet 
multiplicity (fig.~\ref{fig:W.1409.8639:04}) largely follows
the same pattern as its analogue in $\Zjs$ production, discussed
above for the measurement of ref.~\cite{Aad:2013ysa} (see
fig.~\ref{fig:Z.1304.7098:03}). If anything, in the present
case the excellent agreement of the merged predictions with the
data extends to larger $\Njet$ values; however, we do not consider
this fact to be particularly significant, since here one is in a regime
completely dominated by MC effects, and thus subject to significant
uncertainties. We point out that we obtain an identical level of
agreement in the case of the inclusive jet multiplicity. This implies
that, for our predictions, the analogues of the scale factors reported
in table~7 of ref.~\cite{Aad:2014qxa} would all be quite consistent
with each other.

As far as the single-jet transverse momenta are concerned, we have
considered that of the leading jet, in events characterised by different
numbers of jets ($\Njet\ge 1$, $2$, $3$ in figs.~\ref{fig:W.1409.8639:05}, 
\ref{fig:W.1409.8639:07}, and~\ref{fig:W.1409.8639:08}, respectively).
The agreement between merged results and data is generally quite good. There 
is no indication of the predictions being softer than the measurements (as 
was marginally the case for the $\Zjs$ analysis of ref.~\cite{Aad:2013ysa});
the clearest evidence of that, the $\Njet\ge 1$ case as predicted
by \HWpp, is much weaker than its analogue in the $\Zjs$ case
(see fig.~\ref{fig:Z.1304.7098:09}). On the other hand, there is possibly
an indication of the theory being lower than data at the smallest $\pt$'s,
especially for $\Njet\ge 2$, $3$, but this is not statistically very 
significant; we note that a similar trend has been observed in 
ref.~\cite{Aad:2014qxa} for several of the theoretical simulations 
considered there.

The rapidities of the three leading jets, displayed in
figs.~\ref{fig:W.1409.8639:13}--\ref{fig:W.1409.8639:27}, are
in good agreement with the predictions of both \HWpp\ and \PYe,
in terms of both shapes and rates (the latter is consistent with
what we have observed for the $\Njet$ distribution). As in the
case of $\Zjs$ production, the similarity between the \HWpp\ and
\PYe\ FxFx results has to be contrasted with their inclusive counterparts,
whose shapes are significantly different from each other. Furthermore,
such a similarity implies an almost complete {\em in}dependence upon
the shower modelling of the merged predictions; this is in contrast
to what is remarked in ref.~\cite{Aad:2014qxa} regarding the leading
and subleading jet rapidities.

Correlations constructed with the two leading jets are reported
in figs.~\ref{fig:W.1409.8639:23}--\ref{fig:W.1409.8639:26}. 
The agreement between theory and data is generally quite good; we do 
not find any of the seemingly large shape discrepancies\footnote{We
remark that, in the first version of this paper, data and theory 
were incompatible in the rightmost bin of the $\Delta\phi_{j_1j_2}$ 
distribution. This was due to an incorrect normalisation of that
bin in the Rivet analysis, which has now been fixed. It is thus important
that the revised version of {\tt ATLAS\_2014\_I1319490} be used in order
to reproduce the results shown in fig.~\ref{fig:W.1409.8639:23}.}
between the measurements and most of the theoretical predictions
observed in ref.~\cite{Aad:2014qxa}, particularly in the cases
of $\Delta y_{j_1j_2}$ and $\Delta R_{j_1j_2}$. Among our results,
the only one whose shape is statistically only marginally compatible 
with that of the data is $m_{j_1j_2}$ as predicted by \HWpp; we
point out that a similar behaviour has been also found in the case
of $\Zjs$ production (see fig.~\ref{fig:Z.1304.7098:22}). Much smaller
differences between the two MCs are present in the cases of rapidity 
and $R$ distances, with \PYe\ slightly harder than \HWpp. 

We conclude the discussion of the results of ref.~\cite{Aad:2014qxa}
with two $H_{\sss T}$ distributions, measured by imposing
$\Njet\ge 1$ (fig.~\ref{fig:W.1409.8639:15}) and
$\Njet\ge 3$ (fig.~\ref{fig:W.1409.8639:19}). The theory-data
agreement is again good, within somewhat large systematics;
our central predictions, taken at face value, have shapes and
rates pretty compatible with those of the measurements.
A possible exception is the very small $H_{\sss T}$ region,
with theory decreasing faster than data -- this is similar to
the trend observed in ref.~\cite{Aad:2014qxa}.

\clearpage

\vskip 0.4truecm
\noindent
$\bullet$ CMS~\cite{Khachatryan:2014uva}
({\tt arXiv:1406.7533}, Rivet analysis {\tt CMS\_2014\_I1303894}).

\noindent
Study of jet and inclusive properties (the latter defined 
by requiring the presence of at least one jet in the final state), 
and of correlations. Based on an integrated luminosity of
5 fb$^{-1}$, using only the muon channel, with $R=0.5$
anti-$\kt$ jets within $\pt(j)>30$ GeV, $\abs{\eta(j)}<2.4$. Further 
cuts: $\pt(\mu)>24$~GeV, $\abs{\eta(\mu)}<2.1$, $\Delta R(j\mu)\ge 0.5$,
$\mt(\mu\nu)>50$~GeV (see ref.~~\cite{Khachatryan:2014uva} for 
the definition of the missing energy and the neutrino transverse 
momentum); events must contain exactly one muon.
We remark that a technical problem has occurred while running this
Rivet analysis with \PYe\ for some merging scale, which we have failed
to isolate and which has thus prevented us from 
reconstructing some of the observables in the simulation of such 
an MC. Since we believe that \PYe\ is already quite well tested in the
comparison to the $\Wjs$ data of ref.~\cite{Aad:2014qxa} discussed 
previously, as well as for $\Zjs$ production, for the observables
in question we have limited ourselves to presenting the \HWpp\ results.

\begin{figure}[!ht]
  \includegraphics[width=0.499\linewidth]{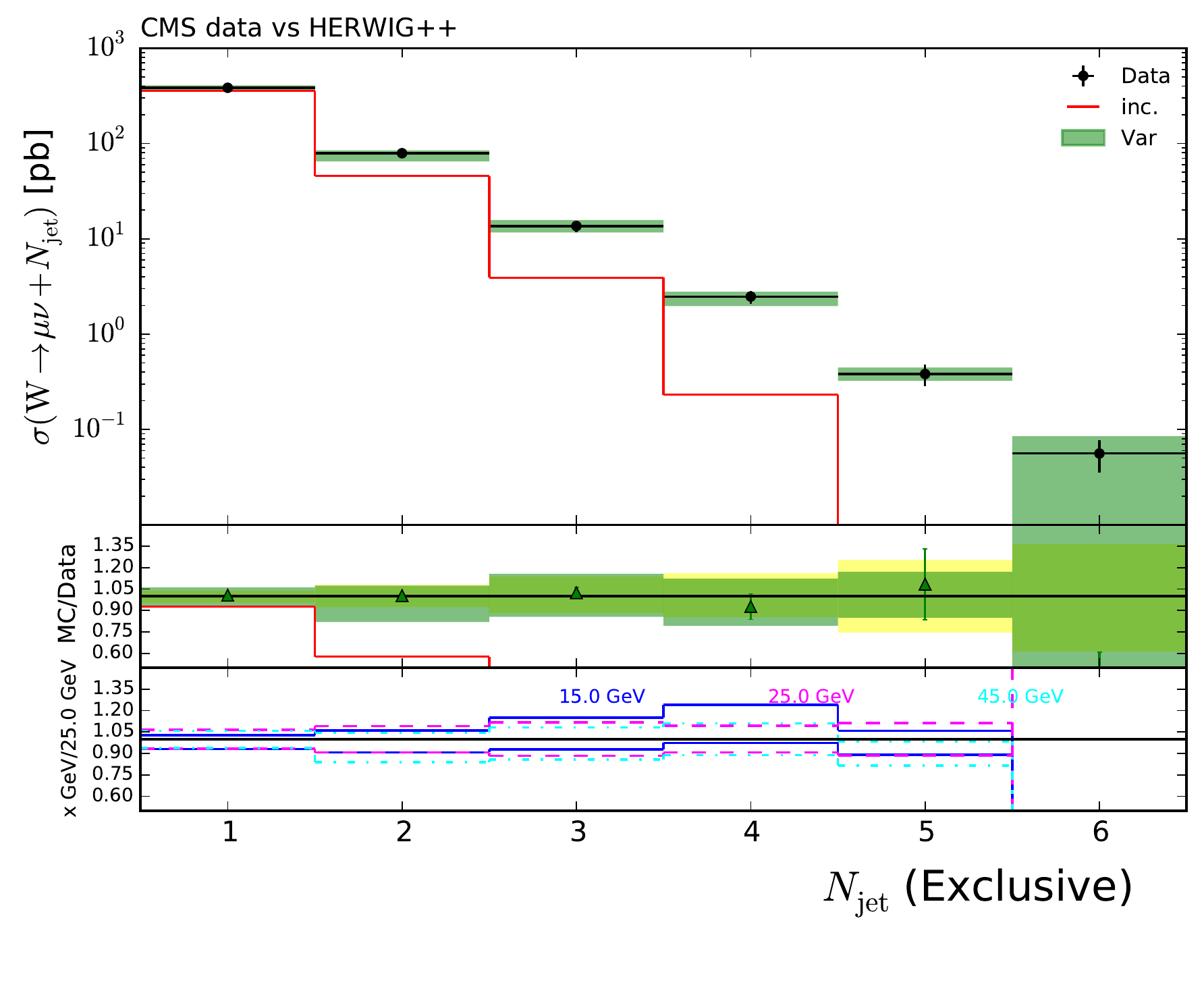}
  \includegraphics[width=0.499\linewidth]{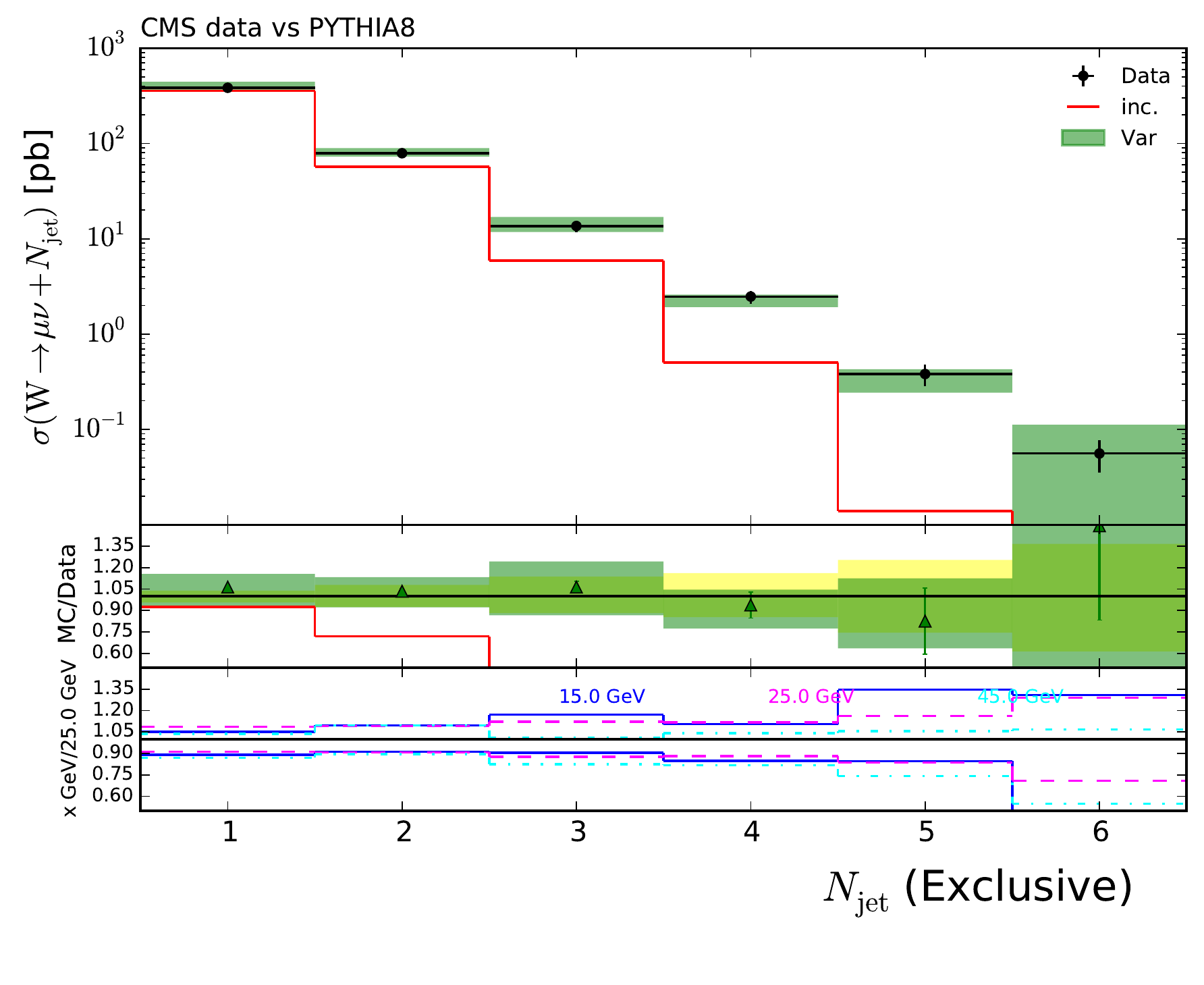}
  \caption{Exclusive jet multiplicity.
 Data from ref.~\cite{Khachatryan:2014uva}, compared to \HWpp\ (left panel) 
 and \PYe\ (right panel) predictions.
 The FxFx uncertainty envelope 
 (``Var'') and the fully-inclusive central result (``inc'') are shown
 as green bands and red histograms respectively. See the end of
 sect.~\ref{sec:tech} for more details on the layout of the plots.
}
  \label{fig:W.1406.7533:13}
\end{figure}

\begin{figure}[!ht]
  \includegraphics[width=0.499\linewidth]{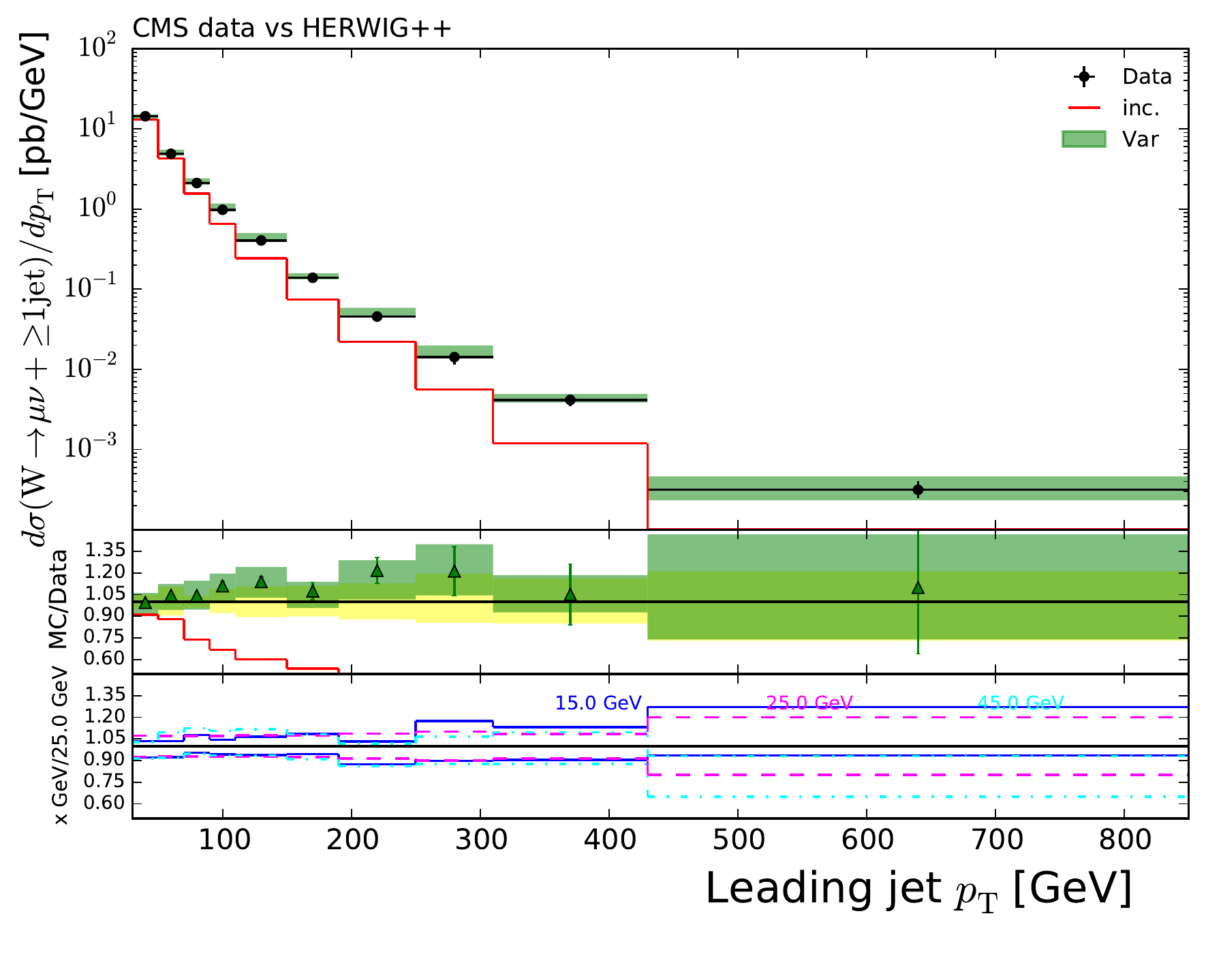}
  \includegraphics[width=0.499\linewidth]{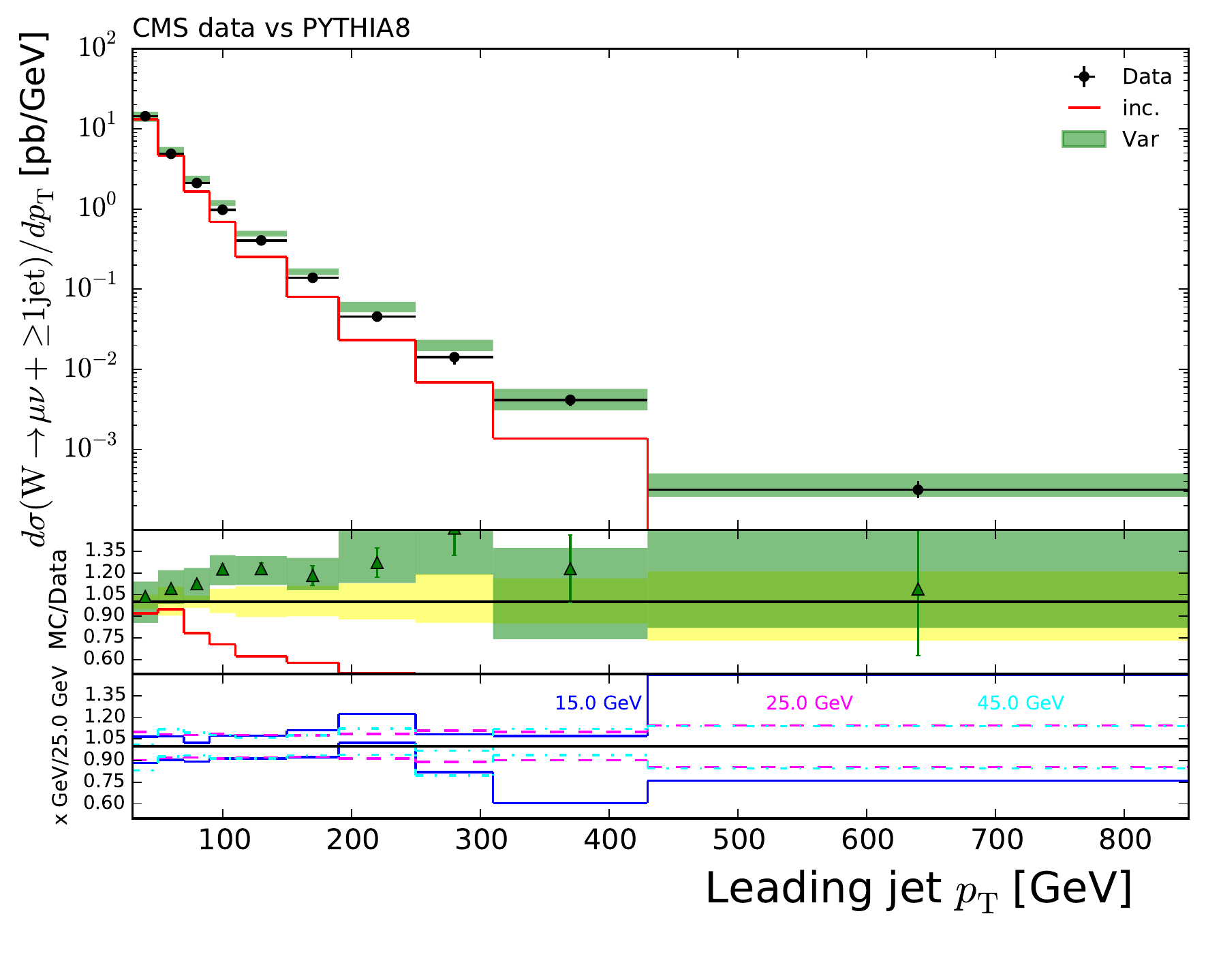}
  \caption{As in fig.~\ref{fig:W.1406.7533:13}, for the 
 transverse momentum of the 1$^{st}$ jet.
}
  \label{fig:W.1406.7533:15}
\end{figure}

\begin{figure}[!ht]
  \includegraphics[width=0.499\linewidth]{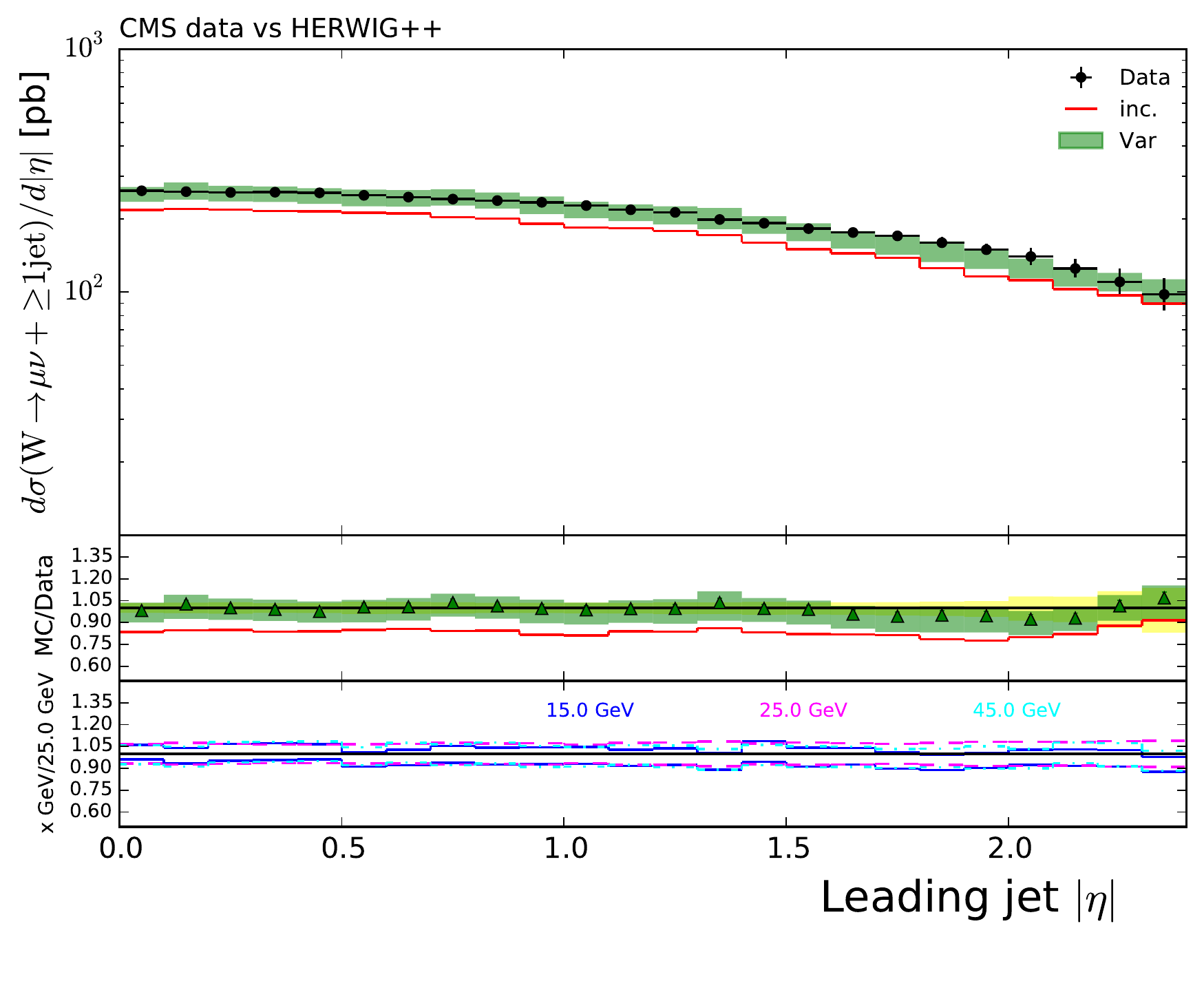}
  \includegraphics[width=0.499\linewidth]{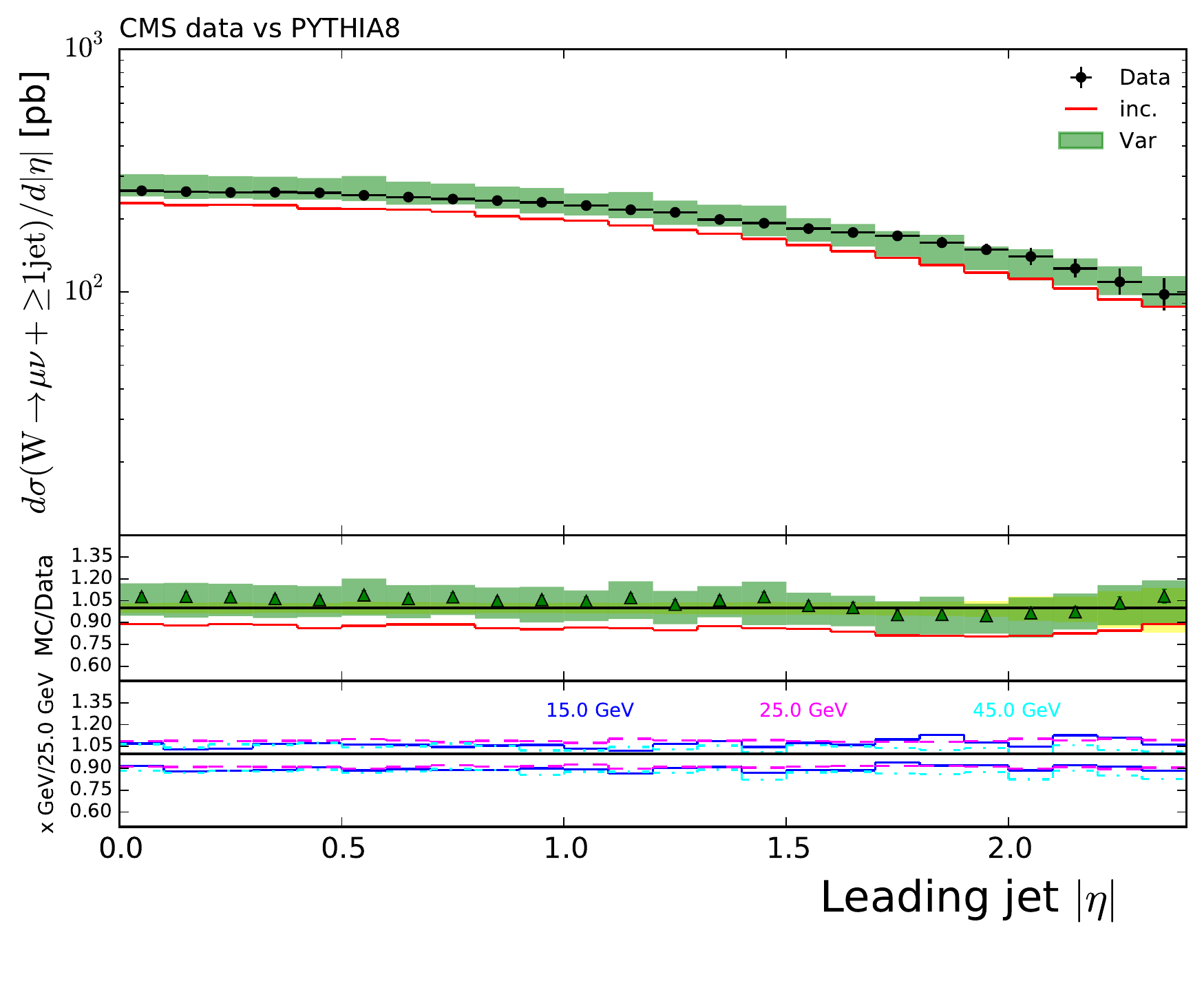}
  \caption{As in fig.~\ref{fig:W.1406.7533:13}, for the 
 pseudorapidity of the 1$^{st}$ jet.
}
  \label{fig:W.1406.7533:05}
\end{figure}

\begin{figure}[!ht]
  \includegraphics[width=0.499\linewidth]{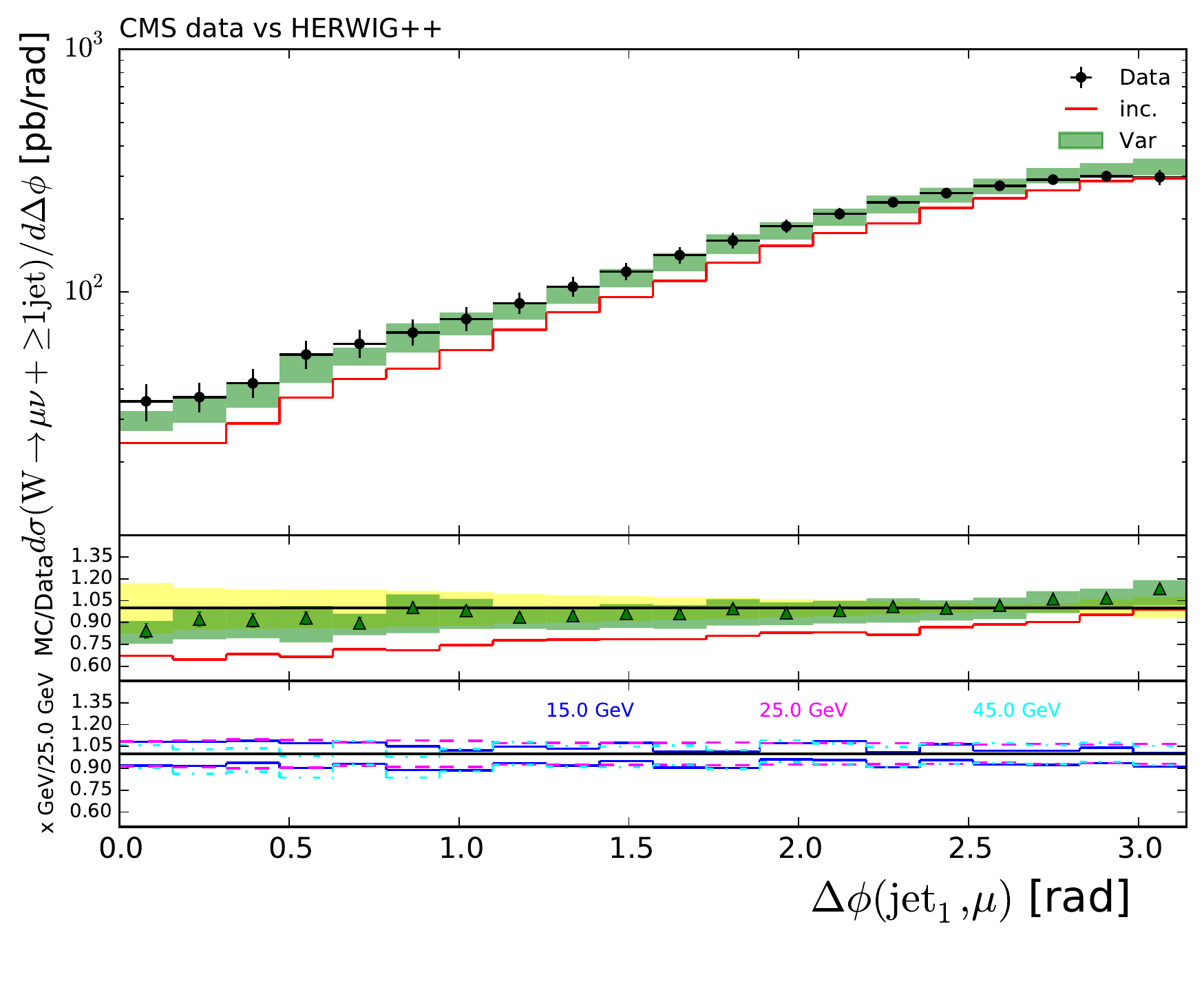}
  \includegraphics[width=0.499\linewidth]{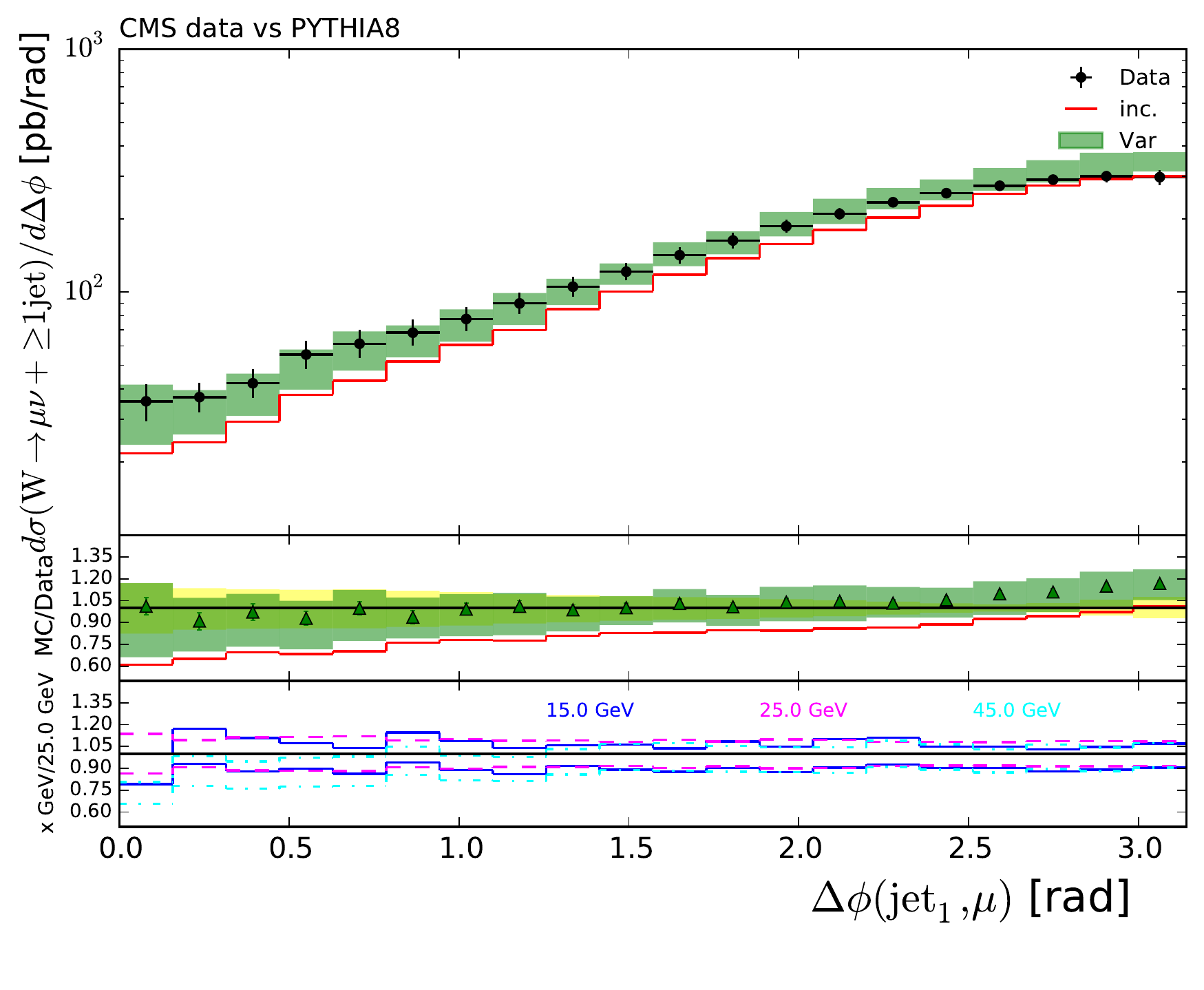}
  \caption{As in fig.~\ref{fig:W.1406.7533:13}, for the 
 azimuthal distance between the $\mu$ and the 1$^{st}$ jet.
}
  \label{fig:W.1406.7533:01}
\end{figure}

\begin{figure}[!ht]
  \includegraphics[width=0.499\linewidth]{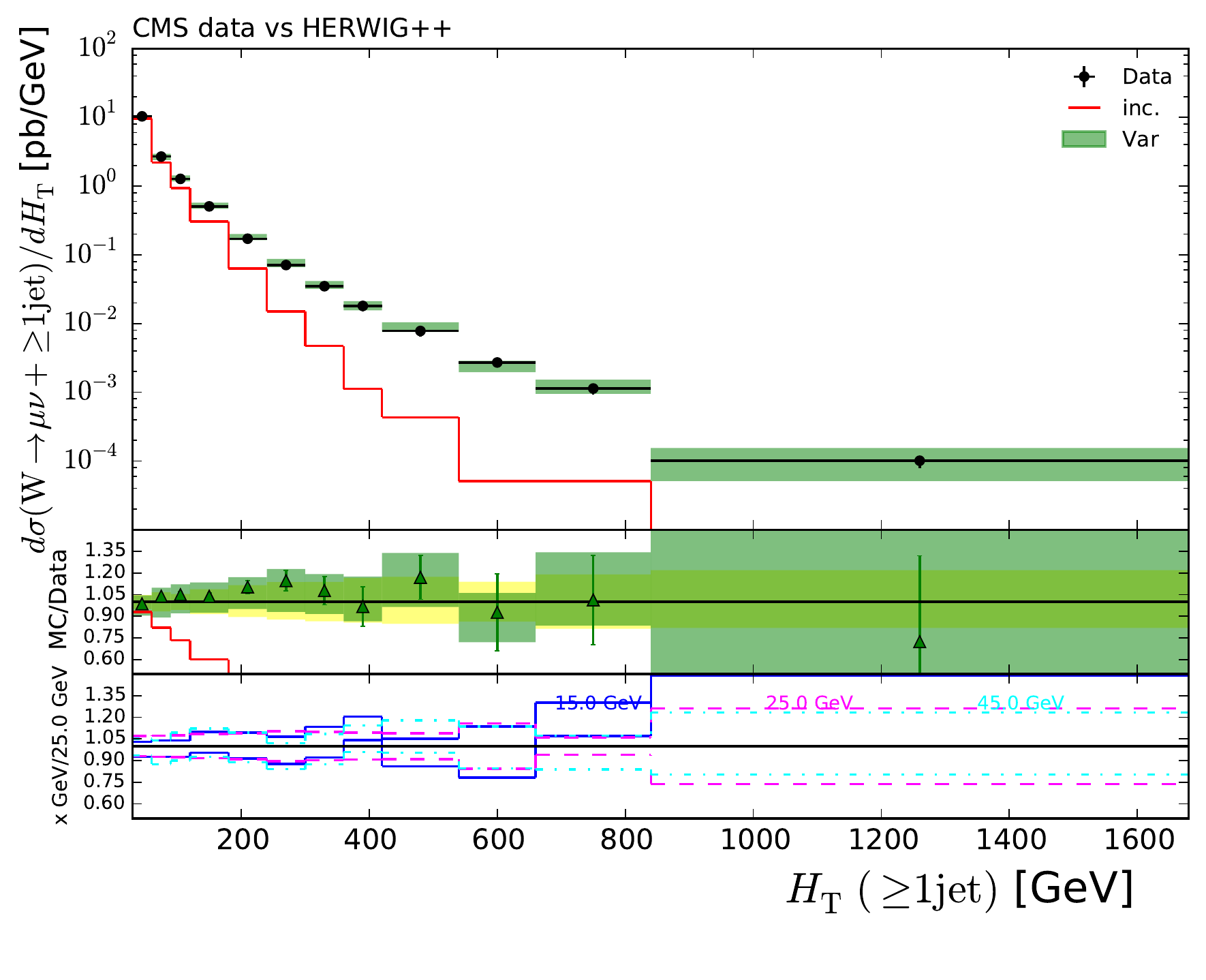}
  \includegraphics[width=0.499\linewidth]{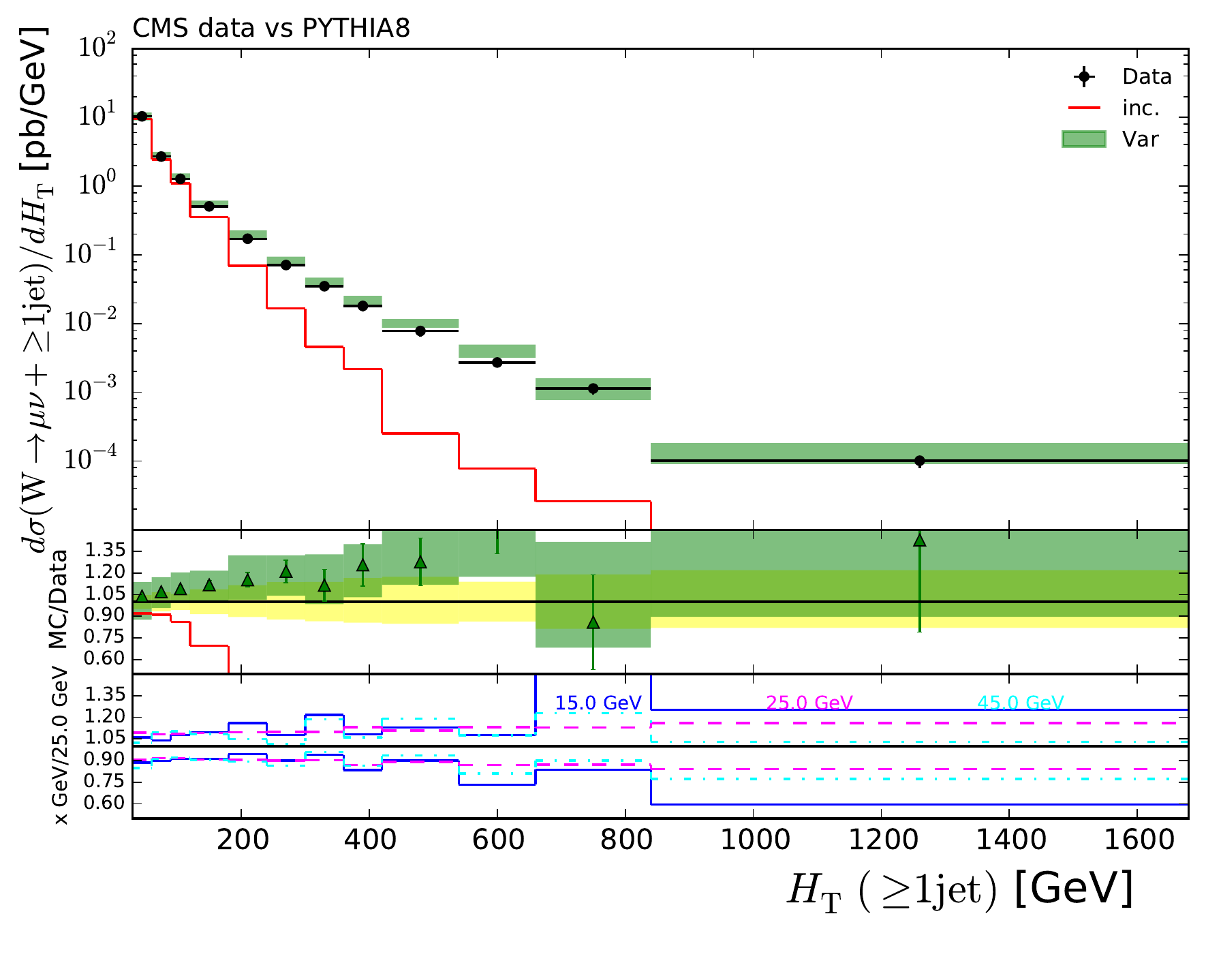}
  \caption{As in fig.~\ref{fig:W.1406.7533:13}, for $H_{\sss T}$.
}
  \label{fig:W.1406.7533:09}
\end{figure}

\begin{figure}[!ht]
  \includegraphics[width=0.499\linewidth]{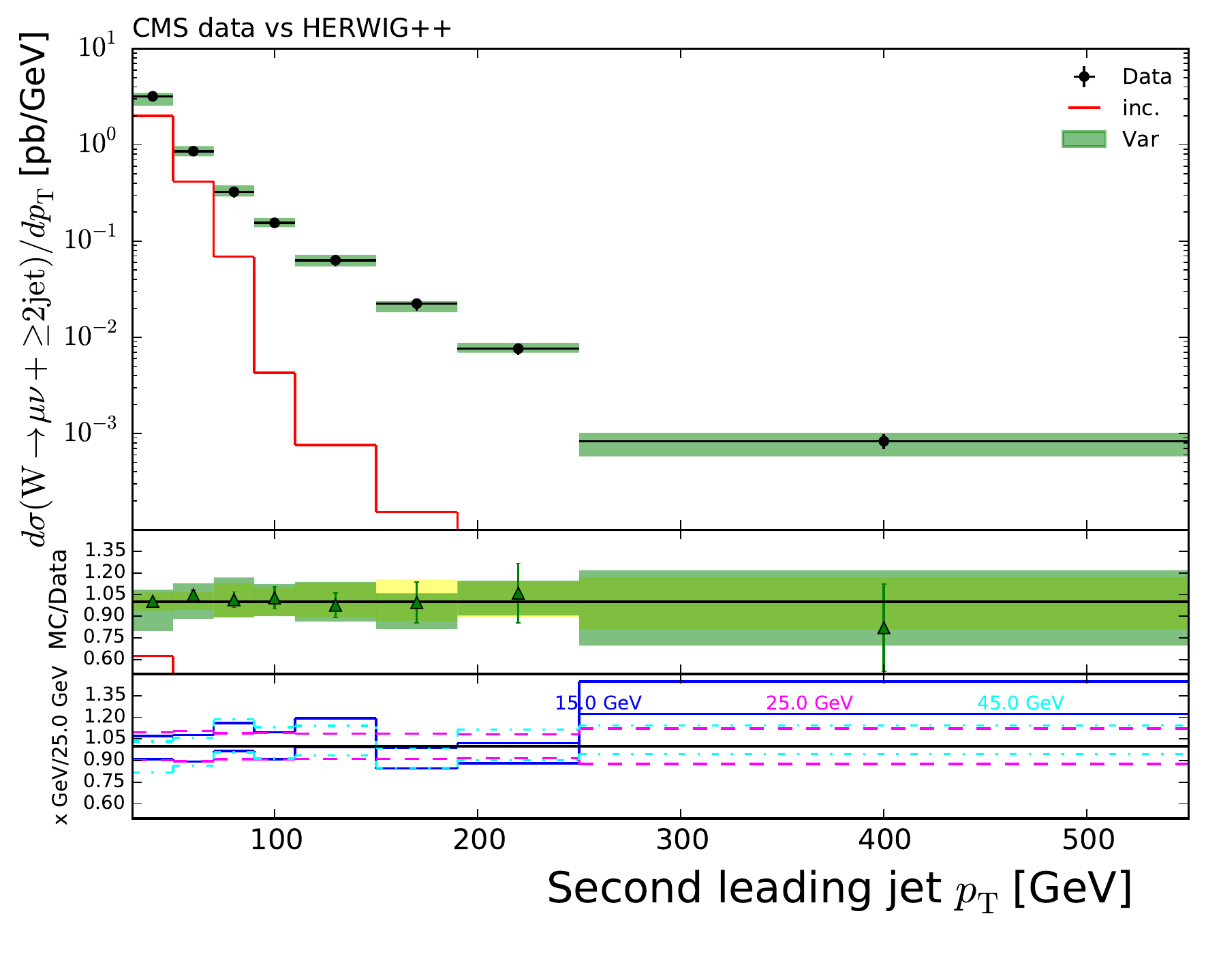}
  \includegraphics[width=0.499\linewidth]{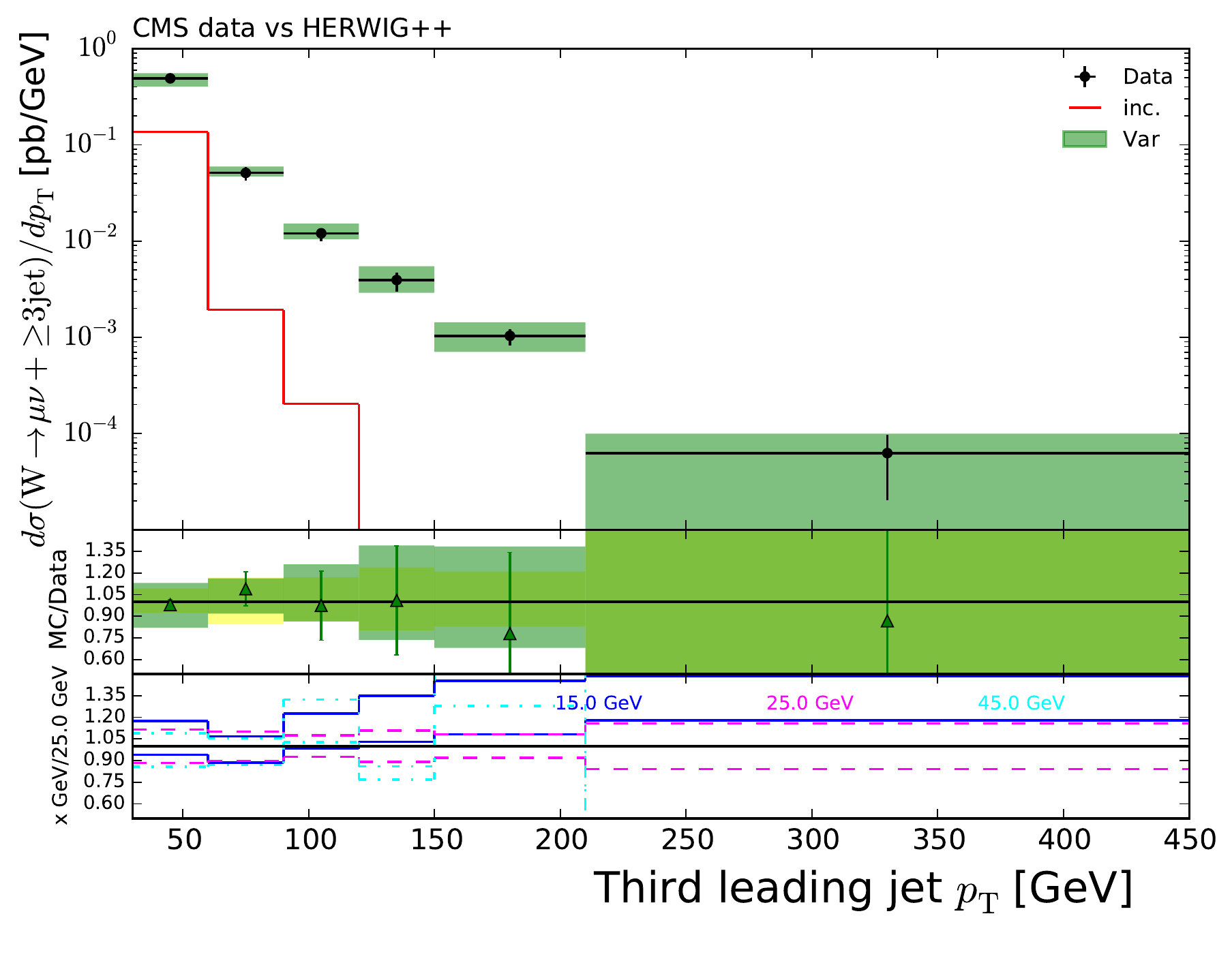}
  \caption{As in fig.~\ref{fig:W.1406.7533:13}, for the 
 transverse momentum of the 2$^{nd}$ (left panel) 
 and of the 3$^{rd}$ (right panel) jet, both
 compared to \HWpp\ predictions.
}
  \label{fig:W.1406.7533:16+17}
\end{figure}

\begin{figure}[!ht]
  \includegraphics[width=0.499\linewidth]{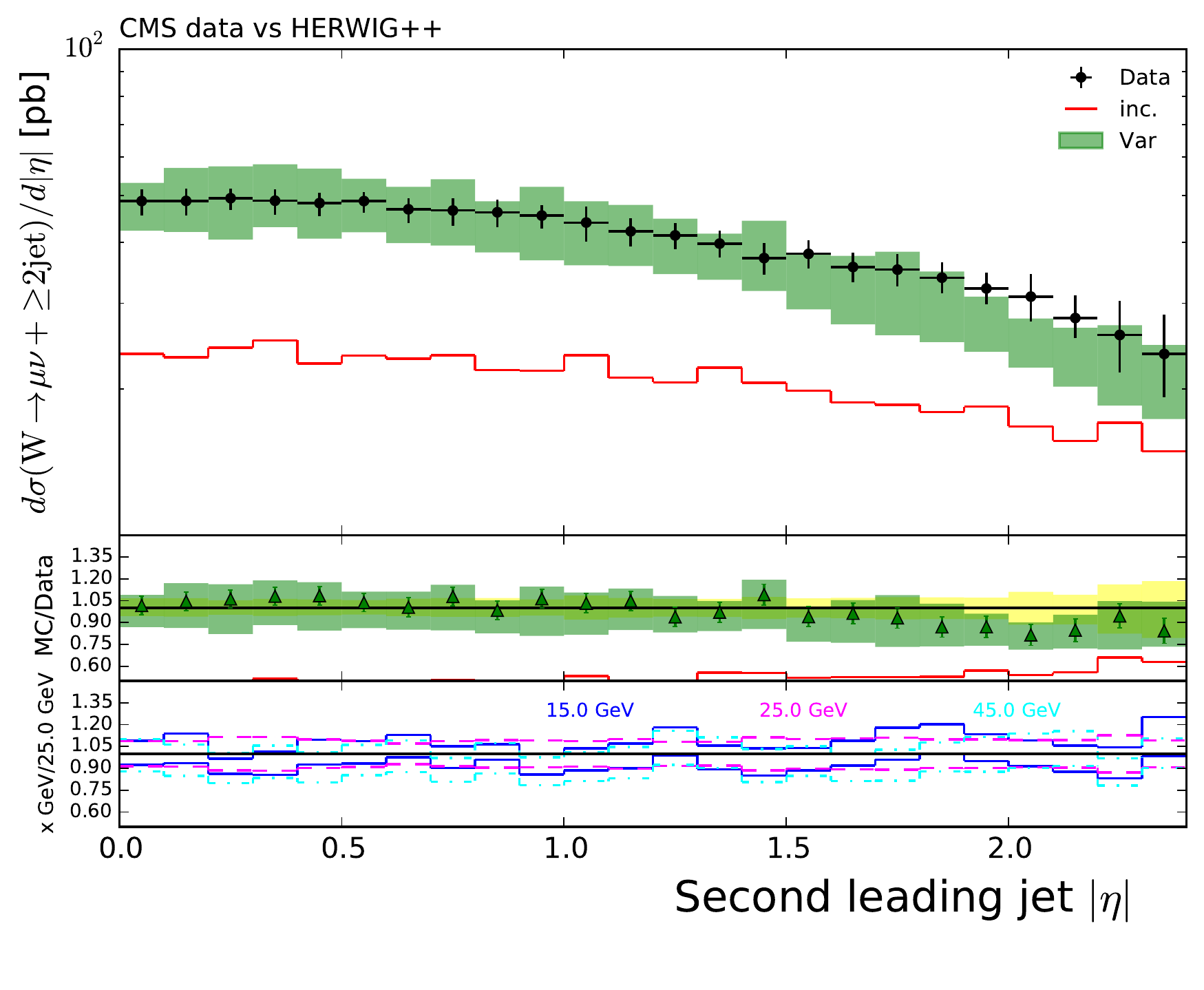}
  \includegraphics[width=0.499\linewidth]{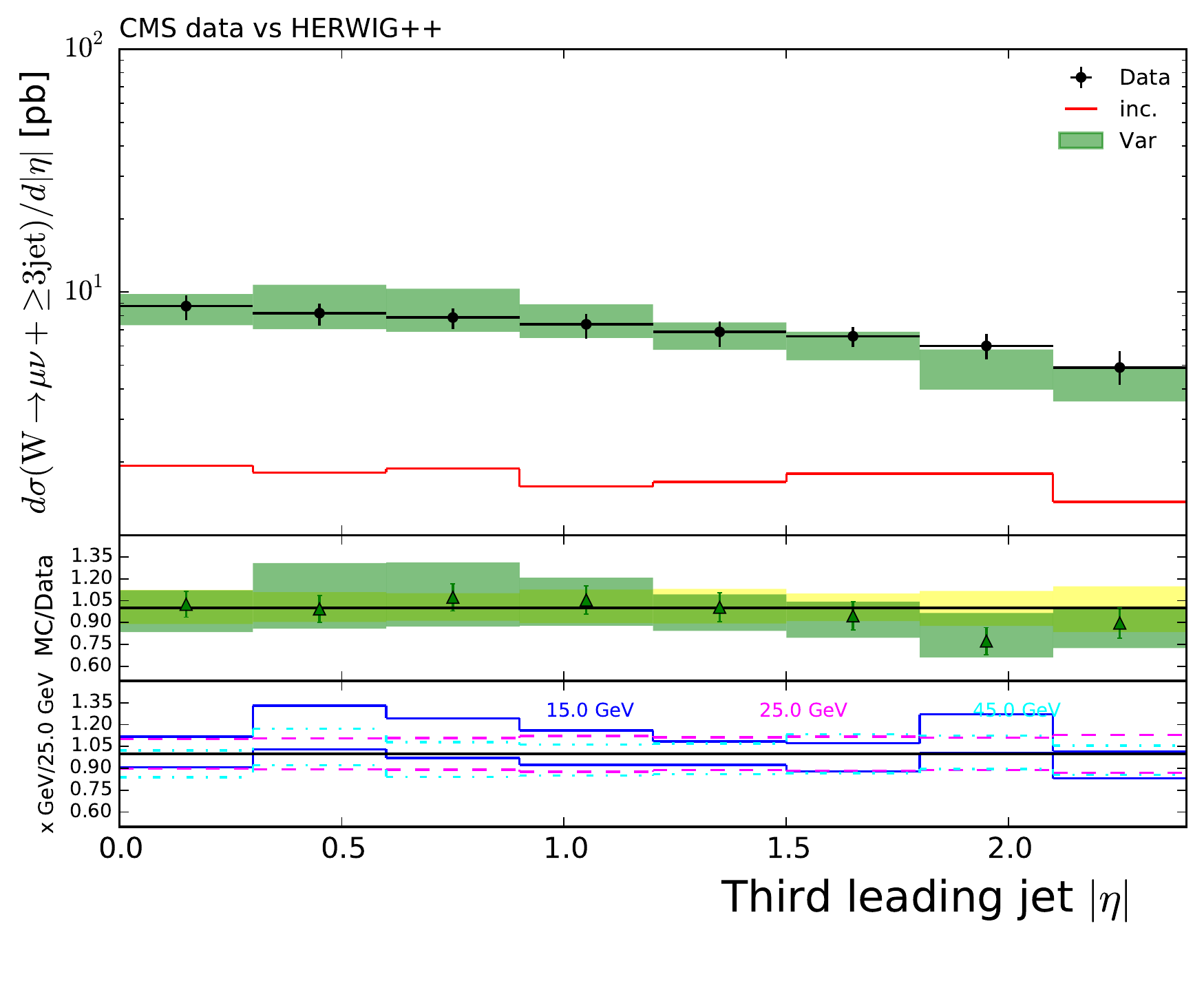}
  \caption{As in fig.~\ref{fig:W.1406.7533:16+17}, for the 
 pseudorapidity of the 2$^{nd}$ (left panel) 
 and of the 3$^{rd}$ (right panel) jet. 
}
  \label{fig:W.1406.7533:06+07}
\end{figure}

\begin{figure}[!ht]
  \includegraphics[width=0.499\linewidth]{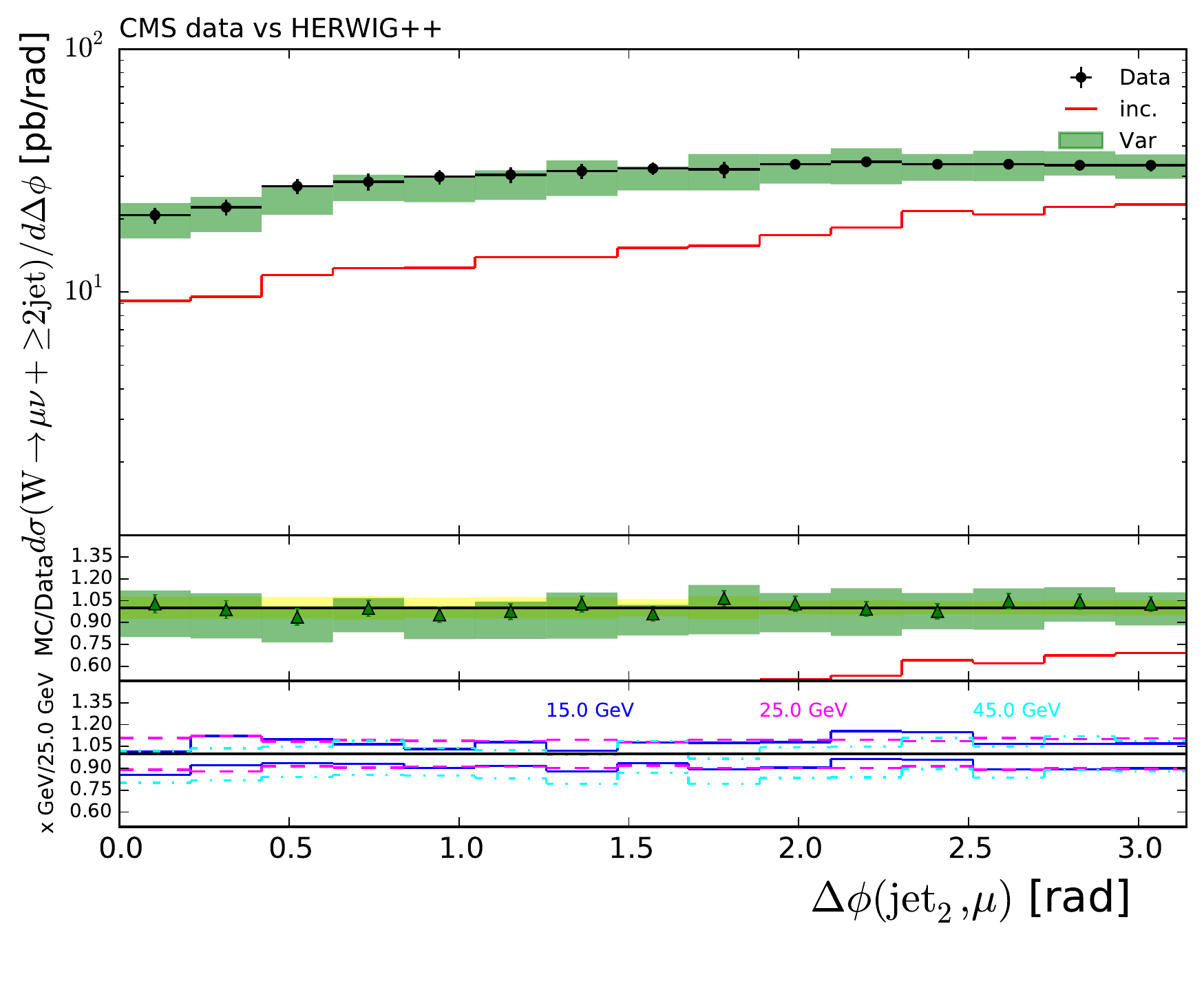}
  \includegraphics[width=0.499\linewidth]{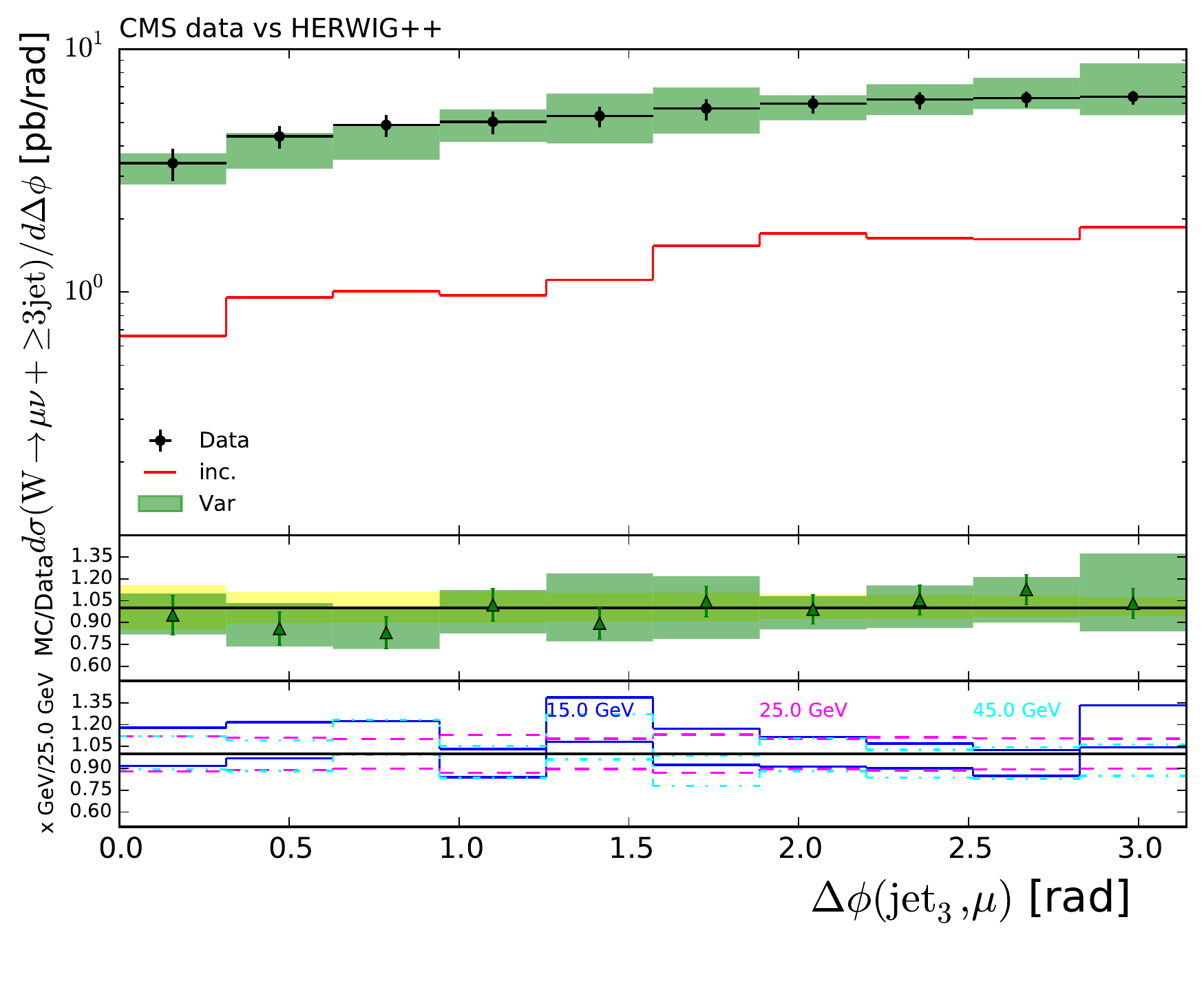}
  \caption{As in fig.~\ref{fig:W.1406.7533:16+17}, for the 
 azimuthal distance between the $\mu$ and the 2$^{nd}$ jet
 (left panel), and between the $\mu$ and the 3$^{rd}$ jet (right panel).
}
  \label{fig:W.1406.7533:02+03}
\end{figure}

\begin{figure}[!ht]
  \includegraphics[width=0.499\linewidth]{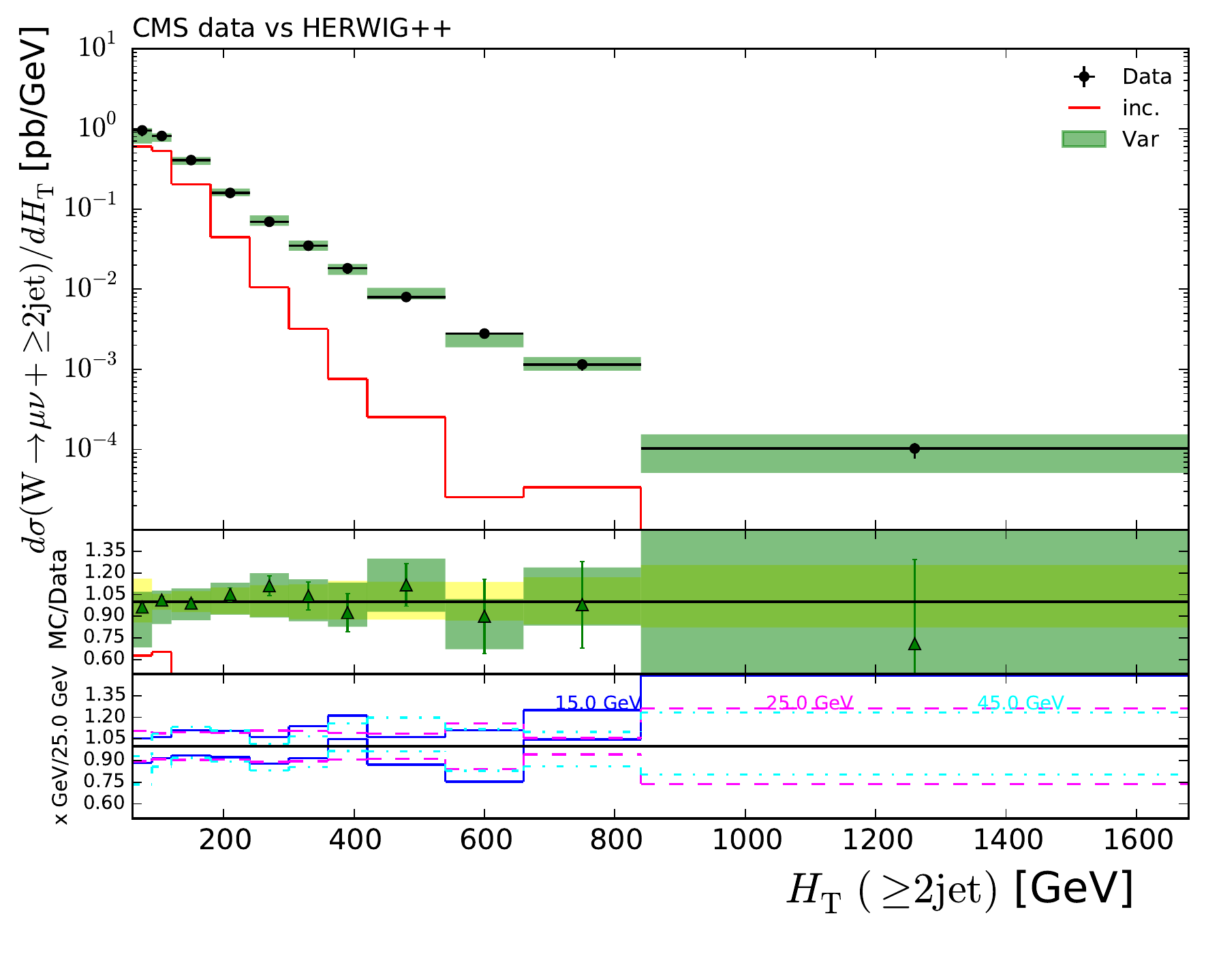}
  \includegraphics[width=0.499\linewidth]{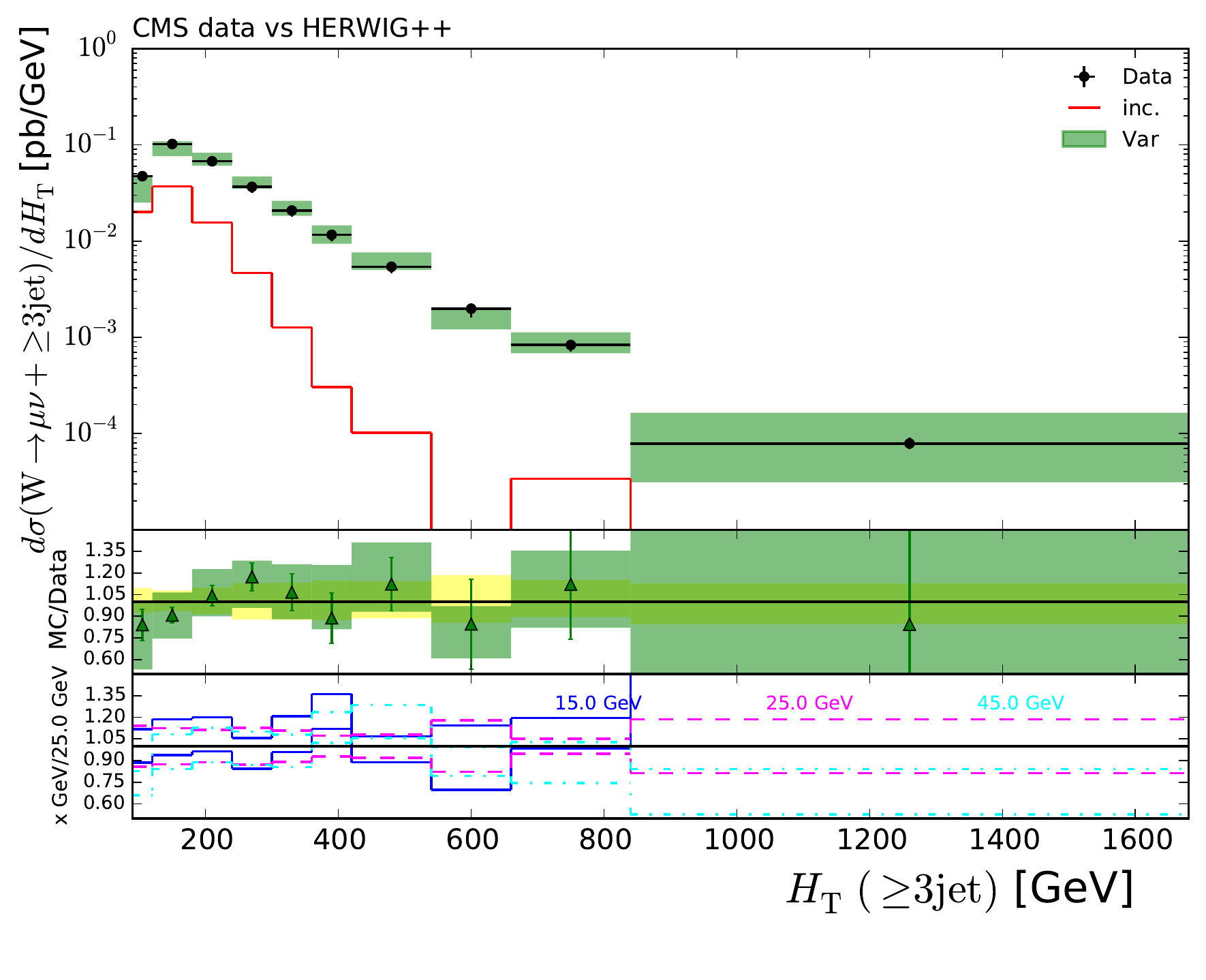}
  \caption{As in fig.~\ref{fig:W.1406.7533:16+17}, for 
 $H_{\sss T}$ in events with at least two (left panel) and
 at least three (right panel) jets. 
}
  \label{fig:W.1406.7533:10+11}
\end{figure}

We have chosen the observables that we consider in the following plots.
Figure~\ref{fig:W.1406.7533:13}: exclusive jet multiplicity;
fig.~\ref{fig:W.1406.7533:15}: transverse momentum of the 1$^{st}$ jet;
fig.~\ref{fig:W.1406.7533:05}: pseudorapidity of the 1$^{st}$ jet;
fig.~\ref{fig:W.1406.7533:01}: azimuthal distance between the $\mu$ and 
the 1$^{st}$ jet;
fig.~\ref{fig:W.1406.7533:09}: $H_{\sss T}$;
fig.~\ref{fig:W.1406.7533:16+17}: transverse momentum of the 2$^{nd}$ 
and of the 3$^{rd}$ jet;
fig.~\ref{fig:W.1406.7533:06+07}: pseudorapidity of the 2$^{nd}$ 
and of the 3$^{rd}$ jet;
fig.~\ref{fig:W.1406.7533:02+03}: azimuthal distance between the $\mu$ and 
the 2$^{nd}$ jet, and between the $\mu$ and the 3$^{rd}$ jet;
fig.~\ref{fig:W.1406.7533:10+11}: $H_{\sss T}$ in events with at least 
two and at least three jets.


The exclusive jet multiplicity (fig.~\ref{fig:W.1406.7533:13}) is
very well predicted by both MCs -- one could repeat almost verbatim
the same comments as for the analysis of ref.~\cite{Aad:2014qxa}
(see fig.~\ref{fig:W.1409.8639:04}).

The inclusive leading-jet observables are reported in
fig.~\ref{fig:W.1406.7533:15} ($\pt$) and fig.~\ref{fig:W.1406.7533:05}
(pseudorapidity). As far as the transverse momentum is concerned, both
MCs tend to be slightly harder than data, an effect which is more
visible in the case of \PYe. This trend, which is statistically not 
very significant (especially in the case of \HWpp), is similar to 
that observed in ref.~\cite{Khachatryan:2014uva}. If one had
to regard our predictions as an NLO-upgraded version of those labelled
``MadGraph'' in ref.~\cite{Khachatryan:2014uva}, one would clearly see a 
significant improvement w.r.t.~the latter. However, we caution against 
taking this comparison too literally, if anything because the LO 
simulations reported in ref.~\cite{Khachatryan:2014uva} have been 
obtained with {\sc\small Pythia6}. For what concerns the leading-jet 
pseudorapidity, both MCs give an excellent description of the
data (and are thus basically identical to each other).

The azimuthal distance between the leading jet and the muon
is shown in fig.~\ref{fig:W.1406.7533:01}, with a fairly good
theory-data agreement. For both MCs, predictions tend to be marginally
steeper than data, although in a way which is not statistically significant. 
A similar trend is observed in ref.~\cite{Khachatryan:2014uva},
and is slightly larger there than here (sometimes much larger,
depending on the specific theory prediction).

We present the $\Njet\ge 1$ $H_{\sss T}$ distribution in 
fig.~\ref{fig:W.1406.7533:09};
we observe a pattern similar to that of the leading jet, with predictions
slightly harder than data, and more markedly so in the case of \PYe.

We now turn to observables which we compare only to \HWpp\ predictions.
We start from single-jet ones, by considering the 2$^{nd}$ and 3$^{rd}$ 
jet transverse momenta (fig.~\ref{fig:W.1406.7533:16+17}) and
pseudorapidities (fig.~\ref{fig:W.1406.7533:06+07}). There is no significant 
discrepancy between theory and data to report in any of these four plots.
As far as the $\pt$'s are concerned, at variance with the case of the leading
jet, our results have essentially the same hardness as the data; this has 
again to be compared to what is found in ref.~\cite{Khachatryan:2014uva}
in the case of the \MadGraphf\ predictions.

The $\Delta\phi$ correlations between the muon and the 2$^{nd}$ and 3$^{rd}$ 
jets are given in fig.~\ref{fig:W.1406.7533:02+03}, with our predictions
again in fairly good agreement with data. Note that here the de-correlation
is stronger than in the case of $\Delta\phi(j_1,\mu)$, but a small 
high-multiplicity matrix-element effect is still present, as can be
seen by comparing the shapes of the merged results with those of
the inclusive ones. Finally, in fig.~\ref{fig:W.1406.7533:10+11}
we present the $H_{\sss T}$ distribution as measured in events with 
$\Njet\ge 2$ and $\Njet\ge 3$. The agreement between theory and
data is good, albeit with non-negligible systematics; in particular,
our simulations are not harder than data, in contrast to what is 
found in ref.~\cite{Khachatryan:2014uva} in the case of LO-based
predictions.

\clearpage

\subsection{Inclusive observables\label{sec:incl}}
We have so far considered observables characterised by hard
hadronic activity in association with a $Z$ or a $W$ boson,
that emphasise the role that multi-parton matrix elements
play in the context of merging techniques such as FxFx.
On the other hand, one expects that physics-wise merged event 
samples improve on inclusive ones in every aspect, being equivalent
to the latter for observables inclusive in QCD radiation, and
able to treat seamlessly the transition between regimes with 
and without well-separated jets. These features are undoubtedly 
beneficial, since for all practical purposes the definition of
what is meant by inclusive, or the occurrence of the transition
between soft and hard kinematics, involve some arbitrariness,
which becomes irrelevant when merged events are employed.
Paradoxically, this might be problematic. In fact, MCs are tuned 
to data, and by far and large the tuning procedures are based
on an LO inclusive picture (typically supplemented by matrix-element
corrections); therefore, the tunes thus obtained, when used in 
a context of merged simulations (especially at the NLO),
may degrade the quality of the theory-data agreement w.r.t.~that 
which results from inclusive LO samples with LO-accurate parameters.

While in principle the solution to this problem is straightforward
(i.e.~to perform tunes with underlying (NLO-)merged samples), in
practice there are several issues that require careful consideration
(e.g.~the possible interplay between multi-parton matrix elements
and multi-particle underlying event (UE henceforth) models; the
impossibility of imposing positivity constraints, simply because
NLO PDFs are not positive definite; and so forth). A discussion of
these issues is beyond the scope of the present work; we shall
limit ourselves to considering the differences between merged
and inclusive predictions for observables that, by construction,
should be rather insensitive to hard radiation, using the measurements
of refs.~\cite{Aad:2012wfa,Chatrchyan:2012tb} as benchmarks, analogously 
to what has been done in sects.~\ref{sec:resZ} and~\ref{sec:resW}.
Such a comparison allows one to gauge the impact of the FxFx merging 
on the typical inclusive observables and thus, indirectly, to start
addressing the question of whether NLO-merged tunes should be considered
sooner than later.

\vskip 0.4truecm
\noindent
$\bullet$ ATLAS~\cite{Aad:2012wfa}
({\tt arXiv:1211.6899}, Rivet analysis {\tt ATLAS\_2012\_I1204784}).

\begin{figure}[!ht]
  \includegraphics[width=0.499\linewidth]{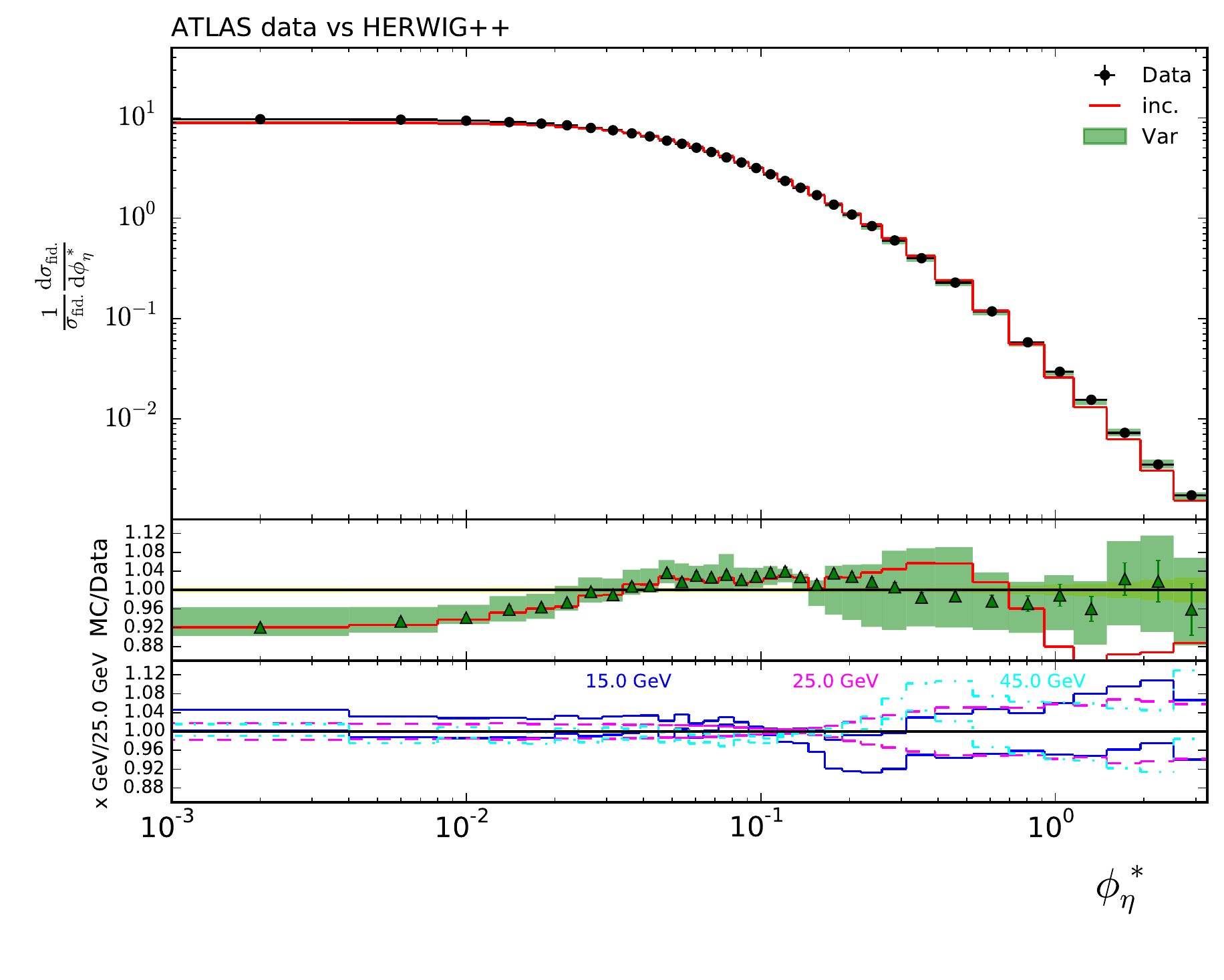}
  \includegraphics[width=0.499\linewidth]{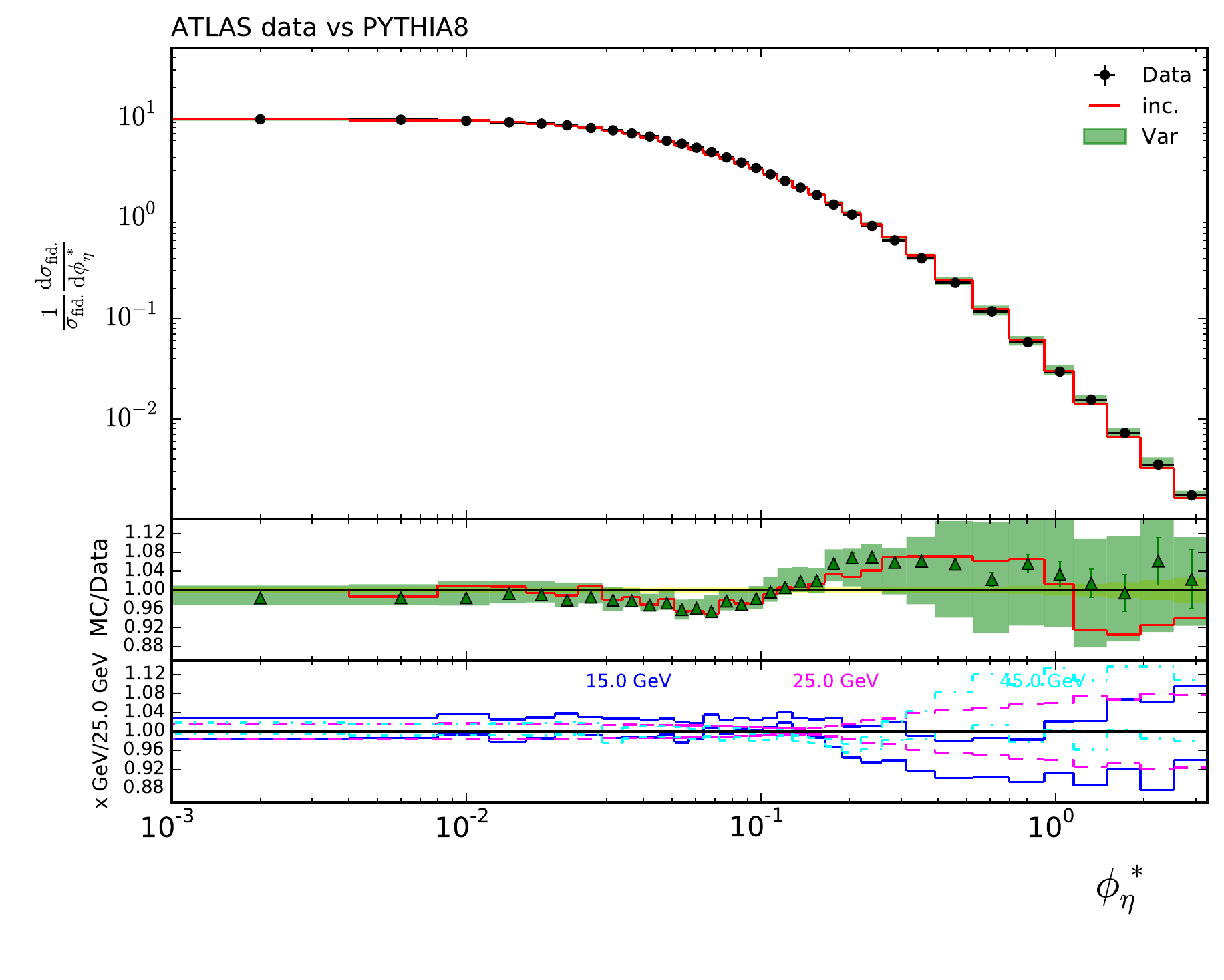}
  \caption{$\phi^\star_\eta$ correlation in the $\epem$ channel, inclusive
 in the $Z$ rapidity. Data from ref.~\cite{Aad:2012wfa}, compared to \HWpp\ 
 (left panel) and \PYe\ (right panel) predictions.
 The FxFx uncertainty envelope 
 (``Var'') and the fully-inclusive central result (``inc'') are shown
 as green bands and red histograms respectively. See the end of
 sect.~\ref{sec:tech} for more details on the layout of the plots.
}
  \label{fig:Z.1211.6899:01}
\end{figure}

\begin{figure}[!ht]
  \includegraphics[width=0.499\linewidth]{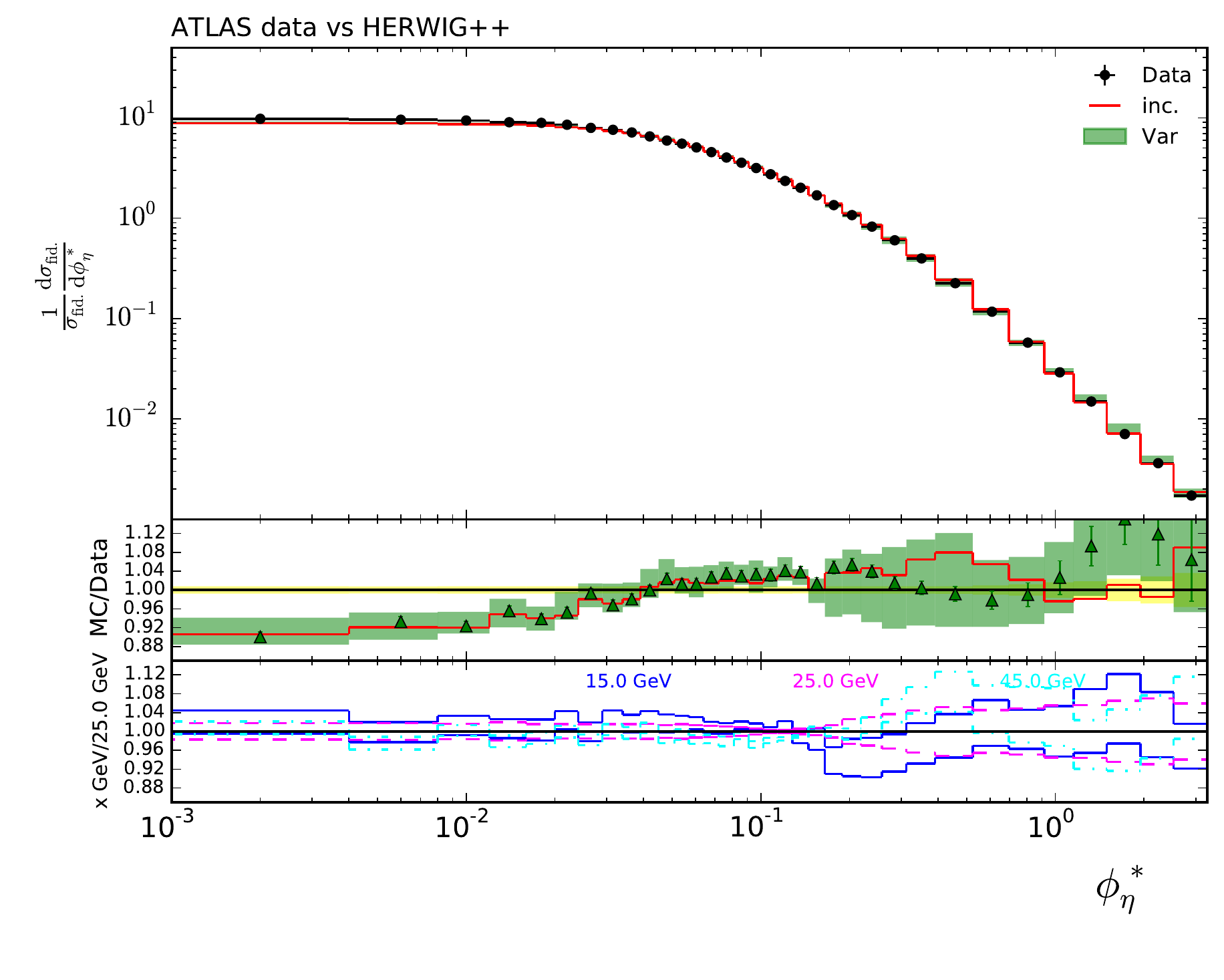}
  \includegraphics[width=0.499\linewidth]{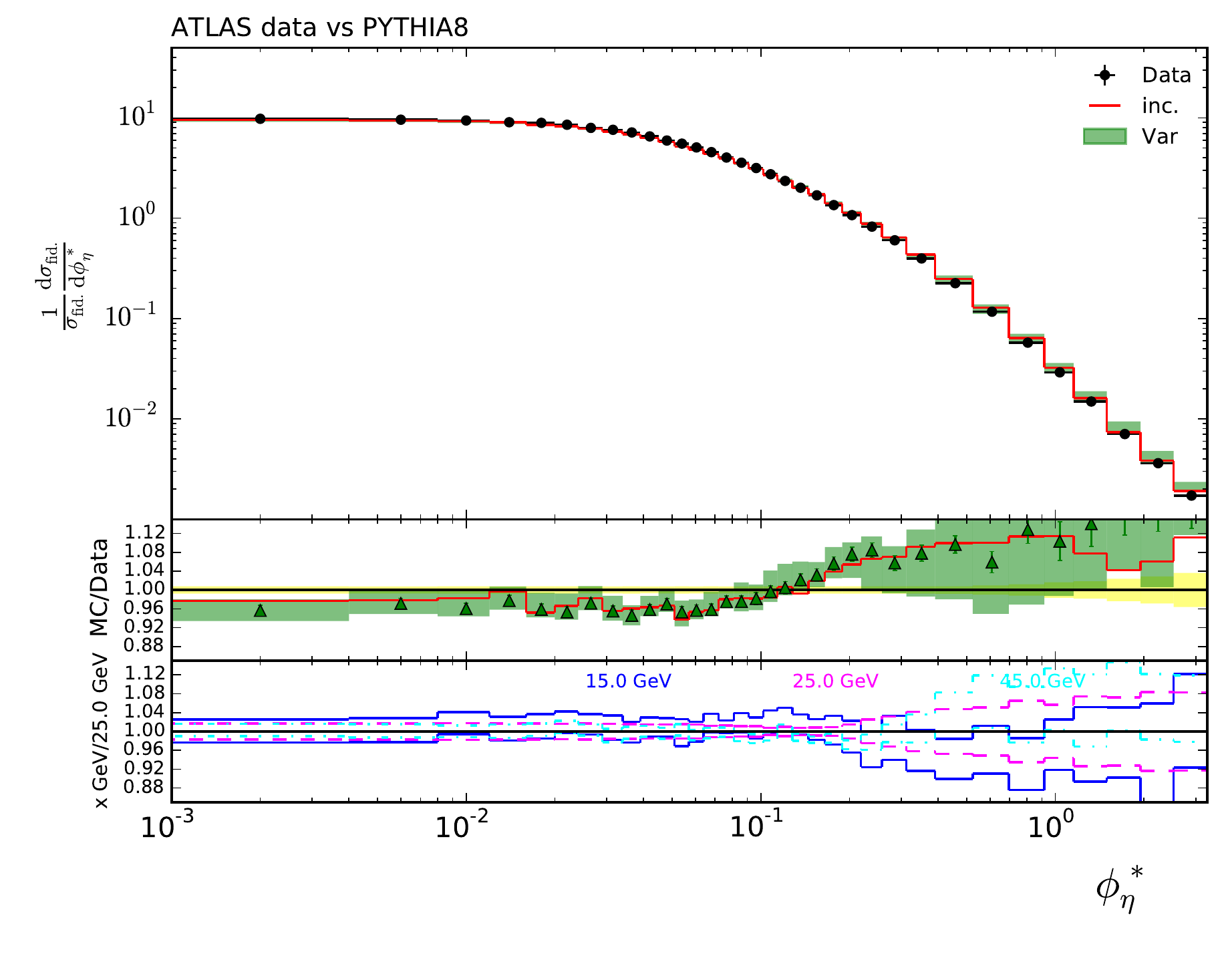}
  \caption{As in fig.~\ref{fig:Z.1211.6899:01}, for the 
 $\abs{y_Z}<0.8$ region.
}
  \label{fig:Z.1211.6899:0201}
\end{figure}

\begin{figure}[!ht]
  \includegraphics[width=0.499\linewidth]{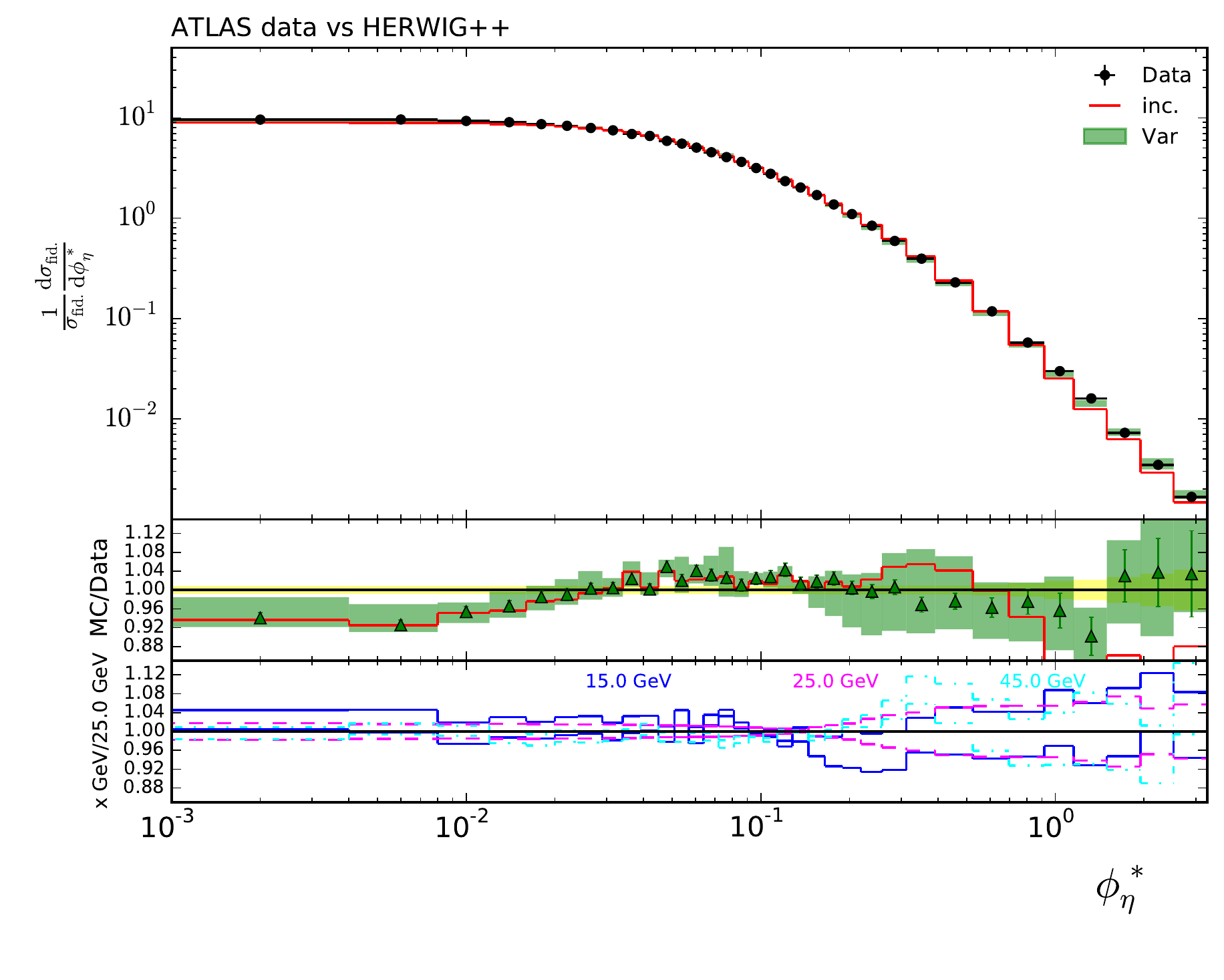}
  \includegraphics[width=0.499\linewidth]{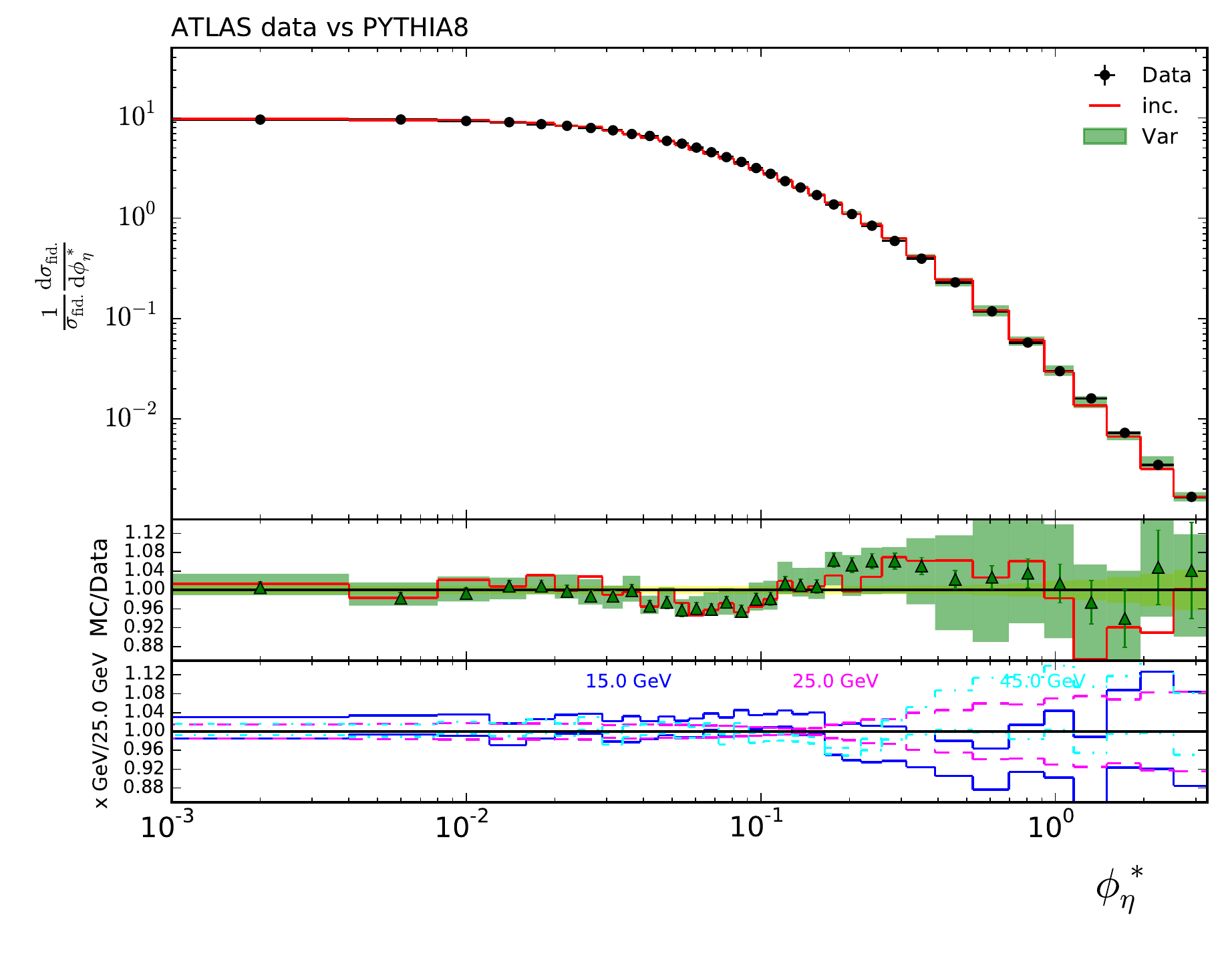}
  \caption{As in fig.~\ref{fig:Z.1211.6899:01}, for the 
 $0.8\le\abs{y_Z}<1.6$ region.
}
  \label{fig:Z.1211.6899:0202}
\end{figure}

\begin{figure}[!ht]
  \includegraphics[width=0.499\linewidth]{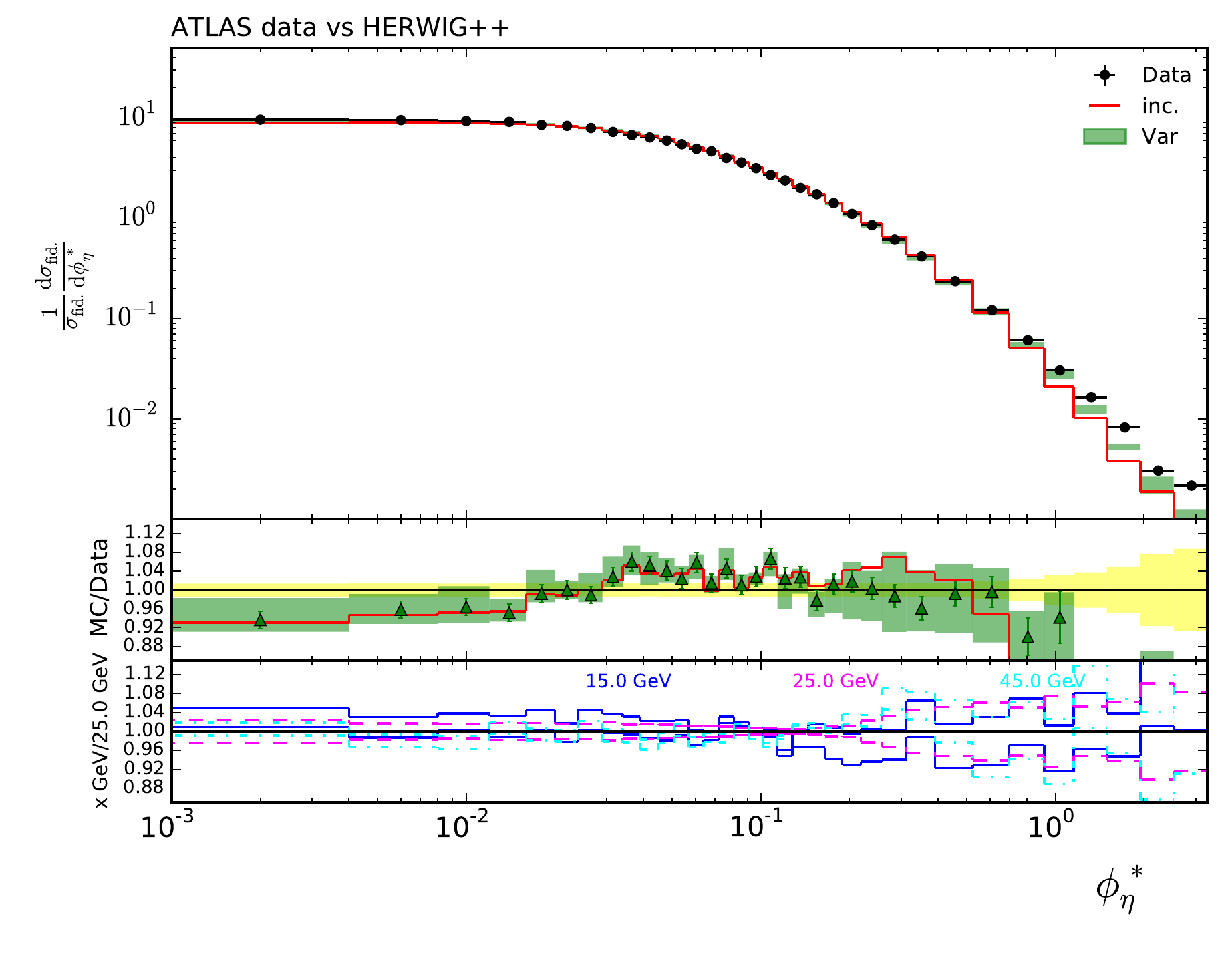}
  \includegraphics[width=0.499\linewidth]{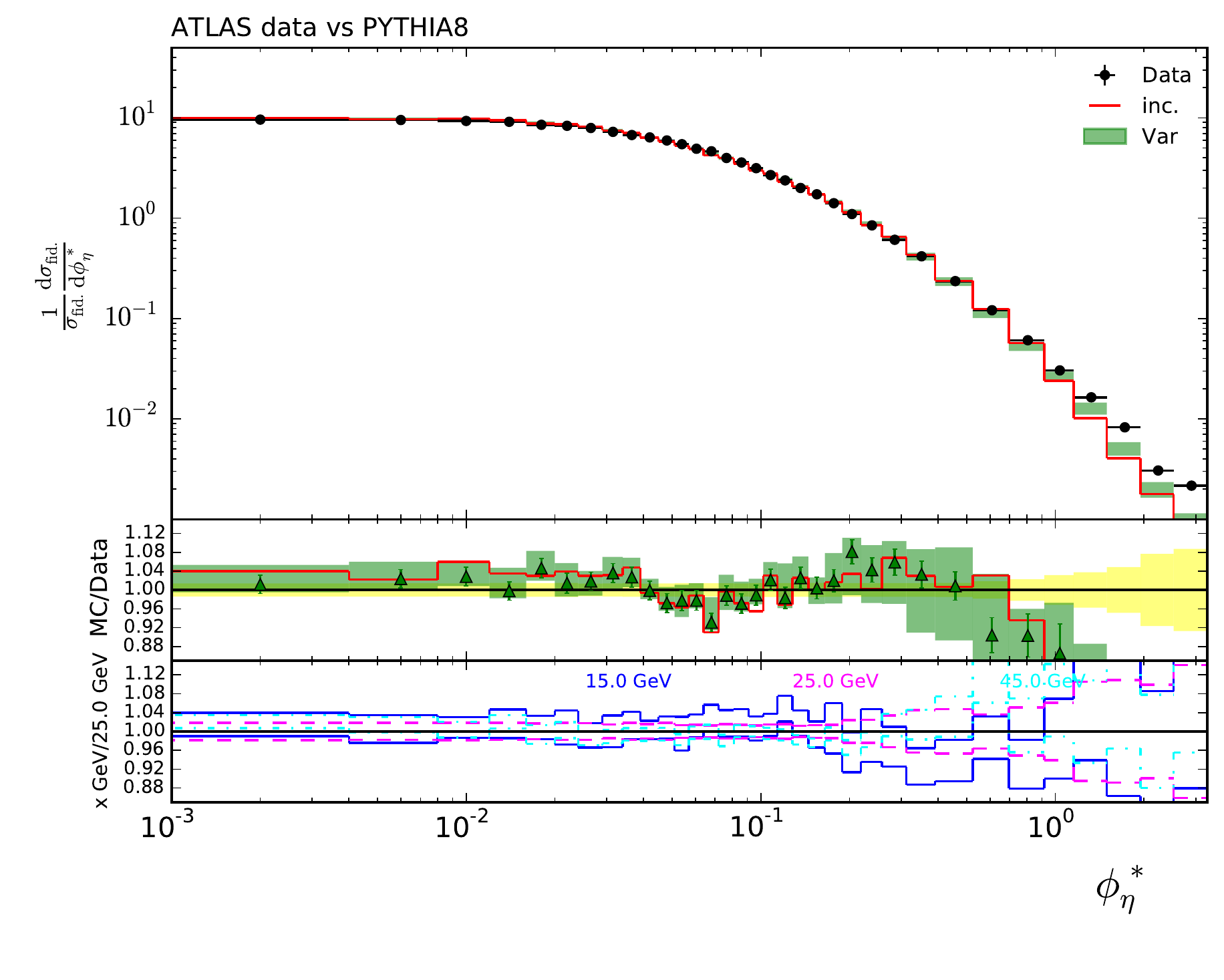}
  \caption{As in fig.~\ref{fig:Z.1211.6899:01}, for the 
 $\abs{y_Z}\ge 1.6$ region.
}
  \label{fig:Z.1211.6899:0203}
\end{figure}

\noindent
Measurement of the $\phi^\star_\eta$ angular correlation 
(see ref.~\cite{Banfi:2010cf} for its definition) in
$\epem$ and $\mpmm$ production. Based on an integrated luminosity 
of 4.6 fb$^{-1}$, within $\pt(\ell)\ge 20$~GeV, 
$66\le M(\ell\ell)\le 116$~GeV, and $\abs{\eta(\ell)}\le 2.4$.
The Rivet analysis outputs the electron and muon results separately.
To be definite, we shall discuss here only the comparison to
$\epem$ data; the situation for $\mpmm$ data is very similar, with
only minor differences which are not significant given the
statistics of our theoretical simulations. The $\phi^\star_\eta$ 
observable is displayed as fully inclusive in the $Z$ rapidity 
(fig.~\ref{fig:Z.1211.6899:01}), and within the
$\abs{y_Z}<0.8$, $0.8\le\abs{y_Z}<1.6$, and $\abs{y_Z}\ge 1.6$
regions (fig.~\ref{fig:Z.1211.6899:0201}, fig.~\ref{fig:Z.1211.6899:0202}, 
and fig.~\ref{fig:Z.1211.6899:0203}, respectively). Note that
$\phi^\star_\eta$, on top of being inclusive, is also designed to 
emphasise the role of low-$\pt(Z)$ production.

The results of figs.~\ref{fig:Z.1211.6899:01}--\ref{fig:Z.1211.6899:0203}
can be summarised as follows. Firstly, they confirm that at small
$\phi^\star_\eta$ (i.e.~at low $\pt$'s) the merged and inclusive 
predictions essentially coincide. However, the region where the two 
start to differ visibly depends on the specific MC employed: 
$\phi^\star_\eta\gtrsim 0.2$ for \HWpp, and $\phi^\star_\eta\gtrsim 0.7$ 
for \PYe, with some dependence on $y_Z$. Secondly, as implied
by the previous item, at small $\phi^\star_\eta$ the level of the theory-data 
agreement is driven by the MC, and therefore affected by the tune of
the latter. Thus, the better description of the low-$\phi^\star_\eta$ data
provided by the \PYe\ simulations w.r.t.~that of \HWpp\ must be 
interpreted in the context of the tunes used in this paper
(see sect.~\ref{sec:tech}). Thirdly, at large $\phi^\star_\eta$ the
merged predictions from the two MCs tend to be rather consistent 
with each other. Since this region can largely be associated with
hard-emission kinematic configurations\footnote{This identification
is hampered by the fact that the $\phi^\star_\eta$ observable squeezes
the large-$\pt$ region in a few bins. From this viewpoint, an analysis
in term of e.g.~the transverse momentum of the lepton pair is superior
w.r.t.~the present one.},
this confirms what has been seen repeatedly in sects.~\ref{sec:resZ} 
and~\ref{sec:resW}. Namely, that in matrix-element-dominated regions
features which are MC-specific matter much less; for the sake of the present 
discussion, this means essentially a significant {\em in}dependence
of tunes (bar perhaps the choice of $\as$ -- see the beginning
of sect.~\ref{sec:res}).

If one wants to use the $\phi^\star_\eta$ data as inputs for tuning,
the implications of the previous discussion are the following.
There exists an MC-specific small-$\phi^\star_\eta$ region where the
inclusion of high-multiplicity NLO matrix elements changes very 
little. In other words, if one tunes here, the results should be
sufficiently ``universal'', and allow one to employ them with
simulations characterised by different perturbative accuracies.
On the other hand, it may be desirable to have a larger lever arm when
tuning, by including in the fits the intermediate-$\phi^\star_\eta$ data
(where again ``intermediate'' is an MC-dependent concept). This
can be acceptable, but the resulting tunes will bear information on
the underlying perturbative description, thus losing universality.
It is this loss of universality which may be responsible for the
degradation of the agreement with data alluded to at the beginning of this 
section. This phenomenon stems from what we may call {\em overtuning}$\,$:
if the natural description of the intermediate region is one that
requires information on higher perturbative orders, the lack of these
in LO-based simulations will tend to be compensated (improperly) by
the tuning, so that when such tunes are used with merged simulations
a certain amount of double counting will be present.

\clearpage

\vskip 0.4truecm
\noindent
$\bullet$ CMS~\cite{Chatrchyan:2012tb}
({\tt arXiv:1204.1411}, Rivet analysis {\tt CMS\_2012\_I1107658}).

\begin{figure}[!ht]
  \includegraphics[width=0.499\linewidth]{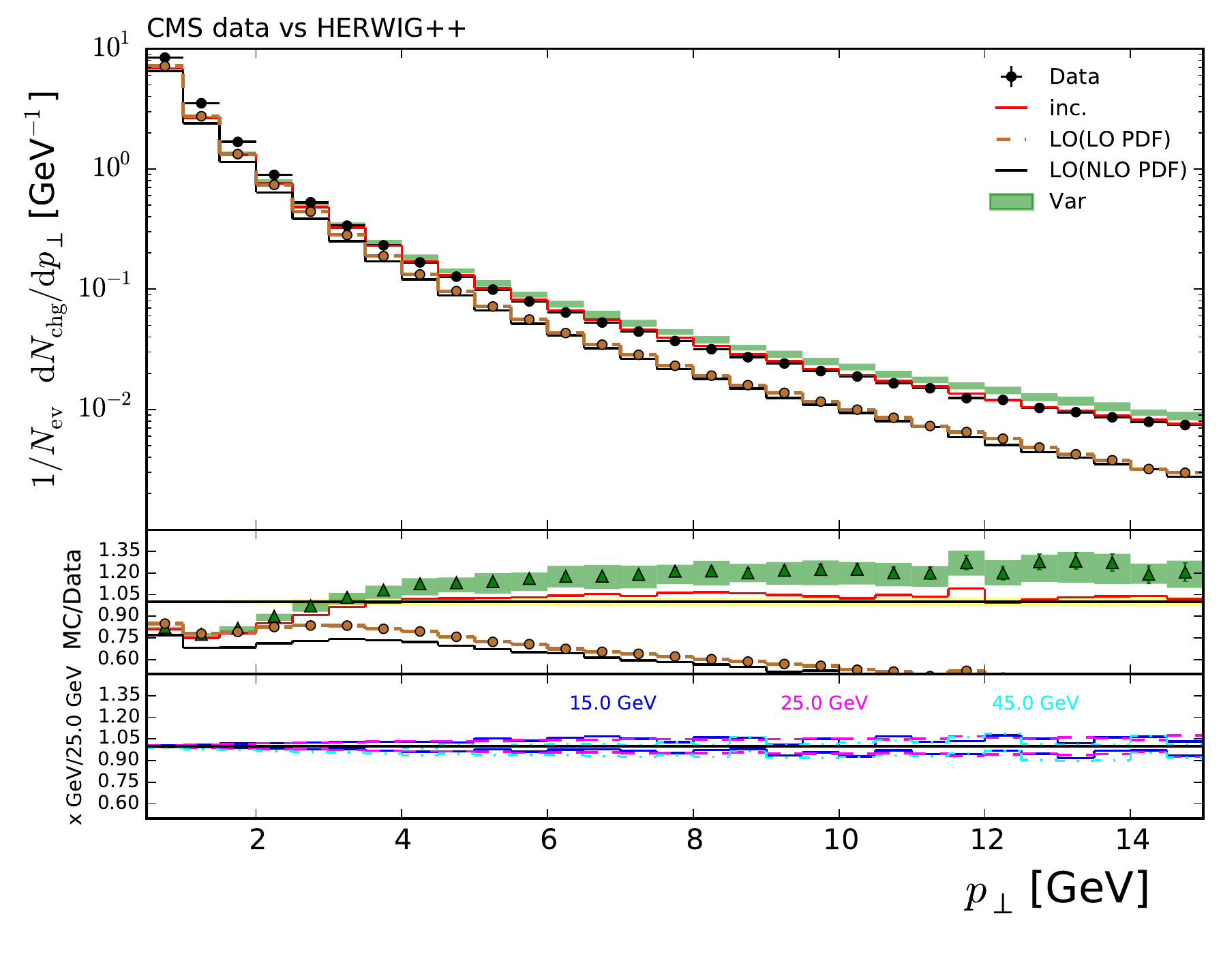}
  \includegraphics[width=0.499\linewidth]{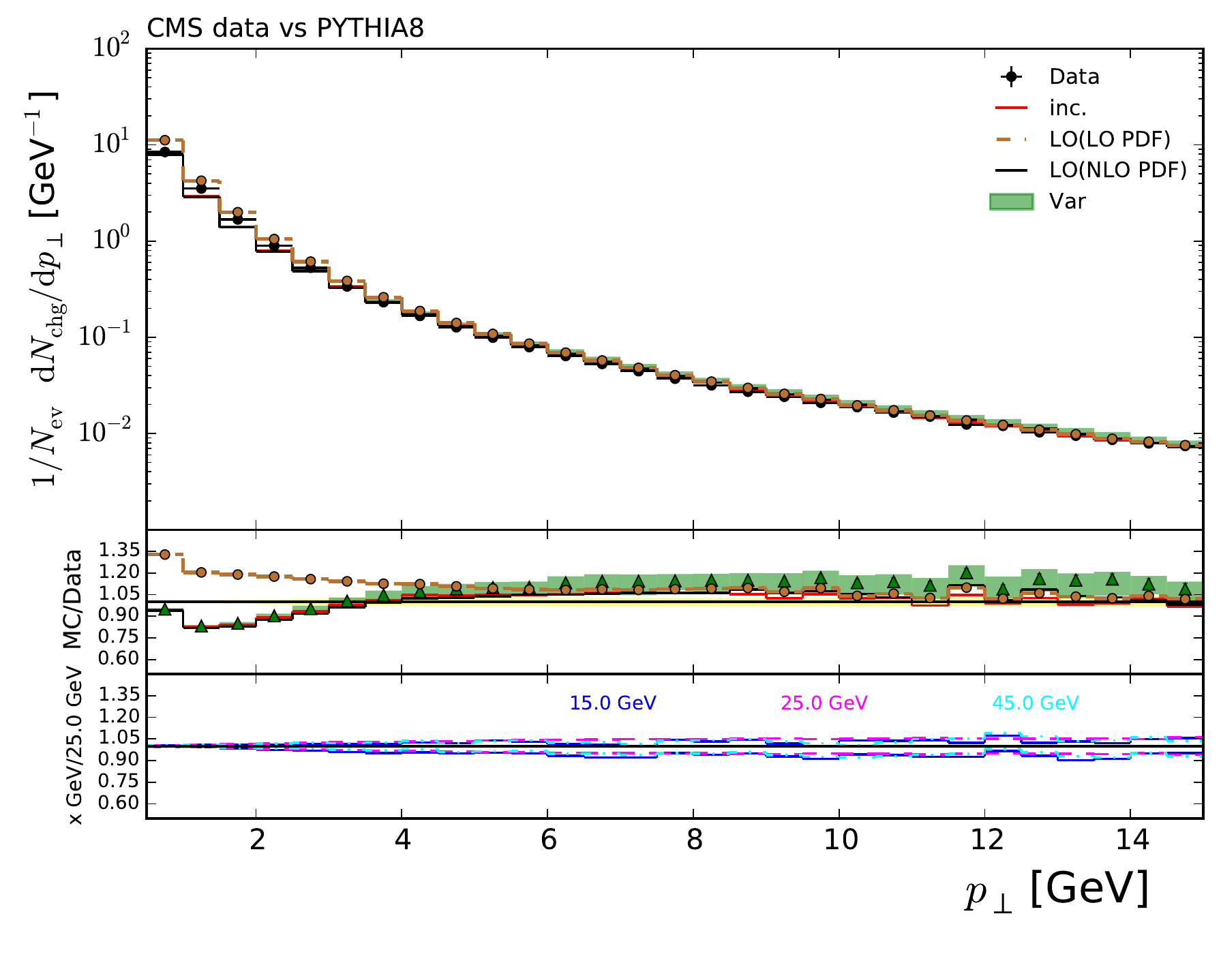}
  \caption{Inclusive charged particle multiplicity as a function 
 of $\pt$ in the away region. Data from ref.~\cite{Chatrchyan:2012tb}, 
 compared to \HWpp\ (left panel) and \PYe\ (right panel) predictions.
 The FxFx uncertainty envelope 
 (``Var'') and the fully-inclusive central result (``inc'') are shown
 as green bands and red histograms respectively. Two LO-accurate
 results are presented as solid black and dashed brown histograms.
 See the end of sect.~\ref{sec:tech} and the text of this section 
 for more details on the layout of the plots.
}
  \label{fig:Z.1204.1411:18}
\end{figure}

\begin{figure}[!ht]
  \includegraphics[width=0.499\linewidth]{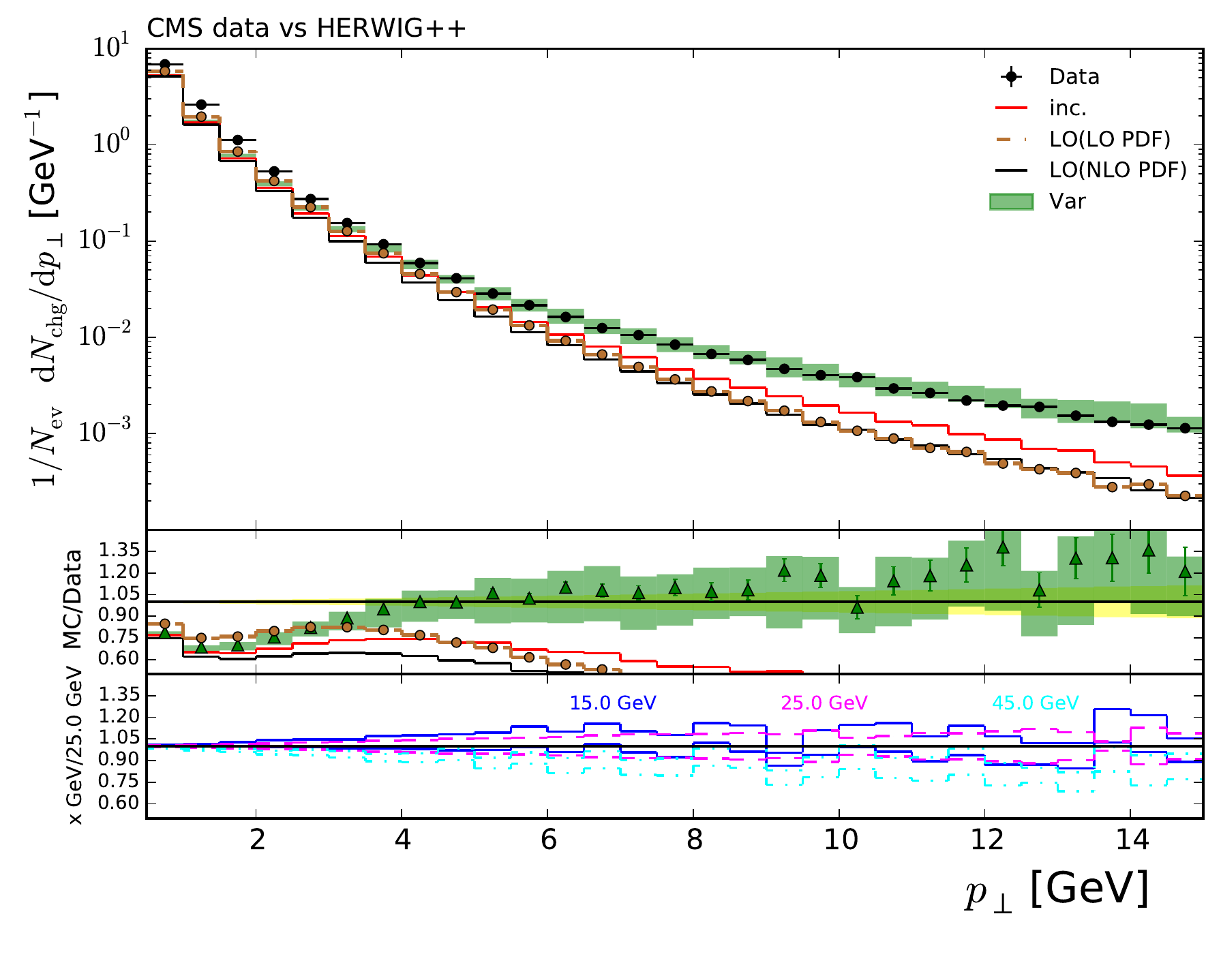}
  \includegraphics[width=0.499\linewidth]{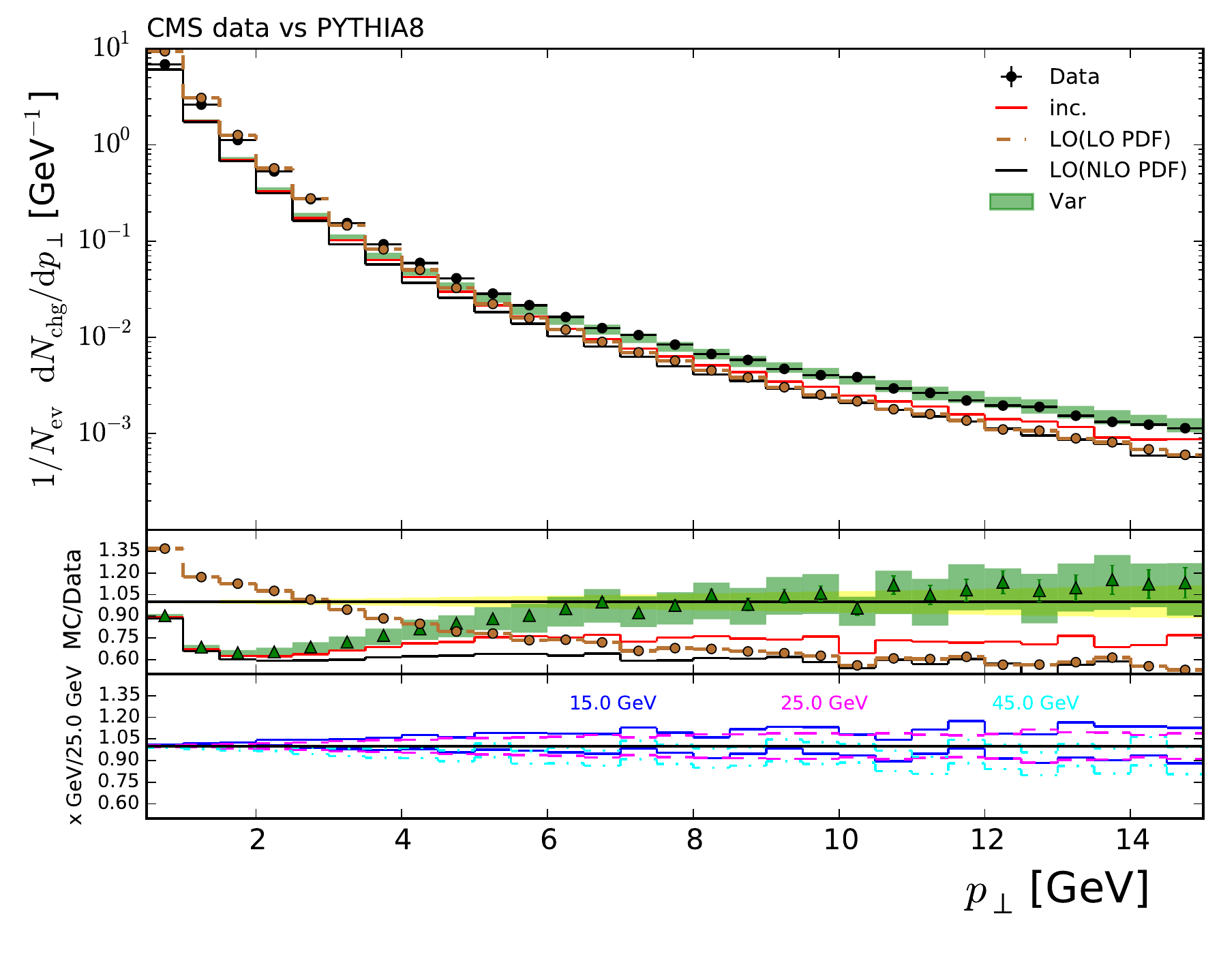}
  \caption{As in fig.~\ref{fig:Z.1204.1411:18}, for the transverse region.
}
  \label{fig:Z.1204.1411:17}
\end{figure}

\begin{figure}[!ht]
  \includegraphics[width=0.499\linewidth]{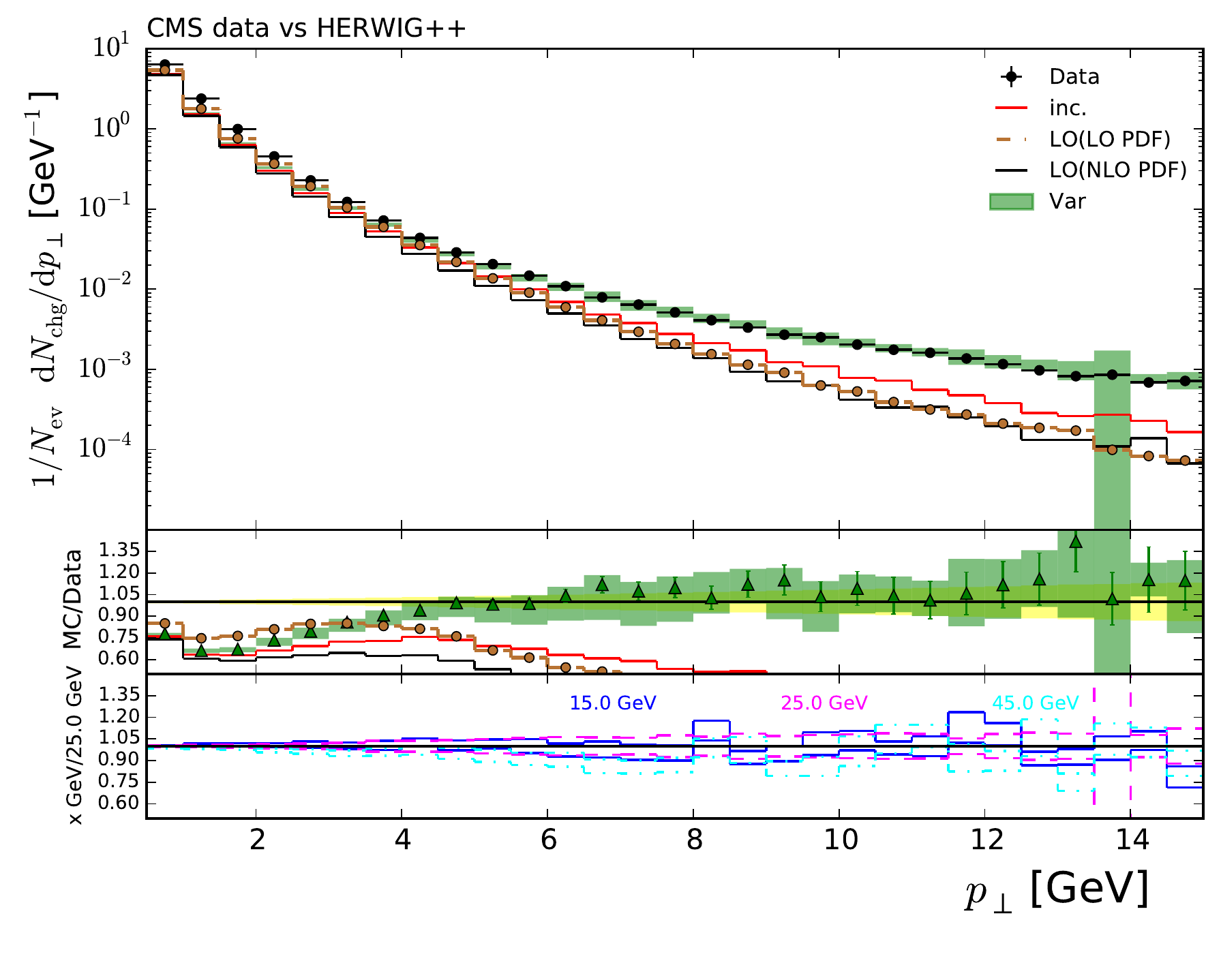}
  \includegraphics[width=0.499\linewidth]{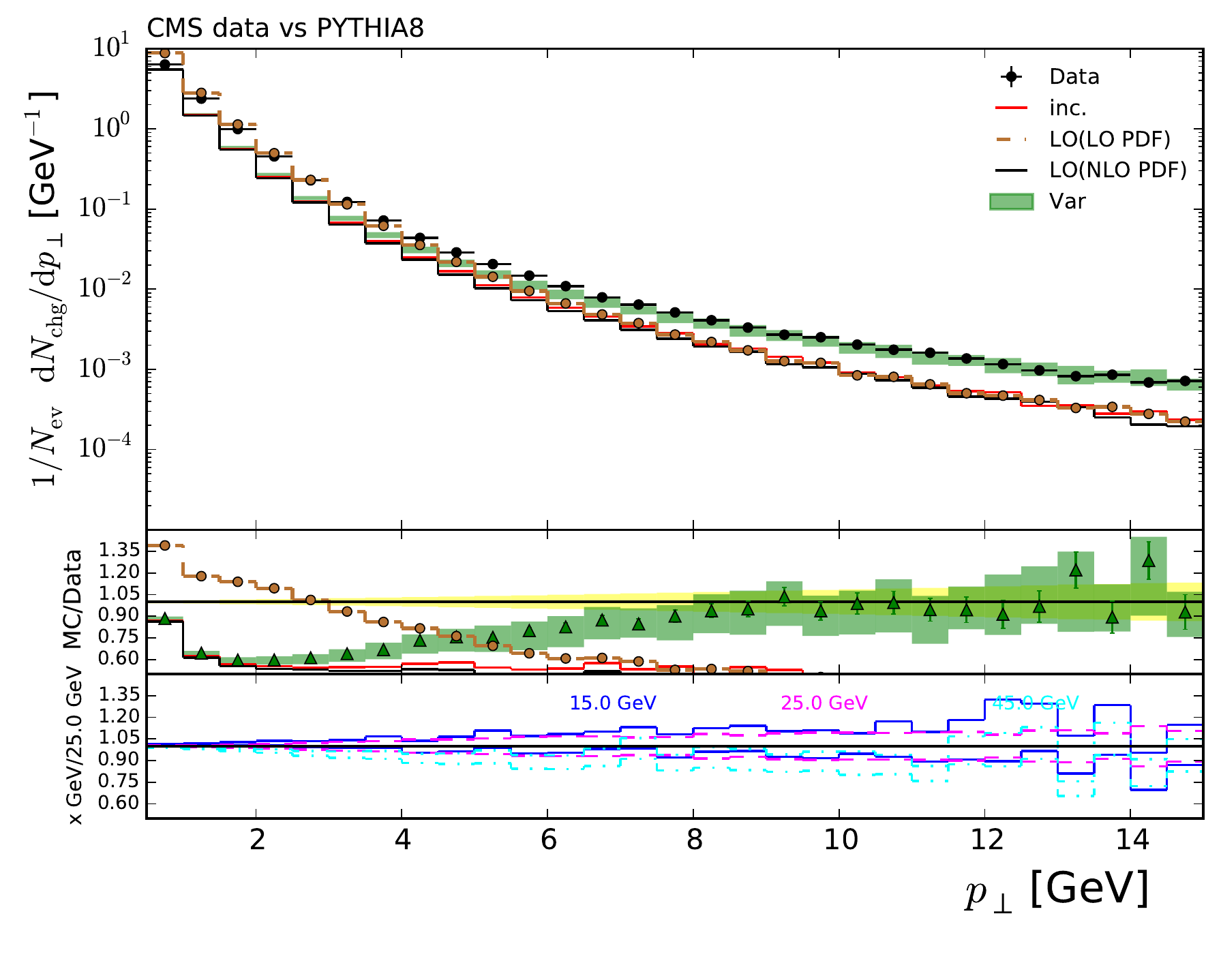}
  \caption{As in fig.~\ref{fig:Z.1204.1411:18}, for the towards region.
}
  \label{fig:Z.1204.1411:16}
\end{figure}

\noindent
Measurement of the underlying event activity in $\mpmm$ production. 
Based on an integrated luminosity of 2.2 fb$^{-1}$, within 
$\pt(\mu)\ge 20$~GeV, $\abs{\eta(\mu)}\le 2.4$, and
either $81\le M(\mpmm)\le 101$~GeV or $\pt(\mpmm)<5$~GeV.
The charged particles considered in the analysis must be within 
$\pt>0.5$~GeV and $\abs{\eta}<2$. We present them, as a function 
of the transverse momentum, for the away, transverse, and towards regions 
in figs.~\ref{fig:Z.1204.1411:18}, \ref{fig:Z.1204.1411:17},
and~\ref{fig:Z.1204.1411:16} respectively, in the
case $81\le M(\mpmm)\le 101$~GeV. These plots have the
same layout as those shown so far, except for the fact that they
contain two additional histograms in the main frame (their ratios
to data are also reported in the upper insets). Such histograms are obtained
with fully-inclusive simulations based on LO matrix elements, convoluted
and showered either with NLO PDFs (solid black), or with LO PDFs
(dashed brown, overlaid with full circles). We point out
that the use of externally-generated LO events is associated with
matrix-element corrections (MECs) in \PYe; on the other hand, these 
corrections are not applied by \HWpp.

By far and large, the towards, transverse, and away regions are
defined as parallel, perpendicular, and anti-parallel to the direction 
of flight of the $Z$, respectively, through conditions on the azimuthal
separation $\Delta\phi$ between the $Z$ and the charged tracks (for more 
details, we refer the reader to the original paper~\cite{Chatrchyan:2012tb}).
Thus, the away region is efficiently filled by the ``first''
hard parton that recoils against the $Z$. This is the reason why
the merged and NLO-inclusive results are close to each other in 
fig.~\ref{fig:Z.1204.1411:18} (with a better agreement between the
two, and the data, in the case of \PYe). Since \PYe\ applies MECs, the 
LO predictions are also in agreement with the NLO ones for such an MC
starting from intermediate $\pt$'s (and actually in the whole $\pt$
range, significantly when NLO PDFs are employed -- see later),
while this is not the case for \HWpp, given that hard matrix elements
are completely lacking in this case (which allows one to see directly the 
dramatic impact of the hard recoil in the away region). Conversely, both the 
transverse and the towards region are designed to remove the contribution
of the hard recoil; furthermore, higher-multiplicity hard configurations 
tend to contribute to the former but not to the latter. We see indeed in 
figs.~\ref{fig:Z.1204.1411:17} and~\ref{fig:Z.1204.1411:16} that the
differences between the merged and the non-merged (both NLO and LO) 
predictions are now significant. Note, in particular, that this
is true also in the towards region (in spite of its being largely
insensitive to hard radiation): this is a consequence of the fact that
multi-parton matrix elements allow the MCs to emit more efficiently
in all corners of the phase space, thanks to shorter colour strings
or narrower radiation cones.

The agreement between data and NLO-merged predictions for 
$\pt\gtrsim 6$~GeV (or lower, depending on the region and/or the MC) 
in all $\Delta\phi$ regions and for both MCs\footnote{The only 
statistically-significant deviation is for \HWpp\ in the away region.
In general, \HWpp\ is systematically slightly higher than \PYe. Inspection 
of figs.~\ref{fig:Z.1204.1411:18}--\ref{fig:Z.1204.1411:16} suggests
that, for these observables, \HWpp\ prefers a merging scale
larger than \PYe's.} has to be compared with the fact that, with 
inclusive NLO or LO+MECs simulations, a satisfactory description 
of the data for such $\pt$'s can only be obtained in the away region.
So while the observables considered here obviously provide one with
excellent inputs for tuning, at the same time the possibility of
overtuning must be carefully assessed. The plots presented above show 
that with NLO-merged results the impact of tuning is mainly limited to 
small transverse momenta, regardless of the $\Delta\phi$ region 
considered, while larger $\pt$'s are essentially unaffected.
We note that this message comes out quite clearly {\em thanks to}
the fact that the tunes we have adopted {\em do not} provide a good
description of the small-$\pt$ data\footnote{We have however verified that
the agreement theory-data significantly improves at small $\pt$'s 
e.g.~by adopting the 4C tune in \PYe\ and the UE-EE-SCR-CTEQ6L1 tune
in \HWpp.}. 

Finally, as far as the small-$\pt$ cross sections are concerned,
one sees immediately the expected good agreement between the two
NLO-accurate results, in all the $\Delta\phi$ regions. For
$\pt\lesssim 2$~GeV (or larger), these two predictions also agree
with the LO one obtained with NLO PDFs. This is interesting, because
at low $\pt$'s we observe a very significant dependence (which disappears 
completely when moving towards larger $\pt$'s) of the LO simulations
on whether the PDFs used are the NLO or the LO ones ; furthermore,
such a dependence is stronger for \PYe\ than for \HWpp\ (although the
pattern NLO PDFs vs LO PDFs is similar in the two MCs). For what 
concerns the latter point, we have found that it is basically an accident: 
by changing the tunes and/or some of the MPI-related parameters in a given
tune, the PDF dependence can be made weaker in \PYe\ and stronger
in \HWpp. This renders it clear that what drives the small-$\pt$ cross
section is not the nominal perturbative accuracy of the PDFs in themselves, 
but rather their small-$x$ behaviours, through the smallest $x$ values
accessed within a given tune. In addition to this, we remark that
at small $\pt$'s there is only a very minimal dependence 
on the perturbative accuracy of the short-distance results, 
when these are all convoluted and showered with the {\em same} PDFs.
The conclusion is that a proper description of the small-$\pt$ data 
can only be achieved in the context of a fully self-consistent treatment 
of non-perturbative ``parameters'' (including the PDFs), and that this is
largely independent of the nature of the hard events used in the simulations.
We point out that such (approximate) universality is subject to the
condition that one does not tune over the soft-hard transition region
(let alone over the hard one): the very same considerations made above for 
the analysis of ref.~\cite{Aad:2012wfa} apply here.
These findings heuristically confirm, in particular, the expectation 
that NLO+PS simulations based on NLO PDFs and tunes obtained with LO PDFs
are liable to worsen the agreement with data at small transverse momenta
{\em for fully-inclusive observables}.
They also imply, however, that in the absence of a proper full-NLO tune
a pragmatic solution is that of tuning using LO(+MECs) matrix elements
with NLO PDFs, provided that one restricts oneself to genuine
low-$\pt$ data.

\section{Summary\label{sec:conc}}
In this paper we have presented a comprehensive comparison
between NLO+PS theoretical predictions obtained by merging
different partonic multiplicities according to the FxFx method,
and several measurements performed by ATLAS and CMS for the leptonic
channels of $\Zjs$ and $\Wjs$ production. In order to deal only 
with well-established quantities and to minimise the number of
operations on experimental results and definitions, we have considered 
7-TeV LHC datasets with high statistical accuracy (based on
an integrated luminosity of about 5 fb$^{-1}$), and associated 
with a public Rivet routine. The primary aim of such a comparison
is the validation of the FxFx procedure in an environment fully
realistic from a phenomenological viewpoint.

Our computations have been carried out in the automated
\aNLO\ framework, with parton showers performed by \HWpp\
and \PYe. Thus, we have validated the relevant computer programs
also in a technical sense: a single sample of unweighted hard events is 
produced simply through input commands (e.g.~there is no {\em a posteriori} 
recombination of different multiplicities), and both \HWpp\ and \PYe\ 
automatically handle FxFx-merged samples.

The results we have obtained are based on underlying NLO 
matrix elements, with Born-level final states characterised
by a lepton pair (thus taking fully into account the off-shellness
of the vector boson, and spin correlations) and either zero, one, or 
two extra QCD partons. Therefore, the largest (tree-level) final-state
multiplicity in our simulations features two leptons and three partons.
We did not merge any LO-accurate samples for multiplicities larger
than those included in our NLO matrix elements. We have used the
same hard-event sample for all the $\Zjs$ analyses (and analogously
for the $\Wjs$ ones); in other words, we have not performed any
analysis-specific event generation.

In the context of merged simulations, and especially for those that
are NLO accurate, the scaling in the number of jets ($\Njet$) must be 
regarded as a genuine theoretical prediction. For this reason, all of 
our differential distributions are presented as they result from the
parton showers, in shapes {\em and} rates. We believe that, at least
when the understanding of production mechanisms is paramount, 
the rescaling of theoretical results by an $\Njet$-dependent
factor, extracted from data, is detrimental, because it entails a loss 
of predictivity, and tends to obscure the various strengths and
weaknesses of the specific combination of matrix elements, merging method,
and MC considered.

The overall agreement between our predictions and the $\Vjs$ data 
is good for both \HWpp\ and \PYe. There is a very limited
number of observables for which minor differences between theory
and data are statistically significant, and these are also the only
cases where the results of the two MCs may visibly differ.
The general agreement between \HWpp\ and \PYe\ implies a significant
reduction of the dependence on MC-specific assumptions, in favour of
an increase in predictivity due to the inclusion of matrix-element
information through merging. We also point out that the theory-data
agreement extends to jet multiplicities which one would not expect
to be predicted with the highest accuracy, given the matrix elements 
we have employed. It thus appears that NLO computations with up to
two extra final-state partons are able to capture the most important
dynamical effects in $\Vjs$ production, providing the MCs with suitable
initial conditions for parton showers, which are never stretched outside
their defining approximations. If taken at face value (i.e.~ignoring
both the theory and experimental uncertainties) certain $\pt$-related
distributions in $\Zjs$ production are softer in our predictions 
than in data; however, given the general picture it seems unlikely
that this is a hint that $Z\!+\!3~{\rm jets}$ NLO matrix elements 
would be necessary (although of course it would be appealing to 
compute them). The only indication of this is in the theory being
lower than data in the $4^{th}$-jet rate.
We note that the typical LO-merged samples
used by the experiments for $\Vjs$ production feature much larger 
parton multiplicities (at least four extra, and typically more).
It is therefore an interesting question that of whether such
high-multiplicity LO matrix elements are truly {\em all} needed
in order to compensate for the lack of genuine NLO effects.
Conversely, one may wonder whether merging procedures at the NLO
are in any case superior to their LO counterparts, irrespective
of the matrix elements included in the computations (obviously,
beyond a minimal threshold).

One aspect where NLO mergings improve by construction on LO ones
is in the reduction of the theoretical systematics. In this paper,
we have assessed three sources of uncertainties, due to the choices
of hard scales, PDFs, and merging scale. We remind the reader that
in \aNLO\ the variations of hard scales and PDFs do not entail
any separate runs, since the results for each individual choice are
stored in the hard-event file as companion weights, correlated with
the main event weight (it is such a correlation that allows one to
use a single shower history for all weights, and hence to significantly
improve the numerical stability of the final results). It is crucial
that compatibility with this feature be systematically incorporated 
in modern simulation and analysis software (such as Rivet).
As far as merging-scale dependence is concerned, we have considered
a rather large range, in order to be as conservative as possible.
Still, this kind of systematics appears to be under good control;
in particular, shape-wise the different choices lead to fairly 
similar results.

We have also presented NLO-merged results for inclusive and
underlying-event observables in $Z$ production. As expected,
in the kinematic regions associated with low-$\pt$ emissions
such results are in agreement with those obtained from a 
fully-inclusive NLO sample; both are in fact ultimately determined
by the underlying MC. In these cases, and for the specific tunes
we have adopted in this paper, the description of
the data is worse, and more MC-dependent, than for the 
$\Zjs$ observables. This is just as well, because it
gives one some further evidence that multi-jet observables
are fairly independent of the general assumptions made in 
individual MCs, and of their tunes in particular.

We have not made any attempt to improve the description of
low-$\pt$ observables by considering different tunes, but have 
limited ourselves to verifying that this is indeed possible. Rather, 
we have employed such observables to give explicit examples of how 
the use of merged predictions could lead to an issue we have called
overtuning, which is basically a double-counting problem.
In essence, one might use the differences between merged
results and those adopted in the context of tuning (typically,
LO plus matrix-element corrections) as a way to assess
the possible impact of high-multiplicity matrix elements
on tunable observables. When this impact is non negligible, 
a new tuning specific for merged samples may be desirable.
Conversely, the differences mentioned above could help 
determine the kinematic regions to be used to obtain ``universal'' 
tunes, that can be safely adopted regardless of the perturbative
accuracy of the simulations. 

We have also shown that at small
transverse momenta the role of the PDFs chosen for tuning is crucial.
This implies that, in such kinematic regions, the use of NLO PDFs
within MCs whose tunes are based on LO PDFs is less than ideal.
Fortunately, our $\Vjs$ results demonstrate that this possible
mismatch is essentially irrelevant in most of the phase space.
While eventually the derivation of proper NLO(-merged) tunes
will offer the best overall option, a lower-cost alternative
for the near future is that of performing MC tunes as they
are done presently, but adopting NLO PDF sets.

Finally, we remark that while the validation of the FxFx merging carried 
out in this paper touches all of the general aspects of the procedure,
it applies directly to the simulations of $WH\!+{\rm jets}$ and 
$ZH\!+{\rm jets}$  production, with $H$ the SM Higgs, owing
to the similarity of these processes with the $\Vjs$ ones.
It is thus compelling to consider such Higgs associated channels
in the context of merged predictions, in particular given the relevance
of jet-multiplicity categorisation in experimental searches.

\section*{Acknowledgements}
We thank Fabio Maltoni for his comments on this manuscript,
Peter Richardson and Torbj\"orn Sj\"ostrand for clarifications on 
and help with \HWpp\ and \PYe, respectively, and Francesco Giuli
for having tested independently \aNLO\ with \Vjs\ production.
We are indebted to Fabio Cossutti, Christian G\"utschow, Joey Huston, 
Jan Kretzschmar, Sascha Savin, and Alessandro Tricoli for having
provided us with information and material relevant to ATLAS and
CMS analyses. 
The work of RF is supported by the
Alexander von Humboldt Foundation, in the framework of the Sofja Kovaleskaja
Award Project ``Event Simulation for the Large Hadron Collider at High
Precision'', endowed by the German Federal Ministry of Education and Research.
SF thanks the CERN TH Unit for hospitality during
the course of this work. 
AP acknowledges support by the MCnet ITN FP7 Marie Curie Initial Training
Network PITN-GA-2012-315877, and a Marie Curie Intra European Fellowship 
within the 7th European Community Framework Programme 
(grant no.~PIEF-GA-2013-622071).
SP is supported by the US Department of Energy under contract
DE-AC02-76SF00515.
The work of PT has received funding from the European Union Seventh
Framework programme for research and innovation under the Marie Curie 
grant agreement N.~609402--2020 researchers: Train to Move (T2M).
This work has been supported in part by the ERC grant 291377 
``LHCtheory: Theoretical predictions and analyses of LHC physics: 
advancing the precision frontier".

\bibliographystyle{JHEP}
\bibliography{WZjets}
\end{document}